\documentclass[a4paper]{article}
\usepackage{authblk}

\usepackage{graphicx} % Required for inserting images
\usepackage{xcolor}

\definecolor{codeblue}{rgb}{0,0.54,0.81}
\definecolor{codegreen}{rgb}{0,0.6,0}
\definecolor{codegray}{rgb}{0.5,0.5,0.5}
\definecolor{codepurple}{rgb}{0.58,0,0.82}
\definecolor{backcolour}{rgb}{0.95,0.95,0.95}

\usepackage{url}
\usepackage{orcidlink}

\usepackage{listings}
\lstdefinestyle{mystyle}{
    backgroundcolor=\color{backcolour},   
    commentstyle=\color{codegreen},
    keywordstyle=\color{codeblue},
    numberstyle=\tiny\color{codegray},
    stringstyle=\color{codepurple},
    basicstyle=\ttfamily\footnotesize,
    breakatwhitespace=false,         
    breaklines=true,
    captionpos=b,
    keepspaces=true,                 
    numbers=left,                    
    numbersep=5pt,                  
    showspaces=false,                
    showstringspaces=false,
    showtabs=false,                  
    tabsize=4,
    aboveskip=\baselineskip,
    belowskip=\baselineskip
}
\usepackage{amsmath}
\usepackage{amssymb}
\usepackage[T1]{fontenc}

\usepackage{graphbox}
\usepackage[section]{placeins}

\usepackage[labelfont=bf]{caption}

\usepackage{hyperref}
\hypersetup{
    colorlinks=true,
    linkcolor=blue,
    filecolor=blue,      
    urlcolor=blue,
    citecolor=blue,
}

\usepackage{tabularray}

\usepackage[authoryear,sort&compress,round]{natbib} 
\newcommand{\pref}[0]{\ensuremath{P_\mathrm{ref}}}
\newcommand{\tref}[0]{\ensuremath{T_\mathrm{ref}}}
\newcommand{\ttra}[0]{\ensuremath{T_0}}
\newcommand{\ttras}[0]{\ensuremath{T_0\mathrm{s}}}
\newcommand{\tlin}[0]{\ensuremath{T_{0,E,\mathrm{lin}}}}
\newcommand{\pert}[0]{\ensuremath{\mathrm{pert}}}
\newcommand{\tran}[0]{\ensuremath{\mathrm{tran}}}
\newcommand{\attv}[0]{\ensuremath{A_\mathrm{TTV}}}
\newcommand{\pttv}[0]{\ensuremath{P_\mathrm{TTV}}}
\newcommand{\logL}[0]{\ensuremath{\log\mathcal{L}}}
\newcommand{\logP}[0]{\ensuremath{\log\mathcal{P}}}
\newcommand{\mttv}[0]{\ensuremath{M_\mathrm{TTV}}}
\newcommand{\mrv}[0]{\ensuremath{M_\mathrm{RV}}}
\newcommand{\rmA}[0]{\ensuremath{\mathrm{A}}}
\newcommand{\rmB}[0]{\ensuremath{\mathrm{B}}}
\newcommand{\rmAB}[0]{\ensuremath{\mathrm{AB}}}
\newcommand{\rmC}[0]{\ensuremath{\mathrm{C}}}
\newcommand{\ltte}[0]{\ensuremath{\Delta_\mathrm{LTTE}}}
\newcommand{\altte}[0]{\ensuremath{\mathcal{A}_\mathrm{LTTE}}}
\newcommand{\kepler}[0]{Kepler}
\title{\textbf{\huge Transit timing variation}}
\author{Luca Borsato\orcidlink{0000-0003-0066-9268}}
\affil{INAF Osservatorio Astronomico di Padova, Vicolo dell'Osservatorio 5, 35122 Padova, Italy}

\date{September 2026}

\providecommand{\keywords}[1]
{
    \begin{center}
    \textbf{\textbf{Keywords}\\} #1
    \end{center}
}

\begin{document}

\maketitle

\begin{abstract}
The transit timing variation (TTV) method is a key approach to the 
confirmation, detection, and characterisation of exoplanetary systems.
This review discusses the detection of TTVs from transit observations and the origins of TTVs, 
mainly due to gravitational interactions with perturbing planets, whether transiting or not.
Additionally, this work details methodologies for deriving planetary parameters, 
employing quasi-global and local optimisation algorithms within a Bayesian framework.
By analysing TTVs, planetary orbital parameters and dynamical properties can be inferred, 
providing valuable insights into system formation and evolution.
The role of resonances and stability in shaping these systems is reviewed, 
along with sensitivity biases that influence mass determination.
Comparative analyses highlight the unique demographic contributions of TTV systems.
TTVs offer a window into the dynamical and physical properties of planetary systems, 
advancing our understanding of their diversity and stability.
\end{abstract}

\keywords{Transit Timing Variations, TTV, multi-planet systems, dynamics, mean-motion resonance, MMR}

\section{Introduction}\label{Intro}

From a historical perspective, the Solar System is the first multiple-planet system ``discovered'' and
the orbits of the rocky planets as well as Jupiter and Saturn were regularly predicted within the precision of the past techniques.
In the first decades of the 19$^{th}$ century, Bouvard stated that new measurements of the orbits of Uranus showed
deviations that could not be clearly understood, requiring larger uncertainties.
However, the later observations of those years continued to show orbital deviations, and there were no reasons to reject them.
By revisiting the observations taken since the discovery of Uranus, 
and testing different models, the perturbed orbit of Uranus allowed both
\citet{Adams1847MmRAS..16..427A} and \citet{LeVerrier1877AnPar..14....1L}
to discover Neptune.
This is the very first case of a hidden planet discovered by its gravitational pull on another planet.

In the field of exoplanets, one can observe the effect of the gravitational interaction among planets within
multi-planet systems by measuring the so-called transit timing variation (TTV).
The gravitational interaction between planets results in transits happening earlier or later than expected,
if assuming only one planet in the system.
Detecting this TTV signal allows one to both discover unknown exoplanet, even if it does not transit,
and to characterise the perturber planet. 
In case of multi-transiting planets it allows one to characterise the system.
Thanks to the \kepler{} mission, the TTV technique showed its potential, providing 
the first clear example of a detected TTV signal in the multiple-transiting planet system 
\kepler-9 \citep{Holman2010Sci...330...51H}, and
the first TTV signal induced by a non-transiting perturber with 
\kepler-19~b \citep{Ballard2011ApJ...743..200B}.
The number of exoplanets and multi-planet systems showing TTV signals 
has since grown,
mainly thanks to the \kepler{} mission, its extension K2, and with the current
Transiting Exoplanet Survey Satellite (TESS) mission.

\section{Method description}\label{Method}

As an exoplanet passes in front of its host star, the stellar flux diminishes, 
reaching a minimum at which one can measure the transit depth, 
which allows the observer to infer the exoplanet's radius. 
This minimum of the stellar flux occurs at the so-called transit time (\ttra)\footnote{In some works it is also indicated as $T_c$.}, 
which is also referred to as the central time of the transit or mid-transit time \citep{Winn2010exop.book...55W}. 
To a first approximation, the transit time represents the centre of symmetry of the transit light curve
when the planetary disc is aligned with the centre of the star.
A more precise definition of the transit time can be found in \citet{Fabrycky2010exop.book..217F},
and it is the time at which the sky-projected distance between the centres 
of the planet and of the star has reached its minimum.

The difference between the \ttra{} of two consecutive transits 
allows one to determine the period ($P$) of the exoplanet. 
However, it is advisable to measure the \ttra{} of at least three consecutive transits
to avoid period-aliases and improve the precision and accuracy of $P$.
Observing three or more transits helps to disambiguate the true orbital period from potential aliases, 
which can occur when only two transits are observed. 
This is particularly important, 
when a significant time gap exists between two isolated transit events, 
as the inferred period, $P$, may correspond either to the time difference between them 
or to an integer sub-multiple (alias) of the true period,
potentially resulting in an incorrect determination of $P$.
A third transit also reduces the probability of false alarms when the transit S/N is low.
Additionally, multiple observations reduce the impact of random errors and systematic effects, 
thereby enhancing the overall accuracy of the period determination.
This can be accomplished by fitting 
the same transit model to all transit observations, 
characterised by a common reference transit time (\tref) and a period (\pref).
This approach is equivalent to phase-folding the transits with respect to a common reference time (\tref) and a period (\pref).
Alternatively, one can fit each single transit\footnote{It can be done etiher by simultaneously fitting all the transits
while fixing the transit model except the \ttras,
or fitting the transit model letting the \ttra{} vary, but fixing the \tref{} and the \pref{} determined from
a previous folded model.}
to determine the \ttras,
then fit a straight line to the \ttras{} 
\citep[for example, using the method of the weighted least squares;][]{NR1992nrfa.book.....P},
computing a reference transit time (\tref) and period (\pref).

\begin{figure}[!htb]
    \centering
    \includegraphics[height=0.33\textheight]{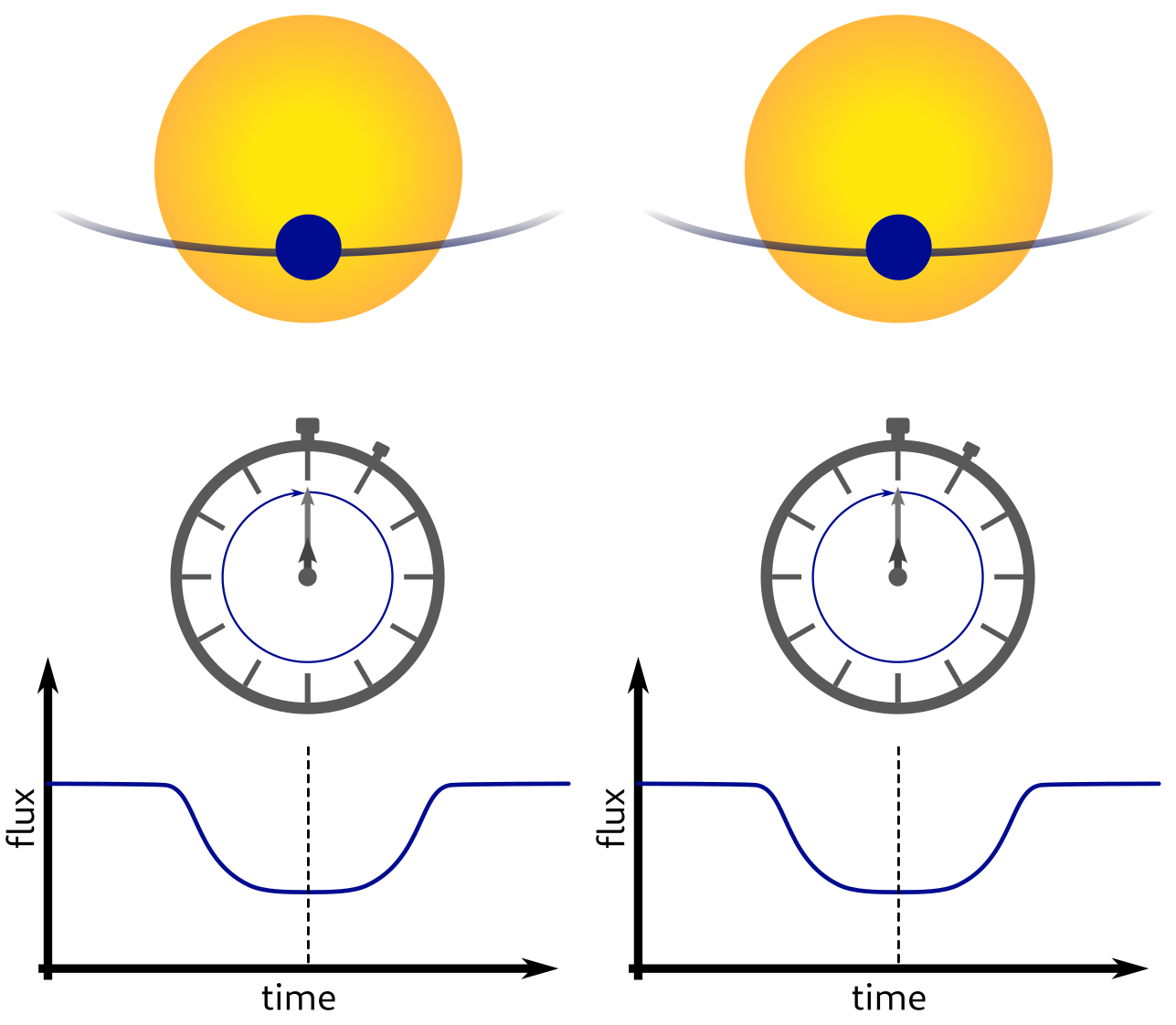}
    \caption{\textbf{Infographic of two periodic transits of an exoplanet.} Schematic illustration of two transit events in the absence of transit timing variations (TTVs).
    In both cases, the exoplanet crosses the centre of the stellar disk (top panels), 
    producing symmetrical light curves (bottom panels) with minima aligned with the expected transit times (dashed vertical lines). 
    The time interval between successive transits corresponds exactly to the orbital period, 
    illustrated by the stopwatch completing a full revolution between events.
    }
    \label{infographic_nottv}
\end{figure}

The \tref{} and \pref{} allow one to predict the expected transit times (\tlin{}) of an exoplanet
with the so-called linear ephemeris:
\begin{equation}
    \label{eq:linephem}
    T_{0,E,\mathrm{lin}} = \tref + E \times \pref,
\end{equation}
where $E$ is the transit number or epoch such that
$E=0$ at \tref{}.
This equation assumes that the transit events are strictly periodic,
and each transit light curve is centred at the \tlin{}, and 
the time between two consecutive transits is exactly \pref{}
(see Fig.~\ref{infographic_nottv}).

By choosing a \tref{} near the middle of the observational time span, 
it minimises the covariance in the linear fit,
effectively reducing the correlation between the uncertainties in \tref{} and \pref.
This approach allows one to improve the precision and accuracy of \tref{} and, even more, of \pref.
The latter is particularly crucial,
as it is the dominant term in the error propagation when predicting future transits. 
A more precise \pref{} leads to more accurate and precise long-term predictions, 
reducing the risk of accumulating significant timing errors over multiple orbital periods.

In the analysis of exoplanets' transits and radial velocities,
it is common to assume that orbital parameters, also known as orbital elements,
follow Kepler's laws, which implies that they are constant in time and
the orbits are periodic, and by definition, closed.
The linear ephemeris is based on this assumption; given a reference transit time,
it predicts future transits with a strictly periodic orbit.
In the analysis of multi-planet systems, 
it is common practice to model the planetary orbits as a simple sum of Keplerian motions, 
regardless of the detection method employed.
However, modelling only with Keplerians would miss the perturbing
or interacting term of the equation of motion
due to additional bodies in the system
\citep{Agol2005MNRAS.359..567A,HolmanMurray2005Sci...307.1288H,Fabrycky2010exop.book..217F,2025haex.book....2A}.
Considering the astrocentric equation of motion
\citep[e.g.][]{MuDe1999book,Fabrycky2010exop.book..217F,2025haex.book....2A} 
of a planet $p$:
\begin{equation}
  \label{eq:force}
  \ddot{\vec r}_{p} =
  -G\left(M_{1}+M_{p}\right) \frac{\vec r _{p}}{r^{3}_{p}}
  +G \sum^{n_\mathrm{b}}_{l=2;l \ne p}{M_{l}\left(\frac{\vec r_{l} - \vec r_{p}}{|\vec r_{l} - \vec r_{p}|^{3}}-\frac{\vec r_{l}}{r^3_{l}}\right)}
\end{equation}
where ${\vec r}_{p}$ is the radius vector of planet $p$,
$M_{1}$ is the mass of the star, 
$M_{p}$ the mass of the planet considered,
$M_{l}$ the masses of the other planets in the system,
$n_\mathrm{b}$ the number of bodies.
The first term of the right-hand side is the Keplerian gravitational acceleration due to the star,
and the second term is responsible for the perturbation of the transiting planet;
this perturbative component consists of two parts:
the \textit{direct} mutual interaction between planet $p$ and the other planets $l$,
and the \textit{indirect} term, which arises because the astrocentric coordinate system is non-inertial, 
reflecting the acceleration of the host star due to the gravitational pull of the other bodies in the system.

\begin{figure}[!htb]
    \centering
    \includegraphics[width=0.33\textheight]{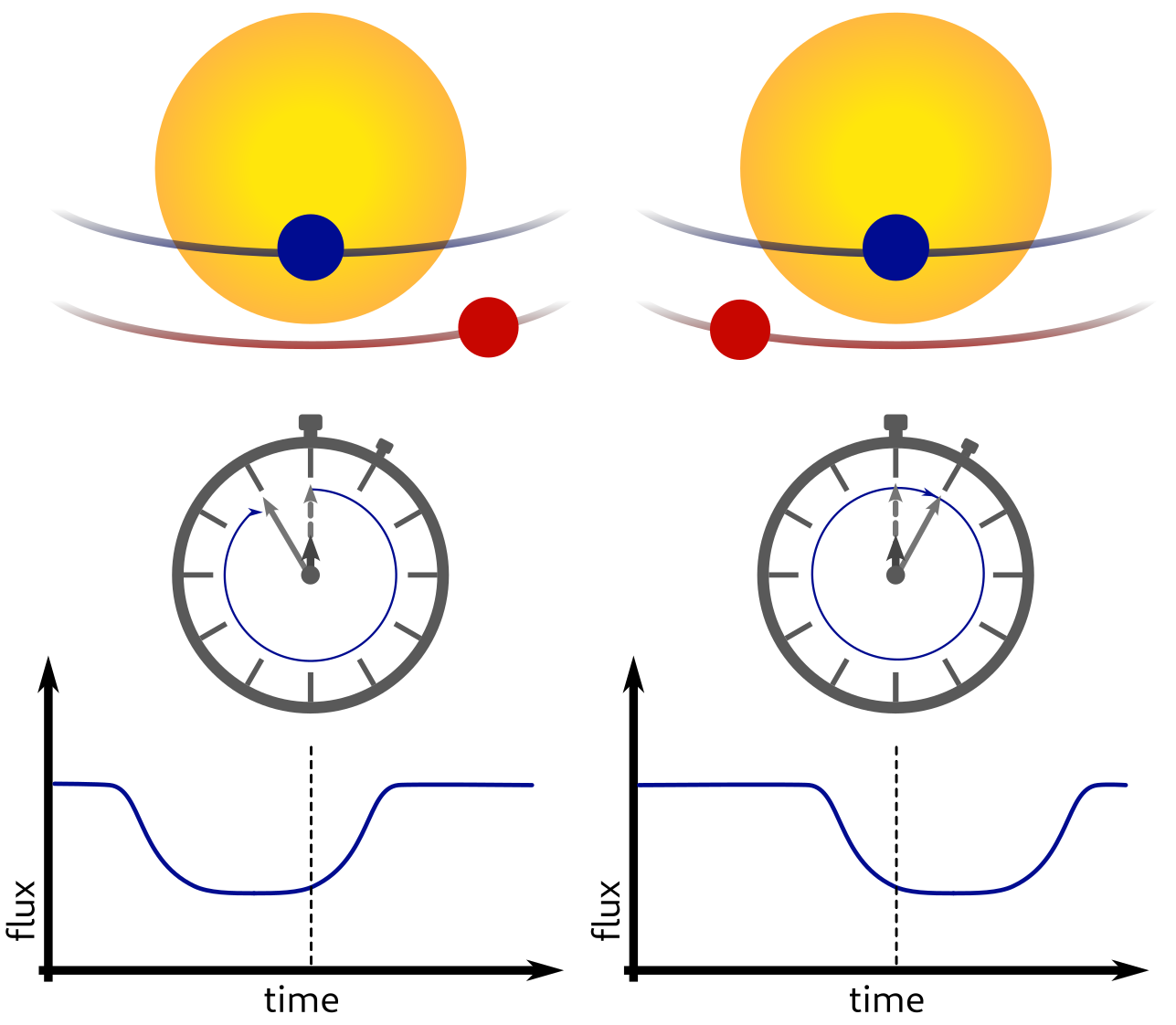}
    \caption{
    \textbf{Infographic of two transits of an exoplanet showing TTV.}
    Schematic illustration of transit timing variations (TTVs) induced by gravitational interactions with an additional planet. 
    The transiting planet (blue) experiences deviations in its expected transit times due to perturbations from a companion (red). 
    In the left panel, the transit occurs earlier than predicted; 
    in the right panel, it occurs later. 
    The light curves (bottom) show the corresponding shifts with respect to the expected mid-transit time (dashed vertical lines), 
    while the stopwatches indicate the offset from the nominal orbital period.
    The orbital geometry shown in the upper panels is illustrative and not intended to represent a physically accurate configuration.
    }
    \label{infographic_ttv}
\end{figure}

\begin{figure}[!htb]
    \centering
    \includegraphics[height=0.66\textheight]{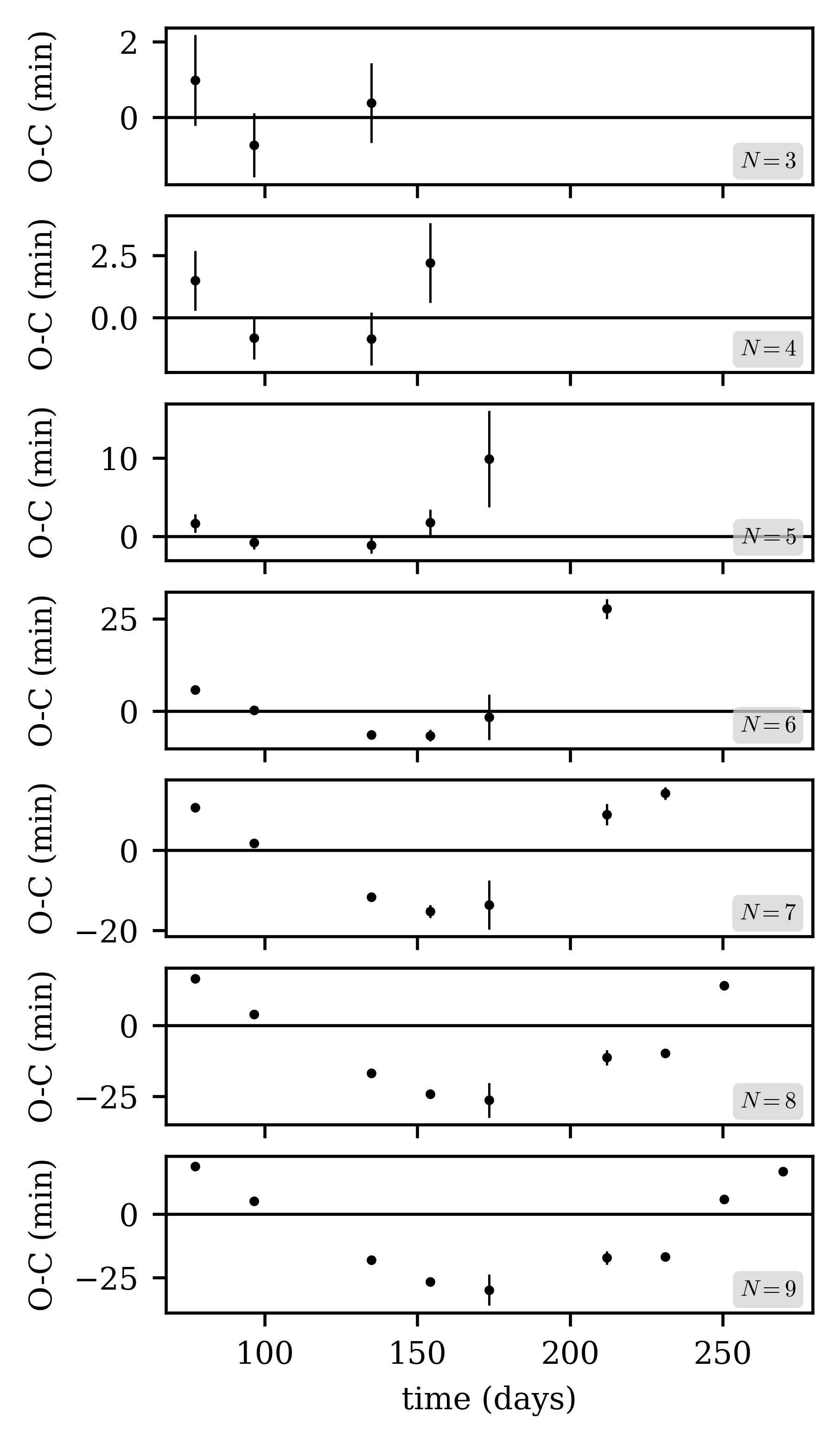}% no path, no extension
    \caption{
    \textbf{$O-C$ plot at increasing number of transits.}
    O-C diagrams at increasing number of transits ($N$, from top to bottom) 
    of \kepler-9~b \citep{Holman2010Sci...330...51H}.
    Transit times from \citet{Borsato2019MNRAS.484.3233B}
    and re-fitted a linear ephemeris for each $N$.
    }\label{OCincreasingN}
\end{figure}

\begin{figure}[!htb]
    \centering
    \includegraphics[width=\columnwidth]{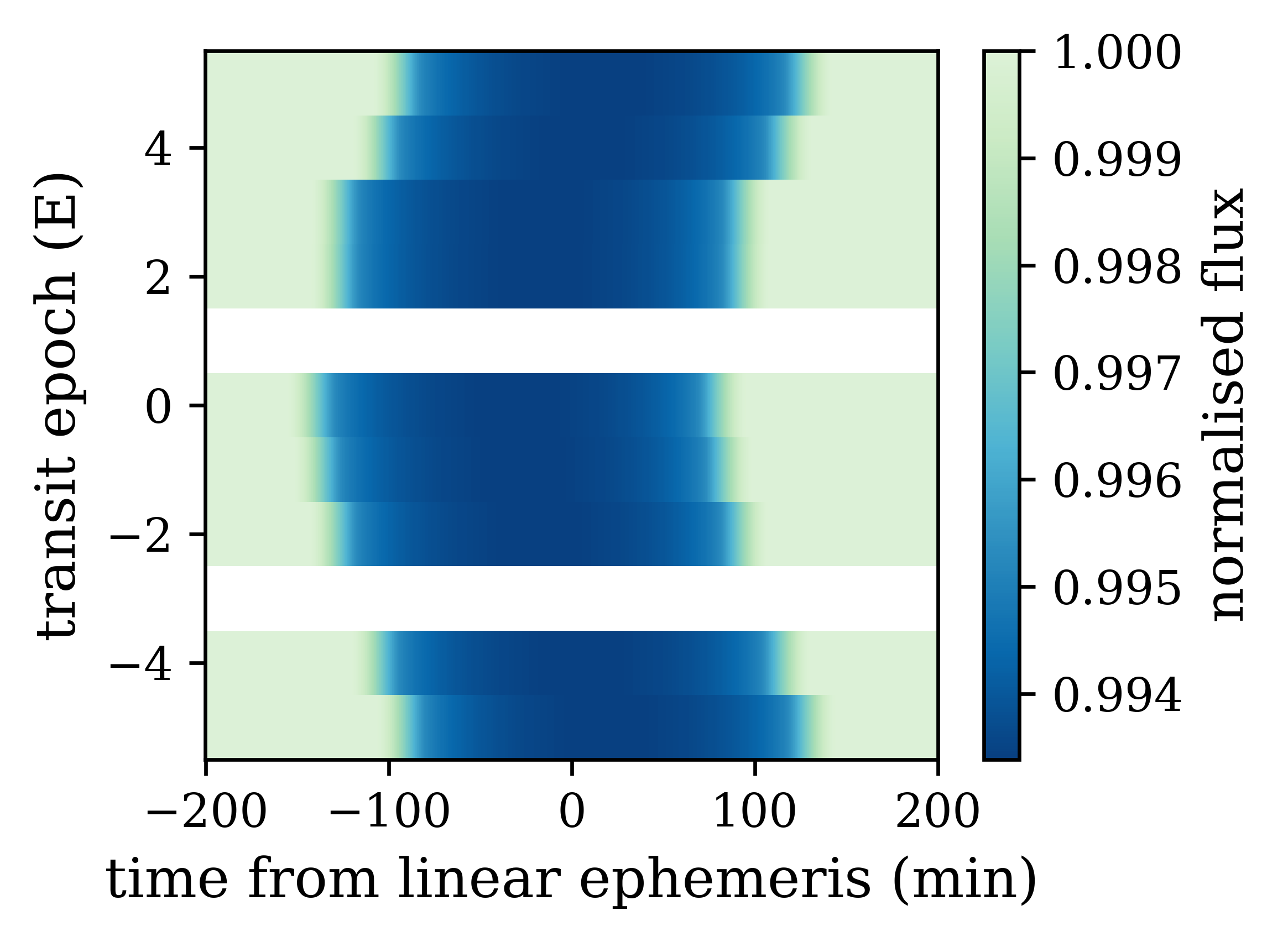}% no path, no extension
    \caption{
    \textbf{TTV through ``River-plot''}
    Simulated ``river plot'' diagram for the first nine observed transits 
    of \kepler-9~b \citep{Holman2010Sci...330...51H}; 
    white rows correspond to not observed transits.
    The X-axis is the time of the light-curve with respect to the expected
    transit time (\tlin) computed with the linear ephemeris in eq.~\ref{eq:linephem}.
    Transit times from \citet{Borsato2019MNRAS.484.3233B}.
    }\label{riverplot}
\end{figure}

The gravitational interaction between planets
induces a transit timing variation (TTV),
meaning that the transits could occur earlier or later
than the predicted, or expected, transit times
(see Fig.~\ref{infographic_ttv}).
This effect is also observable as a deviation of the
transit times (\ttras) from the predicted linear ephemeris (\tlin).
The most common way to visualise the TTV effect is
to plot an Observed-minus-Calculated ($O-C$) diagram,
where $O-C = T_{0,E,\mathrm{obs}} - \tlin$,
as a function of transit epoch ($E$) or time (usually the observed \ttras).
A negative $O-C$ value indicates that the transit occurs earlier than predicted by the linear ephemeris,
while a positive value means a later occurrence.
See in Fig.~\ref{OCincreasingN} some examples of $O-C$ of 
\kepler-9 b \citep{Holman2010Sci...330...51H}
at increasing number ($N$) of observed \ttra{} 
published in \citet{Borsato2019MNRAS.484.3233B}. 
Another diagram used to visualize the TTV effect is the so-called
``river plot'' \citep{CarterAgol2013ApJ...765..132C},
where the normalised stellar flux is plotted as a function of both
the transit number ($E$ in the Y-axis) and 
the time with respect to the expected transit times (\tlin;
see a simulated example in Fig.~\ref{riverplot}).
By observing this TTV signal, it is possible to infer the presence
of an additional planet in the system,
even if it is not transiting the host star
\citep{Agol2005MNRAS.359..567A, Ballard2011ApJ...743..200B,Nesvorny2013ApJ...777....3N}.

There is no universal method for claiming the detection of a TTV signal;
each case requires careful consideration.
For $N$ transit observations, one could compute a reduced chi-squared-like statistic
\begin{equation}
    \label{eq:chisquare}
    \begin{split}
    \chi_\mathrm{r}^{2} & = \frac{\chi^2}{\mathrm{dof}}\ ,\\
    \chi^{2}            & = \sum_{i=1}^{N}\left[ \frac{(O-C)_{i}}{\sigma_{T_{0,i}}}\right]^{2}\ ,\\
    \mathrm{dof}        & = N - 2\ ,
    \end{split}
\end{equation}
where dof are the degrees of freedom.
Checking how far the value of $\chi_\mathrm{r}^{2}$ is from unity 
can provide information about the TTV signal 
and its deviation from the linear ephemeris. 
Another way to assess the TTV signal is to calculate, 
firstly, the amplitude of the TTV ($A_\mathrm{TTV}$),
secondly the signal-to-noise term ($\mathrm{SN_{TTV}}$).
The \attv{} has been defined by \citet{Agol2005MNRAS.359..567A} as the standard deviation of the $O-C$,
but it can also be computed as the semi-amplitude of the $O-C$ or 
the $68.27^\mathrm{th}$ percentile of the absolute value of the $O-C$;
each of these methods has different sensitivities.
For example, the percentile is less sensitive to outliers and, 
in general, yields smaller values than the semi-amplitude method. 
The $\mathrm{SN_{TTV}}$ term is defined as the ratio between 
the \attv{} and the mean (or median) of the \ttras{} uncertainties \citep{Agol2005MNRAS.359..567A}:
\begin{equation}
    \label{eq:snttv}
    \mathrm{SN_{TTV}} = \frac{A_\mathrm{TTV}}{<\sigma_{\ttra}>}\ .
\end{equation}
The higher this ratio, the stronger the TTV signal.
There are no threshold values for these $\mathrm{SN_{TTV}}$ statistics to claim the TTV,
so checking the $\chi_\mathrm{r}^{2}$ and the different $\mathrm{SN_{TTV}}$ values
is the only way to gain a clear picture of the situation. 
As a rule of thumb, if $\chi_\mathrm{r}^{2} \gtrsim 3$ and 
$\mathrm{SN_{TTV}}$ statistics are consistently above 3--5, 
the system deserves further detailed investigation to characterise the nature of the perturbations.

\begin{figure}[!htb]
    \centering
    \includegraphics[height=0.5\textheight]{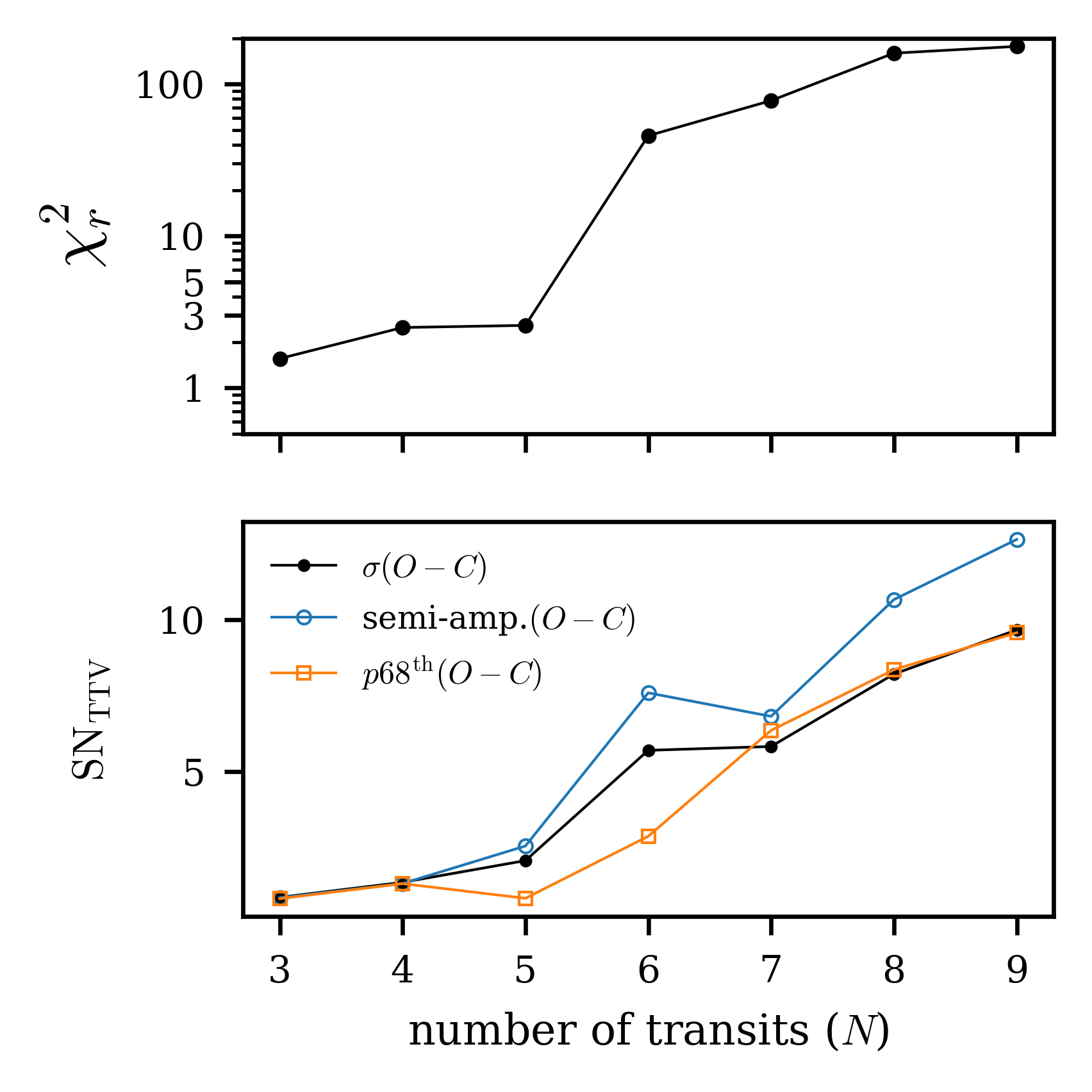}% no path, no extension
    \caption{
    \textbf{TTV detection statistics at increasing number of transits.}
    Statistics of the $O-C$ at increasing number of transits ($N$) of Fig.~\ref{OCincreasingN}.
    \textit{upper-panel}: reduced chi-square ($\chi_\mathrm{r}^{2}$) of the fitted linear ephemeris as function of
    the number of transits.
    \textit{lower-panel}: $\mathrm{SN_{TTV}}$ as TTV amplitude ($A_\mathrm{TTV}$) divided by
    the mean \ttra{} uncertainty ($<\sigma_{\ttra}>$) 
    for three methods to compute the $A_\mathrm{TTV}$.
    }\label{StatsincreasingN}
\end{figure}

Given that the main concept is to measure the timing variation from the linear ephemeris,
the uncertainty of the single transit time ($\sigma_{\ttra}$)
is a key parameter to detect and characterise the TTV signal,
as shown in the eq.~\ref{eq:snttv}.
The determination of the \ttra{} and associated $\sigma_{\ttra}$ 
is mainly driven by the transit ingress and egress phases,
which timescale is usually defined as $\tau_\mathrm{ing,egr}$ \citep{Winn2010exop.book...55W}.
These transit phases occur when the stellar flux decreases and 
they contain most of the timing information \citep{2025haex.book....2A}.
Therefore, high-cadence, high-precision photometry 
with well-sampled ingress/egress phases are necessary to improve the timing error $\sigma_{\ttra}$, 
which can be measured more precisely than $\tau_\mathrm{ing,egr}$
\citep{Winn2010exop.book...55W,2025haex.book....2A}.
Another key aspect for the TTV detection is the number of transits ($N$) accounted for.
In fact, an increasing number of transits ($N$) not only allows for improved
precision and accuracy on the linear ephemeris,
but it is also crucial for being able to claim the existence of a TTV signal,
as can be seen in Fig.~\ref{OCincreasingN} and \ref{StatsincreasingN}.
It is important to note that there is no fixed threshold for the number of transits
required to robustly claim the presence of a TTV signal, 
as this can vary depending on the system and the data quality. 
However, in general, three measured transit times are insufficient to make such a claim, 
as the apparent variations could be due to intrinsic scatter, stellar activity, 
poor transit modelling, or short-term features that can mimic or mask true TTV signals.
To mitigate these issues, observing at least five to six transits is often considered necessary 
to begin identifying a coherent TTV signal and to mitigate the risk of false positives 
(as can be seen in Fig.~\ref{OCincreasingN} and \ref{StatsincreasingN}).
Nonetheless, even with five or more transit times, a linear trend could be observed, 
not because of the absence of a TTV, but rather due to the system being sampled during
a segment of the TTV cycle where the signal is slowly varying 
(e.g., a small, nearly linear portion near a turning point or long-period slope).
This highlights the fact that, in addition to the number of transits, 
the temporal distribution and coverage across the TTV phase are also critical 
-- although the TTV phase is generally unknown in advance.
Another fundamental aspect to keep in mind is that all transit times involved in the analysis
must have the same time standard, and in particular, 
following \citet{Eastman2010PASP..122..935E}, it would advisable to adopt the
Barycentric Julian Day (BJD) in the Barycentric Dynamical Time (TDB), 
$\mathrm{BJD_{TDB}}$, for the all the transit times.

In this review, I focus on TTV signals of planetary origin, 
caused by gravitational interactions among planets within a multiple-planet system. 
However, the methods used to detect such TTVs can be extended to 
other sources of transit timing variations, 
such as 
apsidal precession, 
tidal decay effects resulting from interactions between
a star and a short-period planet
\citep[e.g. see ][]{Rasio1996ApJ...470.1187R,Levrard2009ApJ...692L...9L,Leonardi2024AA...686A..84L},
or spot-crossing events that 
distort the transit shape (manifesting as bumps) and 
introduce variations in the timing measurements
\citep[e.g. see ][]{Oshagh2013AA...556A..19O,Ioannidis2016AA...585A..72I}.

\begin{figure}[!htb]
    \centering
    \includegraphics[width=\textwidth]{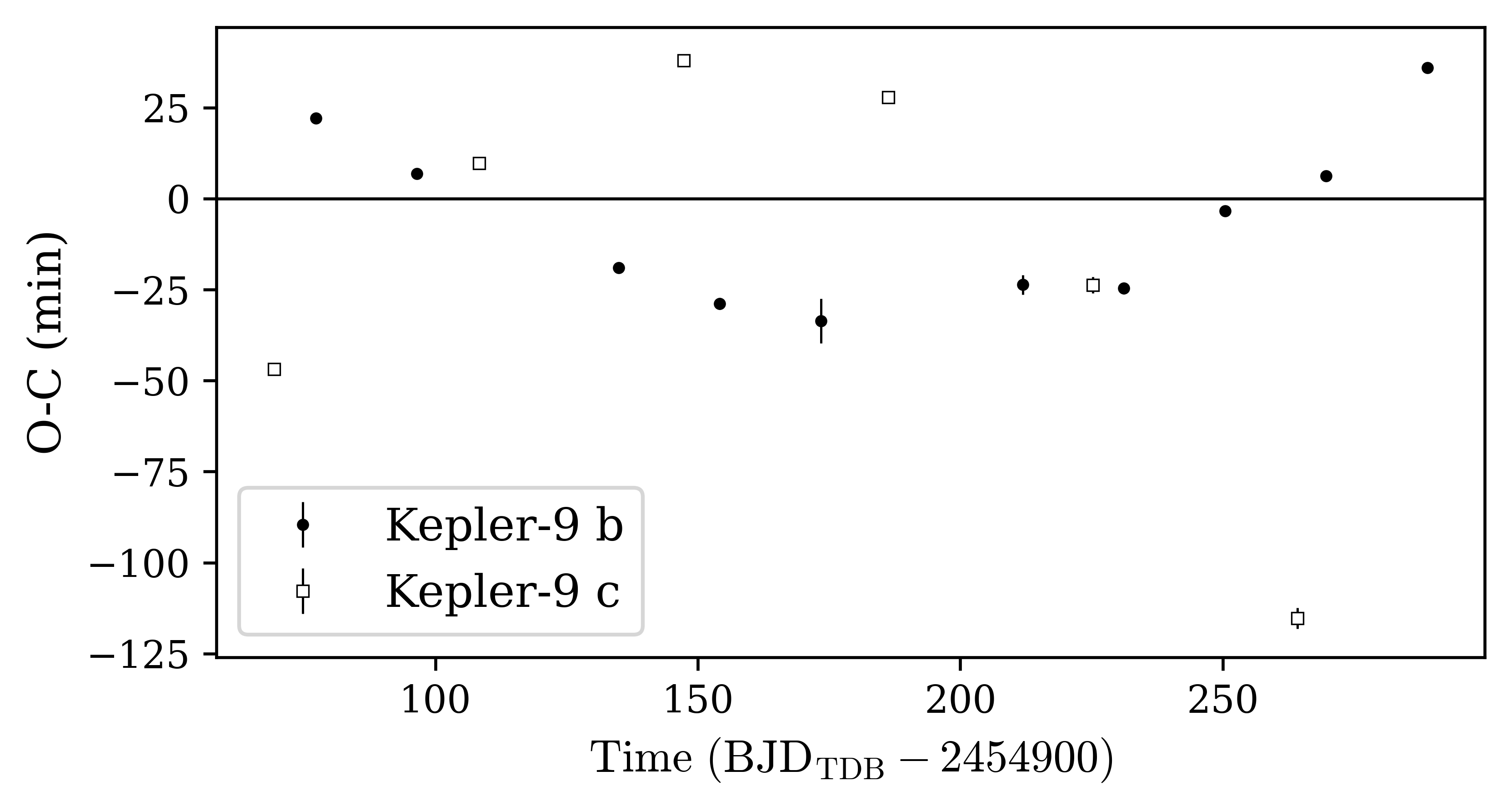}% no path, no extension
    \caption{
    \textbf{\kepler-9 $O-C$ plot for three quarters.}
    O-C diagram of \kepler-9~b (black dots) and c (open squares).
    Transit times from \citet{Borsato2019MNRAS.484.3233B},
    but selected to match first three \kepler{} quarters
    analysed in the discovery paper by \citet{Holman2010Sci...330...51H}
    and after refitting a linear ephemeris.
    }\label{Kepler9oc_H10}
\end{figure}

\begin{figure}[!htb]
    \centering
    \includegraphics[width=\textwidth]{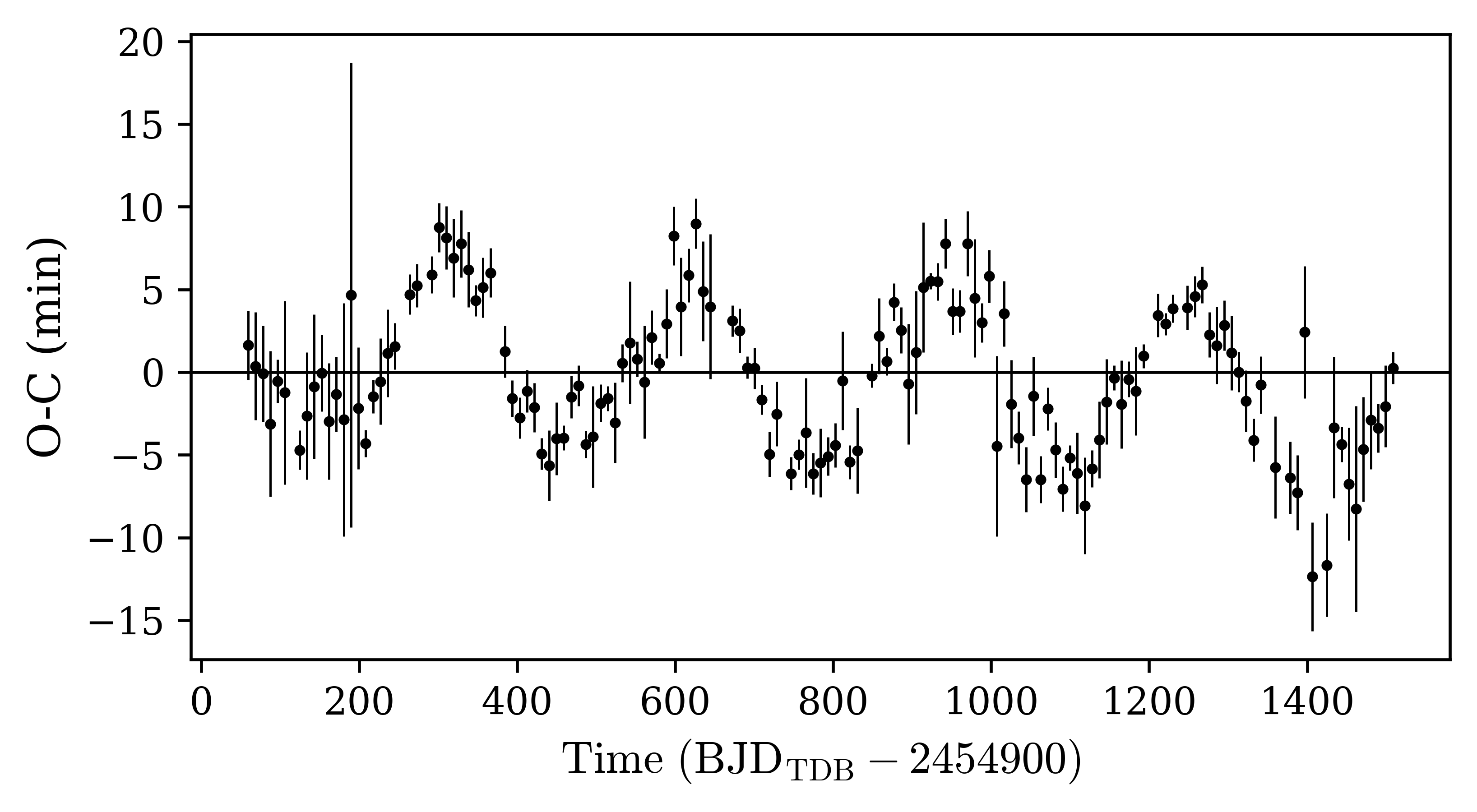}% no path, no extension
    \caption{
    \textbf{\kepler-19~b $O-C$ plot.}
    O-C diagram of \kepler-19~b \citep{Ballard2011ApJ...743..200B}.
    Transit times from \citet{Malavolta2017AJ....153..224M} spanning the full
    \kepler{} mission.
    }\label{Kepler19oc_full}
\end{figure}

The $O-C$ of \kepler-9~b and c is shown in Fig.~\ref{Kepler9oc_H10}
spanning the first three quarters of \kepler{} data (approximately 300 days)
used
to identify the
first planets and the first multiple-planet system showing TTV signals.
In comparison, the $O-C$ diagram spanning the full \kepler{} observations 
of \kepler-19~b can be seen in Fig.~\ref{Kepler19oc_full},
showing the first TTV signal discovered to be induced by non-transiting 
perturber planets.

\subsection{Impact of Planetary Parameters and Orbital Configurations on TTV Characteristics}\label{Method_params}

Observing the TTV signal of a transiting planet gives information
about the scaled mass of the perturber (pert), and not about the mass of the transiting one.
In general, it can be shown that the amplitude computed over the TTV period of the TTVs
of a transiting exoplanet depends, to first order, 
on the perturber's period ($P_\pert$), 
on the planet-to-star mass ratio ($M_\pert/M_\star$),
on the eccentricity ($e_\pert$),
and on a function ($f_{\tran,\pert}$) of the semi-major axis ($a$) ratio 
$\alpha_{\tran,\pert} = \mathrm{min}(a_\tran/a_\pert,a_\pert/a_\tran)$.
Moreover, the TTV phase, amplitude, and modulation is also 
a function of the orbital elements of both planets
\citep{Agol2005MNRAS.359..567A,HolmanMurray2005Sci...307.1288H,2025haex.book....2A}.
Following \citet[][eq. 6]{2025haex.book....2A}, 
the TTV signal of a transiting planet at a transit $i$ can be written as:
\begin{equation}
    \label{eq:ttv}
    \delta t_{\tran,i} = P_\tran \frac{M_\pert}{M_\star} f_{\tran,\pert}(\alpha_{\tran,\pert}, \theta_{\tran,\pert})\ 
\end{equation}
with $\theta_{\tran,\pert}$ the orbital elements of both planets;
$\delta t_{\tran,i}$ can be interpreted as the $(O-C)_{i}$.
Therefore, the analysis and detection of TTV signals make it possible 
to determine the mass and orbital parameters of the perturbing body and the overall system.

\paragraph*{Commensurability and mean-motion resonance.}

When the ratio of the orbital periods 
(or the mean motion, $n=2\pi / P$) 
of the transiting and of perturber planet are close
to the ratio of two small integers, it is said that they exhibit a commensurability.
This is often an indication that the system, or more specifically the planet pair,
is in, or near, a mean-motion resonance (MMR).
Commensurabilities, or MMRs, are typically denoted as $j+q:j$ MMR, 
where the order of the resonance is defined as $q$.
This means that 
a conjunction between the two planets occurs
every $j$ orbits of the inner planet 
and every $j+q$ orbits of the outer one.
For example, planets of the \kepler-9 system are near a 2:1 MMR,
where planet c has $P_\mathrm{c} \simeq 38.962$~days and 
b has $P_\mathrm{b} \simeq 19.243$~days,
that is $\frac{P_\mathrm{c}}{P_\mathrm{b}} \simeq \frac{2}{1} = 2$,
so planet b completes two orbits while planet c completes one;
and K2-19~c ($P_\mathrm{c} \simeq 11.8993$~days)
and K2-19~b ($P_\mathrm{b} \simeq 7.9222$~days)
are near a 3:2 MMR, 
that is $\frac{P_\mathrm{c}}{P_\mathrm{b}} \simeq \frac{3}{2} = 1.5$
and the inner planet b does three orbits while planet c completes two.

To determine whether a pair of planets is in a mean-motion resonance (MMR), 
further dynamical analysis is required. 
This typically involves integrating the orbits of the system 
over a sufficient timescale to track the evolution of the so-called ``critical resonant angles'' ($\phi_1$, $\phi_2$).
For a planet pair, $\phi_1$ and $\phi_2$ can be computed during the evolution for the inner and outer planet as:
\begin{equation}
    \label{eq:resonant_angles}   
    \begin{split}
    \phi_1 & = (j+q) \lambda_\mathrm{outer} - j \lambda_\mathrm{inner} - q \varpi_\mathrm{inner}\ ,\\
    \phi_2 & = (j+q) \lambda_\mathrm{outer} - j \lambda_\mathrm{inner} - q \varpi_\mathrm{outer}\ ,
    \end{split}
\end{equation}
where 
$\varpi=\omega + \Omega$ is the longitude of the pericentre,
$\omega$ the argument of the pericentre,
$\Omega$ the longitude of the ascending node,
$\lambda = \mathcal{M} + \varpi = \mathcal{M} + \omega + \Omega$ represents the mean longitude,
with $\mathcal{M}$ the mean anomaly, and
$j$ and $q$ define $j+q$:$j$ MMR of order $q$.
If at least one of these angles exhibits ``libration'',
meaning it oscillates around a fixed value without circulating
(i.e., without rotating through all possible angles in the range of 0 to $2\pi$),
it indicates that the pair of planets may be in resonance \citep{Veras2011ApJ...727...74V}.
Also, the analysis of the relative apsidal angle (or apsidal difference)
\begin{equation}
    \Delta \varpi = \varpi_\mathrm{inner} - \varpi_\mathrm{outer} = \frac{\phi_1-\phi_2}{q}\ ,    
\end{equation}
provides additional information about the process that leads the planets into the resonant state.
It is important to note, however, that this angle can sometimes librate even in systems outside of 
mean-motion resonance due to secular interactions.
A value $\Delta \varpi \simeq 0^{\circ}$ means that the pericentre of the planets is aligned (apsidal alignment) and
this is usually the result of disk interactions which lead to eccentricity increases.
On the other hand, there is apsidal anti-aligned if $\Delta \varpi \simeq 180^{\circ}$,
meaning that the pair underwent eccentricity damping with protoplanetary disk
interaction during the MMR capture phase \citep{Beauge2003ApJ...593.1124B,Laune2022MNRAS.517.4472L}.
\citet{Deck2013ApJ...774..129D}, followed by \citet{Nesvorny2016ApJ...823...72N} and
\citet{Leleu2021AA...655A..66L_rivers}, developed a one-degree-of-freedom model to
determine whether a pair of planets is in mean-motion resonance (MMR).

The resonance configuration can be extended to a three-planet system,
where the resonance links the dynamics of the three planets together.
In this case, one can define resonant angles similar to the two-body resonance.
Following \citet{Siegel2021AJ....161..290S},
one can assume three planets, identified as $p=1,2,3$, with increasing orbital periods ($P_1 < P_2 < P_3$).
For such a system, lying in a chain of resonance of kind $(j+q : j , k+q : k)$ 
the following critical angles are defined:
\begin{equation}\label{eq:mmr_3b}
    \begin{split}
        \phi_{12} = (j+q) \lambda_2 - j\lambda_1 - q\varpi_2\ , \\
        \phi_{23} = (k+q) \lambda_3 - k\lambda_2 - q\varpi_2\ , \\
        \phi = \phi_{23} - \phi_{12} =  j\lambda_1 - (j+k+q)\lambda_2 + (k+q)\lambda_3\ .
    \end{split}    
\end{equation}
The last equation is the general three-body angle.
When both $\phi_{12}$ and $\phi_{23}$ librate around a specific value,
so does $\phi$.
However, $\phi$ can librate even if both $\phi_{12}$ and $\phi_{23}$ circulate.

Being near or in MMR leads to a strong interaction between the planets and
it has the effect of greatly enhancing the \attv{},
as can be seen in the ``Flames of Resonance'' plot \citep[Figure 2 of][]{Veras2011ApJ...727...74V}.
Such a plot also shows that a configuration in MMR can produce a strong-detectable TTV,
but lower in amplitude than in configurations that are near MMR, but out of it \citep{Veras2011ApJ...727...74V}.
It has been shown, 
for example by \citet{Agol2005MNRAS.359..567A} and \citet{HolmanMurray2005Sci...307.1288H},
that it is possible to detect Earth-like planets orbiting nearby to Jupiter-like planets,
without using the RV method.
For first-order mean motion resonances ($q=1$),
the dominant component of the TTV signal is
driven by changes in the mean motions of the planets
\citep{Agol2005MNRAS.359..567A,HolmanMurray2005Sci...307.1288H}.
This is also due to the fact that the region near the first-order resonance
is larger at low eccentricity \citep{Deck2016ApJ...821...96D}.
In higher-order resonances ($q>1$), 
the contribution of the perturber’s eccentricity,
in the form of the eccentricity vectors 
($e \cos \varpi$, $e \sin \varpi$),
becomes increasingly important, 
leading to more complex TTV signatures 
\citep{Boue2012MNRAS.422L..57B,Deck2016ApJ...821...96D}.

Another effect of being close to a commensurability,
is that the shape of the TTV signal is sinusoidal \citep{Lithwick2012ApJ...761..122L}.
Furthermore, if one detects two transiting planets which are close to commensurability,
two TTV signals that are sinusoidal and anti-correlated to each other are observed.
Due to energy exchanges and conservation, 
\citet{2025haex.book....2A} shown that, in this case, 
the TTV signal of planet 1, at transit $i$, is anti-correlated and proportional to the ratio of the masses of the planets
$M_2/M_1$, removing the dependence with stellar mass:
\begin{equation}
    \label{eq:ttv_mmr}
    \delta t_{1,i} = -\delta t_{2,i} \left(\frac{M_2}{M_1}\right) \left(\frac{P_1}{P_2}\right)^{2/3}\ .
\end{equation}
For planets of equal mass, the outer planet exhibits a larger TTV amplitude. 
This arises from the conservation of Keplerian orbital energy: 
to balance the energy exchanged during mutual gravitational interactions, 
the outer planet, on a larger orbit, must undergo more significant changes
in orbital size compared to the inner planet.
\citet{Lithwick2012ApJ...761..122L} analysed the case of planet pair close to 
a first order ($q=1$) MMR,
and introduced the concept of the TTV super-period (usually called \pttv)
and of the normalised distance from the exact resonance ($\Delta$).
A general form of the \pttv{} and of $\Delta$ can be written as:
\begin{equation}
    \label{eq:pttv}
    \pttv = |\frac{j+q}{P_\mathrm{out}} - \frac{j}{P_\mathrm{in}}|^{-1}    
\end{equation}
and
\begin{equation}
\label{eq:delta}
    \Delta = \frac{P_\mathrm{out}}{P_\mathrm{in}}\frac{j}{j+q} - 1\ .
\end{equation}

As $\Delta$ approaches 0, the planet pair is closer to a MMR state,
increasing the $\attv$.
When one observes two planets transiting, the periods can be measured,
and it is possible to determine the $\Delta$ and, so,
one can also calculate the super-period of this TTV signal.
This is very important because if the observations span a temporal baseline
shorter than \pttv{}, it means that it could be very difficult to infer the perturber parameters,
which would also lead to wrong parameter determination.
In case a single transiting planet showing a sinusoidal TTV covering a full \pttv{} is observed,
the general expression in eq.~\ref{eq:pttv} based on \citet{Lithwick2012ApJ...761..122L}'s formulae,
can be used to compute possible perturber periods,
assuming some period commensurabilities,
effectively reducing the parameter space of the analysis.

\paragraph*{Observational baseline.}

\begin{figure}[!htb]
    \centering
    \includegraphics[width=\textwidth]{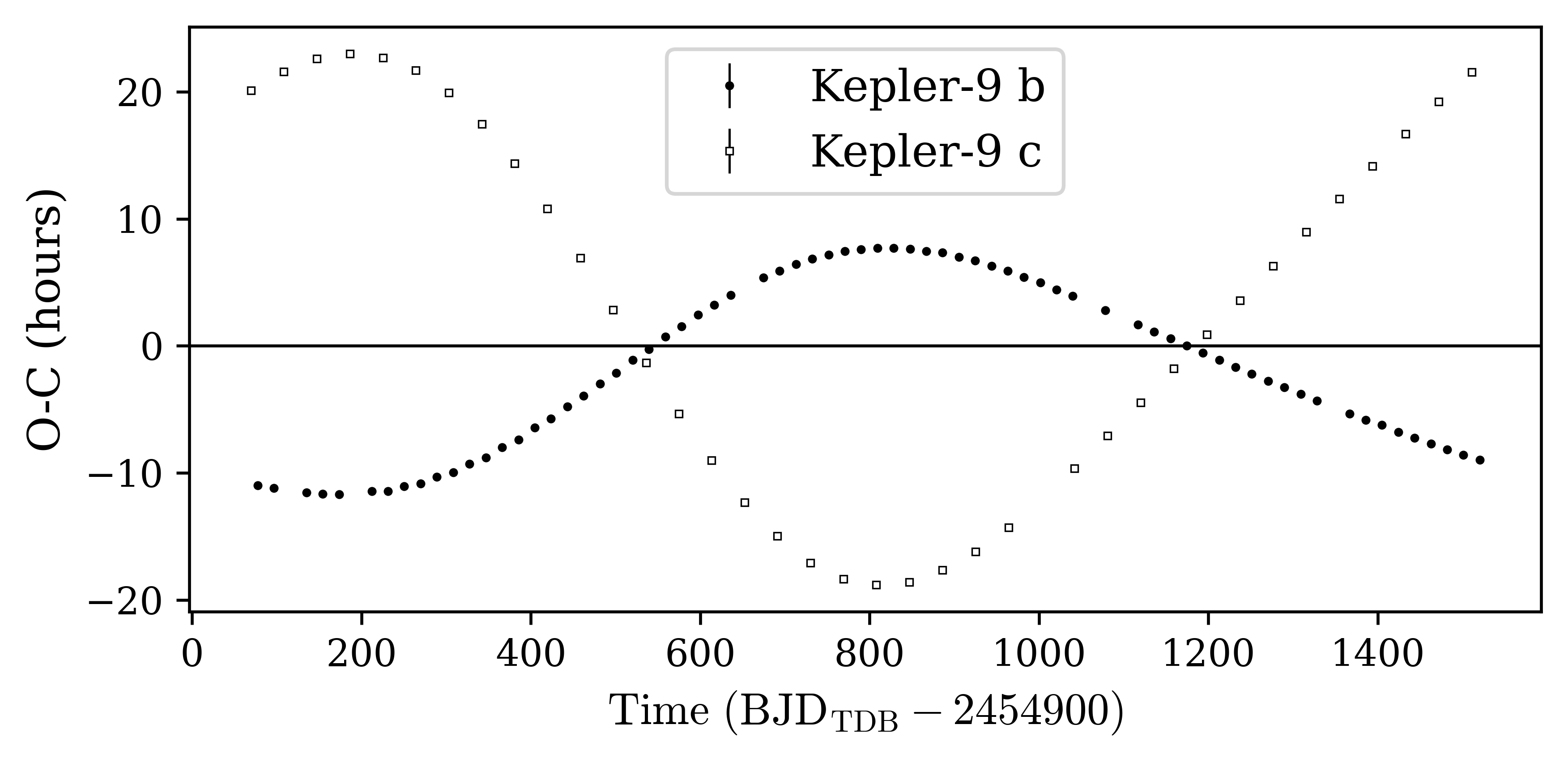}% no path, no extension
    \caption{
    \textbf{\kepler-9 $O-C$ plot for all \kepler{} quarters.}
    O-C diagram of \kepler-9~b (black dots) and c (open squares).
    Transit times from \citet{Borsato2019MNRAS.484.3233B} spanning all \kepler{} quarters.
    }\label{Kepler9oc_full}
\end{figure}

\kepler-9 is not only the first discovered system showing a TTV signal,
but it was also the first system with two transiting planets exhibiting
an anti-correlated TTV curvature (see Fig.~\ref{Kepler9oc_H10}).
The full \pttv{} of \kepler-9 b and c were not fully recovered until four years later, 
when more Kepler data were available.
With a longer baseline, it became possible to retrieve a different TTV pattern and
thoroughly revise the parameters of the planets \citep{Borsato2014AA...571A..38B},
almost halving the masses and eccentricities of both planets. 
This was a clear example of how a long temporal baseline, or observing multiple transits, 
is needed to achieve a sufficient sampling of the TTV signal 
in order to measure accurate and precise orbital parameters and masses. 
I present in Fig.~\ref{Kepler9oc_full} the $O-C$ diagram computed from the \ttras{}
of Kepler-9 b and c,
covering the full Kepler observations and showing the long and substantial anti-correlated TTV signals,
with \attv{} of approximately 10 and 20 hours for b and c, respectively.

\begin{figure}[!htb]
    \centering
    \includegraphics[width=\textwidth]{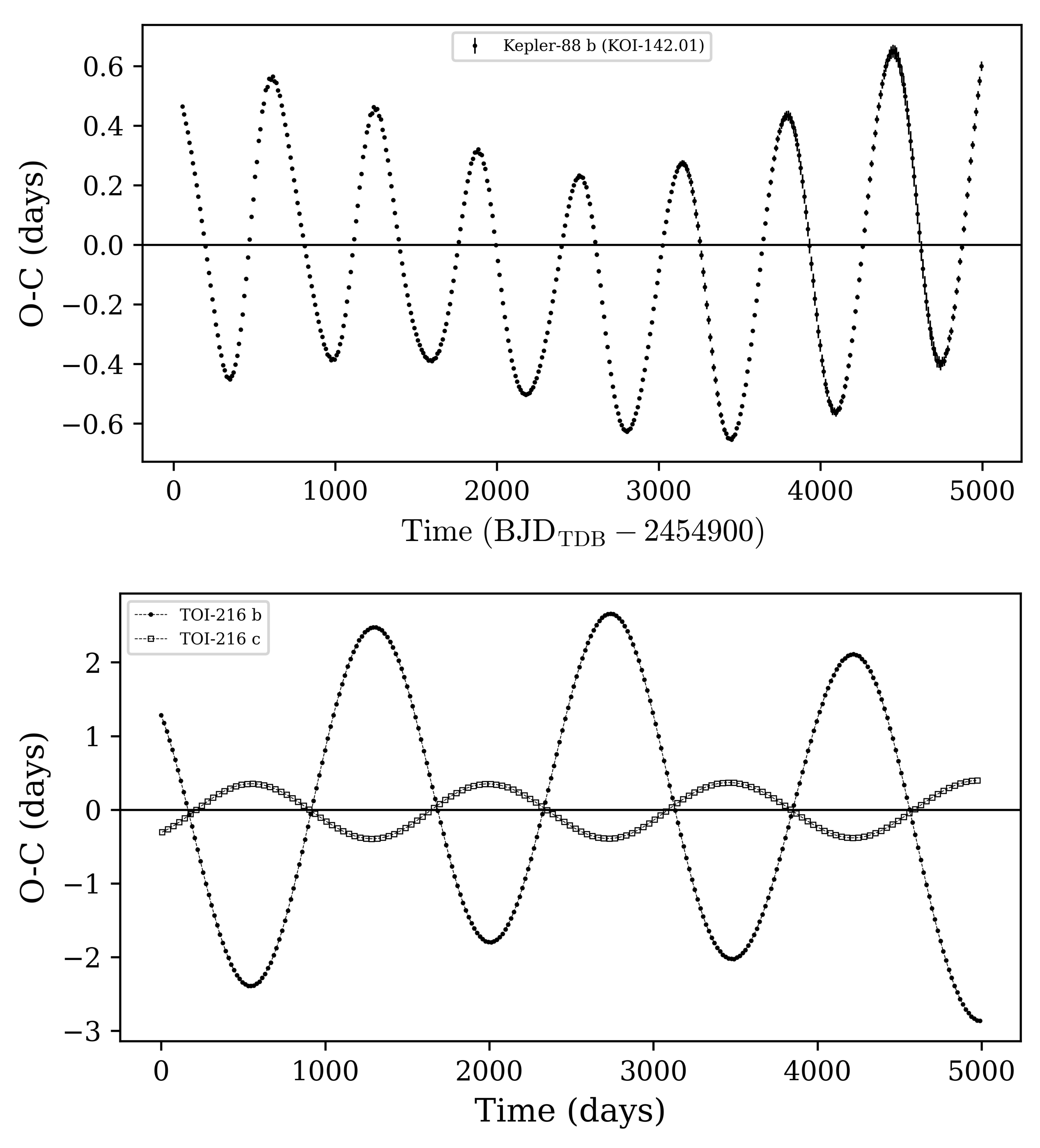}% no path, no extension
    \caption{
    \textbf{$O-C$ plots of \kepler-88 and TOI-216.}
    \textit{Upper panel}: O-C diagram of \kepler-88~b \citep{Nesvorny2013ApJ...777....3N}, the ``King of TTVs'' of the \kepler{} era;
    transit times from \citet{Weiss2020AJ....159..242W}.
    \textit{Lower panel}: synthetic O-C diagrams of TOI-216~b (black circles) and TOI-216~c (open squares), the ``King of TTVs'' at time of writing; parameters of the system from \citet{McKee2023AJ....165..236M}.
    }\label{Kepler88_TOI216_oc}
\end{figure}

During the \kepler{} mission, another exoplanet showed a substantial \attv{} of about 12 hours:
\kepler-88~b \citep[KOI-142;][]{Nesvorny2013ApJ...777....3N}, 
which has been long referred to as ``The King of TTV'' (see Fig.~\ref{Kepler88_TOI216_oc}).
The \kepler-88 system has a sub-Neptune-sized transiting planet on an 11-day orbit 
showing a sinusoidal TTV (see Fig.~\ref{Kepler88_TOI216_oc}) with $\attv \sim 0.6$~days,
that was predicted to be induced by a non-transiting perturber close to a 2:1 MMR 
\citep[period of about 22 days;][]{Nesvorny2013ApJ...777....3N}.
Subsequently, \citet{Barros2014AA...561L...1B} confirmed this perturber, \kepler-88~c, through an RV survey; 
this was the first confirmation of a non-transiting planet predicted from TTV analysis.
Furthermore, with additional RV observations, 
\citet{Weiss2020AJ....159..242W} discovered an additional massive planet ($M_\mathrm{d}\sim 3\, M_\mathrm{Jup}$),
\kepler-88~d, on a more distant orbit of about 1340 days. 

\citet{Kipping2019MNRAS.486.4980K} analysed TESS data of TOI-216 and 
identified two transiting planets near a 2:1 MMR, 
exhibiting anti-correlated TTV curvature ($\attv \sim 40$ and $\sim 10$ minutes for planet b and c, respectively),
reminiscent of the \kepler-9 system.
Subsequently, \citet{Dawson2021AJ....161..161D} extended the observational baseline 
by incorporating additional TESS sectors and ground-based follow-up observations 
over a span of approximately 850 days. 
They reported significantly larger TTV amplitudes, on the order of 33 and 7 hours for planets b and c, respectively, 
with a super-period exceeding the duration of the observations.
These revised TTV amplitudes effectively positioned TOI-216~b as the new ``King of TTVs'', 
previously attributed to \kepler-88~b.
Moreover, numerical integrations of the system's orbital parameters over a timespan of approximately 5\,000 days 
suggest that TOI-216~b could exhibit TTV amplitudes as large as  $\attv \gtrsim 2$~days (see Fig.\ref{Kepler88_TOI216_oc}).

\paragraph*{Degeneracies and ``chopping'' TTV.}

Observing two transiting planets in the same system for a sufficient time to cover the full \pttv{} and \attv{}
allows one to characterise the planetary system and its architecture.
However, some degeneracies between parameters still exist, such as the mass-eccentricity degeneracy, 
where the same TTV can be observed for different combinations of mass and eccentricity of the perturber.
When a pair of planets is close to a MMR, their interaction grows strongest at minimum separation.
If the planets are coplanar and their orbits have low eccentricities, 
this strong interaction occurs at conjunction, which could be far from the transit events. 
This conjunction repeats with a synodic period $P_\mathrm{syn} = |1/P_1 - 1/P_2|^{-1}$.
This periodicity implies an additional effect to consider in the TTV signal: 
small alternating TTVs, on top of the longer TTV, 
on short-timescale and occurring at $P_\mathrm{syn}$ and its harmonics
\citep{Nesvorny2014ApJ...790...58N,Deck2015ApJ...802..116D,2025haex.book....2A}.
This effect is small because the terms in the harmonics do not add coherently.
\citet{Deck2015ApJ...802..116D} defined this effect as ``chopping'' TTV and
showed that its amplitude depends on the perturber's mass-to-star ($M_\pert/M_\star$) and 
the semi-major axis ratio ($\alpha_{\tran,\pert} = a_\tran/a_\pert$).

\begin{figure}[!htb]
    \centering
    \includegraphics[width=\textwidth]{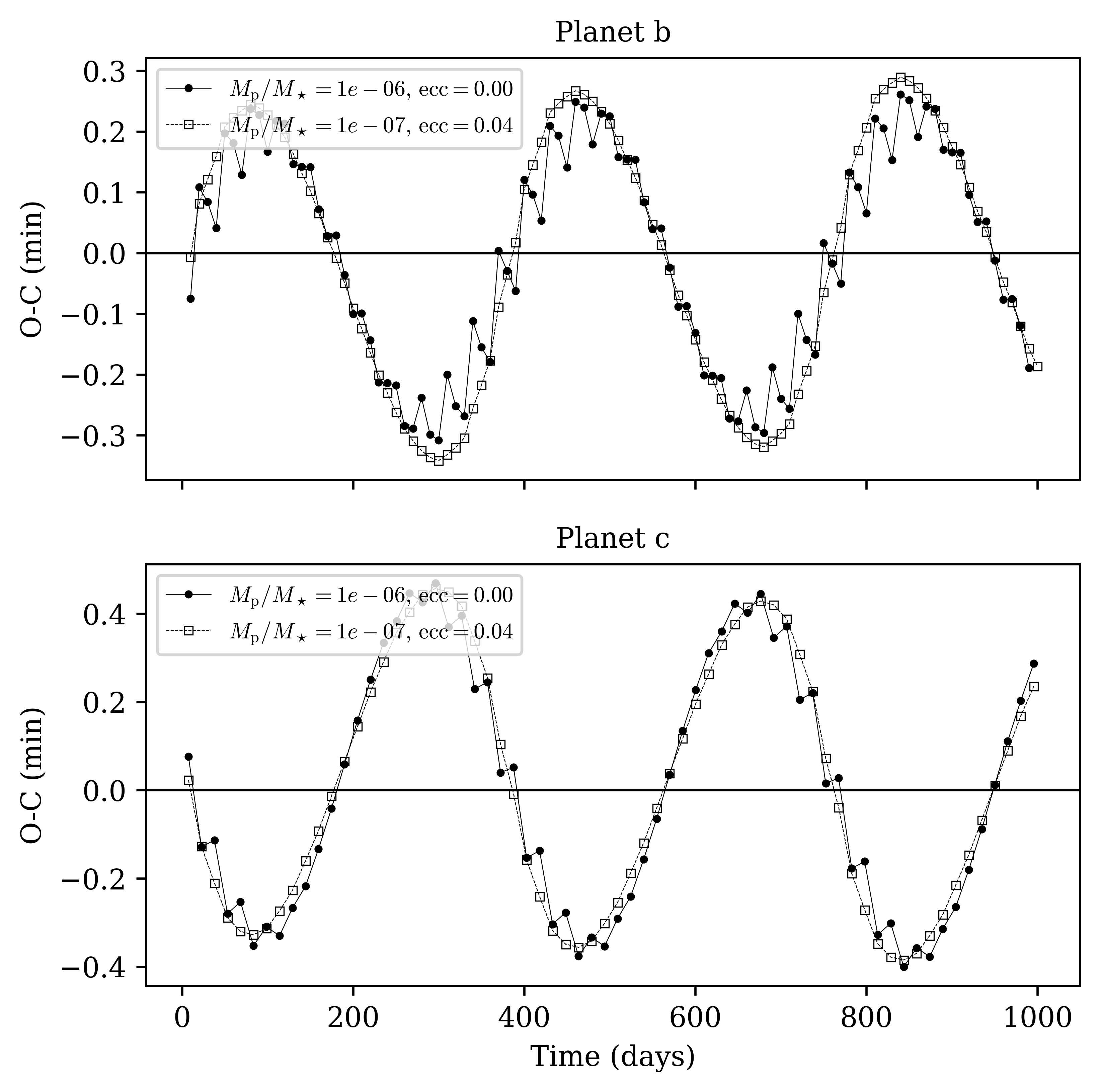}% no path, no extension
    \caption{
    \textbf{$O-C$ plots showing the TTV ``chopping'' effect.}
        O-C diagrams of two hypothetical planets with periods of
        10 days (\textit{top}: planet b) and 15.2 days (\textit{bottom}: planet c), yielding a period ratio of 1.52.
        Two different orbital configurations are shown based on planet-to-star mass ratio ($M_\mathrm{p}/M_\star$) and eccentricity (ecc):
        $M_\mathrm{p}/M_\star = 10^{-6}$ and $\mathrm{ecc}=0$ (black circles);
        $M_\mathrm{p}/M_\star = 10^{-7}$ and $\mathrm{ecc}=0.04$ (open squares).
        The ``chopping'' effect appears as a zig-zag pattern superimposed on the TTV signal
        only for the configuration with $\mathrm{ecc}=0$ (black circles and solid line).
        Note that these configurations yield the same TTV amplitude when 
        the difference in the planetary arguments of pericentre 
        of the two planets within the system is $180^\circ$.
    }\label{ChoppingTTV}
\end{figure}

An example of the ``chopping'' signal can be seen in Fig.~\ref{ChoppingTTV},
which reproduces the simulation presented in \citet{2025haex.book....2A} 
of a two-planet system with a period ratio of 1.52 (inner planet b with $P=10.0$ days).
In this case, by changing the planet-to-star mass ratio ($M_\mathrm{p}/M_\star = 10^{-6}$ and $10^{-7}$)
and eccentricity ($\mathrm{ecc} = 0$ and $0.04$),
it can be observed the same amplitude and modulation of the TTV signals.
However, in one scenario, the ``chopping'' effect is clearly visible as zigzag pattern,
which helps to distinguish between the two configurations.
It is important to note that in these two configurations, 
the difference between the arguments of pericentre ($\omega$) of the two planets in the system
needs to be 180 degrees;
otherwise, one would observe completely different patterns and amplitudes.
If the required precision to observe this effect is attained,
it allows for the breaking of some degeneracy between the mass and other parameters of the perturber 
\citep{Nesvorny2016ApJ...823...72N},
especially the mass-eccentricity degeneracy \citep{Deck2015ApJ...802..116D}.

\subsection{Analytical vs direct integration methods}\label{Method_analytical_integration}

Characterising multi-planetary systems through TTV
analysis presents a challenging inverse problem. 
For each planet, six orbital parameters, along with mass and radius, 
must be determined, resulting in a high-dimensional parameter space. 
Two primary approaches have been developed to tackle this challenge: 
the use of analytical formulae and direct N-body numerical integration.

\paragraph*{Analytical approach.}
Analytical developments, typically based on perturbation theory and
often assuming near-circular, coplanar orbits, and specific orbital configurations (e.g. low-order MMR), 
offer computational efficiency and provide intuitive insights into the system's dynamics. 
However, they may lack accuracy for highly eccentric orbits or systems exactly in,
or far from, resonance. 
Various studies have focused on developing this analytical approach
to overcome different limitations, improving the accuracy and precision of
mass and orbital parameter determinations. 
For further details, see: 
\citet{Agol2005MNRAS.359..567A,Nesvorny2008ApJ...688..636N,Lithwick2012ApJ...761..122L,
Boue2012MNRAS.422L..57B,Nesvorny2014ApJ...790...58N,
Deck2015ApJ...802..116D,Agol2016ApJ...818..177A,Hadden2016ApJ...828...44H,
Linial2018ApJ...860...16L}.

\paragraph*{Numerical integration method.}
Conversely, direct numerical integration of an N-body system, while computationally intensive,
provides a more comprehensive solution applicable to a wider range of orbital configurations.
This approach can accurately model complex multi-body interactions
and inherently accounts for the chopping effect.
However, it may struggle with degeneracies in the parameter space and
requires significant computational resources. 
A very useful and fundamental guide on how to set up the direct N-body approach
to compute the transit times during orbital integration can be found in \citet{Fabrycky2010exop.book..217F}.
It is crucial to note that the gravitational interactions 
between bodies cause orbital elements to oscillate over time, 
necessitating the specification of a reference epoch when fitting system parameters.

The choice between analytical and direct integration methods
often depends on the specific system's characteristics and
the desired balance between computational speed and model accuracy.
To overcome the required computational cost of the N-body integrator,
\citet{Deck2014ApJ...787..132D} developed \texttt{TTVFast}\footnote{\url{https://github.com/kdeck/TTVFast}},
a code that combines an N-body integrator with a Keplerian approximation.

Combining transits, TTV signals, and RV data enables 
full characterisation of planetary systems,
breaking parameter degeneracies, and refining parameter estimates.
This approach also provides the planets' bulk densities, 
which are crucial for constraining composition and evolution models.
The general approach typically requires several iterative steps. 
Initially, one must determine the period and radius of planets and establish a common model of the transits. 
Subsequently, each \ttra{} is fitted while fixing the period and transit parameters;
then deriving the TTV signal.
Finally, a TTV inversion method is applied to obtain the system parameters. 
This iterative process allows for a comprehensive analysis of the planetary system, 
gradually refining the model and parameters with each step.
It is important to note that transit analysis provides information about the scaled radius $R_\tran/R_\star$,
while TTV analysis yields the scaled mass $M_\pert/M_\star$, and in some cases $M_\pert/M_\tran$ (see Section~\ref{Method_params}).
Consequently, one does not have direct access to the absolute radius and mass of the planets,
and the uncertainty is hence dominated by the stellar parameters.

\paragraph*{Photo-dynamical model.}
An alternative approach is the photo-dynamical method, 
which involves the direct modelling of the transit photometry
during the integration of the planetary orbits.
This makes it possible to directly determine the bulk (or mean) density of the planets,
as well as the planet-to-star radius and mass ratios.
However, the Newtonian $MR^{-3}$ degeneracy is still present;
this can be resolved by measuring the light-time travel effect, 
or by fitting photometry and radial velocity data simultaneously 
to determine absolute masses and radii, 
thereby eliminating dependence on stellar evolution models.
This yields absolute radii and masses 
with a precision that is potentially better than that of the host star
\citep{Almenara2015MNRAS.453.2644A}.
One of the first works to introduce this term, as `photometric-dynamical', 
was \citet{Carter2011Sci...331..562C} in their study of KOI-126, 
a hierarchical triple stellar system.
The drawback of this method is that it combines 
the computational time required for direct N-body integration with
that of the photometric modelling of the transits. 
An additional aspect that can further slow down the photo-dynamical analysis
is the simultaneous detrending process of the light curves and
the activity signal in the RV, commonly modelled with Gaussian Process (GP) kernels,
which significantly increase computational complexity.

A non-exhaustive list of publicly available software packages and codes used for 
analytical, numerical N-body, and photo-dynamical modelling
of transit timing variations (TTVs) is provided in Table~\ref{tab:ttv_tools}. 

\paragraph*{Global and local optimisation.}

Practically, the first step in solving this inverse problem is to define the model, 
for example by combining orbital integration, transit times or photometry, 
radial velocity, decorrelation, Gaussian processes, etc. 
Next, one must define an objective function to be optimised. 
This is typically the likelihood ($\mathcal{L}$) function, 
or preferably its logarithm (\logL). 
However, a more comprehensive approach incorporates prior information, 
leading to the log-probability (\logP) as the objective function. 
This \logP{} can be expressed as the sum of the log-likelihood and 
the logarithm of the prior probabilities (log-priors):
$\logP \approx \logL + \sum{\log(\mathrm{prior})}$
where the priors can be uniform, Gaussian, or any other appropriate probability distribution function. 
Combining different models, such as numerical integration, transit modelling, decorrelation, etc., 
will increase the number of parameters, 
introducing correlations among them and potentially reducing the efficiency of many algorithms.
The general approach is to use a quasi-global optimiser, 
that
can be utilised to either maximise the \logP{} or minimise the negative log-probability,
depending on the specific implementation. 
They are defined by some mathematics rules,
sometimes inspired by natural, biological, genetic, or social behaviour,
which allow for the search for
optimal solutions (if they exist) in a wide parameter space, 
by evolving a defined number of parameter combinations until certain criteria are met
(see Table~\ref{tab:opti_tools} for a non-exhaustive list of quasi-global algorithms).
However, these methods come with their own challenges. 
They are prone to early convergence, potentially settling on local optima 
rather than the global optimum \citep{yang2010nature}. 
The solutions found are not guaranteed to be the global optimum, 
especially in complex, multimodal parameter spaces. 
Additionally, the results can be sensitive to initial conditions, 
and finding the right balance between exploration of the parameter space and 
exploitation of promising solutions is an ongoing challenge \citep{yang2010nature}.
Each of these algorithms has various hyper-parameters that can be tweaked, 
and their outputs typically do not include statistical information necessary 
to quantify parameter uncertainties, such as confidence or credible intervals.
However, they are generally less sensitive to correlations among parameters 
compared to traditional optimisation methods and 
they can be easily parallelize to improve computational speed. 
Despite these limitations, quasi-global optimisation algorithms remain valuable tools 
in tackling complex inverse problems in exoplanet research, 
particularly when dealing with high-dimensional parameter spaces and multiple data types.

It is good practice to refine an initial orbital configuration, 
derived from literature, prior knowledge, or from a quasi-global optimiser, 
using a local maximiser or minimiser 
(see a list of possible algorithms in Table~\ref{tab:opti_tools}).
While the local optimisation methods are effective for refining solutions, 
it is crucial to note their limitations in estimating parameter uncertainties and correlations. 
Methods such as Levenberg-Marquardt can provide an estimate of the parameter covariance matrix, 
offering some insight into local correlations. 
Similarly, quasi-Newton methods like BFGS construct an approximation of the Hessian matrix, 
which can yield information about local parameter relationships.
However, these estimates are based on local approximations of the objective function, 
often assuming a quadratic form. 
In complex problems, particularly those with strong non-linearities or
multimodal probability landscapes common in exoplanet data analysis, 
such local approximations may not fully capture the true parameter correlations. 
Consequently, the uncertainties derived from these methods are often underestimated, 
especially in the presence of significant non-linearities. 
Therefore, it is important to interpret these uncertainty estimates with caution and
to be aware of their limitations when dealing with complex astrophysical models.

\paragraph*{Bayesian approach.}
Given the limitations of both global and local optimisation methods, 
researchers often turn to Bayesian frameworks to sample the posterior probability distribution of model parameters.
This approach allows for a more comprehensive exploration of the parameter space and
provides a robust quantification of uncertainties and correlations. 
Two primary classes of algorithms (see a list of packages and implementations in Table~\ref{tab:bayesian_tools})
are commonly employed in this context,
Markov Chain Monte Carlo (MCMC) methods and Nested Sampling,
the latter offer the additional advantage of computing the Bayesian evidence, 
which is crucial for model comparison.

In TTV studies, Bayesian methods such as MCMC and Nested Sampling 
have proven particularly valuable for analysing multi-planet systems and
scenarios characterised by significant parameter degeneracies. 
These approaches excel in exploring complex, high-dimensional parameter spaces, 
naturally accounting for parameter correlations. 
A parameter space is often considered high-dimensional when 
involving approximately 20–30 parameters, a threshold where traditional MCMC algorithms
may suffer from reduced efficiency and convergence issues due to the curse of dimensionality. 
For Nested Sampling, efficiency can remain satisfactory up to 100–200 parameters, 
but this comes at the cost of a substantial increase in computational demand 
\citep{Sharma2017ARAA..55..213S, Feroz2008MNRAS.384..449F, Feroz2009MNRAS.398.1601F}.

It is crucial to note that while these Bayesian methods provide
a comprehensive representation of the posterior distribution, 
they can become computationally intensive for complex models with a high number of parameters. 
Strong parameter correlations further slow down the computational process, 
making it less efficient to probe the parameter space 
\citep{Ford2006ApJ...642..505F,Hogg2018ApJS..236...11H}.
 
To address these challenges, alternative algorithms like Hamiltonian Monte Carlo (HMC)
and Dynamic Nested Sampling have been introduced. 
HMC, which uses gradient-based information to guide sampling, 
is particularly effective for high-dimensional problems with smooth and well-behaved posteriors. 
Specifically, a posterior is considered smooth when it is continuously differentiable, 
and well-behaved when it lacks sharp discontinuities, extreme curvatures, or strong multi-modalities.
These properties ensure that gradient-based methods can efficiently explore 
the parameter space without becoming trapped or producing unstable trajectories.
By exploiting the geometry of the parameter space, HMC can achieve faster convergence and
improved sampling efficiency compared to standard MCMC 
\citep{Neal2011hmcm.book..113N, Betancourt2017arXiv170102434B}. 
Dynamic Nested Sampling, an extension of the traditional Nested Sampling approach, 
adapts its strategy dynamically to focus computational resources on regions of high posterior probability, 
offering a significant performance boost for complex, multi-modal posteriors 
\citep{Higson2019DynamicNested, Handley2015MNRAS.453.4384H}.

These advanced techniques provide promising alternatives 
for overcoming computational challenges in Bayesian inference,
particularly for high-dimensional TTV analyses, 
where they enable a more efficient exploration of the parameter space and
accurate determination of posterior distributions.

Moreover, the efficiency and convergence of these algorithms can be sensitive to
the choice of priors and proposal distributions, requiring careful setup and monitoring.
In case of MCMC algorithms it is somehow required to identify, and discard, 
the so-called burn-in period of the chains\footnote{A chain, or walker, 
in an MCMC analysis is a parameter set that evolves at each step of the algorithm.},
or walkers, and this can be usually achieved testing the convergence with
the Gelman-Rubin $\hat{R}$ \citep{gelman1992}, 
the Geweke's statistic \citep{geweke1991},
the auto-correlation function \citep[see e.g.][]{GoodmanWeare2010CAMCS...5...65G},
and visual inspection (e.g. a trace plot of each parameter showing the evolution of each walker).

\paragraph*{Orbital elements and parameterisation.}

An N-body planetary system, in an astrocentric reference frame,
is described by two stellar parameters, the mass $M_\star$ and radius $R_\star$,
and eight parameters for each $p$-th planet, the mass $M_{p}$, radius $R_{p}$, 
and six osculating orbital elements, which are 
the orbital period $P_{p}$, eccentricity $e_{p}$, argument of pericentre $\omega_{p}$\footnote{
In this context, the argument of pericentre $\omega$ refers to the planet ($\omega_{p}$), 
not the star ($\omega_\star$) as in radial velocity studies;
the relation between the two is $\omega = \omega_{p}=\omega_{\star} + 180^{\circ}$ 
\citep{Eastman2013PASP..125...83E}.}, 
mean anomaly $\mathcal{M}_{p}$\footnote{The mean anomaly $\mathcal{M}_{p}$
is defined as a linear function of time $t$,
defining the $p$-th planet's position on its orbit relative to the argument of pericentre $\omega_{p}$.
It is given by $\mathcal{M}_{p}=n_{p}(t-\tau_{p})$ where $n_{p}=2\pi/P_{p}$ is the mean motion and
 $\tau_{p}$ the time of pericentre passage \citep{MuDe1999book}.},
orbital inclination $i_{p}$, and longitude of the ascending node $\Omega_{p}$.
I stress again that the orbital elements have to be defined at a reference time or epoch of choice of the user.
While planetary radii are not directly involved in fitting, 
they are necessary for transit and collision checks. 
A full list, definitions, and possible alternatives to these parameters
can be found, for example, in \citet{MuDe1999book}.
Directly fitting these parameters can introduce degeneracies and strong correlations,
slowing down the analysis and reducing the efficiency of the algorithms 
in exploring the parameter space.
As I have already mentioned, in cases of TTV analysis
there is not information about the absolute masses,
so it is recommended to fit for the scaled-masses ($M_{p}/M_\star$ or $M_\pert/M_\tran$);
however, the photo-dynamical approach allows for fitting densities $\rho_{p}$,
as well as stellar density $\rho_{\star}$,
as described by \citet{Sozzetti2007ApJ...664.1190S} and in \citet{Perryman2018exha.book.....P},
especially when modelling multiple transiting planets.
Additionally, as pointed by \citet{Eastman2013PASP..125...83E}, 
in a circular orbit ($e_{p}=0$) the argument of pericentre $\omega_{p}$
is undefined and it is conventionally set to $90^{\circ}$.
For this reason, the mean anomaly $\mathcal{M}_{p}$ loses meaning.
Hence it is convenient to use and fit the mean longitude $\lambda_{p} = \mathcal{M}_{p} + \varpi_{p}$,
where $\varpi_{p} = \omega_{p} + \Omega_{p}$ is the longitude of the pericentre.
Since the parameter pair $(e_{p}, \omega_{p})$ is strongly correlated and 
$\omega_{p}$ becomes ill-defined for low-eccentricity
\citep{Ford2006ApJ...642..505F},
it would be better to fit for the eccentricity vectors $(e_{p}\cos{\omega_{p}}$, $e_{p}\sin{\omega_{p}})$,
and sometimes a more suitable parameterisation, $(\sqrt{e_{p}}\cos{\omega_{p}}$, $\sqrt{e_{p}}\sin{\omega_{p}})$,
is often used.
These two combinations of $(e_{p}, \omega_{p})$  allow for uniform priors on $e$ and
mitigates issues as $e \rightarrow 0$ 
\citep{Anderson2011ApJ...726L..19A,Eastman2013PASP..125...83E}.

In transit analysis, it is common to fit 
the inclination $i_{p}$, or the impact parameter $b_{p}$,
treating it as consistent across all transits. 
The symmetry of the transit light curve with respect to the chord through the star's centre means 
$b \geq 0$, or equivalently $i_{p} \leq 90^{\circ}$, and
there is no way to determine which hemisphere\footnote{Assuming $i_{p}=90^\circ$, the stellar disk is halved in two hemisphere.}
of the stellar disk a planet cross.
For example, a planet with inclination $i_{p} = 89^{\circ}$ is indistinguishable from $i_{p} = 91^{\circ}$.

In a multiple-planet system, orbital inclination can vary from one transit to another 
due to gravitational perturbations,
varying the transit geometry and leading to transit duration variation (TDV).
It also has to be taken into account that 
the interaction between two planets lying the same stellar hemisphere
is slightly different from the two planets passing on the two hemispheres.
However, the TDV is quite difficult to measure 
from transit analysis due to correlation and degeneracies with
transit depth and the scaled semi-major axis $a_{p}/R_\star$,
so, it is common to fix (at the reference time)
the inclination of the planets from the transit analysis.

It is advisable to be careful when deciding to fit or fix the inclinations of planets,
because inclination (of the perturber) variation could, also, change the shape of the TTV signal.
So, it would be possible to fix the inclination of a transiting, reference, planet 
and let the inclinations of other(s) planet(s) vary.
If measurements of transit duration are available,
with a dynamical approach and following \citet{Fabrycky2010exop.book..217F}
it is quite easy to compute and fit the total duration, the time passed from first to fourth contact time,
or the full duration, the time from the second and third contact time, 
or both.
For the definition of such contact times of a transit see, for example, \citet{Winn2010exop.book...55W} and \citet{Fabrycky2010exop.book..217F}.
The TDV degeneracy and parameter correlations can be overcome with a photo-dynamical method that
naturally accounts for those effects.
It is also important to note that two transiting planets could have the same inclination,
but have a non-zero mutual inclination $\Delta I$, 
as it depends on the longitude of the ascending nodes of the planets:
\begin{equation}
    \label{eq:DeltaI}
    \Delta I = \arccos(\cos i_p \cos i_l + \sin i_p \sin i_l \cos(\Omega_p-\Omega_l))\ , \mathrm{with}\ p \neq l.
\end{equation}
For an exoplanetary system, the longitudes of the planetary ascending nodes, $\Omega_p$, are undefined;
therefore, it must be defined with a reference coordinate system and a line of nodes. 
Following \citet{Winn2010exop.book...55W}, the X-Y plane is made coincident with the sky plane, 
with the Z-axis pointing towards the observer and aligning the line of nodes with the X-axis;
assuming that the descending node lies on the +X-axis.
For a reference transiting planet $p$, 
it can be set $\Omega_p = 0^\circ$ or $180^\circ$ and
determine all others $\Omega_{l}$, with $l \neq p$,
or the difference $\Delta \Omega_{p,l}$, with $ l \neq p$.

The posterior distribution obtained from Bayesian analysis, 
such as through MCMC or Nested Sampling, 
provides critical insights into the parameter space. 
A corner plot, also referred to as a triangle diagram, 
offers a comprehensive visual representation of the posterior distribution. 
It simultaneously displays the marginalised distributions for each parameter 
along with their pairwise correlations, providing an intuitive grasp of the parameter space's structure.
The marginalised posterior for each parameter can be characterised using 
its median and the highest density interval (HDI), or high posterior density (HPD). 
The HDI, a Bayesian analogue of the frequentist confidence interval (CI), 
captures the region of parameter space with the highest probability density,
referred to as a credible interval,
thereby conveying the range of plausible values\footnote{The HDI at 
$68.25\%$, 
$95.45\%$, and 
$99.73\%$
is the equivalent of the CI at 
$[-1\sigma,+1\sigma]$, 
$[-2\sigma,+2\sigma]$, and 
$[-3\sigma,+3\sigma]$, 
respectively, for a normal distribution.}.
While the median and HDI summarise the posterior, they do not necessarily correspond to
a set of parameters that reproduce the observed data effectively. 
For this reason, it is often advisable to report the maximum-likelihood estimation (MLE) or 
the maximum-a-posteriori probability (MAP) set as the best-fit parameters. 
These represent the parameter values that maximise the likelihood or posterior probability, respectively.
Care must be taken when converting between fitted parameters and their physical counterparts. 
For instance, fitting parameters such as $\sqrt{e}\cos{\omega}$ and $\sqrt{e}\sin{\omega}$ 
must be transformed back to $e$ and $\omega$ before calculating median, HDI, MLE, and MAP values. 
Directly interpreting the median of the transformed parameters is inadvisable, 
as it may yield results inconsistent with physical reality. 
This issue is particularly acute for parameters involving non-linear transformations.
To illustrate the uncertainty in observables such as transit-timing variations (TTVs) 
one can plot $O-C$ percentiles derived from a sample of parameter sets randomly drawn from the posterior.
This approach shows the range of plausible model predictions while respecting the posterior distribution.

Despite these challenges, the ability of Bayesian methods to provide comprehensive uncertainty estimates and
to handle complex parameter relationships makes them indispensable tools in modern exoplanet research, 
particularly in the analysis of TTV data where parameter correlations and multi-modal solutions are common.

In terms of the number of transits required for a robust dynamical characterisation, 
a general fitting principle -- such as in least-squares optimisation -- 
is that the number of free parameters in the model should not exceed 
the number of independent data points. 
In other words, meaningful constraints on a system can only be obtained if
the number of observations exceeds the number of fitted parameters.
For a transiting planet whose dynamical characterisation involves typically six to eight parameters, 
it is therefore necessary to have a larger number of transit time measurements. 
The required number also depends on the precision of those measurements. 
As a guideline, according to \citet{Nesvorny2008ApJ...688..636N}, 
approximately 20 transits are typically needed to characterise a perturbing planet, 
although a higher number is advisable, especially when photometric or timing precision is limited.
The inclusion of RV data can significantly improve the fit by breaking model degeneracies. 
Even if the perturbing planet is too low-mass to be detected directly in the RV signal, 
these observations remain highly valuable. 
Nevertheless, RV measurements can still improve the determination of the transiting planet’s mass, 
which in turn helps constrain the perturber’s mass more accurately via the TTVs.
Ultimately, there is no universal rule for the minimum number of transit times required; 
it depends on the complexity of the system, the precision of the data, 
and the nature of the dynamical interactions under investigation. 
A case-by-case assessment is necessary, considering 
the quality and quantity of the available data,
the signal-to-noise ratio of the TTVs, 
and the complexity of the planetary system.

\section{Demographics of characterised planetary systems}\label{Demographic}

The number of confirmed exoplanets, and hence of planetary systems,
is continuously evolving. 
At the time of writing\footnote{April 2026} there are 6153 confirmed exoplanets\footnote{
Source: composite data at \url{https://exoplanetarchive.ipac.caltech.edu/}}
hosted in 4585 systems, 
of which 1050 are multi-planetary systems.
Among all the confirmed exoplanets, 
1476 have measured masses and radii,
with relative errors below 99\%.
There are 486 planets flagged with TTV signals, hosted in 311 systems, 
of which 147 systems contain 268 exoplanets with measured masses and radii.

\subsection{Multi-planetary systems}\label{Demographic_multiplanets}

\begin{figure}[!htb]
    \centering
    \includegraphics[width=\textwidth]{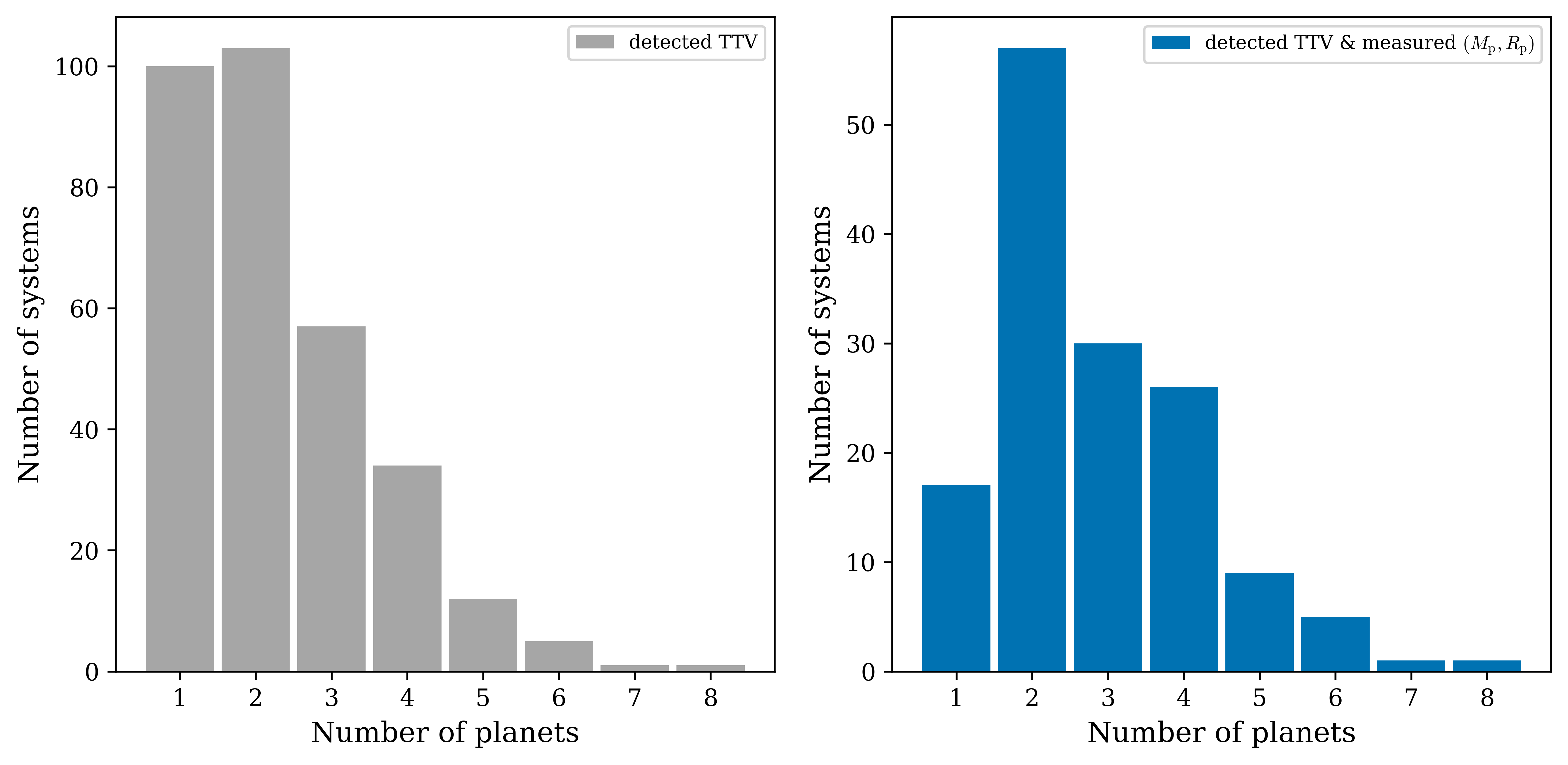}
    \caption{
    \textbf{Number of planets for stars with TTV flag.}
        Distribution of the number of planets per star for systems with (\textit{left panel}) detected TTV signals
        and with (\textit{right panel}) measured masses ($\frac{\sigma_{M_\mathrm{p}}}{M_\mathrm{p}}<99\%$)
        and radii ($\frac{\sigma_{R_\mathrm{p}}}{R_\mathrm{p}}<99\%$). 
        The histogram shows the absolute frequency of stars (systems) hosting different numbers of planets, 
        with the majority of systems containing two planets, followed by two, three, and four planets.
        The number of two-planet systems drops below the four-planet systems considering both mass and radius as well-mesaured.
        This distribution likely reflects a selection effect, where transit surveys like \kepler{} and TESS are 
        more sensitive to short-period planets, while additional, undetected planets may exist in these systems.
    }\label{multiplicity}
\end{figure}

It is notable that the current number of stars flagged with TTV signals
peaks at a multiplicity of two planets, followed by systems with one, three, and four planets, 
as shown in Fig.~\ref{multiplicity}, which illustrates the distribution of 
the number of planets per star with detected TTV signals (on left panel) and
with TTV and both measured mass and radius (on the right panel).
The latter shows a drops in the number of systems with one planet with both well-measured mass and radius.
This trend is likely influenced by selection effects inherent to the detection method. 
Specifically, transit surveys, such as \kepler{} and TESS, 
are more sensitive to short-period planets, while additional undetected planets may exist within these systems. 
Extended baseline surveys and upcoming missions 
may reveal additional planets in already known systems.
Currently, 19 systems hosting more than five planets showing TTVs have been discovered;
in particular, 12 stars host five planets, which are
HD 108236, 
Kepler-102,
Kepler-122, 
Kepler-139,
Kepler-238, 
Kepler-32, 
Kepler-33, 
Kepler-48, 
Kepler-55, 
Kepler-82, 
Kepler-84,
L~98-59;
among these nine have both mass and radius measured
(
Kepler-102, 
Kepler-122, 
Kepler-139,
Kepler-238, 
Kepler-33,
Kepler-48, 
Kepler-82,
Kepler-84
and L~98-59
)
Only five systems have six planets, that is
HIP 41378, 
Kepler-11,
Kepler-80, 
TOI-1136,
and TOI-178;
all of them
have measured masses and radii.
The unique systems with seven and eight planets (showing TTV) are 
TRAPPIST-1 and KOI-351 (or \kepler-90), respectively; 
both systems have measured masses and radii.
See the references of each system in Table~\ref{tab:systems}.

\begin{figure}[!htb]
    \centering
    \includegraphics[width=\textwidth]{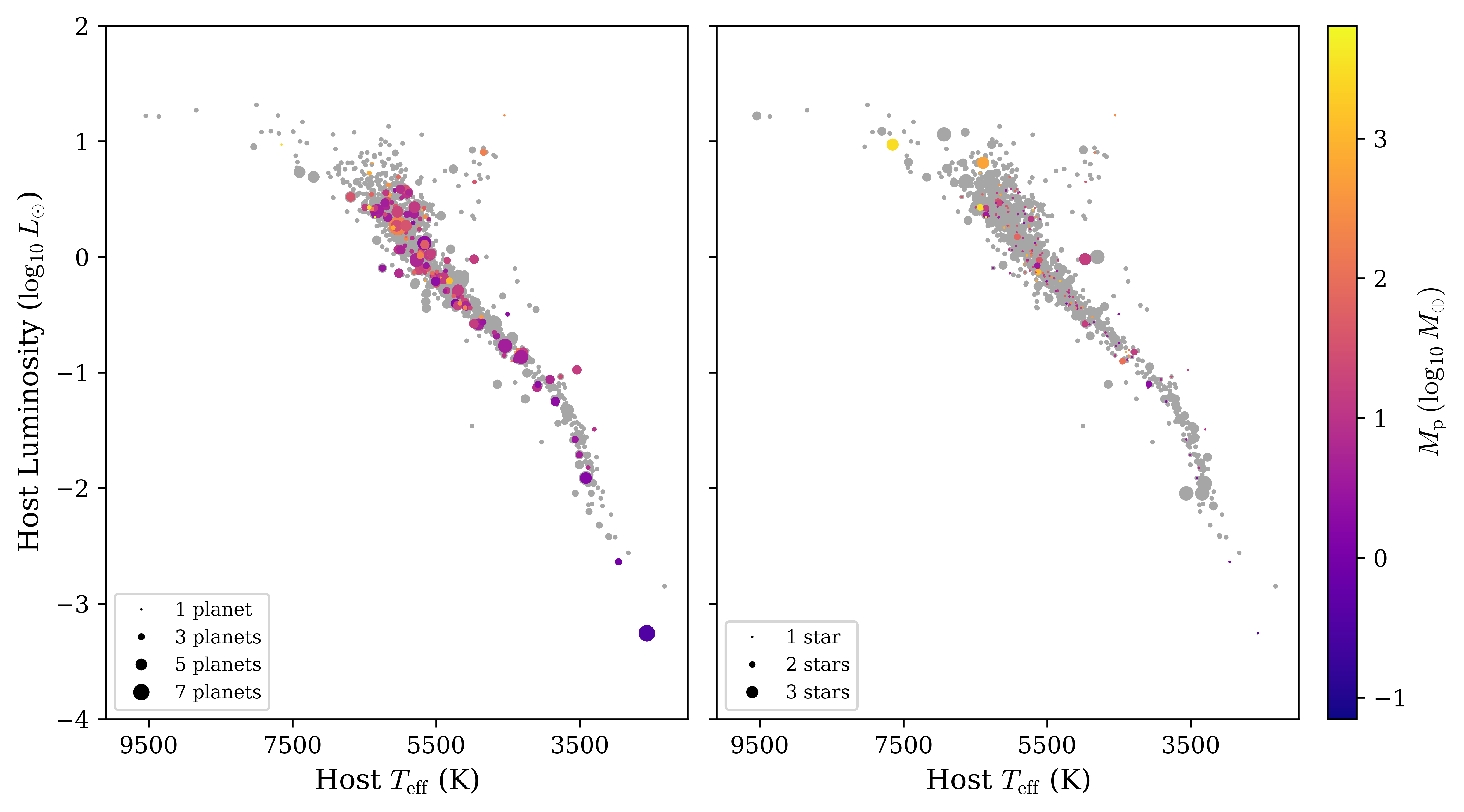}% no path, no extension
    \caption{Hertzsprung-Russel diagram of stars hosting planets with 
    \textbf{Hertzsprung--Russel diagram of host stars with planets showing or not TTV signals.}
    measured mass ($\frac{\sigma_{M_\mathrm{p}}}{M_\mathrm{p}}<99\%$) and
    radius ($\frac{\sigma_{R_\mathrm{p}}}{R_\mathrm{p}}<99\%$).
    Point size indicates the number of the planets (\textit{left panel}) and
    the number of stars (\textit{right panel}) 
    in the system.
    Systems with no TTV signal are plotted in gray, 
    while TTV systems are colour-coded with mass measurements in $\log_{10}$ scale.
    }\label{HR_diagram}
\end{figure}

One can plot a Hertzsprung--Russell (H-R) diagram for systems with measured masses and radii, 
distinguishing between those with and without TTV signals (see Fig.~\ref{HR_diagram}).
Both systems with and without TTVs display a broad distribution along the Main Sequence,
showing TTV systems detected down to M dwarf stars.
This distribution likely reflects a combination of observational biases and intrinsic factors. 
In particular, transiting planets (on which TTV measurements rely, even if the perturber does not transit) 
are more easily detected around stars of intermediate brightness and low variability.
These stars are also well-suited for high-precision photometric monitoring, 
such as that provided by \kepler{}, TESS, and CHEOPS, 
making them ideal targets for identifying small timing deviations in transit curves.

One can observe that TTV systems that host planets across a broad range of masses, 
with some of the more massive planets (orange/yellow hues in Fig.~\ref{HR_diagram})
also present among the TTV sample.
This is notable, as TTVs are often associated with low-mass, near-resonant planets.
The presence of higher-mass planets suggests that TTV signals may arise from
a variety of configurations, including massive planets interacting with other,
potentially unseen and/or lower-mass perturber bodies.

Furthermore, the system with seven planets, TRAPPIST-1, 
is located at low $T_\mathrm{eff}$ and low luminosity, 
showing that highly populated planetary systems (ideal for TTV detection) 
can be found around smaller, cooler stars.
This trend may be due to geometric factors (transits have deeper signals around small stars)
or reflect differences in planet formation and efficiency across stellar types 
to not disrupt the planetary system.
The fact that both TTV and non-TTV systems are largely confined to the Main Sequence
indicates that the majority of host stars are relatively young to middle-aged, 
typically ranging from hundreds of millions to a few billion years old.
However, recent analyses have expanded this census to more evolved targets. 
Specifically, three Red Giant Branch (RGB) stars, Kepler-56, Kepler-91, and Kepler-278,
have been identified as hosting planets with TTVs, 
alongside 14 slightly evolved (sub-giant) hosts including 
HAT-P-13, 
HAT-P-7, 
Kepler-23, 
Kepler-33, 
Kepler-87, 
Kepler-92, 
Kepler-101, 
Kepler-128, 
Kepler-145, 
Kepler-223, 
Kepler-277, 
Kepler-338, 
and TOI-2180.
The low number of evolved stars hosting planets (with or without TTV)
may result from a combination of 
stellar evolution effects 
(e.g., planetary engulfment, orbital disruption) and 
observational challenges 
(e.g., pulsations and noise in giant stars obscuring transit timing variations).
But, asteroseismology can be used to better characterise host star,
hence planets, such as for Kepler-36 and Kepler-56.
Regarding the temporal evolution of these systems, 
\citet{Dai2024AJ....168..239D} found that 
the fraction of systems with at least one pair close to commensurability (either first- or second-order) 
is $86\% \pm 13\%$ among young systems ($<100$~Myr), 
falling to $38\% \pm 12\%$ for adolescent systems ($0.1$--$1$~Gyr), 
and $23\% \pm 3\%$ for mature systems ($>1$~Gyr). 
Consistent with this, \citet{Murillo2026AJ....171...63L} found that 
$28.3\% \pm 6.2\%$ of the young planets in their sample (age $<800$~Myr) show evidence of TTVs. 
This observed decline in MMR occurrence rate as systems age 
serves as a diagnostic signature of the post-formation dynamical processes 
that reshape planetary architectures. 
While most compact multi-planet systems are thought to emerge from the gas-disk phase in resonant chains, 
these configurations are fundamentally delicate. 
Over secular timescales, mechanisms such as disk-dispersal, planetesimal scattering, 
and long-term tidal effects act to destabilise and eventually break these resonant locks. 
Consequently, 
the high incidence of resonance in young systems likely represents the primordial initial conditions of planetary systems, 
which subsequently evolve into the more disordered, 
non-resonant states that characterise the mature population as these disruptive processes take hold.
Understanding these early states is key to uncovering the evolutionary processes 
that lead to the lower occurrence of Mean-Motion Resonance (MMR) and 
TTV signals observed in mature systems, 
although the current sample size remains too small to draw definitive conclusions.

Systems with multiple stars showing TTV signals appear to lie 
to the upper end of the main sequence.
This trend is not very clear and it is probably due to the low number of systems
in the sample and the result of observational biases.

\subsection{Mean-motion resonances and chains}\label{Demographic_mmr}

As already mentioned in Section~\ref{Method}, when pairs of planets 
show period commensurability, that is
when they are in or near a MMR state,
the amplitude of the TTV is enhanced, allowing for the detection of Earth-like planets.
Furthermore, detecting resonant, or quasi-resonant, configurations gives further information
about the architecture of the system, and also on the possible
formation and evolution processes 
\citep[for additional details and further readings see][]{Beauge2003ApJ...593.1124B,Laune2022MNRAS.517.4472L}.

\begin{figure}[!htb]
    \centering
    \includegraphics[width=\textwidth]{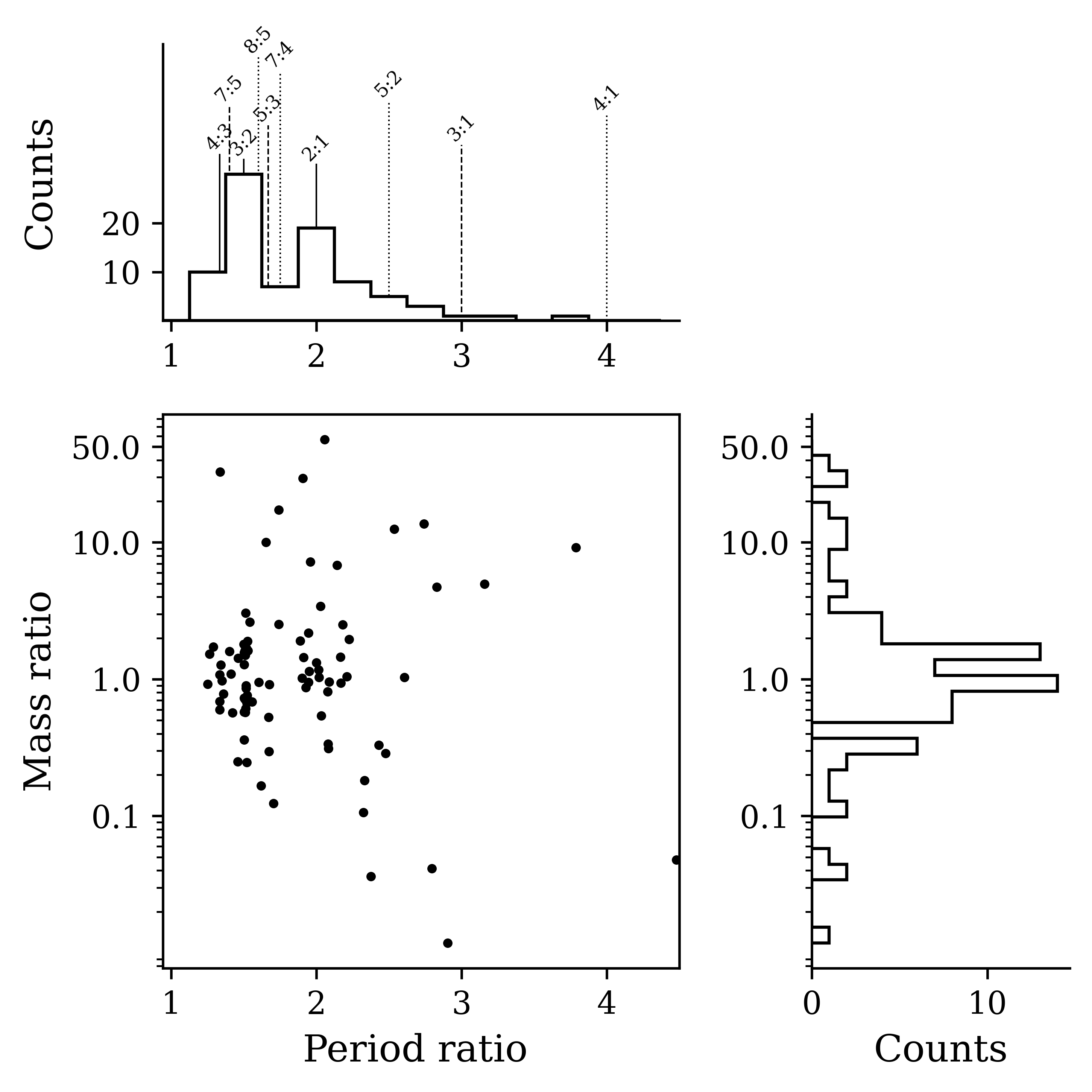}% no path, no extension
    \caption{
    \textbf{Mass and Period ratios of planets flagged with TTV.}
        The $P$-ratio (outer/inner period) versus $M$-ratio (outer/inner mass) 
        for planetary pairs in systems with detected TTV signals. 
        The scatter plot (\textit{lower-left panel}) illustrates the relationship between these ratios, 
        with a noticeable clustering around resonant $P$-ratios, particularly near the 3:2 and 2:1 first order commensurabilities (\textit{top panel}). 
        The histograms in the \textit{top-left} and \textit{bottom-right} show the distributions of the $P$-ratio and $M$-ratio, respectively.
        These distributions highlight the prevalence of near-resonant configurations for $P$-ratios, 
        while the $M$-ratios span a wider range, indicating the variety of planetary masses observed in these systems. 
        This suggests the importance of resonance trapping mechanisms and the potential influence of migration processes on planetary system architecture.
    }\label{Mratio_vs_Pratio}
\end{figure}

The analysis presented in Fig.~\ref{Mratio_vs_Pratio} offers key insights into the relationships 
between period ratios ($P$-ratio) and mass ratios ($M$-ratio), of outer to inner planets,
in planetary systems exhibiting TTVs.
The majority of planetary pairs are concentrated near $P$-ratio values around 1.5–2,
corresponding to common first order mean-motion commensurabilities (e.g., 3:2 or 2:1).
The distribution of $M$-ratio spans several orders of magnitude, 
from values near 0.1 up to beyond 50, highlighting the wide range of mass ratios within these systems. 
Notably, most planetary pairs have $M$-ratio values below 10, 
suggesting a preference for systems with planets of comparable mass. 
This concentration could reflect the increased detectability of TTV signals in systems with similar mass.
The $P$-ratio distribution in Fig~\ref{Mratio_vs_Pratio} shows 
clear peaks at period ratios associated with first order commensurability, that is 3:2, 2:1. 
This reinforce the basis that TTV signals are often associated with planets near mean-motion resonance configurations. 
These resonant interactions likely amplify the gravitational perturbations, 
making these systems prime candidates for TTV detections.

The histogram for $M$-ratio (Fig~\ref{Mratio_vs_Pratio}) reveals a peak at low values, close to one, 
with a slight skew towards values below one.
This suggests that many systems consist of an inner more massive planet paired with an outer less massive companion.
The peak at $M$-ratio\ $=1$ indicates that most planetary pairs exhibiting TTVs have similar masses.
In particular, there are 28 pairs of planets with $M$-ratios between 0.5 and 1.5.
Among these, one can identify 
\kepler-11~e, d;
TRAPPIST-1~c, b;
and finally TRAPPIST-1~g, f, 
which are pairs of planets hosted in a system with high multiplicity. 
This could suggest a formation scenario in which 
planets with similar masses are more likely to occur in multi-planet systems.
However, the long tail extending to $M$-ratio values exceeding 50 suggests a significant fraction of systems
where one planet is substantially more massive, such as a gas giant paired with a smaller terrestrial planet.
So far, short-period gas giants, also known as hot Jupiters ($P \lesssim 10$ days, $M \gtrsim 0.2\, M_\mathrm{Jup}$),
were widely believed to be solitary within their systems
\citep{Latham2011ApJ...732L..24L, Steffen2012PNAS..109.7982S, Huang2016ApJ...825...98H},
and formed by high-eccentricity migration 
\citep[HEM; ][]{Weidenschilling1996Natur.384..619W, Eggleton2001ApJ...562.1012E, Wu2003ApJ...589..605W}.
For example, a system hosting only one gas giant might be the result of a multi-planet system
disrupted by planet-planet scattering, 
but determining the precise dynamical process the system experienced is challenging.
However, recent studies have revealed a growing number of cases where
hot Jupiters are accompanied by smaller, predominantly inner companions.
One of the first identified systems of this kind was WASP-47,
an example of a hot Jupiter with close companions.
To date, seven of such systems have been confirmed to host gas giants alongside at least one smaller, inner companion, which are
WASP-47,
Kepler-730,
TOI-1130, 
WASP-132,
TOI-2000, 
WASP-84,
and 
TOI-5398 (see references in Table~\ref{tab:systems}).
These systems could reflect the long tail at high $M$-ratio distribution.
The discovery and characterisation of such planetary systems impose
significant constraints on models of planetary formation and migration,
providing critical insights into the complex dynamical and evolutionary processes shaping these systems.
The clustering of planetary pairs near resonances in the $P$-ratio distribution 
highlights the role of resonance in shaping system dynamics. 
Such configurations are consistent with theoretical predictions of 
migration-driven resonance trapping during planetary system formation
\citep[e.g. see ][]{Beauge2006MNRAS.365.1160B,Chambers2009AREPS..37..321C,Lykawka2013ApJ...773...65L}.
The observed distribution of both $P$-ratio and $M$-ratio is likely influenced by detection biases. 
As I have already said, transit surveys such as \kepler/K2 and TESS are 
inherently more sensitive to shorter-period planets and those with larger mass ratios, 
as these produce stronger TTV signals.
Again, with extended observational baselines and next-generation missions, 
additional planets will be uncovered in these systems, 
in particular in those configurations where undetected planets 
on longer orbital period may be driving TTV signals.
These findings underline the complex interplay of planetary masses and orbital resonances in shaping TTV signals, 
offering a rich opportunity for further exploration of planetary system architectures. 
This reinforce the idea that TTV analysis is a powerful tool for planet detection, characterisation,
and system architecture studies.

\subsection{Orbital Stability}\label{Demographic_stability}

Planetary systems currently observed in resonance must have either 
remained largely unperturbed throughout their history or 
experienced significant dissipative processes that restored or 
maintained the resonant configuration. 
Resonances, however, can also act as a stabilising mechanism, 
particularly for massive planets in close orbits. 
Notably, the exoplanetary population includes systems whose existence
is only possible due to the protective effects of strong resonant interactions
\citep{Ketchum_2013}.
In fact, the analysis of the apsidal difference $\Delta \varpi$ helps determine whether the system is 
in a librating state (oscillating around a stable point) or 
a circulating state (rotating through all possible angles).
Systems in resonance often show libration of $\Delta \varpi$,
which contributes to stabilising the configuration by reducing long-term chaotic interactions.
In addition to what I already said in Section~\ref{Method_params},
if the resonant critical angles librates around a fixed value and
also the apsidal difference librates around 0 or 180 degrees, 
the system is an aligned/anti-aligned symmetric stationary configuration,
indicating a region that exhibits enhanced protection against orbital instabilities
\citep{Beauge2003ApJ...593.1124B, Correia2018, Laune2022MNRAS.517.4472L}.
This is particularly relevant in systems with massive planets, or with planets in close orbits,
where gravitational interactions are strong.
Additionally, a one-degree-of-freedom model used to evaluate whether a system is 
in resonance can provide valuable insights into its dynamical and stability properties.

However, it is essential to verify the system's stability, 
chaotic nature or instability, using complementary methods.
A common approach to assess the stability of multi-planet systems is 
to perform long-term numerical integrations of planetary orbits,
often spanning millions of years, 
typically using a symplectic integrator like the Wisdom-Holman algorithm
\citep{WisdomHolman1991AJ....102.1528W}.
While this method is less prone to false positives,
it is computationally demanding and sensitive to the chosen time step.
Alternative methods, requiring shorter integration times,
can provide insights into the system's chaotic behaviour.
For instance, frequency analysis techniques 
\citep{Laskar1990Icar...88..266L, Laskar1993PhyD...67..257L}
examine the dominant frequencies of the system to infer its dynamical state.
Another example is the Mean Exponential Growth factor of Nearby Orbits \citep[MEGNO; ][]{Cincotta2000AAS..147..205C}, 
which estimates the Lyapunov characteristic number over integration timescales of approximately $\sim 10^3$ orbital periods.
This approach can efficiently identify chaotic or stable systems and
is computationally more efficient than direct long-term integrations
\citep{Reichl1993sptn.book...47R}.
See in Table~\ref{tab:stability} a list of possible methods and indicators 
to determine if a planetary system is stable, chaotic, or unstable.

\subsection{Sensitivity biases of planetary mass determination}\label{Demographic_sensitivitybias}

\begin{figure}[!htb]
    \centering
    \includegraphics[width=\textwidth]{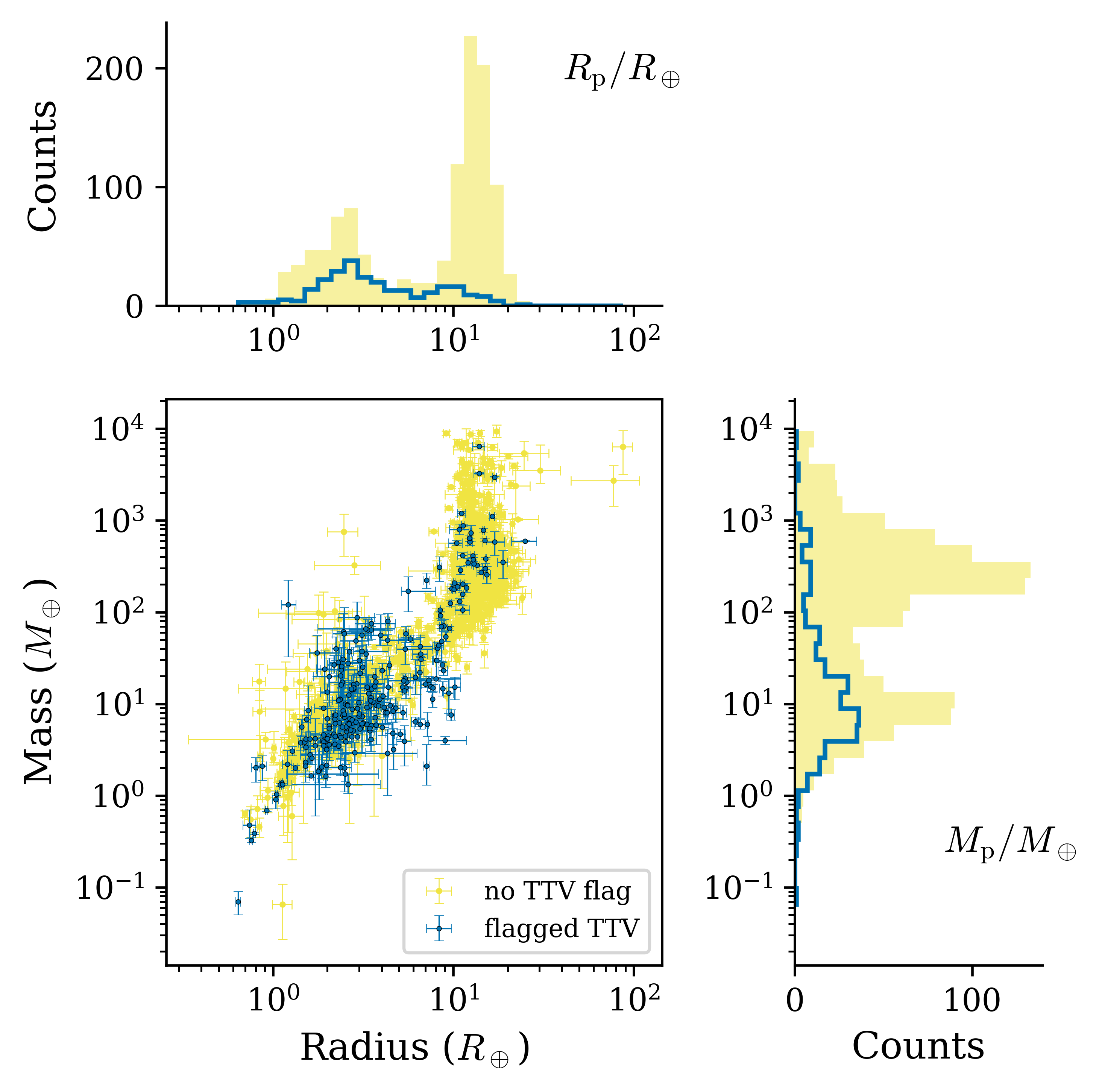}% no path, no extension
    \caption{
    \textbf{Mass-Radius diagram.}
    Mass-Radius diagram of exoplanets with and without TTV signals.
    All the planets shown have a measured mass and radius with 
    relative error ($\frac{\sigma_{M_\mathrm{p}}}{{M_\mathrm{p}}}$ and 
    $\frac{\sigma_{R_\mathrm{p}}}{{R_\mathrm{p}}}$) below 99\%.
    The \textit{lower-left panel} shows the scatter plot of planetary mass versus radius, both in $\log_{10}$-scale.
    Planets flagged with TTV signals are represented by 
    blue and black circles, 
    while those without TTV flags are shown in 
    light yellow.
    The \textit{top panel} displays the radius distribution, 
    and the \textit{right panel} shows the mass distribution. 
    In these histograms, 
    the blue outline represents TTV-flagged planets, 
    while the yellow-filled area corresponds to planets without TTV flags.
    There is evidence of the sensitivity of the TTV method towards
    lower density, 
    visible at $M_\mathrm{p}=2-10\, M_\oplus$,
    where TTV span a wider range in radius compared with planets characterised with other methods.
    The over-density in the radius distribution at around $10 - 20\,R_\oplus$
    can be attributed to the hot Jupiter planets.
    }\label{MR_diagram}
\end{figure}

The \kepler{} mission observed approximately 150 thousand Sun-like stars,
but their brightness was insufficient for mass measurement 
through the RV method with available facilities.
The TTV technique emerged as the primary alternative 
for obtaining mass estimates of exoplanets. 
Surprisingly, this method yielded mass values that began to populate 
a less dense region in the mass-radius diagram compared to RV analysis (see Fig.~\ref{MR_diagram}).
These planets have been referred to as `puff' or `super-puff' planets \citep{Liang2021AJ....161..202L}, 
characterised by their relatively low masses but unexpectedly large radii. 
Initially, there seemed to be some discrepancy between mass measurements derived from TTV (\mttv) 
and RV (\mrv) methods. 
However, \citet{Steffen2016MNRAS.457.4384S}, using Monte Carlo simulations, 
demonstrated that this apparent discrepancy was not a methodological issue but rather a result of sensitivity bias.
This work demonstrated that \mrv{} measurements tend to be higher for a given planetary radius, 
whereas \mttv{} values are more evenly distributed. 
As a result, the two methods are complementary, probing distinct regions of the mass-radius ($M$-$R$) plane.

\begin{figure}[!htb]
    \centering
    \includegraphics[width=\textwidth]{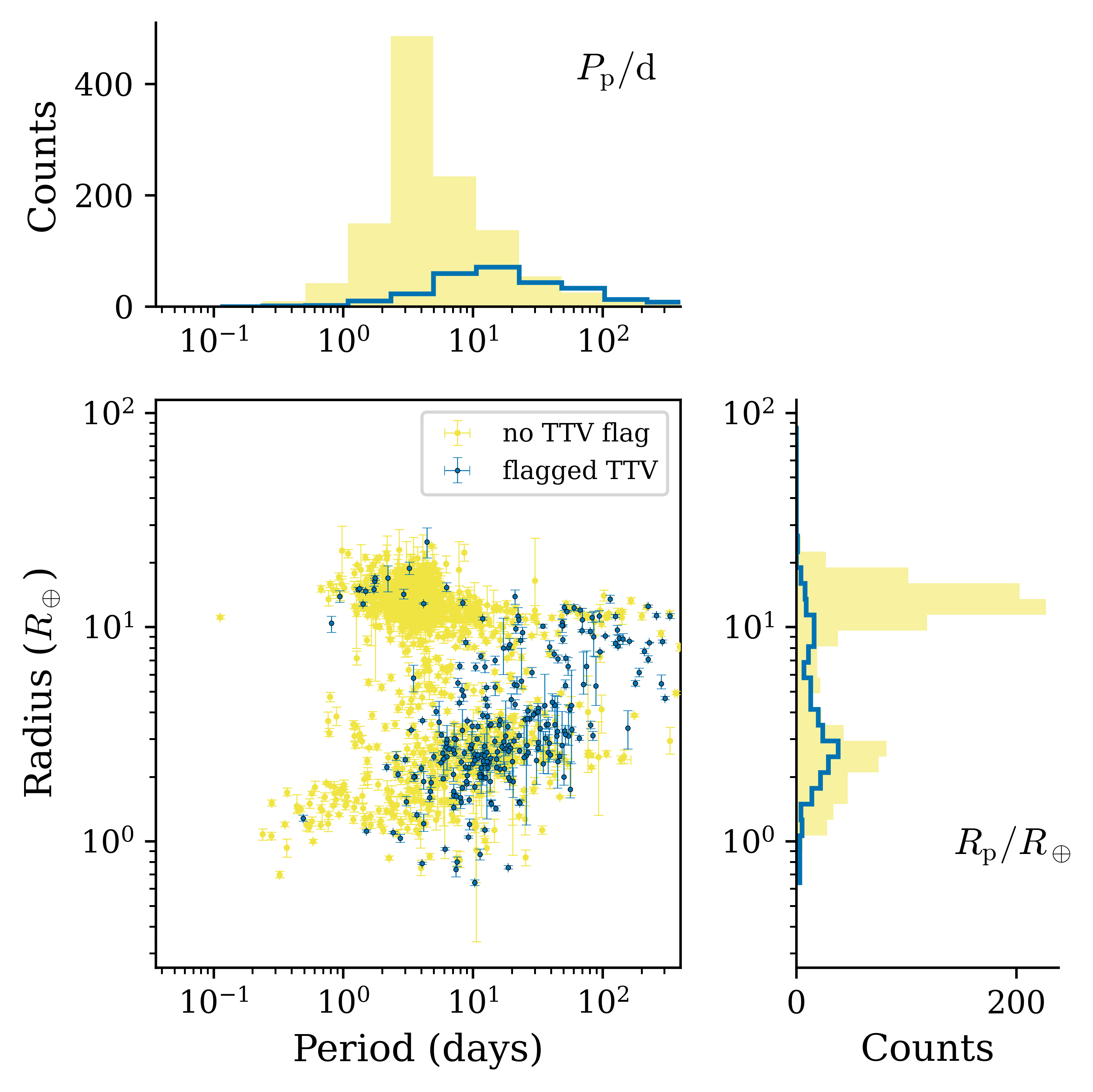}% no path, no extension
    \caption{
    \textbf{Radius-Period diagram.}
    Radius-Period diagram of exoplanets with and without TTV signals.
    The plot has same markers and colour code as in Fig.~\ref{ChoppingTTV}.
    The \textit{lower-left panel} shows the scatter plot of planetary radius versus orbital period, both in $\log_{10}$-scale.
    The triangular-shaped and almost empty region between around 1 and $10\,R_\oplus$ and at period shorter than 2-3 days is 
    the so-called Neptunian desert.
    The \textit{top panel} displays the period distribution, 
    and the \textit{right panel} shows the radius distribution.
    }\label{RP_diagram}
\end{figure}

This sensitivity effect remains quite evident in current $M$-$R$ diagrams, as shown in Fig.~\ref{MR_diagram}.
Both distributions in mass and radius of planets not showing TTVs are double peaked,
with the most predominant peak at gas giants in the range 
$M_\mathrm{p}=100 - 1000\, M_\oplus$ and $R_\mathrm{p} = 10 - 20\, R_\oplus$,
that is associated with the hot Jupiter over-density,
easily discernible in the period-radius diagram (in Fig.~\ref{RP_diagram}),
particularly in the upper-left region ($R_\mathrm{p} = 10 - 20\, R_\oplus$, $P \lesssim 10$ days),
due to the higher sensitivity of the transit methods at short period planets with large radii.
The second peak in the mass distribution of Fig.~\ref{MR_diagram} is
located at the super-Earth and sub-Neptune regimes 
$M_\mathrm{p}=5 - 15\, M_\oplus$ and $R_\mathrm{p} = 2 - 4\, R_\oplus$.
The planets with TTV show a different behaviour,
with a double-peaked distribution in radius, but with the highest peak
towards the super-Earth and Neptune range and
a less prominent peak around $8-9\, R_\oplus$.
In mass, the distribution of TTV-flagged planets shows only a peak around
Earth- and Neptune-like planets at $5-15\, M_\oplus$,
declining at greater masses.

This suggests that the TTV method is particularly effective in 
characterising Neptune-like planets and super-Earths. 
Additionally, TTV-flagged planets appear to fill gaps in the $M$-$R$ diagram,
particularly in the super-Earth and mini-Neptune regions, and at lower densities,
indicating the method's crucial role in obtaining a more complete characterised planetary sample.
It is worth noting that there are numerous cases where RV and TTV measurements agree, such as 
\kepler-9 \citep{Borsato2019MNRAS.484.3233B}, \kepler-18 \citep{Cochran2011ApJS..197....7C}, 
and more recently, WASP-47 \citep{Dai2015ApJ...813L...9D,Nascimbeni2023AA...673A..42N} and
TOI-1130 \citep{Korth2023AA...675A.115K, Borsato2024AA...689A..52B}. 
These examples demonstrate that while sensitivity biases exist, 
the \mttv{} values are increasingly aligning with other methods of planetary mass measurement, 
contributing to a more comprehensive understanding of exoplanetary systems.

The radius-period ($R$-$P$) diagram in Fig.~\ref{RP_diagram} provides additional key insights 
into the distribution and properties of planets with and without TTV signals.
As I already mentioned, in this diagram the hot Jupiter population is clearly visible in the 
region between $10 - 15\, R_\oplus$ and $P \lesssim 10$ days.
Another interesting feature visible in this diagram,
is the sparsely populated region known as the Neptunian desert \citep{Szabo2011ApJ...727L..44S},
located at short orbital periods ($P < 3$ days) and intermediate radii ($R \sim 2-10\, R_\oplus$). 
This desert reflects the scarcity of Neptune-like planets at close distances to their stars. 
This phenomenon is thought to result from a combination of photo-evaporation stripping the atmospheres of sub-Neptunes 
and the inability of intermediate-mass planets to migrate and survive in such tight orbits
\citep{Owen2018MNRAS.479.5012O}.
The population of TTV planets appears to be more uniformly distributed in the $P-R$ diagram, 
avoiding the Neptunian desert.
This is a clear mark of different formation and evolution process of the hot-single Jupiter systems,
possibly through HEM process,
from the disk-drive migration responsible to form TTV multi-planet systems.

The $M-R$ and $R-P$ diagrams and parameter distributions of TTV and non-TTV planetary population
illustrate the complementary strengths of different planet detection techniques. 
While the transit method preferentially identifies hot Jupiters and highly irradiated short-period planets, 
the RV method is more sensitive to massive planets (also at longer periods).
TTV provides extended access to multi-planet systems and regions of parameter space dominated by low-density and intermediate-mass planets;
importantly, the planet inducing the TTV does not necessarily transit.
Together, these methods paint a more comprehensive picture of planetary system architectures and formation processes.

\subsection{Circumbinary planets}\label{Demographic_circumbinary}

Binary and multiple stars are quite common in the Galaxy,
and they host planets, just as for single stars (see Fig.~\ref{HR_diagram}).
Focusing on binary stars, one can identify two main categories of exoplanetary orbital configurations,
the S-type and P-type 
\citep{Schwarz2011MNRAS.414.2763S, MarzariThebault2019Galax...7...84M, Columba2023AA...675A.156C}.

\begin{figure}[!htb]
    \centering
    \includegraphics[width=\textwidth]{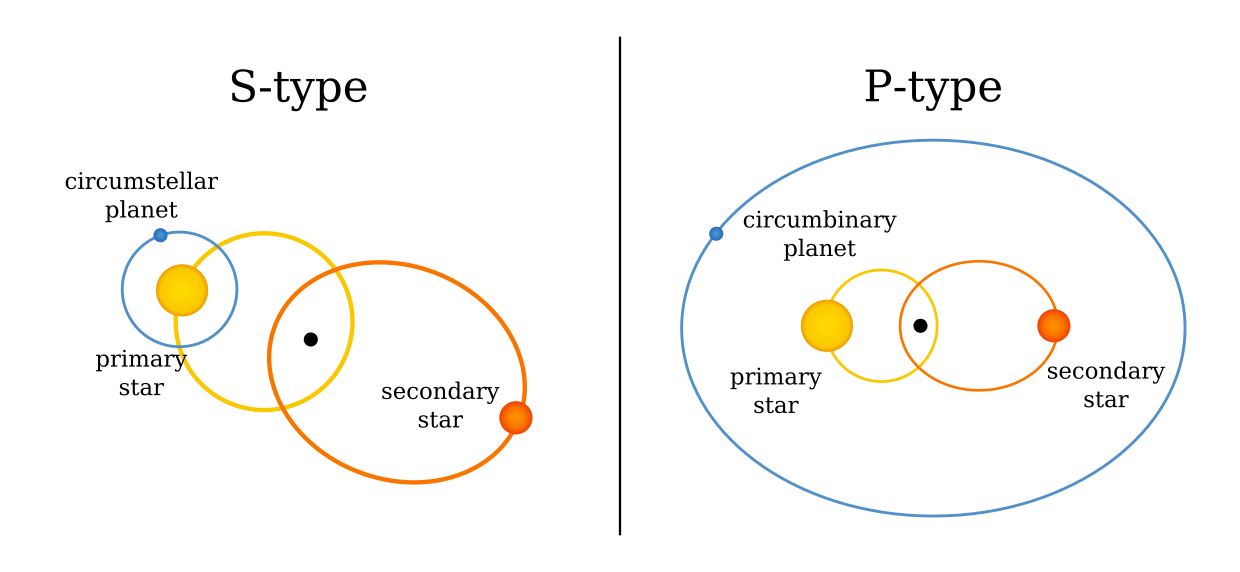}% no path, no extension
    \caption{
    \textbf{Exoplanets in binary systems.}
    Infographic of an exoplanet in a stellar binary system.
    The \textit{lower-left panel} shows the S-type configuration, 
    with the circumstellar planet (in blue) orbiting, for example, the primary star (yellow) of the system.
    The \textit{right panel} shows the P-type configuration, with the circumbinary planet (in blue) orbiting the
    entire stellar system (composed of a primary and a secondary star, in yellow and red, respectively).
    The black dot indicates the centre of mass of the binary system.
    }\label{binary}
\end{figure}

The first, S-type configuration, identifies the so-called circumstellar planet,
which orbits only one of the two binary components (see left panel of Fig.~\ref{binary}).
They have been mainly detected with transit and radial velocity observations, 
such as in the case of XO-2 \citep{Damasso2015AA...575A.111D},
a wide-binary system hosting two circumstellar planets, one for each stellar component, 
where one was detected with transit and the other through radial velocities.

In the case of P-type class, the planet is defined as circumbinary,
meaning it orbits around the stellar binary system as a whole (see right panel of Fig.~\ref{binary}). 
Most massive and long-period circumbinary planets have been discovered through direct imaging,
or around eclipsing binaries as either transiting bodies or 
by inducing eclipse timing variation (ETV, also known as eclipse transit timing variation, ETTV).
While the usual methods have detected a majority of circumbinary planets around Main Sequence binaries, 
ETTV detected several evolved circumbinary systems hosting multiple planets. 
In these systems, one star has evolved off the main sequence to either B-type subdwarf or white dwarf stages 
\citep[e.g.,][]{Baran2015AA...577A.146B,Potter2011MNRAS.416.2202P},
and in some systems, the presence of planets is still under discussion, 
thus requiring a longer observation baseline \citep{Pulley2022MNRAS.514.5725P}.

As with TTVs, creating an O-C of the eclipse timing makes it possible
to identify timing variations (ETV).
There are different possible sources for this ETV that do not involve a circumbinary planet,
such as apsidal motion, mass transfer or ejection within the system, and the Applegate effect
\citep{Sterken2005ASPC..335....3S,Applegate1992ApJ...385..621A,Brown-Sevilla10.1093/mnras/stab1843}.

In case of additional bodies orbiting around the binary,
the motion of the system's barycentre is perturbed,
causing apparent variation in the binary's period;
this is known as the eclipse timing variation (ETV).
One cause of ETV is due to the light-travel time effect 
\citep[LTTE, also known as Roemer effect;][]{Irwin1952ApJ...116..211I,Irwin1959AJ.....64..149I} 
and results from the earlier or delayed time of arrival of light
due to the finite value of the speed of light 
\citep[the LTTE also needs to be taken into account in the standard TTV numerical analysis, for further details, see][]{Fabrycky2010exop.book..217F}.

Let me define an eclipse binary system composed by primary (A) and secondary (B) star,
and assume a third circumbinary body (C, as in Fig.~\ref{binary}).
Measuring the eclipse timing of the binary ($T_{0,\rmAB}$) and
subtracting a linear ephemeris ($\tref + E \times P_\rmAB$),
as for the TTV case,
allows for the identification of the unknown body through the $O-C$ diagram:
\begin{equation}\label{eq:oc_binary}
    (O-C)_{E} = T_{0,\rmAB,E} - (\tref + E \times P_\rmAB) = \ltte\ ,
\end{equation}
with $P_\rmAB$ the binary period and $E$ the eclipse number, where $E=0$ for $T_{0,\rmAB,E} = \tref$.
The non-flat $O-C$ is the fingerprint of the LTTE caused by the gravitational perturbation
of the circumbinary body.

Before the advent of personal computers, the equation mainly used for the LTTE
was developed by \cite{Irwin1952ApJ...116..211I, Irwin1959AJ.....64..149I}.
An updated version has been suggested by \citet{Borkovits2015MNRAS.448..946B, Borkovits2016MNRAS.455.4136B}.
The two equations used an offset for the origin of the reference system,
resulting in a constant difference given by
$a_\rmAB e_\rmC \sin \omega_\rmC \sin i_\rmC$,
where $a_\mathrm{AB}$ is the semi-major axis of the binary,
$e_\rmC$, $\omega_\rmC$, and $i_\rmC$
the eccentricity, the argument of periastron, and the inclination of body C,
respectively.
The proposed equation for the \ltte{} given by
\citet{Borkovits2015MNRAS.448..946B,Borkovits2016MNRAS.455.4136B} is:
\begin{equation}\label{eq:ltte}
    \ltte = - \frac{a_\rmAB \sin i_\rmC}{c} \frac{(1-e^{2}_\rmC) \sin (\nu_\rmC + \omega_\rmC) }{1+e_\rmC \cos \nu_\rmC}\, ,
\end{equation}
with $c$ beeing the speed of light and
$\nu_\rmC$ the true anomaly.
The negative sign is due to the relation $\omega_\rmAB = \omega_\rmC + 180^{\circ}$,
and the amplitude of the LTTE is given by:
\begin{equation}\label{eq:altte}
    \altte = \frac{a_\rmAB \sin i_\rmC}{c} \sqrt{1 - e^{}_\rmC \cos^{2}\omega_\rmC}\, ,
\end{equation}
that can be linked to the mass through the mass function:
\begin{equation}\label{eq:massfunc_binary}
    f(M_\rmC) =  \frac{M^{3}_\rmC \sin^{3} i_\rmC}{M_{\rmAB\rmC}} = \frac{4 \pi^{2}a^{3}_\rmAB \sin^{3} i_\rmC}{G P^{2}_\rmC}\, ,
\end{equation}
where $M_{\rmAB \rmC} = M_\rmA + M_\rmB + \rmC$ is the sum of the masses of each component of the system
(the binary \rmAB{} and the third body \rmC),
$P_\rmC$ is the orbital period of \rmC,
and $G$ is the gravitational constant.

For the additional body C, the unknown parameters involved in the modelling process are
the inclination $i_\rmC$,
the eccentricity $e_\rmC$,
the argument of pericentre $\omega_\rmC$ (or a combination of $e$ and $\omega$),
the mass $M_\rmC$,
the period $P_\rmC$,
and the time of passage at the pericentre $T_{\mathrm{peri},\rmC}$.
The last parameter identifies when the body is at the pericentre and 
it is needed to compute the true anomaly $\nu_\rmC$ at each observation time $T_{0,\rmAB,E}$,
through the known relations:
\begin{equation}\label{eq:peric_to_trueanom}
    \begin{split}
    \mathcal{M}_{\rmC,E} = \frac{2\pi}{P_\rmC} (T_{0,\rmAB,E} - T_{\mathrm{peri},\rmC})\ ,\\
    \mathcal{E}_{\rmC,E} - e \sin \mathcal{E}_{\rmC,E} = \mathcal{M}_{\rmC,E}\ (\mathrm{Kepler's equation}),\\
    \tan \frac{\nu_{\rmC,E}}{2} = \sqrt{\frac{1+e_\rmC}{1-e_\rmC}} \tan \frac{\mathcal{E}_{\rmC,E}}{2}\ ,
    \end{split}
\end{equation}
where $\mathcal{M}_{\rmC,E}$ is the mean anomaly of body C at the eclipse $E$,
$\mathcal{E}_{\rmC,E}$ the eccentric anomaly\footnote{Eccentric anomaly is usually identified as $E$, 
but it has been used the $\mathcal{E}$ notation to avoid confusion with the eclipse/transit number $E$.}.
As already described in Section~\ref{Method_analytical_integration}, 
it can be possible to determine orbital parameters of the additional body 
applying quasi-global and local optimiser, and Bayesian frameworks.
This term for a third body can be repeated to include additional bodies in the model.
As for the TTV case, for ETV systems also the timing precision and the observational baseline
are fundamental to understand the nature of the ETV signal and
measure the model parameters.
Examples of eclipse timing variation cases and survey using ground and space-based observations can be found in \citet{Brown-Sevilla10.1093/mnras/stab1843} and \citet{Bours2016MNRAS.460.3873B},
and with a focus on
\kepler{} and TESS data in 
\cite{Marcadon2024ApJ...976..242M} and \citet{Borkovits2025AA...695A.209B}.

\section{Example of data analysis}\label{Example}

In this section, I will use publicly available \texttt{python} codes to simulate
a multiple-planet system, generate synthetic transit times,
compare parameter sensitivity and different tools.
The following analysis is publicly available as a full \texttt{python notebook} at
the \texttt{github} repository \url{https://github.com/lucaborsato/TTV_DataAnalysis}.
I based the synthetic system on the parameters of TOI-216 presented in 
Table~2 of \citet{McKee2023AJ....165..236M}.
I will show how to generate transit times from three different codes,
such as \texttt{trades}, \texttt{rebound}, and \texttt{TTVFast},
and plot the synthetic Observed--Calculated ($O-C$) diagram,
by fitting a linear ephemeris on the transit times and
I will show the impact of different values for a few parameters on the TTV signal (amplitude and phase).
Then, I will create a synthetic and noisy dataset that will be used for next analysis,
showing how to define fitting parameters, priors, log-likelihood, log-probability, and
how to run \texttt{PyDE}, \texttt{emcee}, and \texttt{nautilus}.

Firstly, I will import all the needed packages and configurations.

\begin{lstlisting}[language=Python]
import numpy as np
import os
import sys
import pickle
os.environ["OMP_NUM_THREADS"] = "1" # needed to avoid issues with parallelization
from multiprocessing import Pool
from scipy.stats import norm, halfnorm, uniform
# TRADES
from pytrades import pytrades
from pytrades import constants as cst
from pytrades import ancillary as anc
from pytrades import plot_oc as poc
from pytrades.convergence import log_probability_trace, full_statistics
# REBOUND
import rebound
# TTVFAST
import ttvfast
# PyDE
from pytransit.utils import de as pyde
# emcee
import emcee
# NAUTILUS
import nautilus
# for corner plots
import pygtc
# plots
import matplotlib as mpl
import matplotlib.pyplot as plt
from matplotlib.gridspec import GridSpec
# use default settings from TRADES
anc.set_rcParams()
# increase dpi
plt.rcParams["figure.dpi"] = 600
plt.rcParams["savefig.dpi"] = 600
\end{lstlisting}

I define the 2-planet system based on the parameters published in \citet{McKee2023AJ....165..236M}, with 
planetary parameters, for TOI-216 b and TOI-216 c,
defined at the dynamical reference time
$ t_\mathrm{ref,dyn} = \mathrm{BJD_{TDB}} - 2457000 = 1325.31 $ days.  
For demonstration purpose, and convenience, I will use zero as reference time $ t_\mathrm{ref,dyn} = 0$.  
Let me define masses in solar masses (\texttt{mass}), 
radii in solar radii (\texttt{radius}), 
periods in days (\texttt{period}), 
arguments of periastron/pericentre (\texttt{argp}), 
mean anomalies (\texttt{meana}), 
mean longitudes (\texttt{meanl}),
inclinations (\texttt{inc}), 
longitudes of ascending nodes (\texttt{longn})
in degrees.

\begin{lstlisting}[language=Python]
body_names = ["star", "b", "c"]

mass   = np.array([0.763, 0.0554*cst.Mjups, 0.525*cst.Mjups])
radius = np.array([0.757, 7.84*cst.Rears, 10.09*cst.Rears])
period = np.array([0.0, 17.0988, 34.5508])
ecc    = np.array([0.0, 0.1593, 0.009])
argp   = np.array([0.0, 292.0, 236.0])
meanl  = np.array([0.0, 82.18, 27.50])
inc    = np.array([0.0, 88.554, 89.801])
# for convenience keep the angles between 0 and 360 degress
longn  = np.array([0.0, 0.0, -0.80])%360.0
meana = (meanl-longn-argp) %360.0
\end{lstlisting}

I will start simulating the system with \texttt{trades}.
Firstly, I need to initialise the code with the number of bodies in the system
(three, a star and two planets), and other parameters that set,
for example, whether it has to return transit duration,
check of close encounters, and others short-term stability flags.

\begin{lstlisting}[language=Python]
n_body = len(mass)
duration_check = 1 # do computation of the transit duration
t_epoch = 0.0 # reference time of integration
t_start = 0.0 # start time of integration
t_int   = 5000.0 # duration of integration in days

pytrades.args_init(
    n_body,
    duration_check,
    t_epoch=t_epoch, # not needed, but good habits to do it here
    t_start=t_start, # not needed, but good habits to do it here
    t_int=t_int, # not needed, but good habits to do it here
    encounter_check=True, # check for close encounters
    do_hill_check=False, # check for stability condition based on Hill radius
    amd_hill_check=False, # check for stability condition based on AMD-Hill criterion
    rv_res_gls=False, # use GLS method on RV residuals to avoid introduction of signals close to the planetary periods
)
\end{lstlisting}

I integrate the orbits of the planets for a specified duration, 
obtaining a sequence of time steps and their corresponding orbital states. 
Each line within the \texttt{orbits} output contains the astrocentric state vector 
(position and velocity) for each body in the system. 
Specifically, the first six values represent the state vector of the central star 
(which are zero in an astrocentric coordinate system), 
followed by six values for planet b, and then a further six values for planet c. 
The boolean variable \texttt{stable} indicates the system's dynamical status: 
1 identifies a stable system, while 0 denotes that an instability condition has occurred, 
such as a close encounter, a planet falling into the star, or planetary ejection.

\begin{lstlisting}[language=Python]
time_steps, orbits, stable = pytrades.kelements_to_orbits_full(
    t_epoch,
    t_start,
    t_int,
    mass,
    radius,
    period,
    ecc,
    argp,
    meana,
    inc,
    longn,
    specific_times=None, # list of additional times at which computing the orbits, as for observed RVs
    step_size=None, # define the output stepsize in unit of days
    n_steps_smaller_orbits=10.0, # define the output stepsize as the minimum period divided this value
)
\end{lstlisting}

\begin{figure}[!htb]
    \centering
    \includegraphics[width=\textwidth]{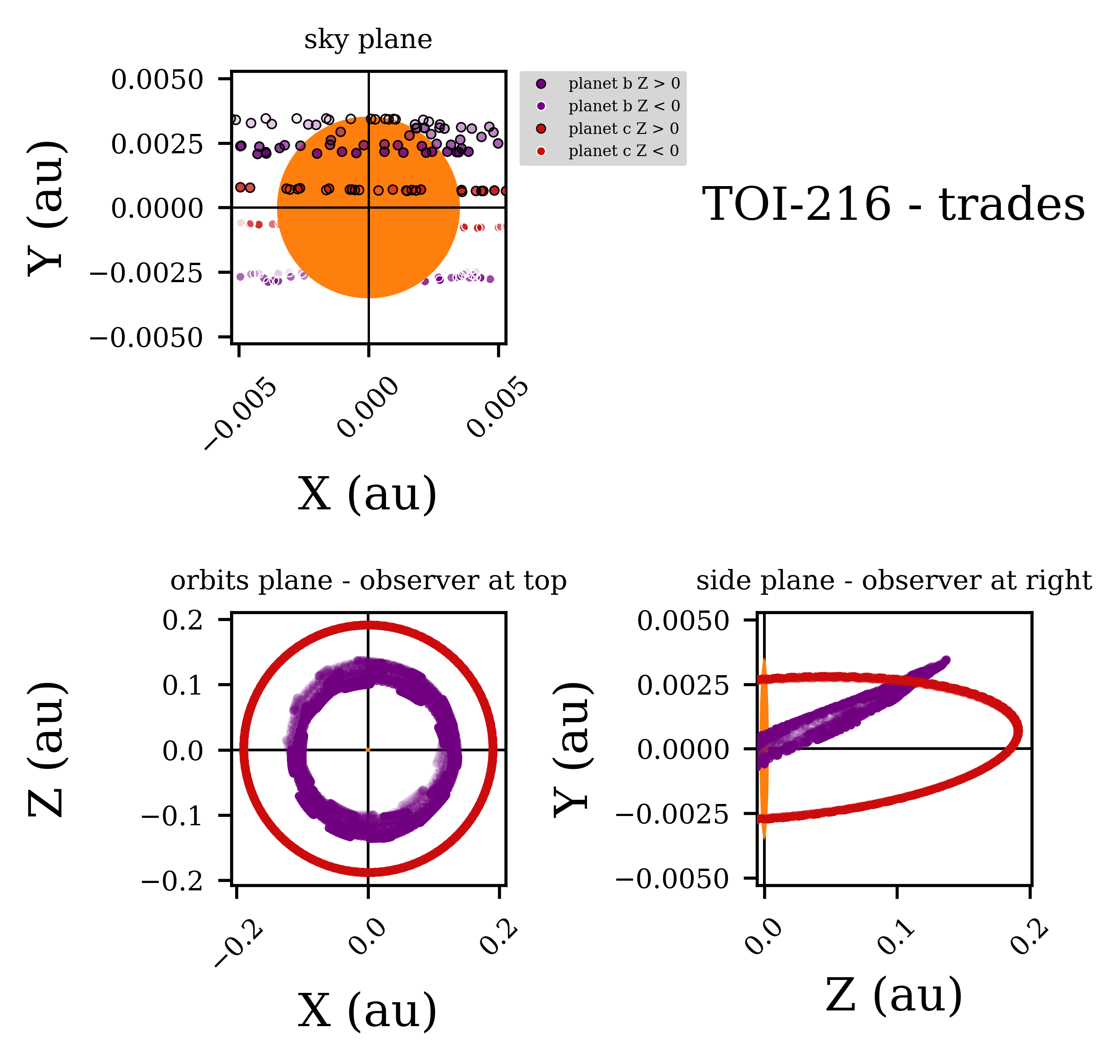}% no path, no extension
    \caption{Orbits of TOI-216 simulated with \texttt{trades}.
    \textbf{Planetary orbits simulated with \texttt{trades}.}
    Planet b is shown as purple circle and planet c as red circles 
    (with opacity increasing with integration time), 
    both with black stroke for $Z>0$ and white stroke for $Z<0$.
    \textit{Upper-left} panel shows the sky-plane $X-Y$, as view by an observer/telescope.
    \textit{Lower-left} panel shows the plane of the orbits $X-Z$ (view from top).
    \textit{Lower-right} panel shows the $Z-Y$ plane, side view as the observer would be placed on the right;
    the x-limits have been defined for only $Z>0$ coordinates to show that planets are not exactly on the orbital plane.
    }\label{da_orbits_trades_default}
\end{figure}

It is possible to visualise the orbits of the planets in the system in three projections in Fig.~\ref{da_orbits_trades_default}, 
the sky plane $(X-Y)$ (what an observer \textit{sees}),
the orbit plane $(X-Z)$ (observer on top), and 
a side view of the orbit $(Z-Y)$ (observer on right side).
I use a function within \texttt{trades} to simplify the process.

\begin{lstlisting}[language=Python]
fig= pytrades.base_plot_orbits(
    time_steps,
    orbits,
    radius,
    n_body,
    body_names,
    figsize=(4, 4),
    sky_scale='star',
    side_scale='positive',
    title="TOI-216 - trades",
    show_plot=True,
)
plt.close(fig)
\end{lstlisting}

Now, I have the orbits at each time step, and I have to extract the transit times.
In \texttt{trades} there is a function that check if a transit event occurs for the planets,
and returns for each transit the transit times in \texttt{transits}, 
the total transit duration \texttt{durations} in min, 
the projected spin-orbit misalignment \texttt{lambda\_rm} (the $\lambda$ of the Rossiter-McLaughlin effect),
the orbital elements \texttt{kep\_elem},
and the identification number of the transiting planet \texttt{body\_flag}.
The function needs to specify the total number of transit events for all planets, 
denoted as \texttt{n\_all\_transits}. 
To avoid numerical and memory issues, 
it would be advisable to define this as the number of time steps multiplied by the number of planets.
However, I will subsequently select only the true transit events and store them in a \texttt{python} dictionary.

\begin{lstlisting}[language=Python]
transiting_body = 1 # Set the transiting body to 1 (all planets)
n_transits = [len(time_steps)-1]*(n_body-1) # prepare a list of the number of transits for each planet
n_all_transits = np.sum(n_transits) # total number of transits for all the planets
# Compute the transits, durations, lambda_rm, Kepler elements, and body flags
transits, durations, lambda_rm, kep_elem, body_flag = pytrades.orbits_to_transits(
    n_all_transits, time_steps, mass, radius, orbits, transiting_body
)
# The function returns:  
# - `transits` array of transit times  
# - `durations` array of transit durations (total duration, $T_{14}$) in minutes  
# - `lambda_rm` array of the projected spin-orbit misalignment angles  
# - `kep_elem` array of Keplerian orbital elements (period in days, semi-majoar axis in au, eccentricity, inclination, mean anomaly, argument of pericenter in degrees)  
# - `body_flag` array of flag that identifies the body (2 for planet the first planet b, 3 for the second planet c), it needs to recover the transit times for different planets.

trades_transits = {}
for pl_letter, pl_number in zip(body_names[1:], [2,3]):
    sel_pl = body_flag == pl_number
    n_transits = np.sum(sel_pl)
    print("planet {} (id {}) with {} transits in {:.0f} days of integration".format(pl_letter, pl_number, n_transits, t_int))
    trades_transits[pl_letter] = {
        "n_transits":  n_transits,
        "transit_times": transits[sel_pl],
        "transit_durations": durations[sel_pl],
        "lambda_rm": lambda_rm[sel_pl],
        "kep_elem": kep_elem[sel_pl],
    }
\end{lstlisting}

I repeat the orbital integration with \texttt{rebound}, 
and then I will extract the transits with the \texttt{trades} function.
First, define a function that set and run a \texttt{rebound} simulation.

\begin{lstlisting}[language=Python]
def run_rebound(
    mass,
    radius,
    period,
    ecc,
    argp,
    meana,
    inc,
    longn,
    time_steps
):

    radius_au = radius * cst.RsunAU
    argp_r   = argp * cst.deg2rad
    meana_r  = meana * cst.deg2rad
    inc_r    = inc * cst.deg2rad
    longn_r  = longn * cst.deg2rad

    sim = rebound.Simulation()
    sim.units = ["msun", "au", "days"] # Units of solar mass, au, and days

    sim.add(m=mass[0], r=radius_au[0]) #star
    nbody = len(mass)
    for ib in range(1, nbody):
        sim.add(
            m=mass[ib],
            r=radius_au[ib],
            P=period[ib],
            e=ecc[ib],
            omega=argp_r[ib],
            M=meana_r[ib],
            inc=inc_r[ib],
            Omega=longn_r[ib],
            primary=sim.particles[0]
        )
    
    sim.move_to_com()

    times = (time_steps - time_steps[0])
    ntime = len(times)

    orbits = np.zeros((ntime, 6*nbody))
    for i, t in enumerate(times):
        sim.integrate(t)
        for ib in range(nbody):
            orbits[i, 0+(6*ib)] = sim.particles[ib].x
            orbits[i, 1+(6*ib)] = sim.particles[ib].y
            orbits[i, 2+(6*ib)] = sim.particles[ib].z
            orbits[i, 3+(6*ib)] = sim.particles[ib].vx
            orbits[i, 4+(6*ib)] = sim.particles[ib].vy
            orbits[i, 5+(6*ib)] = sim.particles[ib].vz

    return orbits
\end{lstlisting}

I use the time steps from \texttt{trades} to ensure
the timestamps of the \texttt{rebound} simulation match. 
After that, I will extract the transit times.

\begin{lstlisting}[language=Python]
orbits_rebound = run_rebound(
    mass,
    radius,
    period,
    ecc,
    argp,
    meana,
    inc,
    longn,
    time_steps
)
# Compute the transits, durations, lambda_rm, Kepler elements, and body flags
transits_rebound, durations_rebound, lambda_rm_rebound, kep_elem_rebound, body_flag_rebound = pytrades.orbits_to_transits(
    n_all_transits, time_steps, mass, radius, orbits_rebound, transiting_body
)
rebound_transits = {}
for pl_letter, pl_number in zip(body_names[1:], [2,3]):
    sel_pl = body_flag_rebound == pl_number
    n_transits = np.sum(sel_pl)
    print("planet {} (id {}) with {} transits in {:.0f} days of integration".format(pl_letter, pl_number, n_transits, t_int))
    rebound_transits[pl_letter] = {
        "n_transits":  n_transits,
        "transit_times": transits_rebound[sel_pl],
        "transit_durations": durations_rebound[sel_pl],
        "lambda_rm": lambda_rm_rebound[sel_pl],
        "kep_elem": kep_elem_rebound[sel_pl],
    }
\end{lstlisting}

Then, I try the \texttt{python} version of \texttt{TTVFast}\footnote{
Original code available at \url{https://github.com/kdeck/TTVFast}
and wrapped in python at
\url{https://github.com/simonrw/ttvfast-python}.
} 
code \citep{Deck2014ApJ...787..132D}.
It will compute the orbits and all the transits of both planets.

\begin{lstlisting}[language=Python]
planet_b = ttvfast.models.Planet(
    mass = mass[1],# mass: Mplanet in units of M_sun
    period = period[1],# period: Period in days
    eccentricity = ecc[1],# eccentricity: E between 0 and 1
    inclination = inc[1],# inclination: I in units of degrees
    longnode = longn[1],# longnode: Longnode in units of degrees
    argument = argp[1],# argument: Argument in units of degrees
    mean_anomaly = meana[1],# mean_anomaly: mean anomaly in units of degrees
)

planet_c = ttvfast.models.Planet(
    mass = mass[2],# mass: Mplanet in units of M_sun
    period = period[2],# period: Period in days
    eccentricity = ecc[2],# eccentricity: E between 0 and 1
    inclination = inc[2],# inclination: I in units of degrees
    longnode = longn[2],# longnode: Longnode in units of degrees
    argument = argp[2],# argument: Argument in units of degrees
    mean_anomaly = meana[2],# mean_anomaly: mean anomaly in units of degrees
)

planets = [planet_b, planet_c]
gravity = cst.Giau # needed to provide the Gravitational constant in proper units
stellar_mass = mass[0]
dt = np.min(np.diff(time_steps)) # provide time steps

results = ttvfast.ttvfast(
    planets, 
    stellar_mass, 
    t_epoch,
    dt, 
    t_int, 
    rv_times=None, 
    input_flag=1 # 0 = Jacobi 1 = astrocentric elements 2 = astrocentric cartesian
)

ttvfast_index, ttvfast_epoch, ttvfast_transits, _, _ = results["positions"]
SEL_OK =  np.array(ttvfast_transits) > -2.0
ttvfast_index = np.array(ttvfast_index)[SEL_OK] # planet id
ttvfast_epoch = np.array(ttvfast_epoch)[SEL_OK] # the epoch or transit number
ttvfast_transits = np.array(ttvfast_transits)[SEL_OK] # transit times
\end{lstlisting}

I have the transit times for both planets from three different programs.
To visualize an $O-C$ diagram, I need to fit a linear ephemeris on the transit times,
but first I have to define some initial guess for the \tref{} and the \pref{}, 
the functions to compute the epoch (or transit number) and the weighted least square.

\begin{lstlisting}[language=Python]
def compute_epoch(Tref, Pref, transit_times):
    dt = transit_times - Tref
    epoch = np.rint(dt / Pref) # it returns the nearest integer, that is if 0.1 => 0, 0.9 => 1.
    return epoch

# Weighted Lest-Squares based on Numerical Recipes
# case without errors on transit times
def linear_fit_no_errors(x, y):
    nx = len(x)
    S = nx
    Sx = np.sum(x)
    Sy = np.sum(y)

    t = x - (Sx / S)
    m = np.dot(t, y)
    St2 = np.dot(t, t)
    m = m / St2
    q = (Sy - (m * Sx)) / S
    err_q = np.sqrt((1.0 + ((Sx*Sx)/(S*St2)))/S)
    err_m = np.sqrt(1.0/St2)

    t = y - (q + m * x)
    chi2 = np.dot(t,t)
    dof = nx-2
    sigdat = np.sqrt(chi2/dof)
    err_q = err_q*sigdat
    err_m = err_m*sigdat
    
    return m, err_m, q, err_q, chi2

# case with errors
def linear_fit_with_errors(x, y, ey):
    nx = len(x)
    w = 1.0 / (ey*ey)
    S = np.sum(w)
    Sx = np.dot(w, x)
    Sy = np.dot(w, y)

    t = (x - (Sx/S))/ey
    m = np.dot(t/ey, y)
    St2 = np.dot(t, t)
    m = m / St2
    q = (Sy - (Sx*m))/S
    err_q = np.sqrt((1.0 + ((Sx*Sx)/(S*St2)))/S)
    err_m = np.sqrt(1.0/St2)

    return m, err_m, q, err_q

# automatic selection of case with or without errors
def linear_fit(x, y, ey=None):
    if(ey is None):
        m, err_m, q, err_q, chi2 = linear_fit_no_errors(x, y)
        # res = y - (q + m * x)
    else:
        m, err_m, q, err_q = linear_fit_with_errors(x, y, ey)
        res = (y - (q + m * x)) / ey
        chi2 = np.dot(res, res)

    return (q, err_q), (m, err_m), chi2

# automatic linear ephemeris from input or fitting
def linear_ephemeris(T0s, eT0s=None, Tref_in = None, Pref_in = None, fit=False):

    nx = len(T0s)

    # let me define a first guess of the reference time
    Tref, Pref = [0.0, 0.0], [0.0, 0.0]
    if Tref_in is None:
        Tref[0] = T0s[nx//2] # ~median value
    elif isinstance(Tref_in, (tuple, list, np.ndarray)):
        Tref = Tref_in
    else:
        Tref[0] = Tref_in

    # let me define a first guess of the reference period
    if Pref_in is None:
        Pref[0] = np.median(np.diff(T0s)) # median of difference of consecutive transits
    elif isinstance(Pref_in, (tuple, list, np.ndarray)):
        Pref = Pref_in
    else:
        Pref[0] = Pref_in

    # let me define the initial epochs
    epoch = compute_epoch(Tref[0], Pref[0], T0s)
    
    if fit:
        # fit the transit times to obtain the linear ephemeris
        Tref, Pref, chi2 = linear_fit(epoch, T0s, ey=eT0s)
        # recompute the epoch (or transit numbers)
        epoch = compute_epoch(Tref[0], Pref[0], T0s)

    # compute the predicted transit times
    Tlin = Tref[0] + Pref[0] * epoch
    # compute the O-C in days
    oc = T0s - Tlin
    if not fit:
        chi2 = np.dot(oc,oc)
    
    return Tref, Pref, chi2, epoch, Tlin, oc
\end{lstlisting}

\begin{figure}[!htb]
    \centering
    \includegraphics[width=\textwidth]{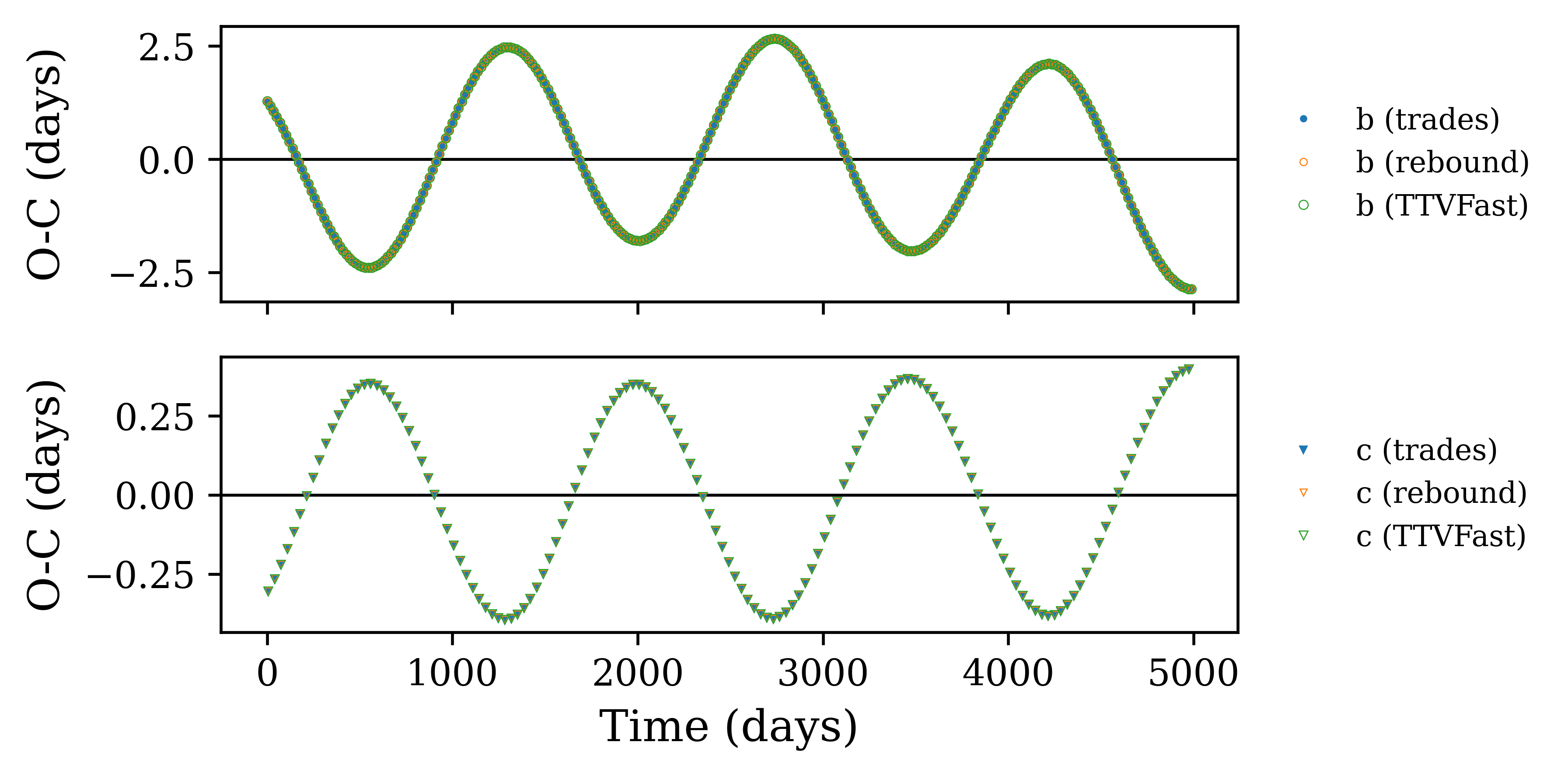}% no path, no extension
    \caption{$O-C$ diagram for planet b (\textit{upper-panel}) and c (\textit{lower-panel}).
    \textbf{$O-C$ diagram with default parameters.}
    Transit times from \texttt{trades} have been plotted as blue markers,
    from \texttt{rebound} as open-orange markers,
    and from \texttt{TTVFast} as open-green markers.
    }\label{da_oc_default}
\end{figure}

I compute the linear ephemeris for each method and do the $O-C$, shown in Fig.~\ref{da_oc_default}.

\begin{lstlisting}[language=Python]
fig, axs = plt.subplots(2, 1, sharex=True, figsize=(5,3))

u = [1.0, "days"]
markers = anc.filled_markers

for i, pl_letter in enumerate(body_names[1:]):
    print("planet {}".format(pl_letter))

    ax = axs[i]
    
    # trades
    tra_tr = trades_transits[pl_letter]["transit_times"]
    n_tra_tr = len(tra_tr)
    Tref_tr, Pref_tr, chi2_tr, epoch_tr, Tlin_tr, oc_tr_days = linear_ephemeris(
        tra_tr, eT0s=None, 
        Tref_in = None, Pref_in = None, 
        fit=True
    )
    print("{:13s}: Tref = {:.5f} +/- {:.5f}, Pref = {:.5f} +/- {:.5f} with chi^2 = {:.2f} ==> chi^2_reduce = {:.4f}".format(
        "trades",
        *Tref_tr, *Pref_tr, chi2_tr, chi2_tr/(n_tra_tr-2))
     )

    ax.axhline(0, color="k", lw=0.8)
    ax.plot(
        tra_tr,
        oc_tr_days*u[0],
        marker=markers[i],
        ms=1,
        color="C0",
        label="{} (trades)".format(pl_letter),
        ls='',
    )

    # rebound
    tra_re = rebound_transits[pl_letter]["transit_times"]
    n_tra_re = len(tra_re)
    Tref_re, Pref_re, chi2_re, epoch_re, Tlin_re, oc_re_days = linear_ephemeris(
        tra_re, eT0s=None, Tref_in = Tref_tr, Pref_in = Pref_tr, fit=False
    )
    print("{:13s}: Tref = {:.5f} +/- {:.5f}, Pref = {:.5f} +/- {:.5f} with chi^2 = {:.2f} ==> chi^2_reduce = {:.4f}".format(
        "rebound",
        *Tref_re, *Pref_re, chi2_re, chi2_re/(n_tra_re-2))
     )

    ax.plot(
        tra_re,
        oc_re_days*u[0],
        marker=markers[i],
        ms=2,
        mfc="None",
        mew=0.3,
        color="C1",
        label="{} (rebound)".format(pl_letter),
        ls='',
    )

    # TTVFast
    sel_ttvf = ttvfast_index == i
    tra_ttvf = ttvfast_transits[sel_ttvf]
    n_tra_ttvf = len(tra_ttvf)
    Tref_ttvf, Pref_ttvf, chi2_ttvf, epoch_ttvf, Tlin_ttvf, oc_ttvf_days = linear_ephemeris(
        tra_ttvf, eT0s=None, 
        Tref_in = Tref_tr[0], 
        Pref_in = Pref_tr[0], 
        fit=True
    )
    print("{:13s}: Tref = {:.5f} +/- {:.5f}, Pref = {:.5f} +/- {:.5f} with chi^2 = {:.2f} ==> chi^2_reduce = {:.4f}".format(
        "TTVFast",
        *Tref_ttvf, *Pref_ttvf, chi2_ttvf, chi2_ttvf/(n_tra_ttvf-2))
     )

    ax.plot(
        tra_ttvf,
        oc_ttvf_days*u[0],
        marker=markers[i],
        ms=2.5,
        mfc="None",
        mew=0.3,
        color="C2",
        label="{} (TTVFast)".format(pl_letter),
        ls='',
    )
    

    ax.legend(loc='center left', bbox_to_anchor =(1.01, 0.5), fontsize=8, frameon=False)
    ax.set_ylabel("O-C ({})".format(u[1]))

axs[0].xaxis.set_tick_params(labelbottom=False)
ax.set_xlabel("Time (days)")

fig.align_ylabels(axs)
plt.show()
plt.close(fig)
# OUTPUT
# planet b
# trades       : Tref = 2495.46865 +/- 0.09570, Pref = 17.21893 +/- 0.00114 with chi^2 = 770.26 ==> chi^2_reduce = 2.6652
# rebound      : Tref = 2495.46865 +/- 0.09570, Pref = 17.21893 +/- 0.00114 with chi^2 = 770.26 ==> chi^2_reduce = 2.6653
# TTVFast      : Tref = 2495.46861 +/- 0.09570, Pref = 17.21893 +/- 0.00114 with chi^2 = 770.26 ==> chi^2_reduce = 2.6652
# planet c
# trades       : Tref = 2490.34650 +/- 0.02199, Pref = 34.50089 +/- 0.00053 with chi^2 = 10.03 ==> chi^2_reduce = 0.0701
# rebound      : Tref = 2490.34650 +/- 0.02199, Pref = 34.50089 +/- 0.00053 with chi^2 = 10.03 ==> chi^2_reduce = 0.0701
# TTVFast      : Tref = 2490.34645 +/- 0.02199, Pref = 34.50089 +/- 0.00053 with chi^2 = 10.03 ==> chi^2_reduce = 0.0701
\end{lstlisting}

I can test the sensitivity on planetary parameters of the TTV signals.
I do this by wrapping some functions in one to compute the orbits
and get the transits.

\begin{lstlisting}[language=Python]
def run_and_get_transits(
    t_e,
    t_s,
    t_i,
    mass_in, 
    radius_in,
    period_in,
    ecc_in,
    argp_in,
    meana_in,
    inc_in,
    longn_in,
    planet_names,
):
    ttt, ooo, stable = pytrades.kelements_to_orbits_full(
        t_e,
        t_s,
        t_i,
        mass_in,
        radius_in,
        period_in,
        ecc_in,
        argp_in,
        meana_in,
        inc_in,
        longn_in,
        specific_times=None,
        step_size=None,
        n_steps_smaller_orbits=10.0,
        )
    sort_ttt = np.argsort(ttt)
    ttt = ttt[sort_ttt]
    ooo = ooo[sort_ttt, :]
    
    tra_body = 1 # Set the transiting body to 1 (all planets)
    n_transits = [len(ttt)-1]*(n_body-1) # prepare a list of the number of transits for each planet
    n_all_transits = np.sum(n_transits) # total number of transits for all the planets
    # Compute the transits, durations, lambda_rm, Kepler elements, and body flags
    tratime, durmin, l_rm, kelem, bd_flag = pytrades.orbits_to_transits(
        n_all_transits, ttt, mass_in, radius_in, ooo, tra_body
    )
    
    out_transits = {}
    for i_pl, pl_letter in enumerate(planet_names):
        pl_number = i_pl + 2
        sel_pl = bd_flag == pl_number
        n_transits = np.sum(sel_pl)
        # print("planet {} (id {}) with {} transits in {:.0f} days of integration".format(pl_letter, pl_number, n_transits, t_i))
        out_transits[pl_letter] = {
            "n_transits":  n_transits,
            "transit_times": tratime[sel_pl],
            "transit_durations": durmin[sel_pl],
            "lambda_rm": l_rm[sel_pl],
            "kep_elem": kelem[sel_pl],
        }

    return ttt, ooo, stable, out_transits
\end{lstlisting}

\begin{figure}[!htb]
    \centering
    \includegraphics[width=\textwidth]{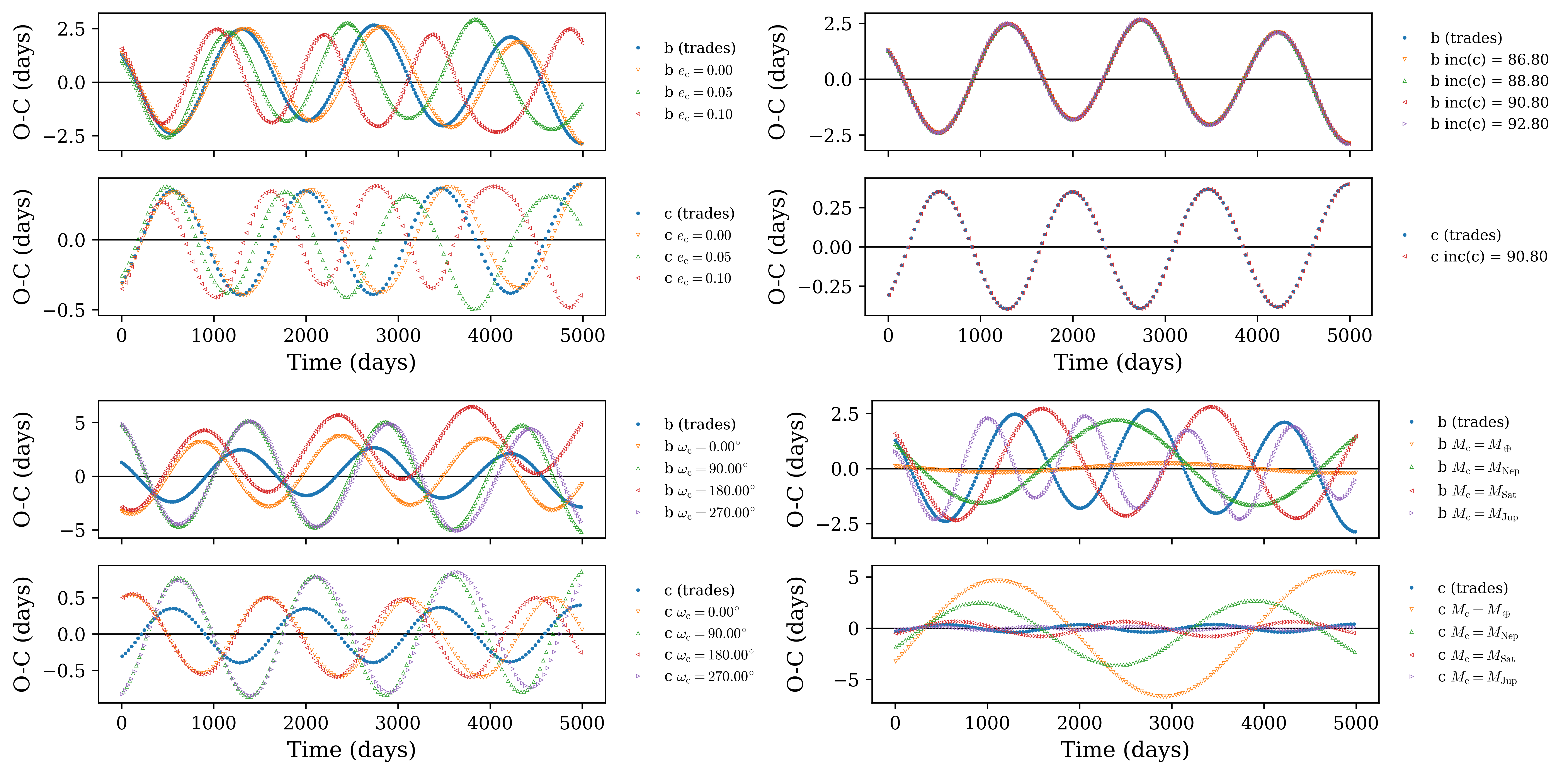}
    \caption{
    \textbf{TTV sensitivity to the planetary parameters.}
    $O-C$ diagrams for different variations of orbital parameters of planet c.
    \textit{Upper-left}: eccentricity variation with values of 0.0, 0.05, and 0.1;
    \textit{Upper-right}: inclination values of 86.80, 88.80, 90.80, and 92.80 degree, in three cases the planet c does not transit anymore;
    \textit{Lower-left}: argument of pericenter values of 0.0, 90.0, 180.0, and 270.0 degress;
    \textit{Lower-right}: mass values of $1 M_\oplus,\, 1 M_\mathrm{Nep},\, 1 M_\mathrm{Sat},\, 1 M_\mathrm{Jup}$.
    }\label{da_oc_par_variation}
\end{figure}

Then I change a value of one parameter of planet c and plot the $O-C$.
Below, I present just the code for the eccentricity case, but it is easily extendible to other parameters.
See in Fig.~\ref{da_oc_par_variation} the output of the variation 
of the eccentricity (upper-left panel),
of the inclination (upper-right panel),
of the argument of pericentre (lower-left panel),
and of the mass (lower-right panel)
of planet c.

\begin{lstlisting}[language=Python]
fig, axs = plt.subplots(2, 1, sharex=True, figsize=(5,3))

u = [1.0, "days"]
markers = anc.filled_markers

lineph_trades = {}

for i, pl_letter in enumerate(body_names[1:]):
    print("planet {}".format(pl_letter))

    ax = axs[i]
    
    # trades original
    tra_tr = trades_transits[pl_letter]["transit_times"]
    n_tra_tr = len(tra_tr)
    Tref_tr, Pref_tr, chi2_tr, epoch_tr, Tlin_tr, oc_tr_days = linear_ephemeris(
        tra_tr, eT0s=None, 
        Tref_in = None, Pref_in = None, 
        # Tref_in = tra_tr[0], Pref_in = period[i+1], 
        fit=True
    )
    print("{:13s}: Tref = {:.5f} +/- {:.5f}, Pref = {:.5f} +/- {:.5f} with chi^2 = {:.2f} ==> chi^2_reduce = {:.4f}".format(
        "trades",
        *Tref_tr, *Pref_tr, chi2_tr, chi2_tr/(n_tra_tr-2))
     )
    lineph_trades[pl_letter] = (Tref_tr, Pref_tr)

    ax.axhline(0, color="k", lw=0.8)
    ax.plot(
        tra_tr,
        oc_tr_days*u[0],
        marker=markers[0],
        ms=1,
        color="C0",
        label="{} (trades)".format(pl_letter),
        ls='',
    )

# here I define values
par_c_test = [0.0, 0.05, 0.1]
for itest, p_test in enumerate(par_c_test):
    ecc_new = ecc.copy() # create a copy of orginal values
    ecc_new[2] = p_test # change only the planet c
    
    time_steps_new, orbits_new, stable_new, trades_transits_new = run_and_get_transits(
        t_epoch,
        t_start,
        t_int,
        mass, radius,
        period,
        ecc_new,
        argp, meana,
        inc, longn,
        body_names[1:]
    )
    
    for i, pl_letter in enumerate(body_names[1:]):
    
        ax = axs[i]

        Tref_tr, Pref_tr = lineph_trades[pl_letter]
        
        # trades updated ecc
        tra_new = trades_transits_new[pl_letter]["transit_times"]
        n_tra_new = len(tra_new)
        Tref_new, Pref_new, chi2_new, epoch_new, Tlin_new, oc_new_days = linear_ephemeris(
            tra_new, eT0s=None,
            Tref_in = Tref_tr,
            Pref_in = Pref_tr,
            fit=True
        )
        print("{:13s}: Tref = {:.5f} +/- {:.5f}, Pref = {:.5f} +/- {:.5f} with chi^2 = {:.2f} ==> chi^2_reduce = {:.4f}".format(
            "rebound",
            *Tref_new, *Pref_new, chi2_new, chi2_new/(n_tra_new-2))
         )
    
        ax.plot(
            tra_new,
            oc_new_days*u[0],
            marker=markers[1+itest],
            ms=2,
            mfc="None",
            mew=0.3,
            color="C{}".format(1+itest),
            label="{} $e_\mathrm{{c}} = {:.2f}$".format(pl_letter, p_test),
            ls='',
        )

for ax in axs:
    ax.legend(loc='center left', bbox_to_anchor =(1.01, 0.5), fontsize=8, frameon=False)
    ax.set_ylabel("O-C ({})".format(u[1]))

axs[0].xaxis.set_tick_params(labelbottom=False)
ax.set_xlabel("Time (days)")

fig.align_ylabels(axs)
plt.show()
plt.close(fig)
\end{lstlisting}

\begin{figure}[!htb]
    \centering
    \includegraphics[width=\textwidth]{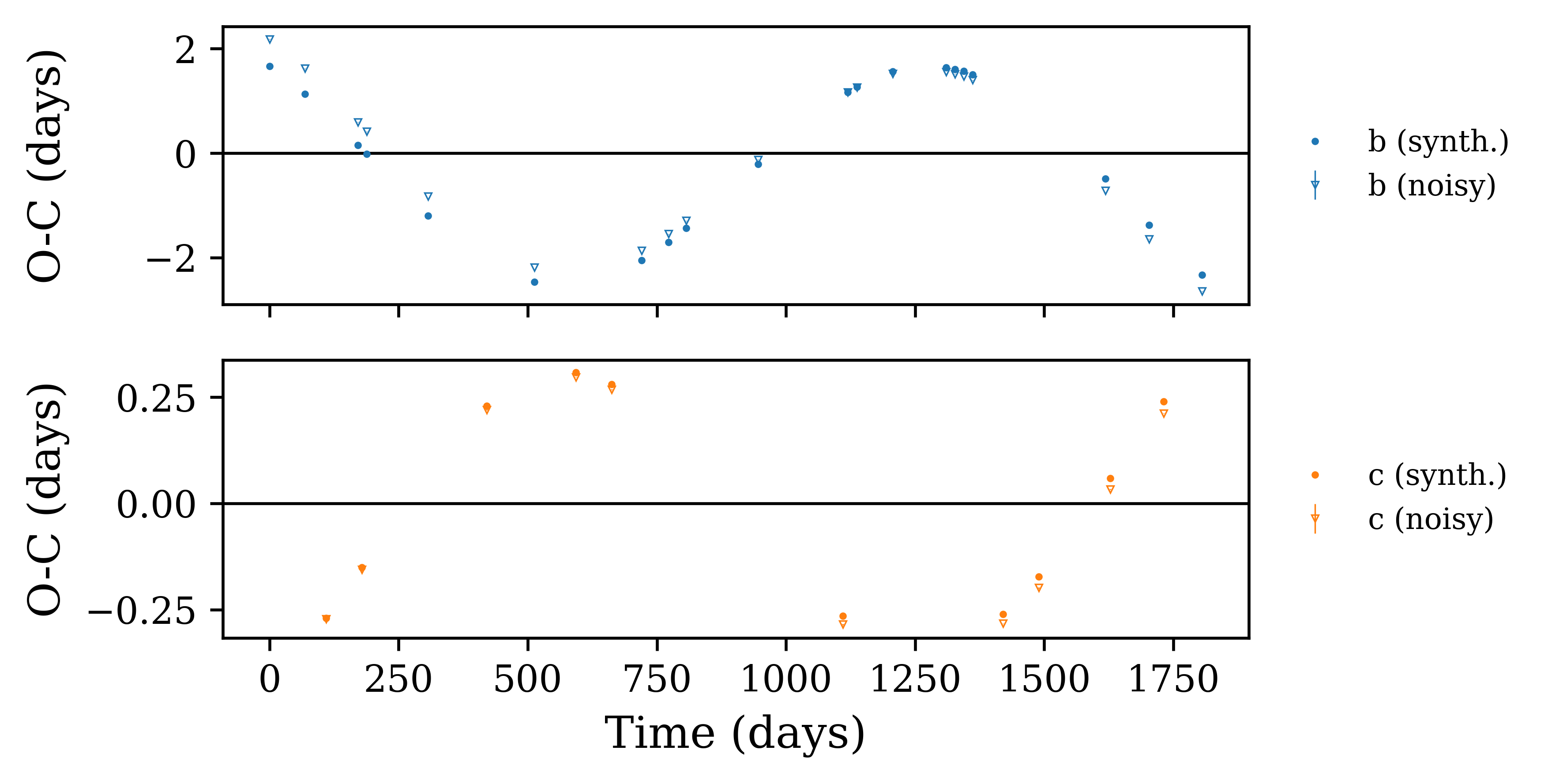}% no path, no extension
    \caption{
    \textbf{Noisy $O-C$ diagram.}
    $O-C$ diagrams of selected transit times with (filled marker) and
    without added noise (open marker).
    }\label{da_oc_noisy}
\end{figure}

I simulate a synthetic-observed dataset, selecting transit times, covering about five years,
from \texttt{TTVFast} output and adding some uncertainty on
the transit times and selecting error to associate to them.
For the uncertainty, I use the unique values provided in \citet{Dawson2021AJ....161..161D}.
Then, I re-compute and store the new ``observed'' linear ephemeris and plot the $O-C$ (see Fig.~\ref{da_oc_noisy}).

\begin{lstlisting}[language=Python]
time_sel_start = 0.0
time_sel_end   = 365.25 * 5
seed = 42
uncertainty = {
    "b": [0.002, 0.003, 0.004], # Dawson+ 2021
    "c": [0.0004, 0.0005, 0.0007, 0.0008, 0.0009, 0.001, 0.0011, 0.0015, 0.002, 0.003], # Dawson+ 2021
}
n_tra_syn_b, n_tra_syn_c = 20, 10

def select_transit_times(
    pl_letter, pl_idx, tra_index, tra_times, err_pool, 
    t1=0.0, t2=365.25, n_tra_syn=20, 
    err_scale_mult=1.0,
    noise_scale_mult=1.0,
    seed=42
):

    np.random.seed(seed=seed)
    
    sel_tra = tra_index == pl_idx
    all_tra_syn = tra_times[sel_tra][np.logical_and(
    tra_times[sel_tra] >= t1,
    tra_times[sel_tra] <= t2,
)]
    n_tra = len(all_tra_syn)

    tra_syn = np.random.choice(all_tra_syn, size=n_tra_syn, replace=False)
    err_tra_syn = np.random.choice(err_pool, size=n_tra_syn, replace=True)
    err_mean, err_std = np.mean(err_pool), np.std(err_pool, ddof=1)
    tra_syn_noisy = tra_syn + np.random.normal(loc=err_mean, scale=err_std, size=n_tra_syn)*noise_scale_mult
    
    return tra_syn, err_tra_syn, tra_syn_noisy

err_scale_mult_b = 1 + np.random.random(n_tra_syn_b)*3
tra_syn_b, err_tra_syn_b, tra_syn_noisy_b = select_transit_times(
    "b", 0, 
    ttvfast_index, ttvfast_transits, 
    uncertainty["b"], 
    t1=time_sel_start, t2=time_sel_end, 
    n_tra_syn=n_tra_syn_b,
    err_scale_mult=err_scale_mult_b,
    noise_scale_mult=1.0,
    seed=seed
)

err_scale_mult_c = 1 + np.random.random(n_tra_syn_c)*3
tra_syn_c, err_tra_syn_c, tra_syn_noisy_c = select_transit_times(
    "c", 1, 
    ttvfast_index, ttvfast_transits, 
    uncertainty["c"],
    t1=time_sel_start, t2=time_sel_end,
    n_tra_syn=n_tra_syn_c,
    err_scale_mult=err_scale_mult_c,
    noise_scale_mult=1.0,
    seed=seed
)

# prepare dictionary to store linear ephemeris with and without added noise
lineph_syn = {}
lineph_syn_noisy = {}

i, pl_letter = 0, "b"
print("planet {}".format(pl_letter))

Tr = tra_syn_b[n_tra_syn_b // 2]
Pr = 17.16073 # Dawson+ 2021
print("input: Tref = ", Tr, "Pref = ", Pr)
Tref_b, Pref_b, chi2_b, epoch_b, Tlin_b, oc_b_days = linear_ephemeris(
    tra_syn_b, eT0s=None, 
    Tref_in = Tr, Pref_in = Pr, 
    fit=True
)
print("{:13s}: Tref = {:.5f} +/- {:.5f}, Pref = {:.5f} +/- {:.5f} with chi^2 = {:.2f} ==> chi^2_reduce = {:.4f}".format(
    "synthetic",
    *Tref_b, *Pref_b, chi2_b, chi2_b/(n_tra_syn_b-2))
 )
lineph_syn[pl_letter] = (Tref_b, Pref_b)

Tr = tra_syn_noisy_b[n_tra_syn_b // 2]
Pr = 17.16073 # Dawson+ 2021
print("input: Tref = ", Tr, "Pref = ", Pr)
Tref_b, Pref_b, chi2_b, epoch_b, Tlin_b, oc_b_days = linear_ephemeris(
    tra_syn_noisy_b, eT0s=err_tra_syn_b, 
    Tref_in = Tr, Pref_in = Pr, 
    fit=True
)
print("{:13s}: Tref = {:.5f} +/- {:.5f}, Pref = {:.5f} +/- {:.5f} with chi^2 = {:.2f} ==> chi^2_reduce = {:.4f}".format(
    "noisy",
    *Tref_b, *Pref_b, chi2_b, chi2_b/(n_tra_syn_b-2))
 )
lineph_syn_noisy[pl_letter] = (Tref_b, Pref_b)

i, pl_letter = 1, "c"
print("planet {}".format(pl_letter))

Tr = tra_syn_c[n_tra_syn_c // 2]
Pr = 34.525528 # Dawson+ 2021
print("input: Tref = ", Tr, "Pref = ", Pr)
Tref_c, Pref_c, chi2_c, epoch_c, Tlin_c, oc_c_days = linear_ephemeris(
    tra_syn_c, eT0s=None, 
    Tref_in = Tr, Pref_in = Pr, 
    fit=True
)
print("{:13s}: Tref = {:.5f} +/- {:.5f}, Pref = {:.5f} +/- {:.5f} with chi^2 = {:.2f} ==> chi^2_reduce = {:.4f}".format(
    "synthetic",
    *Tref_c, *Pref_c, chi2_c, chi2_c/(n_tra_syn_c-2))
 )
lineph_syn[pl_letter] = (Tref_c, Pref_c)

Tr = tra_syn_noisy_c[n_tra_syn_c // 2]
Pr = 34.525528 # Dawson+ 2021
print("input: Tref = ", Tr, "Pref = ", Pr)
Tref_c, Pref_c, chi2_c, epoch_c, Tlin_c, oc_c_days = linear_ephemeris(
    tra_syn_noisy_c, eT0s=err_tra_syn_c, 
    Tref_in = Tr, Pref_in = Pr, 
    fit=True
)
print("{:13s}: Tref = {:.5f} +/- {:.5f}, Pref = {:.5f} +/- {:.5f} with chi^2 = {:.2f} ==> chi^2_reduce = {:.4f}".format(
    "noisy",
    *Tref_c, *Pref_c, chi2_c, chi2_c/(n_tra_syn_c-2))
 )
lineph_syn_noisy[pl_letter] = (Tref_c, Pref_c)
\end{lstlisting}

Now that I have a synthetic dataset generated with \texttt{TTVFast}, 
I can further proceed with a dynamical analysis of the system.
I use \texttt{trades} to integrate the orbits and compute the transit times.
So, firstly I have to initialise a flag to say which planet transits (both b and c in this case),
and load the data into memory.
Then, I will define the fitting labels, parameters and boundaries, and 
functions needed to convert back from fitting to physical parameters,
to compute the log-prior, log-likelihood $\ln \mathcal{L}$ and log-probability $\ln \mathcal{P}$.
The TTV signal depends on the ratio $M_\mathrm{p}/M_\star$, and I use this parameter for both planets,
instead of the mass.

\begin{lstlisting}[language=Python]
transit_flag = [0, 1, 1] # 0 = not transiting (star), 1 transiting (b & c)
# load transits into memory
b_sources_id = np.ones(n_tra_syn_b).astype(int)
pytrades.set_t0_dataset(2, epoch_b, tra_syn_noisy_b, err_tra_syn_b, sources_id=b_sources_id)
c_sources_id = np.ones(n_tra_syn_c).astype(int)
pytrades.set_t0_dataset(3, epoch_c, tra_syn_noisy_c, err_tra_syn_c, sources_id=c_sources_id)

# this will be a global variable, it will not be passed to log-like/prob function
fit_labels = [
    "M_b/M_star", 
    "M_c/M_star",
    "P_b", 
    "P_c",
    "e_b", 
    "e_c",
    "w_b", 
    "w_c",
    "meana_b", 
    "meana_c",
]
n_fit = len(fit_labels)
print("Number of fitting parameters = {}".format(n_fit))

# let me define an initial set of parameters to test next functions
fit_pars_initial = [
    mass[1]/mass[0],
    mass[2]/mass[0],
    period[1],
    period[2],
    ecc[1],
    ecc[2],
    argp[1],
    argp[2],
    meana[1],
    meana[2]
]

# this will be a global variable, it will not be passed to log-like/prob function
# using tight boundaries just for convenience
fit_boundaries =[
    [0.01*cst.Mjups, 1.0*cst.Mjups],
    [0.01*cst.Mjups, 1.0*cst.Mjups],
    [period[1]-1.0, period[1]+1.0],
    [period[2]-1.0, period[2]+1.0],
    [0.0, 0.5],
    [0.0, 0.5],
    [0.0, 360.0],
    [0.0, 360.0],
    [0.0, 360.0],
    [0.0, 360.0],
]

# let me define a set of parameter to test with random values
np.random.seed(seed=123456)
fit_pars_test = [
    uniform.rvs(loc=fit_boundaries[i][0], scale=np.ptp(fit_boundaries[i])) for i in range(0, n_fit)
]

def fitting_to_physical_params(fit_pars):

    m_x = np.zeros((n_body))
    r_x = radius.copy() # fixed
    p_x = m_x.copy()
    e_x = m_x.copy()
    w_x = m_x.copy()
    ma_x = m_x.copy()
    i_x = inc.copy() # fixed
    ln_x = longn.copy() # fixed
    
    m_x[0] = mass[0] # not fitting stellar mass, global variable
    ifit = 0
    m_x[1] = fit_pars[ifit] * mass[0]
    ifit += 1
    m_x[2] = fit_pars[ifit] * mass[0]
    
    ifit += 1
    p_x[1] = fit_pars[ifit]
    ifit += 1
    p_x[2] = fit_pars[ifit]
    
    ifit += 1
    e_x[1] = fit_pars[ifit]
    ifit += 1
    e_x[2] = fit_pars[ifit]
    
    ifit += 1
    w_x[1] = fit_pars[ifit]
    ifit += 1
    w_x[2] = fit_pars[ifit]
    
    ifit += 1
    ma_x[1] = fit_pars[ifit]
    ifit += 1
    ma_x[2] = fit_pars[ifit]
    
    return m_x, r_x, p_x, e_x, w_x, ma_x, i_x, ln_x 

def check_fitting_boundaries(fit_pars):

    for ifit, ibound in enumerate(fit_boundaries):
        p = fit_pars[ifit]
        if (p < ibound[0]) or ( p > ibound[1]):
            return False

    return True

def fitting_to_observables(fit_pars):

    m_fit, r_fit, p_fit, e_fit, w_fit, ma_fit, i_fit, ln_fit = fitting_to_physical_params(fit_pars)
    
    (
        body_flag_sim,
        epo_sim,
        transits_sim,
        durations_sim,
        lambda_rm_sim,
        kep_elem_sim,
        stable,
    ) = pytrades.kelements_to_observed_t0s(
        t_epoch,
        t_start,
        t_int,
        m_fit,
        r_fit,
        p_fit,
        e_fit,
        w_fit,
        ma_fit,
        i_fit,
        ln_fit,
        transit_flag
    )
    # print("kel to T0s")
    return (
        body_flag_sim,
        epo_sim,
        transits_sim,
        durations_sim,
        lambda_rm_sim,
        kep_elem_sim,
        stable,
    )

def fitting_to_observables_dict(fit_pars):

    (
        body_flag_sim,
        epo_sim,
        transits_sim,
        durations_sim,
        lambda_rm_sim,
        kep_elem_sim,
        stable,
    ) = fitting_to_observables(fit_pars)

    transits = {}
    for pl_letter, pl_number in zip(body_names[1:], [2,3]):
        sel_pl = body_flag_sim == pl_number
        n_tra = np.sum(sel_pl)
        print("planet {} (id {}) with {} transits in {:.0f} days of integration".format(pl_letter, pl_number, n_tra, t_int))
        transits[pl_letter] = {
            "n_transits":  n_tra,
            "transit_times": transits_sim[sel_pl],
            "transit_durations": durations_sim[sel_pl],
            "lambda_rm": lambda_rm_sim[sel_pl],
            "kep_elem": kep_elem_sim[sel_pl],
        }
    
    return transits
\end{lstlisting}

MCMC algorithms and MAP optimization are concerned with the shape of the posterior distribution and finding its mode(s),
not its absolute normalization.  
When you work with log-probabilities, 
adding a constant term to the log-posterior does not change the location of the maximum or 
the relative probabilities between different points in parameter space.  
So, in case of \texttt{emcee} the uniform priors 
($\ln_\mathrm{prior}(\theta) = \ln ( 1 / (\theta_\mathrm{max} - \theta_\mathrm{min}) )$) can be discarded,
and only Normal-Gaussian (or other types) priors can be computed in log-probability.  
It is mandatory to compute the log-prior for uniform priors in case of
Nested Sampling and/or model selection (Bayes factor) analysis.

\begin{lstlisting}[language=Python]
ln_const = -0.5*(n_tra_syn_b+n_tra_syn_c)*np.log(2.0*np.pi) # compute this only once

def log_boundaries(fit_pars):

    check_bounds = check_fitting_boundaries(fit_pars)
    if not check_bounds:
        return -np.inf
    return 0.0

def log_likelihood(fit_pars):

    lnL = ln_const
    
    (
        body_flag_sim,
        epo_sim,
        transits_sim,
        durations_sim,
        lambda_rm_sim,
        kep_elem_sim,
        stable,
    ) = fitting_to_observables(fit_pars)
    if not stable:
        return -np.inf

    res_b = tra_syn_noisy_b - transits_sim[body_flag_sim ==2]
    wres_b = res_b / err_tra_syn_b
    lnL_b = -0.5*np.sum(np.log(err_tra_syn_b)) - 0.5*np.sum(wres_b*wres_b)
    
    res_c = tra_syn_noisy_c - transits_sim[body_flag_sim ==3]
    wres_c = res_c / err_tra_syn_c
    lnL_c = -0.5*np.sum(np.log(err_tra_syn_c)) - 0.5*np.sum(wres_c*wres_c)
    
    lnL += lnL_b + lnL_c
    
    return lnL

def log_probability(fit_pars):

    ln_prior = log_boundaries(fit_pars)
    if np.isinf(ln_prior):
        return ln_prior

    lnP = log_likelihood(fit_pars)
    if np.isinf(lnP):
        return lnP
    lnP += ln_prior
    return lnP

# test them
lnP_0 = log_probability(fit_pars_initial)
print("logP = ",lnP_0) # 21.2320468
lnP = log_probability(fit_pars_test)
print("logP = ",lnP) # -23485070127.44914
\end{lstlisting}

One starts probing the parameter space with the Differential Evolution (DE) algorithm
implemented within \texttt{pyde}.
I have to set some hyper-parameters for DE and for print purpose.

\begin{lstlisting}[language=Python]
# f: the difference amplification factor. Values of 0.5-0.8 are good in most cases.
de_f = 0.5
# c: The cross-over probability. Use 0.9 to test for fast convergence, and smaller values (~0.1) for a more elaborate search.
de_c = 0.5
# -maximise (True) or minimise (False)
de_maximize = True
de_fit_type = -1 if de_maximize else 1
# n_pop: number of population, that is the number of different configuration to test at each generation (a step, or a run)
n_pop = n_fit * 4 # suggested n_pop(min) = n_fit * 2, n_pop(ok) = n_fit * 4, n_pop(good)= n_fit * 10
# n_gen: numer of generations, that is the number of evolution of the configurations, it stops when reach this number
n_gen = 5000 # < 10000 is a very low number of generation, just to show how it works
iter_print = n_gen // 10 # print and store every 10% of the total number of steps

seed = 42
n_threads = min(n_pop // 1, len(os.sched_getaffinity(0))) # just a test, usually // 2 is ok
# probably using newer version of python and os has the function os.process_cpu_count()

load_de = True # if you want just to load previous analysis without repeating, e.g. for plotting or getting the best-fit or last-population

print(
    "number of the population = {}".format(n_pop),
    "number of the generation = {}".format(n_gen),
    "number of the threads = {}".format(n_threads),
    "seed = {}".format(seed),
    sep="\n"
)
\end{lstlisting}

I am ready to run it as an iterator allowing it to store evolution of the parameters and
print progress.

\begin{lstlisting}[language=Python]
de_pop = np.zeros((n_gen, n_pop, n_fit))
de_fitness = np.zeros((n_gen, n_pop)) - np.inf
de_pop_best = np.zeros((n_gen, n_fit))
de_fitness_best = np.zeros((n_gen)) - np.inf

if load_de: # load previous analysis
    de_pop = pickle.load(open('de_pop.pkl', 'rb'))
    de_fitness = pickle.load(open('de_fitness.pkl', 'rb'))
    de_pop_best = pickle.load(open('de_pop_best.pkl', 'rb'))
    de_fitness_best = pickle.load(open('de_fitness_best.pkl', 'rb'))
else:
    with Pool(n_threads) as pool: # using parallel implementation with Pool, it works on shared memory (so no clusters)
        de_obj = pyde.DiffEvol(
            log_probability,
            fit_boundaries,
            n_pop,
            f=de_f,
            c=de_c,
            seed=seed,
            maximize=de_fit_type,
            pool=pool
        )
        for iter_de, res_de in enumerate(de_obj(n_gen)):
            de_pop[iter_de, :, :]    = de_obj.population.copy()
            de_fitness[iter_de, :]   = de_fit_type * de_obj._fitness.copy()
            de_pop_best[iter_de, :]  = de_obj.minimum_location.copy()
            de_fitness_best[iter_de] = de_fit_type * de_obj.minimum_value
        
            if ((iter_de + 1) % iter_print) == 0:
                print("Completed iter {:5d} / {:5d} ({:5.1f}%)".format(iter_de+1, n_gen, 100*(iter_de+1)/n_gen))
    
if not load_de: # store to files
    pickle.dump(de_pop, open('de_pop.pkl', 'wb'))
    pickle.dump(de_fitness, open('de_fitness.pkl', 'wb'))
    pickle.dump(de_pop_best, open('de_pop_best.pkl', 'wb'))
    pickle.dump(de_fitness_best, open('de_fitness_best.pkl', 'wb'))
\end{lstlisting}

\begin{figure}[!htb]
    \centering
    \includegraphics[width=\textwidth]{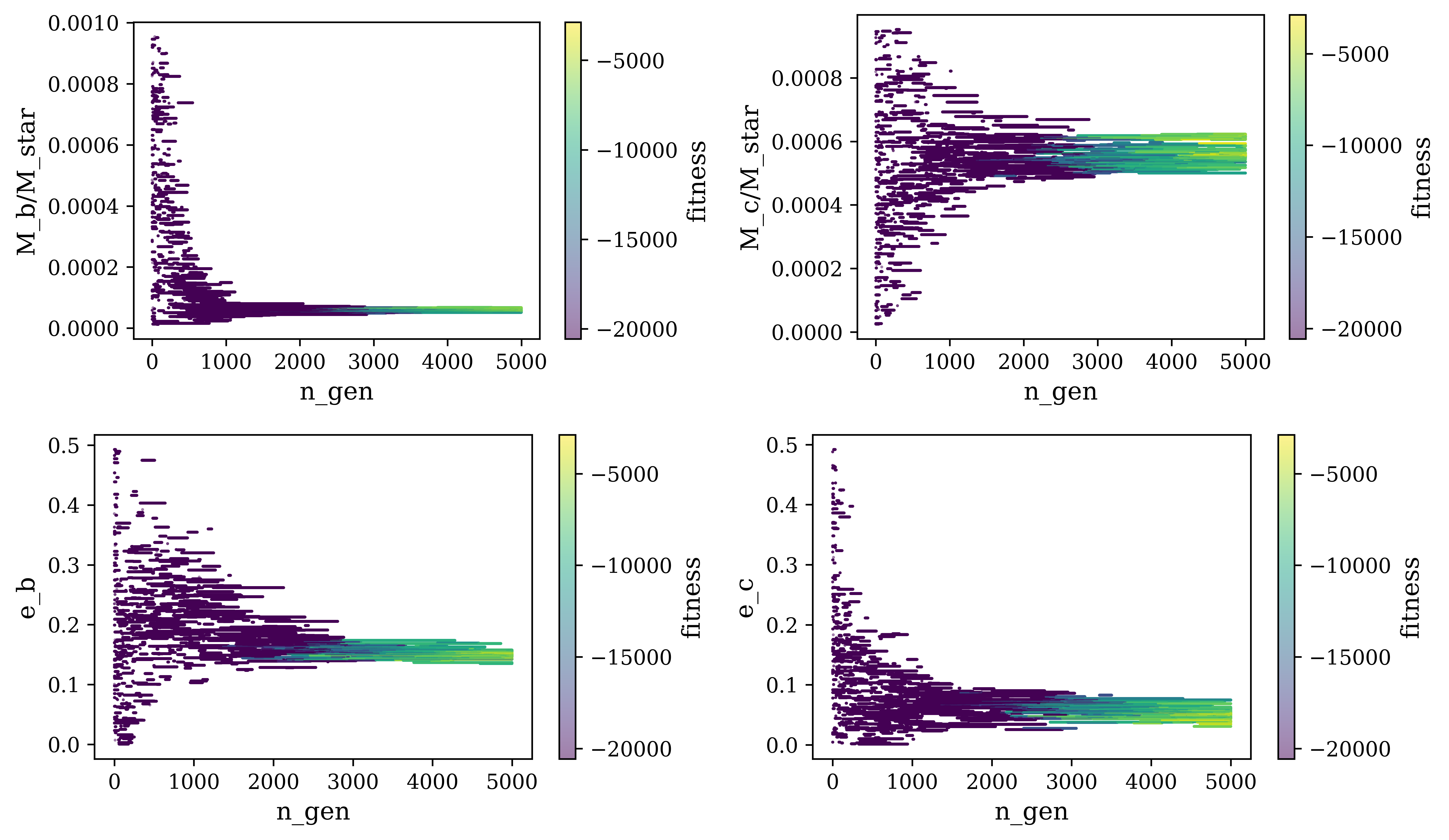}% no path, no extension
    \caption{
    \textbf{DE evolution.}
    Parameter evolution with DE implemented in \texttt{pyde}:
    \textit{upper-left}: $M_\mathrm{b}/M_\star$;
    \textit{upper-right}: $M_\mathrm{c}/M_\star$;
    \textit{lower-left}: $e_\mathrm{b}$;
    \textit{lower-right}: $e_\mathrm{c}$.
    }\label{da_pyde_evolution_1}
\end{figure}

I can check the evolution and the convergence of each parameter, 
see an example in Fig~\ref{da_pyde_evolution_1} of the evolution of the scaled masses and the eccentricities.

\begin{lstlisting}[language=Python]
skip_this = False # if you want to skip this, set it to True

if not skip_this:
    xx = np.repeat(np.arange(n_gen), n_pop)
    cc = de_fitness.reshape((n_gen * n_pop))
    
    for ifit in range(0,n_fit):
        fig = plt.figure(figsize=(5,3))
        yy = de_pop[:,:,ifit].reshape((n_gen * n_pop))
        plt.scatter(
            xx,
            yy,
            s=2,
            c=cc,
            vmin=np.percentile(cc, 50), # the minimum value will be the median of the fitness
            vmax=np.max(cc),
            alpha=0.5,
            edgecolors='None',
            linewidths=0.0,
        )
        plt.colorbar(label="fitness")
        plt.ylabel(fit_labels[ifit])
        plt.xlabel("n_gen")
        
        plt.tight_layout()
        plt.show()
        plt.close(fig)
\end{lstlisting}

Once \texttt{pyde} completed the run, it is possible to grab the population (the series of configurations)
of the last generation (or iteration) and pass it to \texttt{emcee} to probe
the posterior distribution in a Bayesian framework.
As for \texttt{pyde}, also \texttt{emcee} has some parameters and keywords to be fine-tuned
and set-up.

\begin{lstlisting}[language=Python]
n_walkers = n_pop # same as `pyde`, this is the number of chains
n_steps = 1000 # number of iteration to run
thin_by = 10 # apply a thinning factor while running, it means that emcee will run for n_steps x thin_by, but it will keep/return only n_steps values. 
# This factor helps reducing correlation of the posterior and can be used as a store parameter

load_emcee = True # as for pyde, if you want to load a previous analysis.

# define the move / sampling step
# default is the Affine-Invariant Ensemble MCMC (A.I.E.M.):
# emcee_move = [
#     (emcee.moves.StretchMove(), 1.0),
# ]
#
# for complex problems the emcee's author suggested sampling with DE80%+DESnooker20%
emcee_move = [
    (emcee.moves.DEMove(), 0.8),
    (emcee.moves.DESnookerMove(), 0.2),
]

# just for progress bar
pka = {
    'ncols': 75,
    'dynamic_ncols': False,
    'position': 0
}
\end{lstlisting}

So, run \texttt{emcee} with the last \texttt{pyde} population and store evolution
in an \texttt{hdf5} file every \texttt{thin\_by} steps.

\begin{lstlisting}[language=Python]
backend_filename = os.path.join(os.path.abspath("./"), "sampler.hdf5")

if load_emcee:
    sampler = emcee.backends.HDFBackend(backend_filename, read_only=True)
else:
    backend = emcee.backends.HDFBackend(backend_filename, compression="gzip")
    backend.reset(n_walkers, n_fit) # needed in case the file already exists and you don't care ...
    
    with Pool(n_threads) as pool:
        sampler = emcee.EnsembleSampler(
            n_walkers,
            n_fit,
            log_probability,
            pool=pool,
            moves=emcee_move,
            backend=backend
        )
    
        sampler.run_mcmc(
            de_pop[-1, :, :], # last population of pyde
            n_steps,
            thin_by=thin_by,
            store=True,
            tune=True,
            skip_initial_state_check=False,
            progress=True,
            progress_kwargs=pka
        )
\end{lstlisting}

Once the run is completed, 
I can extract the posterior distribution (after discarding a suitable burn-in period) and 
perform statistical analysis, 
such as obtaining the Maximum A Posteriori (MAP) estimate and 
Highest Density Interval (HDI), 
as well as comparing the results with existing literature.

\begin{lstlisting}[language=Python]
n_burnin = 400 # remove first n_burnin steps, change it accordingly to convergence
full_chains = sampler.get_chain() # (n_steps, n_walkers, n_fit)
full_chains_flat = sampler.get_chain(flat=True) # (n_steps*n_walkers, n_fit)
post_chains = sampler.get_chain(discard=n_burnin) # (n_steps-n_burnin, n_walkers, n_fit)
post_chains_flat = sampler.get_chain(discard=n_burnin, flat=True) # (n_post, n_fit)
full_lnprob = sampler.get_log_prob()
full_lnprob_flat = sampler.get_log_prob(flat=True)
post_lnprob = sampler.get_log_prob(discard=n_burnin)
post_lnprob_flat = sampler.get_log_prob(discard=n_burnin, flat=True)
n_post, _ = np.shape(post_chains_flat)  # dimension of the posterior distribution
map_idx = np.argmax(post_lnprob_flat) # index of the Maximum-a-Posteriori, that is the index of the max-lnP
map_lnprob = post_lnprob_flat[map_idx] # get the max of the lnP
map_pars = post_chains_flat[map_idx, :] # and the corrisponding parameter set
\end{lstlisting}

\begin{figure}[!htb]
    \centering
    \includegraphics[width=\textwidth]{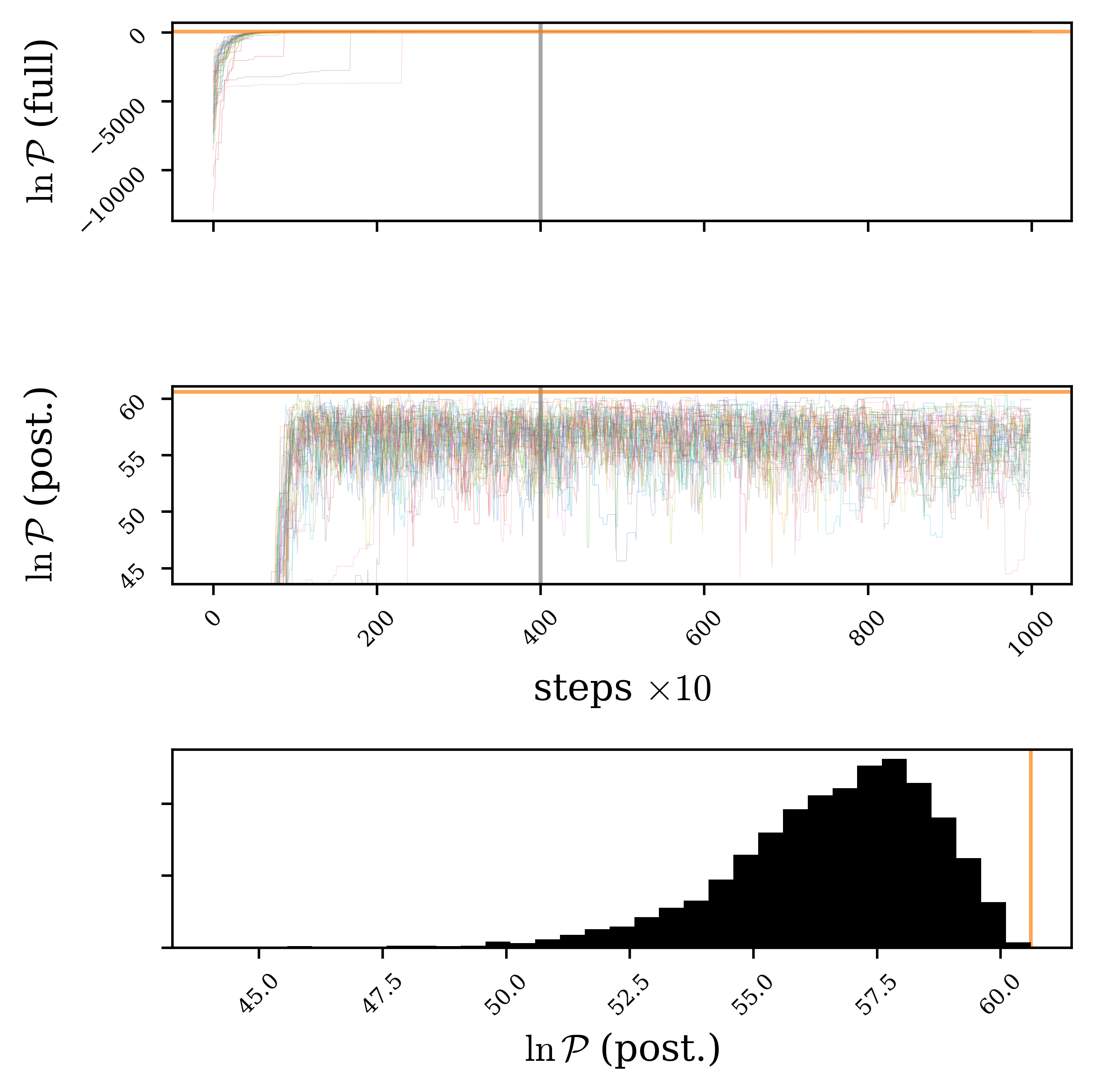}% no path, no extension
    \caption{
    \textbf{$\ln\mathcal{P}$ of \texttt{emcee}}
    Trace plots of the logarithm of the posterior probability ($\ln\mathcal{P}$).
    The top panel displays the full trace for all MCMC walkers.
    The middle panel shows the trace plot with the y-axis limited to the posterior phase, 
    following the burn-in period (indicated by the grey vertical line).
    The bottom panel presents the histogram of the $\ln\mathcal{P}$ distribution, 
    with its maximum value highlighted by an orange vertical line.
    }\label{da_emcee_lnP_1}
\end{figure}

It is important to check the convergence and behaviour of the $\ln\mathcal{P}$ and the chains of the parameters,
to fine-tuning the burn-in parameter and if the analysis needs more steps to reach convergence and stability.
I use some functions within \texttt{trades} to plot, firstly, the $\ln\mathcal{P}$, and monitor its evolution,
as can be seen in Fig.~\ref{da_emcee_lnP_1}.

\begin{lstlisting}[language=Python]
log_probability_trace(
    full_lnprob,
    post_lnprob_flat,
    None,
    n_burn=n_burnin,
    n_thin=thin_by,
    show_plot=True,
    figsize=(5, 5),
    olog=None,
)
\end{lstlisting}

\begin{figure}[!htb]
    \centering
    \includegraphics[width=\textwidth]{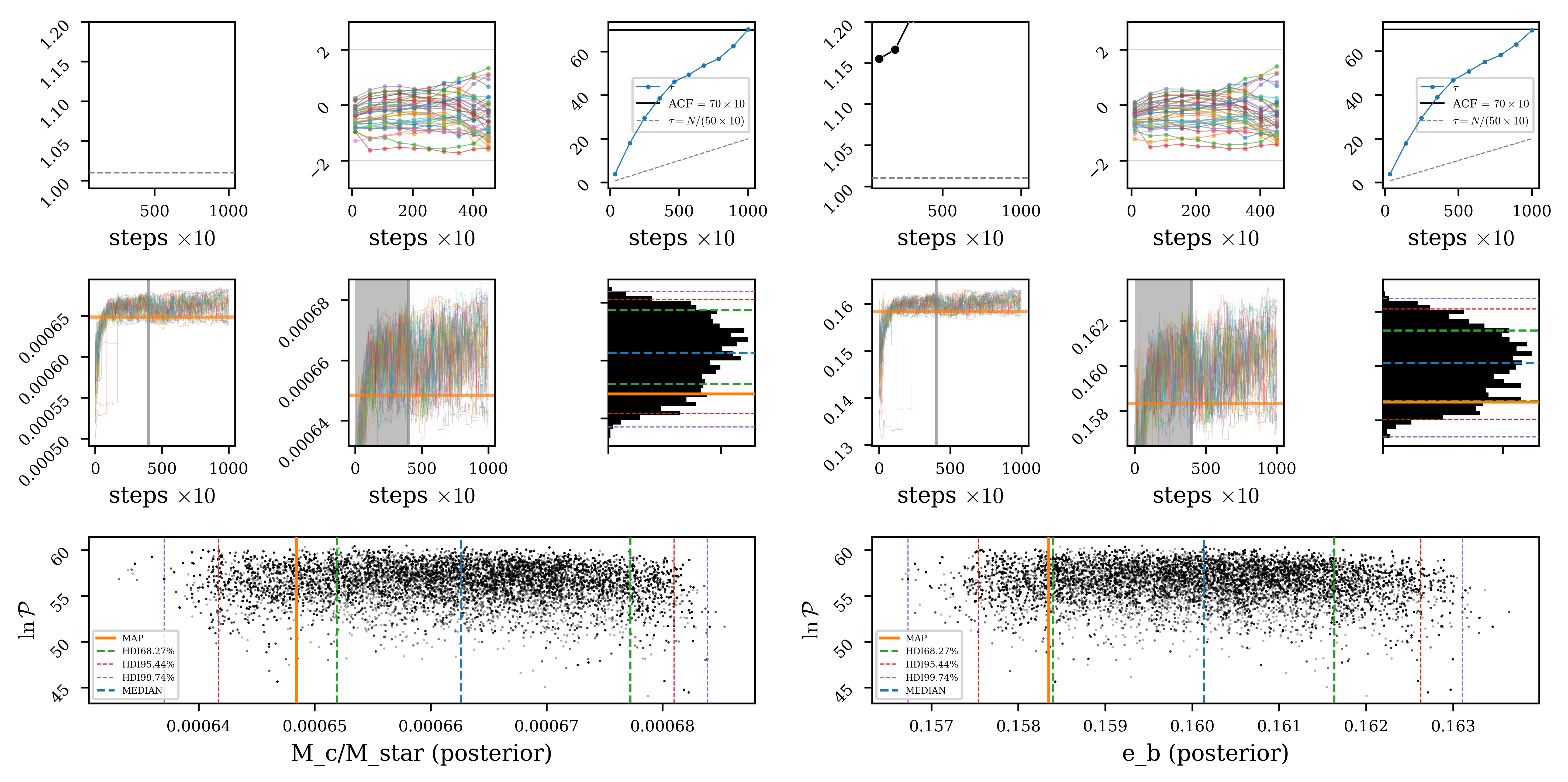}
    \caption{
    \textbf{Trace and convergence plots of the scaled mass of planet c and of the eccentricity of planet b.}
    An example of convergence and trace plots for 
    the scaled mass of planet c ($M_\mathrm{c}/M_\star$, \textit{left}) and
    the eccentricity of planet b ($e_\mathrm{c}$, \textit{right})
    from the \texttt{emcee} run.
    Each panel has sub-panels, divided in
    \textit{top sub-panels} (from left to right) with 
    the Gelman-Rubin statistic (with a convergence threshold of 1.01 indicated by the dashed line); 
    the Geweke diagnostic (where the y-axis represents the $z$-score, indicating deviation from a stationary distribution); 
    and the Auto-Correlation Function (ACF);
    \textit{middle sub-panels} showing the trace plot of fitted parameter in full (left),
    the trace plot limited to the posterior samples after the burn-in as the grey shaded area (centre);
    and the histogram of the posterior distribution (right);
    \textit{bottom sub-panel} illustrating the logarithm of the posterior probability ($\ln\mathcal{P}$) as a function of the fitted parameter, 
    with the vertical lines indicating the Highest Density Intervals (HDIs) at various confidence levels, 
    and the Maximum A Posteriori (MAP) estimate. 
    Note that in some cases, the MAP estimate may fall outside the 68.27\% HDI.
    }\label{da_emcee_traces_1}
\end{figure}

There are different way to monitor the convergence of the chains of the different parameters.
The main methods are the Gelman-Rubin statistics \citep{gelman1992}, 
Geweke statistics \citep{geweke1991},
the auto-correlation function \citep[ACF,][]{GoodmanWeare2010CAMCS...5...65G},
and a visual inspection.
However, keep in mind that these method are indicative of the convergence,
for example the Gelman-Rubin statistics has almost no meaning when using 
the affine-invariant and the differential-evolution samplers.
So, comparing the evolution and the stability of the $\ln\mathcal{P}$ and
the evolution of the parameters, is the only way to determine a proper burn-in and when halt the analysis.
As an example I show in Fig.~\ref{da_emcee_traces_1} the trace plot with different statistics for 
$M_\mathrm{c}/M_\star$ and $e_\mathrm{b}$.

\begin{lstlisting}[language=Python]
skip_this = False # if you want to skip the plots
if not skip_this:
    # let me use `trades` function to plot all the chains/trace plot of each parameter
    exp_acf_fit, exp_steps_fit = full_statistics(
        full_chains,
        post_chains_flat,
        fit_labels,
        map_pars,
        post_lnprob_flat,
        None,
        olog=None,
        ilast=0,
        n_burn=n_burnin,
        n_thin=thin_by,
        show_plot=True,
        figsize=(5, 5),
    )
\end{lstlisting}

Once the convergence of the MCMC chains has been evaluated, 
a process that often requires iterative refinement of the burn-in phase and 
potentially additional sampling (e.g., more steps or walkers), 
I extract the best-fit parameter (MAP) values and 
their associated Highest Density Intervals (HDIs) from the posterior distribution.

\begin{lstlisting}[language=Python]
# HDI or credible intervals
perc = np.array([68.27, 95.44, 99.74]) / 100.0
for i in range(n_fit):
    fitn, fitp, fitpost = fit_labels[i], map_pars[i], post_chains_flat[:, i]
    credint = [anc.hpd(fitpost, c) for c in perc]
    l = "{:12s}: MAP {:10.6f} ".format(fitn, fitp)
    for j, pc in enumerate(credint):
        l += "HDI@{:.2f}% [{:10.6f} , {:10.6f}] ".format(perc[j]*100, pc[0], pc[1])
    print(l)
print()

# conversion of parameters to physical: planetary masses

err_Mstar = 0.021 # Msun -- McKee
Mstar_gaussian = norm.rvs(loc=mass[0], scale=err_Mstar, size=n_post)

post_Mb2s_flat = post_chains_flat[:, 0] # posterior of Mb/Mstar
post_Mb_Me = post_Mb2s_flat * mass[0] * cst.Msear # physical posterior of Mb (in Mearth), multiplying only by Mstar, it is needed to take the MAP value
map_Mb_Me = post_Mb_Me[map_idx] # best-fit value of Mb
post_Mb_Me_noisy = post_Mb2s_flat * Mstar_gaussian * cst.Msear # needed to take into account the uncertainty on Mstar, only for the credible interval
credint = [anc.hpd(post_Mb_Me_noisy, c) for c in perc]
l = "{:12s}: MAP {:10.6f} ".format("M_b", map_Mb_Me)
for j, pc in enumerate(credint):
    l += "HDI@{:.2f}% [{:10.6f} , {:10.6f}] ".format(perc[j]*100, pc[0], pc[1])
print(l)
err_Mb_Me = np.ptp(credint[0])*0.5 # uncertainty on Mb as semi-interval, but it is not the only way

post_Mc2s_flat = post_chains_flat[:, 1] # Mc/Mstar
post_Mc_Me = post_Mc2s_flat * mass[0] * cst.Msear
map_Mc_Me = post_Mc_Me[map_idx]
post_Mc_Me_noisy = post_Mc2s_flat * Mstar_gaussian * cst.Msear
credint = [anc.hpd(post_Mc_Me_noisy, c) for c in perc]
l = "{:12s}: MAP {:10.6f} ".format("M_c", map_Mc_Me)
for j, pc in enumerate(credint):
    l += "HDI@{:.2f}% [{:10.6f} , {:10.6f}] ".format(perc[j]*100, pc[0], pc[1])
print(l)
err_Mc_Me = np.ptp(credint[0])*0.5

print()
print("This example:")
print("M_b = {:10.6f} +/- {:10.6f} Mearth".format(map_Mb_Me, err_Mb_Me))
print("M_c = {:10.6f} +/- {:10.6f} Mearth".format(map_Mc_Me, err_Mc_Me))

print("McKee values:")
print("M_b = {:10.6f} +/- {:10.6f} Mearth".format(0.0554*cst.Mjups*cst.Msear, 0.0020*cst.Mjups*cst.Msear))
print("M_c = {:10.6f} +/- {:10.6f} Mearth".format(0.525*cst.Mjups*cst.Msear, 0.019*cst.Mjups*cst.Msear))
# OUTPUT
# M_b/M_star  : MAP   0.000068 HDI@68.27% [  0.000069 ,   0.000071] HDI@95.44% [  0.000068 ,   0.000072] HDI@99.74% [  0.000067 ,   0.000072] 
# M_c/M_star  : MAP   0.000648 HDI@68.27% [  0.000652 ,   0.000677] HDI@95.44% [  0.000642 ,   0.000681] HDI@99.74% [  0.000637 ,   0.000684] 
# P_b         : MAP  17.099349 HDI@68.27% [ 17.098348 ,  17.099394] HDI@95.44% [ 17.097900 ,  17.099987] HDI@99.74% [ 17.097472 ,  17.100434] 
# P_c         : MAP  34.550492 HDI@68.27% [ 34.550569 ,  34.551060] HDI@95.44% [ 34.550239 ,  34.551185] HDI@99.74% [ 34.550025 ,  34.551420] 
# e_b         : MAP   0.158348 HDI@68.27% [  0.158395 ,   0.161635] HDI@95.44% [  0.157540 ,   0.162629] HDI@99.74% [  0.156731 ,   0.163107] 
# e_c         : MAP   0.011184 HDI@68.27% [  0.004435 ,   0.009974] HDI@95.44% [  0.003778 ,   0.012883] HDI@99.74% [  0.003012 ,   0.014204] 
# w_b         : MAP 292.402423 HDI@68.27% [291.348712 , 292.413278] HDI@95.44% [290.870440 , 292.800727] HDI@99.74% [290.575716 , 293.152739] 
# w_c         : MAP 238.254508 HDI@68.27% [226.302516 , 240.275229] HDI@95.44% [214.522333 , 243.078384] HDI@99.74% [208.168564 , 244.544909] 
# meana_b     : MAP 149.650477 HDI@68.27% [149.628592 , 150.876985] HDI@95.44% [149.235514 , 151.557372] HDI@99.74% [148.735685 , 151.905512] 
# meana_c     : MAP 150.122714 HDI@68.27% [148.379471 , 162.200700] HDI@95.44% [145.334463 , 173.789502] HDI@99.74% [143.750534 , 179.998041] 
# 
# M_b         : MAP  17.346028 HDI@68.27% [ 17.176905 ,  18.321979] HDI@95.44% [ 16.617973 ,  18.887549] HDI@99.74% [ 16.043096 ,  19.377235] 
# M_c         : MAP 164.726021 HDI@68.27% [162.903680 , 173.690149] HDI@95.44% [157.700301 , 179.112357] HDI@99.74% [152.198945 , 183.723807] 
# 
# This example:
# M_b =  17.346028 +/-   0.572537 Mearth
# M_c = 164.726021 +/-   5.393235 Mearth
# McKee values:
# M_b =  17.611338 +/-   0.635788 Mearth
# M_c = 166.894450 +/-   6.039990 Mearth
\end{lstlisting}

\begin{figure}[!htb]
    \centering
    \includegraphics[width=\textwidth]{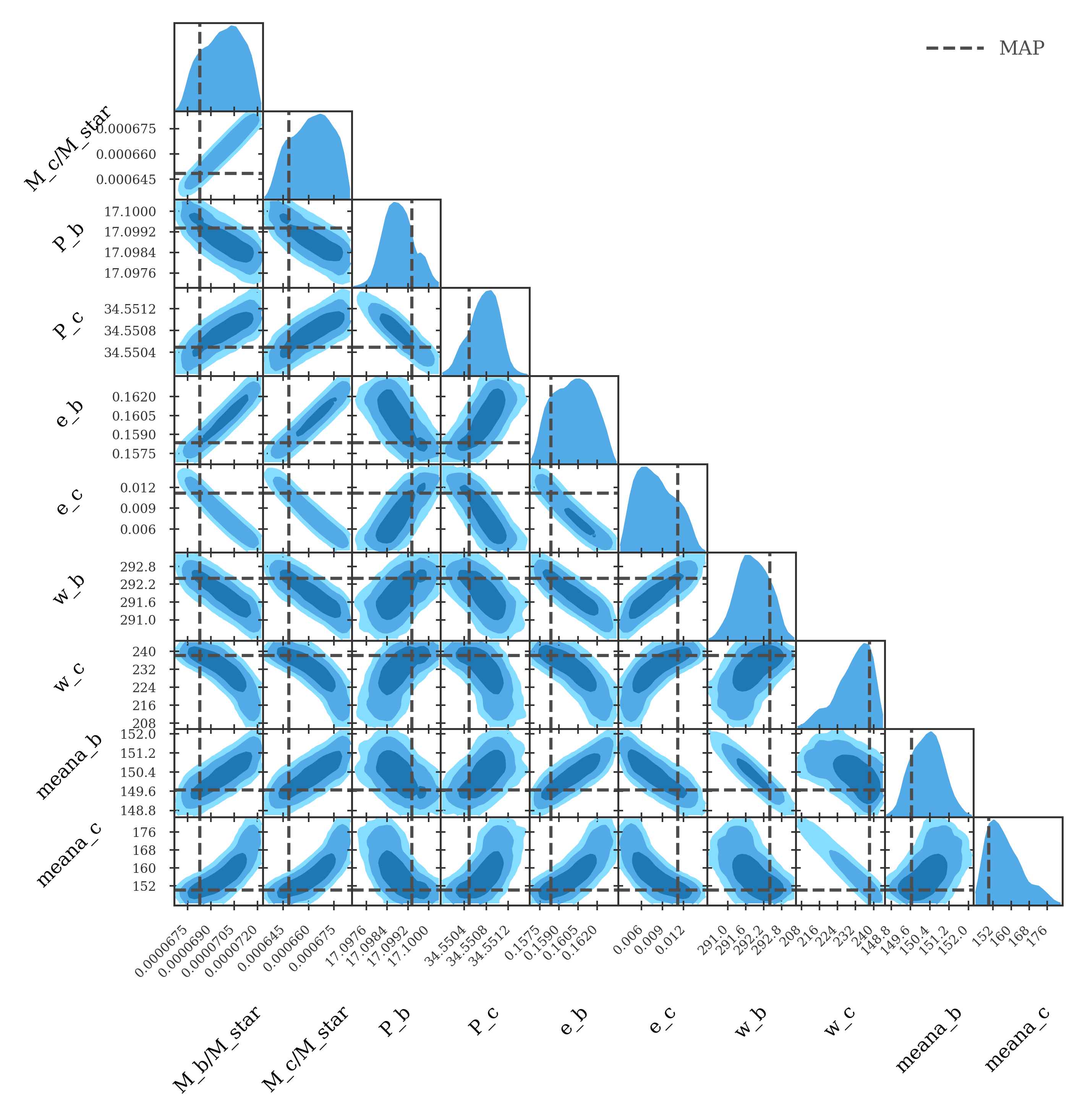}% no path, no extension
    \caption{
    \textbf{Corner plot of \texttt{emcee} run.}
    Triangle or corner plot of the posterior distribution from \texttt{emcee}
    showing possible correlation and the space probed by the analysis.
    }\label{da_corner_emcee_1}
\end{figure}

I can check correlation among parameters and which regions of the parameter space have been explored through the triangle or corner plot,
as in Fig.~\ref{da_corner_emcee_1}.

\begin{figure}[!htb]
    \centering
    \includegraphics[width=\textwidth]{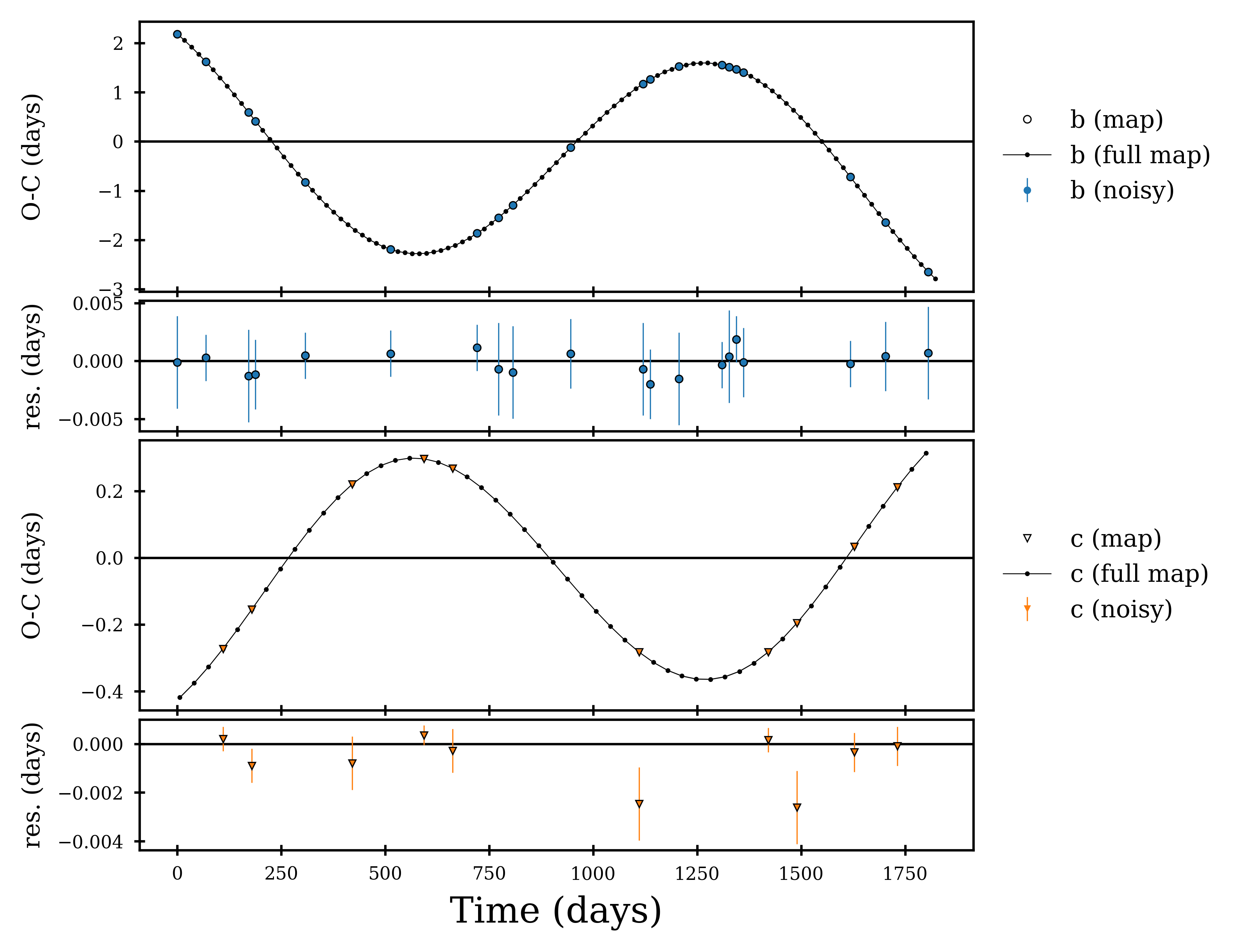}% no path, no extension
    \caption{
    \textbf{$O-C$ diagram of \texttt{emcee} run.}
    Observed-Calculated ($O-C$) diagrams for planet b (top two panels) and planet c (bottom two panels), 
    featuring models derived from the Maximum A Posteriori (MAP) configuration obtained from the \texttt{emcee} run.
    For each planet there are two panels, 
    the upper one displays the $O-C$ diagram: 
    synthetic-observed transit times are shown as blue-filled circles (for planet b) and orange-filled triangles (for planet c); 
    the corresponding MAP model values are depicted as open markers; 
    and the full analytical model (encompassing all transits, including unobserved ones) is represented by a black line with black dots.
    Although 1, 2, and 3$\sigma$ confidence regions from the random samples were plotted as shades of grey, 
    they are not visibly discernible here. 
    This is due to the parameters being well-constrained and 
    the intrinsic amplitude of the associated $O-C$ being very low compared to the amplitude of the TTV signal.
    The lower panel for each planet shows the residuals, representing the difference between
    the synthetic-observed transit times and those computed using the MAP solution.
    }\label{da_oc_emcee_1}
\end{figure}

I want to plot the results in an $O-C$ diagram (see Fig.~\ref{da_oc_emcee_1}) 
with the MAP as the best-fit model
and a draw of random samples to show shaded area corresponding to 1, 2, and 3$\sigma$ regions.
Firstly, I need to extract the transit times for the MAP solution (for the observed transit times,
and all the transits in the integration time).

\begin{lstlisting}[language=Python]
map_transits = fitting_to_observables_dict(map_pars) # get transits exactly at the synthetic-observed times
(
    mass_map, 
    radius_map,
    period_map, 
    ecc_map, 
    argp_map, 
    meana_map,
    inc_map,
    longn_map,
) = fitting_to_physical_params(map_pars) # convert the parameters into physical ones
t_int_syn = time_sel_end # just integrates for the time span by the observations
_, _, _, map_transits_full = run_and_get_transits(
    t_epoch, 
    t_start, 
    t_int_syn,
    mass_map,
    radius_map,
    period_map, 
    ecc_map,
    argp_map, 
    meana_map,
    inc_map,
    longn_map,
    body_names[1:],
) # get all the transits of all planets from t_start and t_start + t_int_syn
\end{lstlisting}

Secondly, I draw random samples from the posterior distribution and get the full models.

\begin{lstlisting}[language=Python]
n_samples = 33 # number of samples to draw
smp_tra = {}
smp_idx = np.random.choice(n_post, n_samples, replace=False) # get indices, without repetitions 
for ismp in smp_idx:
    smp_pars = post_chains_flat[ismp, :]
    (
        mass_smp, 
        radius_smp,
        period_smp, 
        ecc_smp, 
        argp_smp, 
        meana_smp,
        inc_smp,
        longn_smp,
    ) = fitting_to_physical_params(smp_pars)
    _, _, _, smp_transits_full = run_and_get_transits(
        t_epoch, 
        t_start, 
        t_int_syn,
        mass_smp,
        radius_smp,
        period_smp, 
        ecc_smp,
        argp_smp, 
        meana_smp,
        inc_smp,
        longn_smp,
        body_names[1:],
    )
    smp_tra[ismp] = smp_transits_full
\end{lstlisting}

Finally, I can plot the $O-C$ diagram combining MAP and random samples, as can be seen in Fig.~\ref{da_oc_emcee_1}.

\begin{lstlisting}[language=Python]
fig = plt.figure(figsize=(5,5))
fig.subplots_adjust(hspace=0.07, wspace=0.25)

c1, c2, c3 = 0.6827, 0.9544, 0.9974
hc1, hc2, hc3 = c1*0.5, c2*0.5, c3*0.5

lfont = 8
tfont = 6

# zorders for plot management
zo_map = 10
zo_obs = zo_map-1
zo_mod = zo_obs -1
zo_1s = zo_mod - 1
zo_2s = zo_1s - 1
zo_3s = zo_2s - 1

cfsm = plt.get_cmap("gray")
gval = 0.6
dg = 0.1

axs = []
nrows = 6 # (2 + 1) * 2
ncols = 1

u = [1.0, "days"]
markers = anc.filled_markers

all_xlims = []

# =================================================================
i, pl_letter = 0, "b"
print("planet {}".format(pl_letter))

# --- O-C
ax = plt.subplot2grid((nrows, ncols), (0, 0), rowspan=2)
poc.set_axis_default(ax, ticklabel_size=tfont, aspect="auto", labeldown=False)
ax.set_ylabel("O-C ({:s})".format(u[1]), fontsize=lfont)
ax.axhline(0, color="k", lw=0.8)

(Tref_b, Pref_b) = lineph_syn_noisy[pl_letter]
_, _, chi2_b, epoch_b, Tlin_b, oc_b= linear_ephemeris(
    tra_syn_noisy_b, eT0s=err_tra_syn_b, Tref_in = Tref_b, Pref_in = Pref_b, fit=False
)
print("Tref_b: {:12.6f} Pref_b: {:12.6f}".format(Tref_b[0], Pref_b[0]))

tra_map_b = map_transits[pl_letter]["transit_times"]
epo_map_b = compute_epoch(Tref_b[0], Pref_b[0], tra_map_b)
tln_map_b = Tref_b[0] + epo_map_b*Pref_b[0]
oc_map_b = tra_map_b - tln_map_b
res_map_b = tra_syn_noisy_b - tra_map_b

tra_map_full_b = map_transits_full[pl_letter]["transit_times"]
epo_map_full_b = compute_epoch(Tref_b[0], Pref_b[0], tra_map_full_b)
tln_map_full_b = Tref_b[0] + epo_map_full_b*Pref_b[0]
oc_map_full_b = tra_map_full_b - tln_map_full_b

ax.errorbar(
    tra_syn_noisy_b,
    oc_b*u[0],
    yerr=err_tra_syn_b*u[0],
    marker=markers[0],
    ms=2.5,
    mec='None',
    mew=0.4,
    color="C0",
    ecolor="C0",
    elinewidth=0.4,
    capsize=0,
    label="{} (noisy)".format(pl_letter),
    ls='',
    zorder=zo_obs
)

ax.plot(
    tra_syn_noisy_b,
    oc_map_b*u[0],
    color="black",
    marker=markers[0],
    ms=2.5,
    mfc='None',
    mew=0.4,
    label="{} (map)".format(pl_letter),
    ls='',
    zorder=zo_map
)

ax.plot(
    tra_map_full_b,
    oc_map_full_b*u[0],
    color="black",
    marker='o',
    ms=0.6,
    ls='-',
    lw=0.3,
    label="{} (full map)".format(pl_letter),
    zorder=zo_mod
)

oc_smp = []
for ksmp, vsmp in smp_tra.items():
    tra_xxx = vsmp[pl_letter]["transit_times"]
    epo_xxx = compute_epoch(Tref_b[0], Pref_b[0], tra_xxx)
    tln_xxx = Tref_b[0] + epo_xxx*Pref_b[0]
    oc_xxx = tra_xxx - tln_xxx
    oc_smp.append(oc_xxx)
oc_smp = np.array(oc_smp).T * u[0]
hdi1 = np.percentile(oc_smp, [50 - (100*hc1), 50 + (100*hc1)], axis=1).T
hdi2 = np.percentile(oc_smp, [50 - (100*hc2), 50 + (100*hc2)], axis=1).T
hdi3 = np.percentile(oc_smp, [50 - (100*hc3), 50 + (100*hc3)], axis=1).T
ax.fill_between(
    tra_map_full_b,
    hdi1[:, 0],
    hdi1[:, 1],
    color=cfsm(gval),
    alpha=1.0,
    lw=0.0,
    zorder=zo_1s,
)
ax.fill_between(
    tra_map_full_b,
    hdi2[:, 0],
    hdi2[:, 1],
    color=cfsm(gval+dg),
    alpha=1.0,
    lw=0.0,
    zorder=zo_2s,
)
ax.fill_between(
    tra_map_full_b,
    hdi3[:, 0],
    hdi3[:, 1],
    color=cfsm(gval+(dg*2)),
    alpha=1.0,
    lw=0.0,
    zorder=zo_3s,
)

all_xlims.append(ax.get_xlim())

axs.append(ax)

# --- RESIDUALS
ax = plt.subplot2grid((nrows, ncols), (2, 0), rowspan=1)
poc.set_axis_default(ax, ticklabel_size=tfont, aspect="auto", labeldown=False)
ax.set_ylabel("res. ({:s})".format(u[1]), fontsize=lfont)
ax.axhline(0, color="k", lw=0.8)

ax.errorbar(
    tra_syn_noisy_b,
    res_map_b*u[0],
    yerr=err_tra_syn_b*u[0],
    marker=markers[0],
    ms=2.5,
    mec='black',
    mew=0.4,
    color="C0",
    ecolor="C0",
    elinewidth=0.4,
    capsize=0,
    # label="{} (noisy)".format(pl_letter),
    ls='',
    zorder=zo_obs
)

axs.append(ax)

# =================================================================
i, pl_letter = 1, "c"
print("planet {}".format(pl_letter))

# --- O-C
ax = plt.subplot2grid((nrows, ncols), (3, 0), rowspan=2)
poc.set_axis_default(ax, ticklabel_size=tfont, aspect="auto", labeldown=False)
ax.set_ylabel("O-C ({:s})".format(u[1]), fontsize=lfont)
ax.axhline(0, color="k", lw=0.8)

(Tref_c, Pref_c) = lineph_syn_noisy[pl_letter]
_, _, chi2_c, epoch_c, Tlin_c, oc_c= linear_ephemeris(
    tra_syn_noisy_c, eT0s=err_tra_syn_c, Tref_in = Tref_c, Pref_in = Pref_c, fit=False
)
print("Tref_c: {:12.6f} Pref_c: {:12.6f}".format(Tref_c[0], Pref_c[0]))

tra_map_c = map_transits[pl_letter]["transit_times"]
epo_map_c = compute_epoch(Tref_c[0], Pref_c[0], tra_map_c)
tln_map_c = Tref_c[0] + epo_map_c*Pref_c[0]
oc_map_c = tra_map_c - tln_map_c
res_map_c = tra_syn_noisy_c - tra_map_c

tra_map_full_c = map_transits_full[pl_letter]["transit_times"]
epo_map_full_c = compute_epoch(Tref_c[0], Pref_c[0], tra_map_full_c)
tln_map_full_c = Tref_c[0] + epo_map_full_c*Pref_c[0]
oc_map_full_c = tra_map_full_c - tln_map_full_c

ax.errorbar(
    tra_syn_noisy_c,
    oc_c*u[0],
    yerr=err_tra_syn_c*u[0],
    marker=markers[1],
    ms=2.5,
    mec='None',
    mew=0.4,
    color="C1",
    ecolor="C1",
    elinewidth=0.4,
    capsize=0,
    label="{} (noisy)".format(pl_letter),
    ls='',
    zorder=zo_obs
)

ax.plot(
    tra_syn_noisy_c,
    oc_map_c*u[0],
    color="black",
    marker=markers[1],
    ms=2.5,
    mfc='None',
    mew=0.4,
    label="{} (map)".format(pl_letter),
    ls='',
    zorder=zo_map
)

ax.plot(
    tra_map_full_c,
    oc_map_full_c*u[0],
    color="black",
    marker='o',
    ms=0.6,
    ls='-',
    lw=0.3,
    label="{} (full map)".format(pl_letter),
    zorder=zo_mod
)

oc_smp = []
for ksmp, vsmp in smp_tra.items():
    tra_xxx = vsmp[pl_letter]["transit_times"]
    epo_xxx = compute_epoch(Tref_c[0], Pref_c[0], tra_xxx)
    tln_xxx = Tref_c[0] + epo_xxx*Pref_c[0]
    oc_xxx = tra_xxx - tln_xxx
    oc_smp.append(oc_xxx)
oc_smp = np.array(oc_smp).T * u[0]
hdi1 = np.percentile(oc_smp, [50 - (100*hc1), 50 + (100*hc1)], axis=1).T
hdi2 = np.percentile(oc_smp, [50 - (100*hc2), 50 + (100*hc2)], axis=1).T
hdi3 = np.percentile(oc_smp, [50 - (100*hc3), 50 + (100*hc3)], axis=1).T
ax.fill_between(
    tra_map_full_c,
    hdi1[:, 0],
    hdi1[:, 1],
    color=cfsm(gval),
    alpha=1.0,
    lw=0.0,
    zorder=zo_1s,
)
ax.fill_between(
    tra_map_full_c,
    hdi2[:, 0],
    hdi2[:, 1],
    color=cfsm(gval+dg),
    alpha=1.0,
    lw=0.0,
    zorder=zo_2s,
)
ax.fill_between(
    tra_map_full_c,
    hdi3[:, 0],
    hdi3[:, 1],
    color=cfsm(gval+(dg*2)),
    alpha=1.0,
    lw=0.0,
    zorder=zo_3s,
)

all_xlims.append(ax.get_xlim())

axs.append(ax)

# --- RESIDUALS
ax = plt.subplot2grid((nrows, ncols), (5, 0), rowspan=1)
poc.set_axis_default(ax, ticklabel_size=tfont, aspect="auto", labeldown=True)
ax.set_ylabel("res. ({:s})".format(u[1]), fontsize=lfont)
ax.axhline(0, color="k", lw=0.8)

ax.errorbar(
    tra_syn_noisy_c,
    res_map_c*u[0],
    yerr=err_tra_syn_c*u[0],
    marker=markers[1],
    ms=2.5,
    mec='black',
    mew=0.4,
    color="C1",
    ecolor="C1",
    elinewidth=0.4,
    capsize=0,
    ls='',
    zorder=zo_obs
)

axs.append(ax)

all_xlims = np.concatenate(all_xlims)
for ax in axs:
    ax.set_xlim(np.min(all_xlims), np.max(all_xlims))
    ax.legend(loc='center left', bbox_to_anchor =(1.01, 0.5), fontsize=lfont, frameon=False)
ax.set_xlabel("Time (days)")

fig.align_ylabels(axs)
plt.show()
plt.close(fig)
\end{lstlisting}

This was a quick-and-dirty analysis, because I ran \texttt{pyde} and \texttt{emcee}
with a low number of steps and the chains were not well mixed and did not reached convergence and stability.
Now, I will try the same example combining \texttt{pyde} and \texttt{emcee}, but using a different parameterisation.
In particular, 
I will also fit for the mass of the star with a Gaussian prior,
a combination of the masses of the planets as 
$\log_{10}M_\mathrm{b}/M_\star$ and $\log_{10}M_\mathrm{c}/M_\mathrm{b}$
to take into account the information from the anti-correlated signal presented in 
eq.~\ref{eq:ttv_mmr} of Section~\ref{Method_params} and in a log-scale to better probe smaller values,
the eccentricity and argument of pericentres as $\sqrt{e}\cos\omega$ and $\sqrt{e}\sin\omega$, and
the mean longitude $\lambda = \mathcal{M} + \omega + \Omega$.
Here, I will present just the parameter definition and conversion to physical ones, the log-priors, log-likelihood,
and the log-probability, because the \texttt{pyde} and \texttt{emcee} are exactly the same.

\begin{lstlisting}[language=Python]
# this will be a global variable, it will not be passed to log-like/prob function
fit_labels_repar = [
    "M_star",
    "log10(M_b/M_star)", 
    "log10(M_c/M_b)",
    "P_b", 
    "P_c",
    "secosw_b", 
    "sesinw_b", 
    "secosw_c",
    "sesinw_c",
    "meanl_b", 
    "meanl_c",
]
n_fit_repar = len(fit_labels_repar)
print("Number of fitting parameters = {}".format(n_fit_repar))

# let me define an initial set of parameters
fit_pars_initial_repar = [
    mass[0],
    np.log10(mass[1]/mass[0]),
    np.log10(mass[2]/mass[1]),
    period[1],
    period[2],
    np.sqrt(ecc[1])*np.cos(argp[1]*cst.deg2rad),
    np.sqrt(ecc[1])*np.sin(argp[1]*cst.deg2rad),
    np.sqrt(ecc[2])*np.cos(argp[2]*cst.deg2rad),
    np.sqrt(ecc[2])*np.sin(argp[2]*cst.deg2rad),
    meanl[1],
    meanl[2]
]

# let me define physical boundaries
M_s_boundaries = [0.01, 2.0] # stellar mass
M_p_boundaries = [0.01*cst.Mjups, 1.0*cst.Mjups] # planetary mass
ecc_boundaries = [0.0, 0.5]
ang_boundaries = [0.0, 360.0] # angles

# this will be a global variable, it will not be passed to log-like/prob function
# using a slightly larger boundaries for the periods (+/-2 days instead of +/-1)
fit_boundaries_repar = [
    M_s_boundaries, # Msun
    [-10, 0], # 10^-15, 1 as Mb/Ms
    [-10, 3], # 10^-15, 1 as Mc/Mb
    [period[1]-2.0, period[1]+2.0],
    [period[2]-2.0, period[2]+2.0],
    [-np.sqrt(np.max(ecc_boundaries[1])), +np.sqrt(np.max(ecc_boundaries[1]))],
    [-np.sqrt(np.max(ecc_boundaries[1])), +np.sqrt(np.max(ecc_boundaries[1]))],
    [-np.sqrt(np.max(ecc_boundaries[1])), +np.sqrt(np.max(ecc_boundaries[1]))],
    [-np.sqrt(np.max(ecc_boundaries[1])), +np.sqrt(np.max(ecc_boundaries[1]))],
    ang_boundaries, # meanl_b
    ang_boundaries, # meanl_c
]

priors_repar = [
    norm(loc=0.763, scale=0.021)
] + [
    uniform(loc=bd[0], scale=np.ptp(bd)) for bd in fit_boundaries_repar[1:]
]

def fitting_to_physical_params_repar(fit_pars):

    m_x = np.zeros((n_body))
    r_x = radius.copy() # fixed
    p_x = np.zeros((n_body))
    e_x = np.zeros((n_body))
    w_x = np.zeros((n_body))
    ma_x = np.zeros((n_body))
    i_x = inc.copy() # fixed
    ln_x = longn.copy() # fixed
    
    ifit = 0
    m_x[0] = fit_pars[ifit]
    ifit += 1
    m_x[1] = 10**(fit_pars[ifit]) * m_x[0]
    ifit += 1
    m_x[2] = 10**(fit_pars[ifit]) * m_x[1]
    
    ifit += 1
    p_x[1] = fit_pars[ifit]
    ifit += 1
    p_x[2] = fit_pars[ifit]
    
    ifit += 1
    e_x[1] = fit_pars[ifit]**2 + fit_pars[ifit+1]**2
    w_x[1] = (np.arctan2(fit_pars[ifit+1], fit_pars[ifit])*cst.rad2deg)%360.0
    
    ifit += 2
    e_x[2] = fit_pars[ifit]**2 + fit_pars[ifit+1]**2
    w_x[2] = (np.arctan2(fit_pars[ifit+1], fit_pars[ifit])*cst.rad2deg)%360.0
    
    ifit += 2
    ma_x[1] = (fit_pars[ifit] - w_x[1] - ln_x[1])%360.0
    ifit += 1
    ma_x[2] = (fit_pars[ifit] - w_x[2] - ln_x[2])%360.0
    
    return m_x, r_x, p_x, e_x, w_x, ma_x, i_x, ln_x 

def check_fitting_boundaries_repar(fit_pars):

    for ifit, ibound in enumerate(fit_boundaries_repar):
        p = fit_pars[ifit]
        if (p < ibound[0]) or ( p > ibound[1]):
            return False
    
    return True
    
def fitting_to_observables_repar(fit_pars):

    m_fit, r_fit, p_fit, e_fit, w_fit, ma_fit, i_fit, ln_fit = fitting_to_physical_params_repar(fit_pars)
    
    (
        body_flag_sim,
        epo_sim,
        transits_sim,
        durations_sim,
        lambda_rm_sim,
        kep_elem_sim,
        stable,
    ) = pytrades.kelements_to_observed_t0s(
        t_epoch,
        t_start,
        t_int,
        m_fit,
        r_fit,
        p_fit,
        e_fit,
        w_fit,
        ma_fit,
        i_fit,
        ln_fit,
        transit_flag
    )
    return (
        body_flag_sim,
        epo_sim,
        transits_sim,
        durations_sim,
        lambda_rm_sim,
        kep_elem_sim,
        stable,
    )

def fitting_to_observables_repar_dict(fit_pars):

    (
        body_flag_sim,
        epo_sim,
        transits_sim,
        durations_sim,
        lambda_rm_sim,
        kep_elem_sim,
        stable,
    ) = fitting_to_observables_repar(fit_pars)

    transits = {}
    for pl_letter, pl_number in zip(body_names[1:], [2,3]):
        sel_pl = body_flag_sim == pl_number
        n_tra = np.sum(sel_pl)
        transits[pl_letter] = {
            "n_transits":  n_tra,
            "transit_times": transits_sim[sel_pl],
            "transit_durations": durations_sim[sel_pl],
            "lambda_rm": lambda_rm_sim[sel_pl],
            "kep_elem": kep_elem_sim[sel_pl],
        }
    
    return transits

ln_const = -0.5*(n_tra_syn_b+n_tra_syn_c)*np.log(2.0*np.pi) # it computes this only once

def log_boundaries_repar(fit_pars):

    check_bounds = check_fitting_boundaries_repar(fit_pars)
    if not check_bounds:
        return -np.inf

    m_x, r_x, p_x, e_x, w_x, ma_x, i_x, ln_x = fitting_to_physical_params_repar(fit_pars)
    if not M_s_boundaries[0] <= m_x[0] <= M_s_boundaries[1]:
        return -np.inf
    if not M_p_boundaries[0] <= m_x[1] <= M_p_boundaries[1]:
        return -np.inf
    if not M_p_boundaries[0] <= m_x[2] <= M_p_boundaries[1]:
        return -np.inf
    if not ecc_boundaries[0] <= e_x[1] <= ecc_boundaries[1]:
        return -np.inf
    if not ecc_boundaries[0] <= e_x[2] <= ecc_boundaries[1]:
        return -np.inf
    
    return 0.0

def log_priors_repar(fit_pars): # this can be extended for uniform priors when using Nested Sampling

    ln_prior = 0.0

    ms = fit_pars[0]
    
    return priors_repar[0].logpdf(ms)

def log_likelihood_repar(fit_pars):

    lnL = ln_const
    
    (
        body_flag_sim,
        epo_sim,
        transits_sim,
        durations_sim,
        lambda_rm_sim,
        kep_elem_sim,
        stable,
    ) = fitting_to_observables_repar(fit_pars)
    if not stable:
        return -np.inf

    res_b = tra_syn_noisy_b - transits_sim[body_flag_sim ==2]
    wres_b = res_b / err_tra_syn_b
    lnL_b = -0.5*np.sum(np.log(err_tra_syn_b)) - 0.5*np.sum(wres_b*wres_b)
    
    res_c = tra_syn_noisy_c - transits_sim[body_flag_sim ==3]
    wres_c = res_c / err_tra_syn_c
    lnL_c = -0.5*np.sum(np.log(err_tra_syn_c)) - 0.5*np.sum(wres_c*wres_c)
    
    lnL += lnL_b + lnL_c
    
    return lnL

def log_probability_repar(fit_pars):

    ln_prior = log_boundaries_repar(fit_pars)
    if np.isinf(ln_prior):
        return ln_prior

    ln_prior += log_priors_repar(fit_pars)

    lnP = log_likelihood_repar(fit_pars)
    if np.isinf(lnP):
        return lnP
    lnP += ln_prior
    return lnP
\end{lstlisting}

As before, from the posterior with 400 steps of burn-in 
I compute the statistics as the MAP and the HDI,
and convert the fitted parameters to physical ones.

\begin{lstlisting}[language=Python]
# HDI
perc = np.array([68.27, 95.44, 99.74]) / 100.0
for i in range(n_fit):
    fitn, fitp, fitpost = fit_labels_repar[i], map_pars_repar[i], post_chains_flat_repar[:, i]
    credint = [anc.hpd(fitpost, c) for c in perc]
    l = "{:18s}: MAP {:10.6f} ".format(fitn, fitp)
    for j, pc in enumerate(credint):
        l += "HDI@{:.2f}% [{:10.6f} , {:10.6f}] ".format(perc[j]*100, pc[0], pc[1])
    print(l)
print()

# conversion of parameters

post_Ms_flat_repar = post_chains_flat_repar[:, 0] # stellar mass

post_l10Mb2s_flat_repar = post_chains_flat_repar[:, 1] # log10(Mb/Mstar)
post_Mb_Me_repar = 10**(post_l10Mb2s_flat_repar) * post_Ms_flat_repar * cst.Msear
map_Mb_Me_repar = post_Mb_Me_repar[map_idx_repar]

credint = [anc.hpd(post_Mb_Me_repar, c) for c in perc]
l = "{:18s}: MAP {:10.6f} ".format("M_b", map_Mb_Me_repar)
for j, pc in enumerate(credint):
    l += "HDI@{:.2f}% [{:10.6f} , {:10.6f}] ".format(perc[j]*100, pc[0], pc[1])
print(l)
err_Mb_Me_repar = np.ptp(credint[0])*0.5 # semi-interval, but it is not the only solution

post_l10Mc2b_flat_repar = post_chains_flat_repar[:, 2] # log10(Mc/Mb)
post_Mc_Me_repar = 10**(post_l10Mc2b_flat_repar) * post_Mb_Me_repar
map_Mc_Me_repar = post_Mc_Me_repar[map_idx_repar]

credint = [anc.hpd(post_Mc_Me_repar, c) for c in perc]
l = "{:18s}: MAP {:10.6f} ".format("M_c", map_Mc_Me_repar)
for j, pc in enumerate(credint):
    l += "HDI@{:.2f}% [{:10.6f} , {:10.6f}] ".format(perc[j]*100, pc[0], pc[1])
print(l)
err_Mc_Me_repar = np.ptp(credint[0])*0.5 # semi-interval, but it is not the only solution

icw = fit_labels_repar.index("secosw_b")
isw = icw + 1
post_secw_b_flat_repar = post_chains_flat_repar[:, icw]
post_sesw_b_flat_repar = post_chains_flat_repar[:, isw]
post_ecc_b_flat_repar = post_secw_b_flat_repar**2 + post_sesw_b_flat_repar**2
map_ecc_b_repar = post_ecc_b_flat_repar[map_idx_repar]

credint = [anc.hpd(post_ecc_b_flat_repar, c) for c in perc]
l = "{:18s}: MAP {:10.6f} ".format("e_b", map_ecc_b_repar)
for j, pc in enumerate(credint):
    l += "HDI@{:.2f}% [{:10.6f} , {:10.6f}] ".format(perc[j]*100, pc[0], pc[1])
print(l)
err_ecc_b_repar = np.ptp(credint[0])*0.5 # semi-interval, but it is not the only solution

post_argp_b_flat_repar = np.arctan2(post_sesw_b_flat_repar, post_secw_b_flat_repar)*cst.rad2deg
map_argp_b_repar = post_argp_b_flat_repar[map_idx_repar]

credint = [anc.hpd(post_argp_b_flat_repar, c) for c in perc]
l = "{:18s}: MAP {:10.6f} ".format("w_b", map_argp_b_repar)
for j, pc in enumerate(credint):
    l += "HDI@{:.2f}% [{:10.6f} , {:10.6f}] ".format(perc[j]*100, pc[0], pc[1])
print(l)
err_ecc_b_repar = np.ptp(credint[0])*0.5 # semi-interval, but it is not the only solution


icw = fit_labels_repar.index("secosw_c")
isw = icw + 1
post_secw_c_flat_repar = post_chains_flat_repar[:, icw]
post_sesw_c_flat_repar = post_chains_flat_repar[:, isw]
post_ecc_c_flat_repar = post_secw_c_flat_repar**2 + post_sesw_c_flat_repar**2
map_ecc_c_repar = post_ecc_c_flat_repar[map_idx_repar]

credint = [anc.hpd(post_ecc_c_flat_repar, c) for c in perc]
l = "{:18s}: MAP {:10.6f} ".format("e_c", map_ecc_c_repar)
for j, pc in enumerate(credint):
    l += "HDI@{:.2f}% [{:10.6f} , {:10.6f}] ".format(perc[j]*100, pc[0], pc[1])
print(l)
err_ecc_c_repar = np.ptp(credint[0])*0.5 # semi-interval, but it is not the only solution

post_argp_c_flat_repar = np.arctan2(post_sesw_c_flat_repar, post_secw_c_flat_repar)*cst.rad2deg
map_argp_c_repar = post_argp_c_flat_repar[map_idx_repar]

credint = [anc.hpd(post_argp_c_flat_repar, c) for c in perc]
l = "{:18s}: MAP {:10.6f} ".format("w_c", map_argp_c_repar)
for j, pc in enumerate(credint):
    l += "HDI@{:.2f}% [{:10.6f} , {:10.6f}] ".format(perc[j]*100, pc[0], pc[1])
print(l)
err_ecc_c_repar = np.ptp(credint[0])*0.5 # semi-interval, but it is not the only solution

print()
print("This example:")
print("M_b = {:10.6f} +/- {:10.6f} Mearth".format(map_Mb_Me_repar, err_Mb_Me_repar))
print("M_c = {:10.6f} +/- {:10.6f} Mearth".format(map_Mc_Me_repar, err_Mc_Me_repar))

print("McKee values:")
print("M_b = {:10.6f} +/- {:10.6f} Mearth".format(0.0554*cst.Mjups*cst.Msear, 0.0020*cst.Mjups*cst.Msear))
print("M_c = {:10.6f} +/- {:10.6f} Mearth".format(0.525*cst.Mjups*cst.Msear, 0.019*cst.Mjups*cst.Msear))
# M_star            : MAP   0.766808 HDI@68.27% [  0.744168 ,   0.778876] HDI@95.44% [  0.728315 ,   0.799392] HDI@99.74% [  0.709748 ,   0.817473] 
# log10(M_b/M_star) : MAP  -4.163072 HDI@68.27% [ -4.176320 ,  -4.159053] HDI@95.44% [ -4.184626 ,  -4.151047] HDI@99.74% [ -4.220545 ,  -4.150187] 
# log10(M_c/M_b)    : MAP   0.977112 HDI@68.27% [  0.976923 ,   0.977725] HDI@95.44% [  0.976513 ,   0.978185] HDI@99.74% [  0.976254 ,   0.978858] 
# P_b               : MAP  17.099257 HDI@68.27% [ 17.098847 ,  17.100092] HDI@95.44% [ 17.098342 ,  17.100959] HDI@99.74% [ 17.098121 ,  17.104623] 
# P_c               : MAP  34.550594 HDI@68.27% [ 34.550187 ,  34.550754] HDI@95.44% [ 34.549801 ,  34.550962] HDI@99.74% [ 34.548262 ,  34.551046] 
# secosw_b          : MAP   0.151048 HDI@68.27% [  0.149803 ,   0.154556] HDI@95.44% [  0.147172 ,   0.156877] HDI@99.74% [  0.145980 ,   0.160277] 
# sesinw_b          : MAP  -0.368704 HDI@68.27% [ -0.369834 ,  -0.363997] HDI@95.44% [ -0.373051 ,  -0.361685] HDI@99.74% [ -0.374338 ,  -0.355651] 
# secosw_c          : MAP  -0.055337 HDI@68.27% [ -0.059966 ,  -0.052926] HDI@95.44% [ -0.063649 ,  -0.048187] HDI@99.74% [ -0.069297 ,  -0.045538] 
# sesinw_c          : MAP  -0.085513 HDI@68.27% [ -0.109819 ,  -0.078055] HDI@95.44% [ -0.124307 ,  -0.060547] HDI@99.74% [ -0.172985 ,  -0.056908] 
# meanl_b           : MAP  82.077951 HDI@68.27% [ 81.958072 ,  82.127306] HDI@95.44% [ 81.894552 ,  82.243690] HDI@99.74% [ 81.782981 ,  82.324701] 
# 
# M_b               : MAP  17.538345 HDI@68.27% [ 16.750583 ,  17.826441] HDI@95.44% [ 16.094226 ,  18.415487] HDI@99.74% [ 14.775819 ,  18.742388] 
# M_c               : MAP 166.379786 HDI@68.27% [159.323227 , 169.453269] HDI@95.44% [152.926292 , 174.857749] HDI@99.74% [140.465051 , 177.769766] 
# e_b               : MAP   0.158758 HDI@68.27% [  0.156455 ,   0.159350] HDI@95.44% [  0.155259 ,   0.160961] HDI@99.74% [  0.152045 ,   0.161359] 
# w_b               : MAP -67.722419 HDI@68.27% [-67.920292 , -66.970786] HDI@95.44% [-68.457879 , -66.555792] HDI@99.74% [-68.687801 , -65.734289] 
# e_c               : MAP   0.010375 HDI@68.27% [  0.008727 ,   0.015040] HDI@95.44% [  0.006063 ,   0.018376] HDI@99.74% [  0.005937 ,   0.034498] 
# w_c               : MAP -122.907760 HDI@68.27% [-124.296124 , -116.675539] HDI@95.44% [-131.423425 , -114.696488] HDI@99.74% [-134.214300 , -111.624442] 
# 
# This example:
# M_b =  17.538345 +/-   0.537929 Mearth
# M_c = 166.379786 +/-   5.065021 Mearth
# McKee values:
# M_b =  17.611338 +/-   0.635788 Mearth
# M_c = 166.894450 +/-   6.039990 Mearth
\end{lstlisting}

\begin{figure}[!htb]
    \centering
    \includegraphics[width=\textwidth]{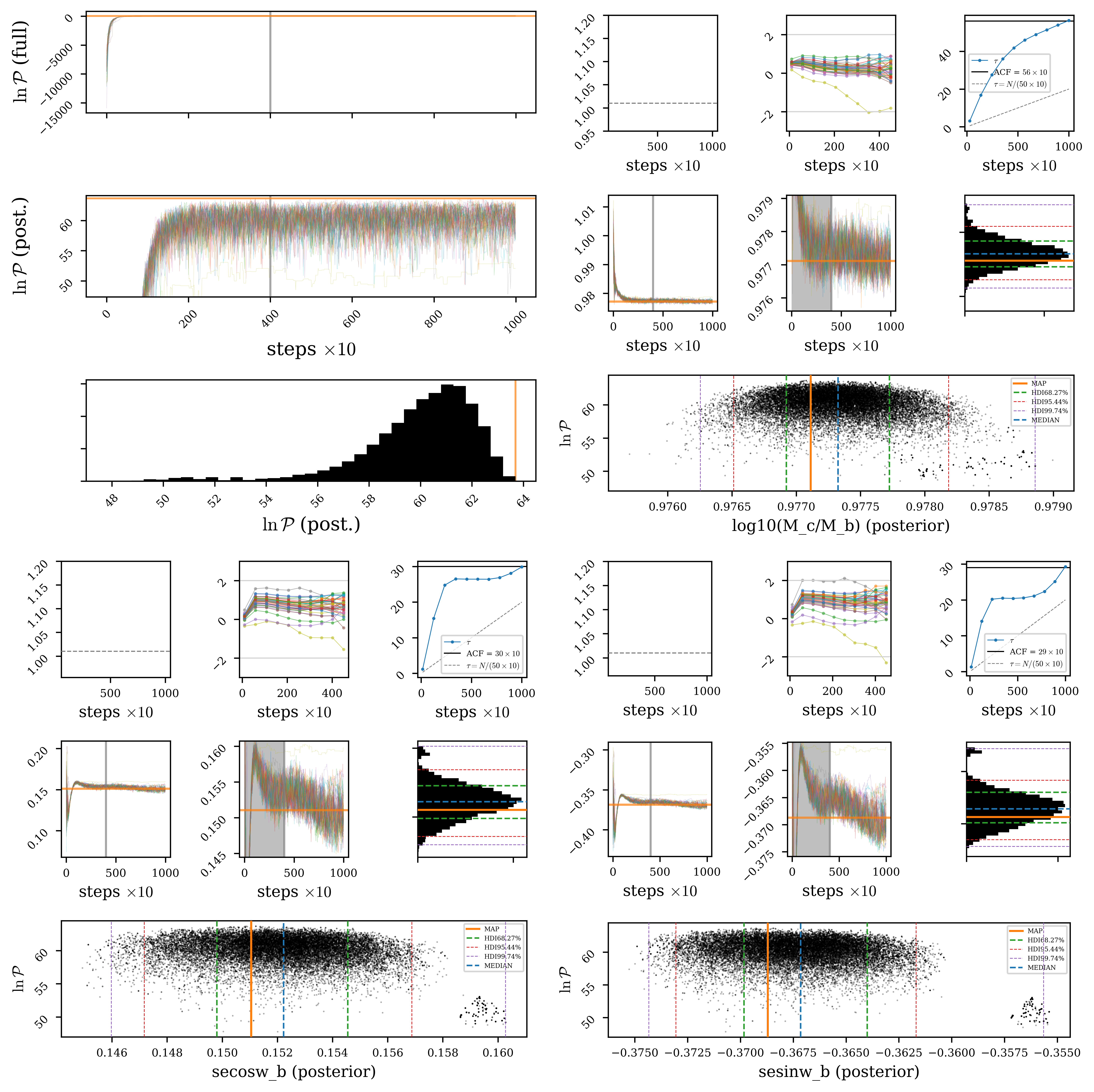}
    \caption{
    \textbf{Trace and convergence plot of $\ln\mathcal{P}$, $\log_{10}(M_\mathrm{c}/M_\mathrm{b})$, $\sqrt{e}\cos\omega$, $\sqrt{e}\sin\omega$)
    of \texttt{emcee} with re-parametrisation.}
    \textit{Upper-left}: trace plot of $\ln\mathcal{P}$ as in Fig.~\ref{da_emcee_lnP_1};
    Each of the other panels shows trace-convergence plots as in Fig.~\ref{da_emcee_traces_1}, 
    but for $\log_{10}(M_\mathrm{c}/M_\mathrm{b})$ (\textit{upper-right}),
    $\sqrt{e}\cos\omega$ (\textit{lower-left}), and 
    $\sqrt{e}\sin\omega$ (\textit{lower-right}) from the re-parametrised analysis with \texttt{emcee}.
   The trace panels for these three parameters indicate an improved level of mixing; 
   however, the trend within the chains suggests that convergence has not yet been fully achieved.
    }\label{da_emcee_traces_2}
\end{figure}

In this case, the MAP value of the eccentricity of b and c are within the 68.27\% HDI, 
contrary to the analysis fitting directly the eccentricity.
It also reached a slightly higher precision on the masses, still perfectly consistent with literature.
Furthermore, if I check the $\ln\mathcal{P}$ and 
the trace plot, for example for $\log_{10}(M_\mathrm{c}/M_\mathrm{b})$ and 
($\sqrt{e}\cos\omega$, $\sqrt{e}\sin\omega$) for planet c in Fig.~\ref{da_emcee_traces_2}
it can seen that the chains have a higher level of mixing and it is closer to the convergence state,
and so, the parameter estimation should be more accurate, precise, and reliable.

\begin{figure}[!htb]
    \centering
    \includegraphics[width=\textwidth, align=c]{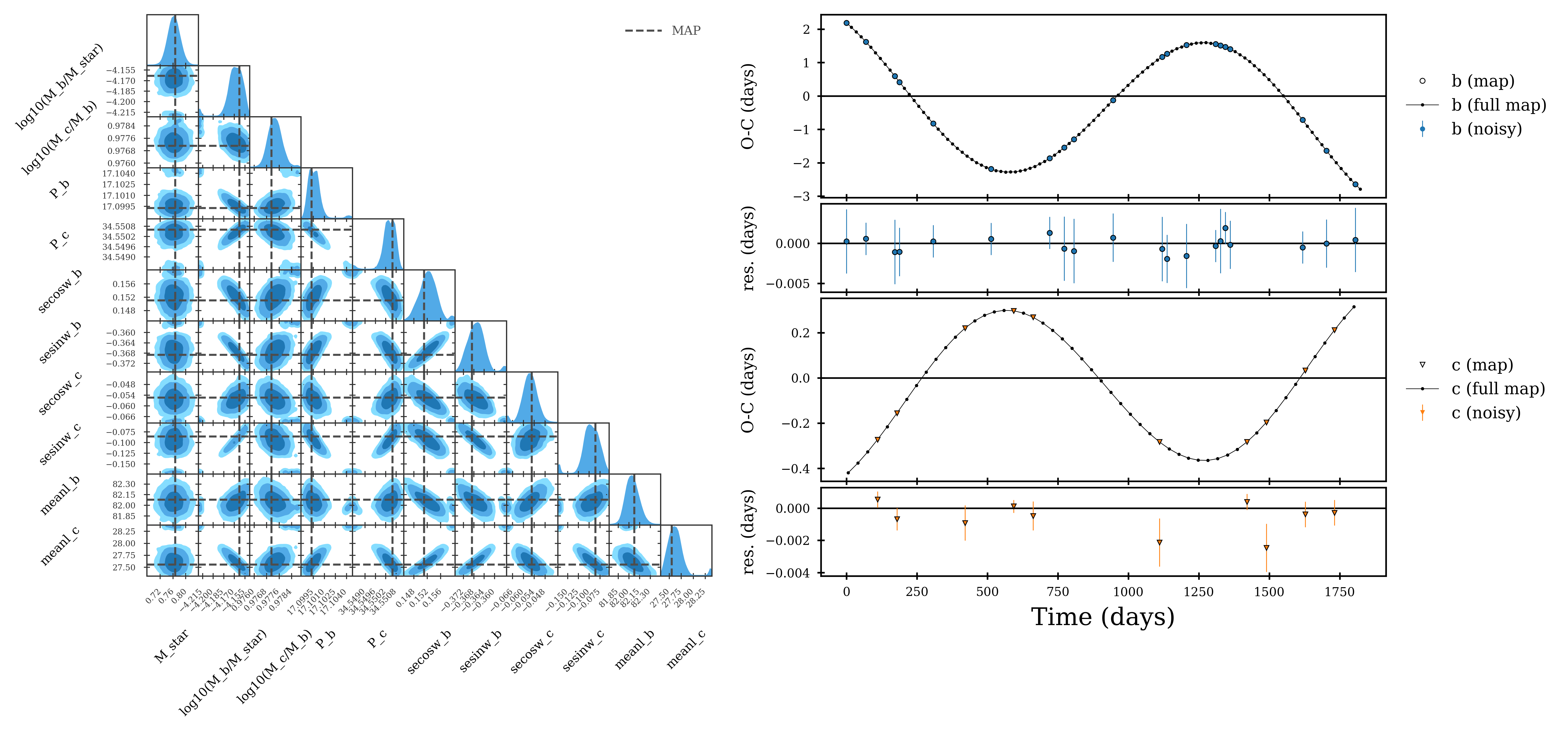}
    \caption{
    \textbf{Corner and $O-C$ plots of \texttt{emcee} with re-parametrisation.)}
    Corner plot (\textit{left}) as in Fig.~\ref{da_corner_emcee_1} and 
    $O-C$ plot (\textit{right}) as in Fig.~\ref{da_oc_emcee_1}, but for the re-parametrised \texttt{emcee} run.
    }\label{da_emcee_corner_oc_2}
\end{figure}

As I have done previously, I can draw the corner plot of the posterior distribution and 
the $O-C$ plot with the MAP model and the random samples (see Fig.~\ref{da_emcee_corner_oc_2}).

The last corner plot in Fig.~\ref{da_emcee_corner_oc_2} shows a small region of the parameter space separate from
the most of the chains and it can be attributed to a chain stuck far from others.
This is due to analysis being very quick, with a total number of iteration 
\texttt{n\_iteration = n\_steps $\times$ thin\_by = 10\,000},
and it would need more iterations or a larger thinning factor,
to get rid of this behaviour.
However, comparing with the first corner plot in Fig.~\ref{da_corner_emcee_1},
it can be seen that the correlation among parameters has been reduced, 
especially for planetary masses and mass-eccentricity pairs,
even if not removed completely for all the parameters.
Keep in mind that there is not a perfect re-parametrisation able to get rid of all correlation,
and letting the MCMC analysis running for more steps is always advisable.
There is not a strict rule for the number of walkers, steps, and thinning factor,
so playing with these numbers and checking the trace-convergence plots during 
the analysis is the only way to progress.\\
Another method for the posterior sampling and parameter estimation, in a Bayesian framework,
is the Nested Sampling algorithm.
There are different publicly available codes and in this quick analysis I am going to use
the \texttt{nautilus} code developed by \citet{nautilus}.
These kind of code needs a different way to provide the priors, usually as a separated function from
the log-likelihood.
I show how to re-formatting the priors, boundaries, and the log-likelihood function to works with \texttt{nautilus}.

\begin{lstlisting}[language=Python]
priors_nautilus = nautilus.Prior() # set the Prior object
# uses previous definition of the parameters: Mstar as Gaussian, others as Uniform within boundaries.
for ilab, lab in enumerate(fit_labels_repar):
    priors_nautilus.add_parameter(lab, dist=priors_repar[ilab])

# creates a variable for the physical boundaries
boundaries_physical = [
    M_s_boundaries,
    M_p_boundaries,
    ecc_boundaries,
]

# and creates a Prior-like object for the physical ones
priors_physical = nautilus.Prior()
priors_physical.add_parameter("M_star", dist=uniform(loc=M_s_boundaries[0], scale=np.ptp(M_s_boundaries)))
priors_physical.add_parameter("M_b", dist=uniform(loc=M_p_boundaries[0], scale=np.ptp(M_p_boundaries)))
priors_physical.add_parameter("M_c", dist=uniform(loc=M_p_boundaries[0], scale=np.ptp(M_p_boundaries)))
priors_physical.add_parameter("ecc_b", dist=uniform(loc=ecc_boundaries[0], scale=np.ptp(ecc_boundaries)))
priors_physical.add_parameter("ecc_c", dist=uniform(loc=ecc_boundaries[0], scale=np.ptp(ecc_boundaries)))

# nautulis prefer to use the logpdf of the prior distribution also for checking the boundaries
# and with fitting parameters as dictionary.
# Tested fitting parameters as list/array but it raises errors/bugs.
# It works in this way
def log_boundaries_nautilus_repar(fit_dict):

    fit_pars = np.array(list(fit_dict.values()))
    
    m_x, r_x, p_x, e_x, w_x, ma_x, i_x, ln_x = fitting_to_physical_params_repar(fit_pars)
    phys = {
        "M_star":m_x[0],
        "M_b":m_x[1],
        "M_c":m_x[2],
        "ecc_b":e_x[1],
        "ecc_c":e_x[2],
    }

    # loop on the `phys` parameters, and if the converted ones are outside the
    # boundaries `phys` it returns -np.inf, otherwise 0.0
    # I do not want to change the value of the log-likelihood/probability,
    # just discard configurations that have not physical meanings (like ecc < 0)
    for k, v in phys.items():
        lnphy = priors_physical.dists[priors_physical.keys.index(k)].logpdf(v)
        if np.isinf(lnphy):
            return -np.inf
    return 0.0

 # log-Priors of fitting parameters
def log_priors_nautulis_repar(fit_dict):

    ln_prior = np.sum(
        [
            priors_nautilus.dists[priors_nautilus.keys.index(k)].logpdf(p) for k, p in fit_dict.items()
        ]
    )
    return ln_prior

def log_likelihood_nautilus_repar(fit_dict):

    ln_prior = log_boundaries_nautilus_repar(fit_dict)
    if np.isinf(ln_prior):
        return ln_prior
    
    lnL = ln_const + ln_prior
    fit_pars = np.array(list(fit_dict.values()))
    (
        body_flag_sim,
        epo_sim,
        transits_sim,
        durations_sim,
        lambda_rm_sim,
        kep_elem_sim,
        stable,
    ) = fitting_to_observables_repar(fit_pars)
    if not stable:
        return -np.inf

    res_b = tra_syn_noisy_b - transits_sim[body_flag_sim ==2]
    wres_b = res_b / err_tra_syn_b
    lnL_b = -0.5*np.sum(np.log(err_tra_syn_b)) - 0.5*np.sum(wres_b*wres_b)
    
    res_c = tra_syn_noisy_c - transits_sim[body_flag_sim ==3]
    wres_c = res_c / err_tra_syn_c
    lnL_c = -0.5*np.sum(np.log(err_tra_syn_c)) - 0.5*np.sum(wres_c*wres_c)
    
    lnL += lnL_b + lnL_c
    
    return lnL
\end{lstlisting}

Let me set the number of live points (\texttt{n\_live}) that \texttt{nautilus} has to use,
and set-up the file to store information.
Then, run the code that will stop when it reaches a $\Delta\ln\mathcal{Z} = 0.01$ and
an effective size $N_\mathrm{eff}$ of at least 10\,000\footnote{
$\Delta\ln\mathcal{Z} = \frac{1}{\sqrt{N_\mathrm{eff}}} = 0.01$}, and
I also set to discard the initial exploration phase.

\begin{lstlisting}[language=Python]
n_live = 3000
n_threads = 100
seed = 42578
delete_file = True
resume = True
nautilus_filename = os.path.join(os.path.abspath("."), "sampler_nautilus.hdf5")
if delete_file:
    if os.path.exists(nautilus_filename):
        os.remove(nautilus_filename)
        
sampler_nautilus = nautilus.Sampler(
    priors_nautilus,
    log_likelihood_nautilus_repar,
    n_dim=n_fit_repar,
    n_live=n_live,
    vectorized=False,
    pass_dict=True,
    pool=n_threads,
    seed=seed,
    filepath=nautilus_filename,
    resume=resume,
)
sampler_nautilus.run(discard_exploration=True, verbose=True)
# RUNNING OUTPUT
# Starting the nautilus sampler...
# Please report issues at github.com/johannesulf/nautilus.
# Status    | Bounds | Ellipses | Networks | Calls    | f_live | N_eff | log Z    
# Finished  | 109    | 2        | 4        | 874100   | N/A    | 10022 | -24.42   
# True
log_z = sampler_nautilus.log_z
n_eff = sampler_nautilus.n_eff
points, log_w, log_l = sampler_nautilus.posterior() # posterior, ln(weights), ln-likelihood
print("log_Z = {:.4f} +/- {:.4f}".format(log_z, 1.0/np.sqrt(n_eff)))
print("N_eff = {:.0f}".format(n_eff))
# OUTPUT
# log_Z = -24.4202 +/- 0.0100
# N_eff = 10022
\end{lstlisting}

The code returns the $\ln\mathcal{L}$ and if I want the $\ln\mathcal{P}$
I need to recompute the $\ln$-prior for the whole posterior.

\begin{lstlisting}[language=Python]
def compute_log_prior_nautilus(pars, labels):
    pars_dict = {lab:val for lab, val in zip(labels, pars)}
    ln_prior = log_priors_nautulis_repar(pars_dict)
    return ln_prior

n_calls, _ = np.shape(points)
log_priors_nau, log_prob_nau = np.zeros(n_calls), np.zeros(n_calls)
for icall in range(n_calls):
    log_priors_nau[icall] = compute_log_prior_nautilus(points[icall], fit_labels_repar)
    log_prob_nau[icall] = log_l[icall] + log_priors_nau[icall]
\end{lstlisting}

If I would have defined the $\ln\mathcal{L}$ function to return additional variables, 
like $\ln\mathcal{P}$, $\ln$-prior, or whatever you want,
the code would returns them as \texttt{blobs}.

\begin{lstlisting}[language=Python]
points, log_w, log_l, blobs = sampler_nautilus.posterior() # posterior, ln(weights), ln-likelihood, blobs (additional outputs)
\end{lstlisting}

\begin{figure}[!htb]
    \centering
    \includegraphics[width=\textwidth, align=c]{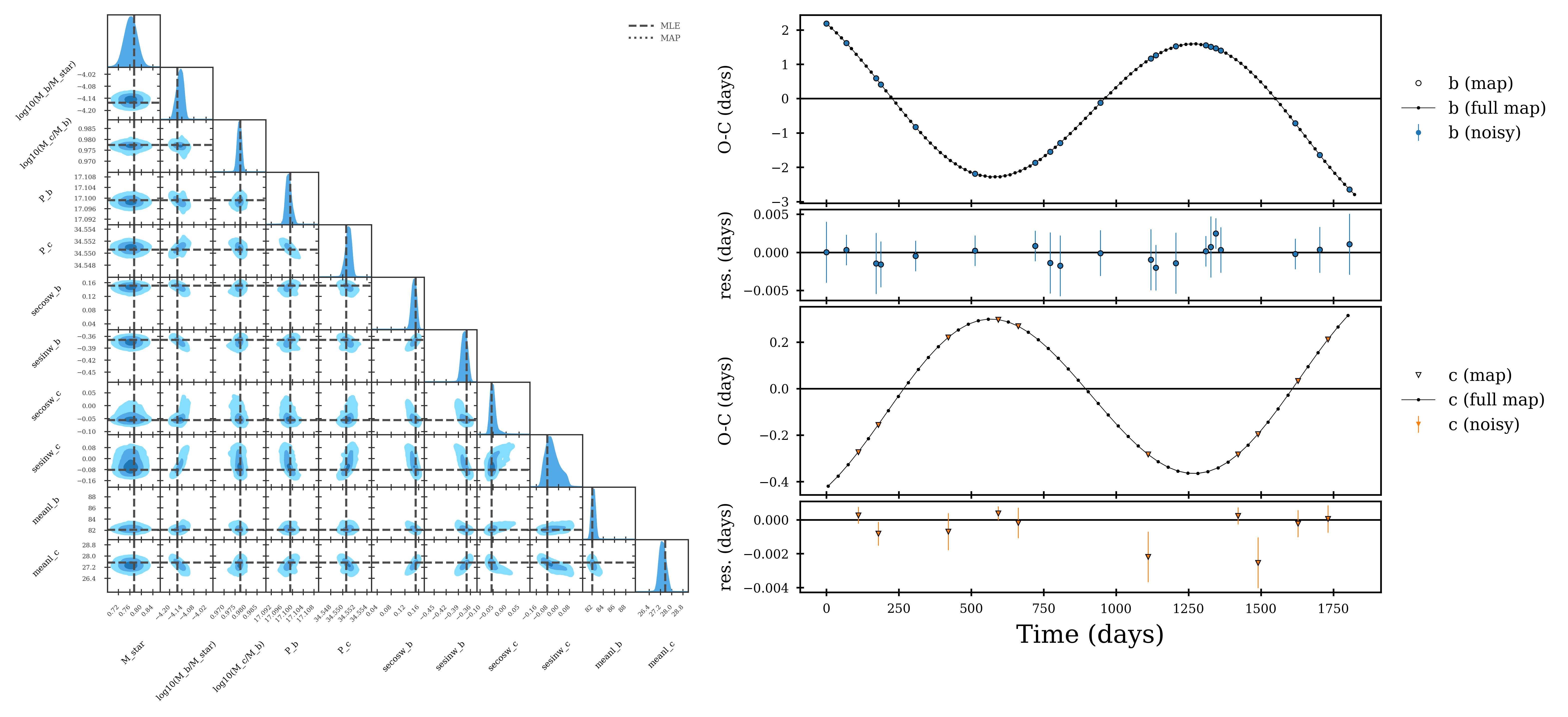}
    \caption{
    \textbf{Corner and $O-C$ plots of \texttt{nautilus} analysis.)}
    Corner plot (\textit{left}) and $O-C$ plot (\textit{right}) as in Fig.~\ref{da_emcee_corner_oc_2},
    but for the \texttt{nautilus} run.
    }\label{da_nautilus_corner_oc}
\end{figure}

I finally can do some statistics on the posterior, obtaining the Maximum-Likelihood-Estimator (MLE),
the Maximum-A-Posteriori (MAP) as best-fit configurations,
the uncertainties as HDI, and creating the corner and the $O-C$ plots (see Fig.~\ref{da_nautilus_corner_oc}).

\begin{lstlisting}[language=Python]
weights = np.exp(log_w)
sel_w = weights > 0.0
pos_points = points[sel_w, :]
pos_weights = weights[sel_w]
pos_log_like = log_l[sel_w]
pos_log_prior = log_priors_nau[sel_w]
pos_log_prob = log_prob_nau[sel_w]
n_nautilus = len(pos_weights)
print("posterior size n_nautilus = ", n_nautilus)
# MLE
idx_mle_nautilus = np.argmax(pos_log_like)
mle_points = pos_points[idx_mle_nautilus, :]
mle_fit_dict = {lab:val for lab, val in zip(fit_labels_repar, mle_points)}
# MAP
idx_map_nautilus = np.argmax(pos_log_prob)
map_points = pos_points[idx_map_nautilus, :]
map_fit_dict = {lab:val for lab, val in zip(fit_labels_repar, map_points)}
# FOR THIS CASE (AND SEED) THE MLE AND MAP ARE EXACTLY THE SAME
# credible intervals
perc = np.array([68.27, 95.44, 99.74]) / 100.0
for i in range(n_fit):
    fitn, fitp, fitpost = fit_labels_repar[i], map_points[i], pos_points[:, i]
    credint = [anc.hpd(fitpost, c) for c in perc]
    l = "{:18s}: MAP {:10.6f} ".format(fitn, fitp)
    for j, pc in enumerate(credint):
        l += "HDI@{:.2f}% [{:10.6f} , {:10.6f}] ".format(perc[j]*100, pc[0], pc[1])
    print(l)
print()

# conversion of parameters

post_Ms_flat_repar = pos_points[:, 0]

post_l10Mb2s_flat_repar = pos_points[:, 1]
post_Mb_Me_repar = 10**(post_l10Mb2s_flat_repar) * post_Ms_flat_repar * cst.Msear
map_Mb_Me_repar = post_Mb_Me_repar[idx_map_nautilus]

credint = [anc.hpd(post_Mb_Me_repar, c) for c in perc]
l = "{:18s}: MAP {:10.6f} ".format("M_b", map_Mb_Me_repar)
for j, pc in enumerate(credint):
    l += "HDI@{:.2f}% [{:10.6f} , {:10.6f}] ".format(perc[j]*100, pc[0], pc[1])
print(l)
err_Mb_Me_repar = np.ptp(credint[0])*0.5 # semi-interval, but it is not the only solution

post_l10Mc2b_flat_repar = pos_points[:, 2]
post_Mc_Me_repar = 10**(post_l10Mc2b_flat_repar) * post_Mb_Me_repar
map_Mc_Me_repar = post_Mc_Me_repar[idx_map_nautilus]

credint = [anc.hpd(post_Mc_Me_repar, c) for c in perc]
l = "{:18s}: MAP {:10.6f} ".format("M_c", map_Mc_Me_repar)
for j, pc in enumerate(credint):
    l += "HDI@{:.2f}% [{:10.6f} , {:10.6f}] ".format(perc[j]*100, pc[0], pc[1])
print(l)
err_Mc_Me_repar = np.ptp(credint[0])*0.5 # semi-interval, but it is not the only solution

icw = fit_labels_repar.index("secosw_b")
isw = icw + 1
post_secw_b_flat_repar = pos_points[:, icw]
post_sesw_b_flat_repar = pos_points[:, isw]
post_ecc_b_flat_repar = post_secw_b_flat_repar**2 + post_sesw_b_flat_repar**2
map_ecc_b_repar = post_ecc_b_flat_repar[idx_map_nautilus]

credint = [anc.hpd(post_ecc_b_flat_repar, c) for c in perc]
l = "{:18s}: MAP {:10.6f} ".format("e_b", map_ecc_b_repar)
for j, pc in enumerate(credint):
    l += "HDI@{:.2f}% [{:10.6f} , {:10.6f}] ".format(perc[j]*100, pc[0], pc[1])
print(l)
err_ecc_b_repar = np.ptp(credint[0])*0.5 # semi-interval, but it is not the only solution

post_argp_b_flat_repar = np.arctan2(post_sesw_b_flat_repar, post_secw_b_flat_repar)*cst.rad2deg
map_argp_b_repar = post_argp_b_flat_repar[idx_map_nautilus]

credint = [anc.hpd(post_argp_b_flat_repar, c) for c in perc]
l = "{:18s}: MAP {:10.6f} ".format("w_b", map_argp_b_repar)
for j, pc in enumerate(credint):
    l += "HDI@{:.2f}% [{:10.6f} , {:10.6f}] ".format(perc[j]*100, pc[0], pc[1])
print(l)
err_ecc_b_repar = np.ptp(credint[0])*0.5 # semi-interval, but it is not the only solution


icw = fit_labels_repar.index("secosw_c")
isw = icw + 1
post_secw_c_flat_repar = pos_points[:, icw]
post_sesw_c_flat_repar = pos_points[:, isw]
post_ecc_c_flat_repar = post_secw_c_flat_repar**2 + post_sesw_c_flat_repar**2
map_ecc_c_repar = post_ecc_c_flat_repar[idx_map_nautilus]

credint = [anc.hpd(post_ecc_c_flat_repar, c) for c in perc]
l = "{:18s}: MAP {:10.6f} ".format("e_c", map_ecc_c_repar)
for j, pc in enumerate(credint):
    l += "HDI@{:.2f}% [{:10.6f} , {:10.6f}] ".format(perc[j]*100, pc[0], pc[1])
print(l)
err_ecc_c_repar = np.ptp(credint[0])*0.5 # semi-interval, but it is not the only solution

post_argp_c_flat_repar = np.arctan2(post_sesw_c_flat_repar, post_secw_c_flat_repar)*cst.rad2deg
map_argp_c_repar = post_argp_c_flat_repar[idx_map_nautilus]

credint = [anc.hpd(post_argp_c_flat_repar, c) for c in perc]
l = "{:18s}: MAP {:10.6f} ".format("w_c", map_argp_c_repar)
for j, pc in enumerate(credint):
    l += "HDI@{:.2f}% [{:10.6f} , {:10.6f}] ".format(perc[j]*100, pc[0], pc[1])
print(l)
err_ecc_c_repar = np.ptp(credint[0])*0.5 # semi-interval, but it is not the only solution

print()
print("This example:")
print("M_b = {:10.6f} +/- {:10.6f} Mearth".format(map_Mb_Me_repar, err_Mb_Me_repar))
print("M_c = {:10.6f} +/- {:10.6f} Mearth".format(map_Mc_Me_repar, err_Mc_Me_repar))

print("McKee values:")
print("M_b = {:10.6f} +/- {:10.6f} Mearth".format(0.0554*cst.Mjups*cst.Msear, 0.0020*cst.Mjups*cst.Msear))
print("M_c = {:10.6f} +/- {:10.6f} Mearth".format(0.525*cst.Mjups*cst.Msear, 0.019*cst.Mjups*cst.Msear))
# OUTPUT
# M_star            : MAP   0.774914 HDI@68.27% [  0.741894 ,   0.783218] HDI@95.44% [  0.722886 ,   0.805598] HDI@99.74% [  0.699399 ,   0.826296] 
# log10(M_b/M_star) : MAP  -4.161391 HDI@68.27% [ -4.161183 ,  -4.132279] HDI@95.44% [ -4.179140 ,  -4.123785] HDI@99.74% [ -4.212080 ,  -4.095137] 
# log10(M_c/M_b)    : MAP   0.977437 HDI@68.27% [  0.976426 ,   0.977597] HDI@95.44% [  0.975693 ,   0.978271] HDI@99.74% [  0.970803 ,   0.982816] 
# P_b               : MAP  17.099169 HDI@68.27% [ 17.097649 ,  17.099206] HDI@95.44% [ 17.097037 ,  17.100527] HDI@99.74% [ 17.093576 ,  17.104235] 
# P_c               : MAP  34.550590 HDI@68.27% [ 34.550618 ,  34.551368] HDI@95.44% [ 34.549991 ,  34.551618] HDI@99.74% [ 34.548422 ,  34.553118] 
# secosw_b          : MAP   0.151004 HDI@68.27% [  0.141527 ,   0.152136] HDI@95.44% [  0.136361 ,   0.155848] HDI@99.74% [  0.113308 ,   0.164413] 
# sesinw_b          : MAP  -0.369190 HDI@68.27% [ -0.381030 ,  -0.368686] HDI@95.44% [ -0.385399 ,  -0.363182] HDI@99.74% [ -0.402403 ,  -0.354453] 
# secosw_c          : MAP  -0.056307 HDI@68.27% [ -0.060305 ,  -0.046214] HDI@95.44% [ -0.066113 ,  -0.018823] HDI@99.74% [ -0.080591 ,   0.048901] 
# sesinw_c          : MAP  -0.081003 HDI@68.27% [ -0.107419 ,  -0.014639] HDI@95.44% [ -0.116741 ,   0.056909] HDI@99.74% [ -0.153026 ,   0.110422] 
# meanl_b           : MAP  82.064652 HDI@68.27% [ 81.997897 ,  82.380717] HDI@95.44% [ 81.861893 ,  82.654408] HDI@99.74% [ 81.246323 ,  84.102942] 
# 
# M_b               : MAP  17.792451 HDI@68.27% [ 17.250286 ,  18.814265] HDI@95.44% [ 16.405858 ,  19.513356] HDI@99.74% [ 15.338770 ,  20.684499] 
# M_c               : MAP 168.916675 HDI@68.27% [163.950658 , 178.616143] HDI@95.44% [155.977623 , 185.163729] HDI@99.74% [146.338879 , 196.354703] 
# e_b               : MAP   0.159103 HDI@68.27% [  0.158841 ,   0.165053] HDI@95.44% [  0.155956 ,   0.167203] HDI@99.74% [  0.151836 ,   0.175625] 
# w_b               : MAP -67.754754 HDI@68.27% [-69.665781 , -67.619441] HDI@95.44% [-70.530912 , -66.812571] HDI@99.74% [-74.427729 , -65.621698] 
# e_c               : MAP   0.009732 HDI@68.27% [  0.001785 ,   0.008103] HDI@95.44% [  0.001264 ,   0.015433] HDI@99.74% [  0.000001 ,   0.022490] 
# w_c               : MAP -124.803707 HDI@68.27% [-157.153390 , -114.446052] HDI@95.44% [-179.257369 , 165.770919] HDI@99.74% [-179.257369 , 179.839635] 
# 
# This example:
# M_b =  17.792451 +/-   0.781989 Mearth
# M_c = 168.916675 +/-   7.332742 Mearth
# McKee values:
# M_b =  17.611338 +/-   0.635788 Mearth
# M_c = 166.894450 +/-   6.039990 Mearth
\end{lstlisting}

From these three independent analysis, I obtained the estimation of the parameters of the system,
consistent with each other and with literature.
The analysis clearly needs further fine tuning,
larger number of population/chains and generations/steps, and hyper-parameters of each code.
It would also be advisable to repeat the same run with different seeds for the random number generator,
allowing the code to sample differently the space of parameters.
In this way one can check if there are multiple solutions, or if the current data can be explained
with the same solution, allowing for the merging of the posteriors.

\clearpage

\section{Tables of some TTV systems, instruments, and tools}\label{Tables}

% ====================================================================
% TTV SYSTEMS
% ====================================================================

\DeclareTblrTemplate{caption-tag}{default}{\textbf{Table\hspace{0.25em}\thetable}}
\DeclareTblrTemplate{caption-sep}{default}{\enskip}
\DeclareTblrTemplate{caption-text}{default}{\InsertTblrText{caption}}
\DeclareTblrTemplate{note-text}{default}{\textit{\InsertTblrNoteText}}

\begin{longtblr}[
    caption = {TTV systems presented},
    label = {tab:systems},
    note{*} = {Confirmed planets in the systems. If indicated as $N+X$, $N$ means confirmed planets and $X$ statistically validated.}
]{
  width = \textwidth,
  colspec = { X[2,l,t] X[1,c,t] X[7,l,t] },
  rowhead = 1,      % Repeats the first row on every page
  % hlines,           % Adds horizontal lines (replaces \hline)
  row{1} = {font=\bfseries}, % Automatically makes header bold
}
    \hline
    % Table Header
    Host star & Number of Planets\TblrNote{*} & References \\
    \hline

    \kepler-9            & $2+1$ & \citet{Holman2010Sci...330...51H,Borsato2014AA...571A..38B,Borsato2019MNRAS.484.3233B}\\
    \kepler-11           & 6     & \citet{Lissauer2013ApJ...770..131L,Borsato2014AA...571A..38B}\\
    \kepler-19           & 3     & \citet{Ballard2011ApJ...743..200B,Malavolta2017AJ....153..224M}\\
    \kepler-23           & 3     & \citet{2023AA...669A.117L}\\
    \kepler-32           & 5     & \citet{Fabrycky2012ApJ...750..114F}\\
    \kepler-33           & 5     & \citet{Lissauer2012ApJ...750..112L}\\
    \kepler-36           & 2     & \citet{Carter2011Sci...331..562C}\\
    \kepler-48           & 5     & \citet{Marcy2014ApJS..210...20M}\\
    \kepler-55           & 5     & \citet{Steffen2013MNRAS.428.1077S}\\ 
    \kepler-56           & 3     & \citet{2013Sci...342..331H}\\ 
    \kepler-80           & 6     & \citet{MacDonald2016AJ....152..105M}\\ 
    \kepler-82           & 5     & \citet{Freudenthal2019AA...628A.108F}\\ 
    \kepler-84           & 5     & \citet{Xie2013ApJS..208...22X}\\
    \kepler-87           & 2     & \citet{2014AA...561A.103O}\\
    \kepler-88           & 3     & \citet{Nesvorny2013ApJ...777....3N,Barros2014AA...561L...1B}\\
    \kepler-91           & 1     & \citet{2014AA...562A.109L,2016ApSS.361...17B}\\
    \kepler-92           & 3     & \citet{Xie2014ApJS..210...25X,2024ApJS..270....8W}\\
    \kepler-101          & 2     & \citet{2014AA...572A...2B} \\
    \kepler-102          & 5     & \citet{Marcy2014ApJS..210...20M,2023AA...677A..33B} \\
    \kepler-122          & 5     & \citet{Rowe2014ApJ...784...45R}\\
    \kepler-128          & 2     & \citet{Hadden2016ApJ...828...44H} \\
    \kepler-136          & 2     & \citet{Rowe2014ApJ...784...45R}\\
    \kepler-145          & 2     & \citet{2025AJ....169...90O}\\
    \kepler-223          & 4     & \citet{phodymm2016Natur.533..509M}\\
    \kepler-238          & 5     & \citet{Xie2014ApJS..210...25X}\\
    \kepler-277          & 2     & \citet{2025AJ....169...90O}\\
    \kepler-278          & 2     & \citet{2020AA...634A..29J}\\
    \kepler-338          & 4     & \citet{Weiss2020AJ....159..242W}\\
    \kepler-730          & 2     & \citet{Canas2019ApJ...870L..17C}\\
    KOI-351 (\kepler-90) & 8     & \citet{Cabrera2014ApJ...781...18C,2014ApJ...784...44L,Liang2021AJ....161..202L,2025AJ....170..146S} \\
    HAT-P-7              & 1     & \citet{2008ApJ...680.1450P,2025ApJ...986..117Y}\\
    HAT-P-13             & 2     & \citet{2012MNRAS.420.2580S} \\
    WASP-47              & 4     & \citet{Hellier2012MNRAS.426..739H,Nascimbeni2023AA...673A..42N} \\
    WASP-84              & 2     & \citet{Maciejewski2023MNRAS.525L..43M} \\
    WASP-132             & 3     & \citet{Grieves2024arXiv240615986G} \\
    HD 108236            & 5     & \citet{Bonfanti2021AA...646A.157B} \\
    HIP 41378            & 5     & \citet{Berardo2019AJ....157..185B,Sulis2024AA...686L..18S,Leonardi2024AA...686A..84L} \\
    TRAPPIST-1           & 7     & \citet{Gillon2017Natur.542..456G,Agol2021PSJ.....2....1A} \\
    TOI-216              & 2     & \citet{Kipping2019MNRAS.486.4980K, Dawson2021AJ....161..161D} \\
    TOI-1130             & 2     & \citet{Huang2020ApJ...892L...7H, Korth2023AA...675A.115K, Borsato2024AA...689A..52B} \\
    TOI-1136             & 6     & \citet{Beard2024AJ....167...70B} \\
    TOI-2180             & 1     & \citet{2022AJ....163...61D,2022RNAAS...6...76D,2024ApJS..271...16D}\\
    TOI-2000             & 2     & \citet{Sha2023MNRAS.524.1113S} \\
    TOI-5398             & 2     & \citet{Mantovan2022MNRAS.516.4432M,Mantovan2024AA...682A.129M} \\
    L~98-59              & 5     & \citet{2019AJ....158...32K,2025AJ....170..154C,2025AJ....170..146S} \\

    \hline
\end{longtblr}

% ====================================================================
% TTV instruments - space
% ====================================================================
\begin{longtblr}[
    caption = {Space-based instruments suitable for TTV study.},
    label = {tab:space_instruments},
]{
  width = \textwidth,
  % X[ratio, alignment, vertical_alignment]
  % Ratios
  colspec = { X[2,l,t] X[6,l,t] X[2,l,t] },
  rowhead = 1,      % Repeats the first row on every page
  % hlines,           % Adds horizontal lines (replaces \hline)
  row{1} = {font=\bfseries}, % Automatically makes header bold
}
    \hline
    Instrument & Notes & References\\
    \hline
    
    \kepler{} &
    survey; detect and characterise TTVs; long baseline; not suitable for ground-based follow-up &
    \citet{Borucki2010Sci...327..977B} \\
    
    \kepler/K2 &
    survey; detect and characterise TTVs; short baseline; suitable for ground-based follow-up &
    \citet{Howell2014PASP..126..398H} \\
    
    TESS      &
    survey; detect TTVs with a few sectors, characterise TTV with multiple sectors; suitable for ground-based follow-up &
    \citet{Ricker2015JATIS...1a4003R} \\

    CHEOPS    &
    single target; refine TTVs; suitable for ground-based follow-up &
    \citet{Benz2021ExA....51..109B} \\

    JWST    &
    single target; refine TTVs; IR band less influenced by stellar activity; suitable for ground-based follow-up &
    \citet{Gardner2023PASP..135f8001G} \\

    PLATO    &
    survey; detect and characterise TTVs; long baseline; suitable for ground-based follow-up; expected launch date end 2026 &
    \citet{Rauer2025ExA....59...26R,Nascimbeni2025AA...694A.313N} \\

    Ariel    &
    single target; refine TTVs; IR band less influenced by stellar activity; suitable for ground-based follow-up; expected launch date in 2029 &
    \citet{Tinetti2018ExA....46..135T,Borsato2022ExA....53..635B} \\
    
    \hline

\end{longtblr}

% ====================================================================
% TTV instruments - ground
% ====================================================================
\begin{longtblr}[
    caption = {Ground-based instruments that have proven to be suitable for TTV study.},
    label = {tab:ground_instruments},
    note{*} = {This is not a full and complete list,
               as many ground-based facilities can be fine tuned and 
               used to observe transits and refine existing TTV signals.
               }
]{
  width = \textwidth,
  % X[ratio, alignment, vertical_alignment]
  % Ratios
  colspec = { X[2,l,t] X[6,l,t] X[2,l,t] },
  rowhead = 1,      % Repeats the first row on every page
  % hlines,           % Adds horizontal lines (replaces \hline)
  row{1} = {font=\bfseries}, % Automatically makes header bold
}
    \hline
    Instrument\TblrNote{*} & Notes & References\\
    \hline
    
    NGTS &
    transit survey; refine strong TTVs of bright K-M stars hosting planets larger than Neptune; follow-up of space-based telescopes &
    \citet{Wheatley2018MNRAS.475.4476W} \\
    
    MUSCAT &
    multi-sites and simultaneous multi-band transit observations; mitigates stellar spots effects; refine strong TTV&
    \citet{narita2019muscat2,Narita2020SPIE11447E..5KN} \\

    ASTEP/ASTEP+ &
    Antarctica site; refine strong TTVs; follow-up of space-based telescopes&
    \citet{Guillot2015AN....336..638G,Mekarnia2016MNRAS.463...45M} \\

    SPECULOOS &
    Cerro Paranal; Near-infrared survey comprises four robotic telescopes; refine strong TTVs of M stars;&
    \citet{2013prpl.conf2K066G} \\

    Copernico (Mount Ekar, Asiago) &
    1.8~m telescope; refine strong TTVs; follow-up of space-based telescopes&
    \citet{Nascimbeni2011AA...527A..85N} \\

    \hline
\end{longtblr}

% ====================================================================
% TTV CODES - ANALYTICAL
% ====================================================================
% TTV CODES - N-BODY
% ====================================================================
% TTV CODES - PHOTO-DYNAMICAL
% ====================================================================
\begin{longtblr}[
    caption = {Analytical, numerical integration, and photo-dynamical tools.},
    label = {tab:ttv_tools},
    note{*} = {Reporting the source if publicly available.}
]{
  width = \textwidth,
  % X[ratio, alignment, vertical_alignment]
  % Ratios
  colspec = { X[2.8,l,t] X[3,l,t] X[6,l,t] },
  rowhead = 1,      % Repeats the first row on every page
  % hlines,           % Adds horizontal lines (replaces \hline)
  row{1} = {font=\bfseries}, % Automatically makes header bold
}
    \hline
    Code & Type & References and Source\TblrNote{*}\\
    
    \hline

    \texttt{TTVFaster} & Analytical & \citet{Agol2016ApJ...818..177A} \mbox{\small \url{https://github.com/ericagol/TTVFaster}} \\
    \texttt{TTV2Fast2Furious} & Analytical & \citet{Hadden2019TTV2Fast2Furious} \mbox{\small \url{https://github.com/shadden/TTV2Fast2Furious}} \\

    \texttt{exostriker} & N-body & \citet{exostriker2019ascl.soft06004T} \mbox{\small \url{https://github.com/3fon3fonov/exostriker}} \\
    \texttt{NbodyGradient.jl} & N-body & \citet{Agol2021MNRAS.507.1582A} \mbox{\small \url{https://github.com/ericagol/NbodyGradient.jl}}\\
    \texttt{Photodynamics.jl} & N-body & \citet{2025MNRAS.540..106L} \mbox{\small \url{https://github.com/langfzac/Photodynamics.jl}}\\
    \texttt{TTVFast} & Analytical/N-body & \citet{Deck2014ApJ...787..132D} \mbox{\small \url{https://github.com/kdeck/TTVFast}}\\
    
    \texttt{TRADES} & N-body/Photo-dynamical & \citet{Borsato2014AA...571A..38B, Borsato2019MNRAS.484.3233B, Borsato2024AA...689A..52B} \mbox{\small \url{https://github.com/lucaborsato/trades}} \\    
    \texttt{pyTTV} & N-body/Photo-dynamical & \citet{Korth2023AA...675A.115K} \\
    \texttt{jnkepler} & N-body/Photo-dynamical & \citet{2024AJ....168..294M} \mbox{\small \url{https://github.com/kemasuda/jnkepler}}\\

    \texttt{photodynam} & Photo-dynamical & \citet{CarterAgol2013ApJ...765..132C} \mbox{\small \url{https://github.com/dfm/photodynam}} \\
    \texttt{PyDynamicaLC} & Photo-dynamical & \citet{Yoffe2021ApJ...908..114Y} \mbox{\small \url{https://github.com/avivofir/PyDynamicaLC}} \\
    \texttt{planet-planet} & Photo-dynamical & \citet{Luger2017ApJ85194L} \mbox{\small \url{https://github.com/rodluger/planetplanet}} \\
    \texttt{PhoDyMM} & Photo-dynamical & \citet{phodymm2016Natur.533..509M} \\

    \hline
\end{longtblr}

% ====================================================================
% ==TABLE:OPTIMISERS==
% ====================================================================
% - GLOBAL
% ====================================================================
% ====================================================================
% - LOCAL
% ====================================================================
\begin{longtblr}[
    caption = {Global and local optimisation tools.},
    label = {tab:opti_tools},
]{
  width = \textwidth,
  % X[ratio, alignment, vertical_alignment]
  % Ratios
  colspec = { X[3,l,t] X[1,l,t] X[7,l,t] },
  rowhead = 1,      % Repeats the first row on every page
  % hlines,           % Adds horizontal lines (replaces \hline)
  row{1} = {font=\bfseries}, % Automatically makes header bold
}
    \hline
    Code & Type & References and Source\\
    \hline
    
    Genetic Algorithm (GA) & Global & \citet{john1992adaptation, Charbonneau1995ApJS..101..309C}; \mbox{e.g. \texttt{PIKAIA}} \mbox{\small \url{https://www.hao.ucar.edu/modeling/pikaia/pikaia.php}} \\
    Differential Evolution (DE) & Global & \citet{StornP97}; \mbox{e.g. \texttt{pyDE}} \mbox{\small \url{https://github.com/hpparvi/PyDE}}\\
    Particle Swarm Optimisation (PSO) & Global & \citet{PSO_KennedyEberheart1995} \\
    Ant Colony & Global & \citet{DorigoStutzle2004ANT} \\

    Levenberg-Marquardt algorithm & Local & \citet{Levenberg1944, Marquardt1963}; \mbox{e.g. MINPACK}\\
    Nelder-Mead Simplex Method & Local & \citet{Nelder-Mead1965} \\
    Broyden-Fletcher-Goldfarb-Shanno (BFGS) & Local & \citet{broyden1970,fletcher1970,goldfarb1970,shanno1970} \\
    Conjugate Gradient Method & Local & \citet{Hestenes1952MethodsOC} \\
    Powell's Method & Local & \citet{Powell1964} \\

    \hline
\end{longtblr}

% ====================================================================
% ==TABLE:BAYESIAN==
% ====================================================================
% - MCMC
% ====================================================================
% ====================================================================
% - NESTEDSAMPLING
% ====================================================================
\begin{longtblr}[
    caption = {Bayesian framework tools},
    label = {tab:bayesian_tools},
]{
  width = \textwidth,
  % X[ratio, alignment, vertical_alignment]
  % Ratios
  colspec = { X[3,l,t] X[4,l,t] X[13,l,t] },
  rowhead = 1,      % Repeats the first row on every page
  % hlines,           % Adds horizontal lines (replaces \hline)
  row{1} = {font=\bfseries}, % Automatically makes header bold
}
    \hline
    Algor./Code & Type & References and Source\\
    \hline
    \textit{Algorithm} & & \\
    
    Metropolis-Hastings algorithm & MCMC & \citet{Metropolis1953JChPh..21.1087M, Hastings1970Bimka..57...97H} \\
    
    Gibbs sampling & MCMC sampler & \citet{Gibbs_sampler} \\

    Affine-invariant ensemble sampler & MCMC sampler & \citet{GoodmanWeare2010CAMCS...5...65G} \\

    Differential evolution sampler & MCMC sampler & \citet{terBraak2008, DENelson2014ApJS..210...11N} \\

    Nested Sampling & Nested Sampling algorithm & \citet{Skilling2004AIPC..735..395S,Skilling10.1214/06-BA127}

    \textit{Code/Implementation} & & \\

    \texttt{emcee} & MCMC code & \citet{Foreman2013,DFM2019JOSS....4.1864F} \\

    \texttt{MultiNest} & Nested-Sampling code & \citet{Feroz2009MNRAS.398.1601F} \mbox{\small \url{https://github.com/JohannesBuchner/MultiNest}} \\

    \texttt{POLYCHORD} & Nested-Sampling code & \citet{Handley2015MNRAS.450L..61H} \mbox{\small \url{https://ascl.net/1502.011}} \\

    \texttt{dynesty}   & Nested-Sampling code & \citet{Speagle2020MNRAS.493.3132S} \mbox{\small \url{https://dynesty.readthedocs.io/en/stable/}} \\
    
    \texttt{ultranest} & Nested-Sampling code & \citet{Buchner2023StSur..17..169B} \mbox{\small \url{https://johannesbuchner.github.io/UltraNest/index.html}} \\
    
    \texttt{nautilus}  & Nested-Sampling code & \citet{nautilus} \mbox{\small \url{https://nautilus-sampler.readthedocs.io/en/stable/index.html}} \\

    \hline
\end{longtblr}

% ====================================================================
% - STABILITY
% ====================================================================
\begin{longtblr}[
    caption = {Stability indicators and tools.},
    label = {tab:stability},
    note{*} = {Chaotic indicators mix numerical integration with mathematical tools to identify long-term instability.}
]{
  width = \textwidth,
  % X[ratio, alignment, vertical_alignment]
  % Ratios
  colspec = { X[4,l,t] X[3,l,t] X[14,l,t] },
  rowhead = 1,      % Repeats the first row on every page
  % hlines,           % Adds horizontal lines (replaces \hline)
  row{1} = {font=\bfseries}, % Automatically makes header bold
}
    \hline
    Criterion or tool & Type & Notes\\
    
    \hline
    Hill stability & Analytical & 
        Determines the minimum orbital separation required for two planets 
        to avoid close encounters, but it assumes circular and coplanar orbits. 
        It doesn't account for long-term chaotic behaviour, resonant interactions, 
        or effects of inclination
        \citep{Gladman1993Icar..106..247G}.\\
    Angular Momentum Deficit (AMD)  & Analytical & 
        The AMD is a measure of how far a planetary system's angular momentum deviates from
        a hypothetical coplanar, circular system with the same masses and semi-major axes. 
        Does not account for resonances.
        A lower AMD generally means higher long-term stability, but some stable systems can
        have relatively high AMD, and some unstable ones can have low AMD
        \citep{Laskar1997AA...317L..75L,Laskar2000PhRvL..84.3240L,LaskarPetit2017AA...605A..72L}.\\
    Normalised Angular Momentum Deficit (NAMD) & Analytical &
        The NAMD is defined as the ratio
        of the AMD to the circular angular momentum (CAM), 
        that is the angular momentum of the planetary
        system with the same masses and semimajor axis values for the
        planets, but with circular and co-planar orbits
        \citep{Chambers2001Icar..152..205C,Turrini2020AA...636A..53T,2026AA...706A.222B}\\
    AMD-Hill stability  & Analytical & 
        A combination of the mutual Hill and AMD stability criteria. 
        It improves upon the two criteria on which it is based, but it still suffers from the same assumptions
        \citet{Petit2018AA...617A..93P}.\\

    N-body integration & Numerical & 
        General and the most robust method to determine system's stability 
        by integrating the whole system for a long time.
        The results could depend on the choice of the numerical integrators, 
        the most suitable for long-term analysis are the symplectic integrators, 
        such as Wisdom-Holmam \citep{WisdomHolman1991AJ....102.1528W}, SYMBA \citep{Chambers1999MNRAS.304..793C}, 
        and Bulirsch-Stoers integrators \citep{NR-Press1996}.
        For example, \texttt{rebound} is an N-body \texttt{C-python} package \citep{rebound}, 
        that implements different integrators,
        such as the Wisdom-Holman, 
        \texttt{whfast} \citep{reboundwhfast}, 
        \texttt{ias15} \citep{rebound},
        and others.\\

    Lyapunov Exponent (LE) & Chaotic indicators\TblrNote{*} &
        Measures the exponential divergence of initially nearby trajectories in phase space.
        A positive LE indicates chaos.\\
    Mean Exponential Growth of Nearby Orbits (MEGNO) & Chaotic indicators & 
        A faster-converging variant of the Lyapunov exponent
        \citep{Cincotta2000AAS..147..205C}.\\
    Frequency Map Analysis (FMA) & Chaotic indicators & 
        Detects chaotic behaviour by looking for changes in the fundamental frequencies of the orbits over time.
         In quasi-periodic motion, the frequencies remain nearly constant, 
         while in chaotic motion, they change over time \citep{Laskar1993PhyD...67..257L}.\\

    \hline
\end{longtblr}

% ====================================================================
% - ACRONYMS
% ====================================================================
\begin{longtblr}[
    caption = {List of acronyms.},
    label = {tab:acronyms},
]{
  width = \textwidth,
  % X[ratio, alignment, vertical_alignment]
  % Ratios
  colspec = { X[3,l,t] X[7,l,t] },
  rowhead = 1,      % Repeats the first row on every page
  % hlines,           % Adds horizontal lines (replaces \hline)
  row{1} = {font=\bfseries}, % Automatically makes header bold
}
    \hline
    Acronym & Meaning\\
    \hline
    
    TTV & Transit Timing Variation or Transit Time Variation \\
    TDV & Transit Duration Variation \\
    MMR & Mean-Motion Resonance \\
    RV & Radial Velocity \\
    BFGS & Broyden–Fletcher–Goldfarb–Shanno \\
    MCMC & Markov Chain Monte Carlo \\
    HMC & Hamiltonian Monte Carlo \\
    HDI & High Density Interval \\
    HPD & High Posterior Density \\
    CI & Confidence Interval \\
    MLE & Maximum-Likelihood Estimation \\
    MAP & Maximum-A-Posteriori Probability \\
    HEM & High-Eccentricity Migration \\
    ETV & Eclipse Timing Variation \\
    ETTV & Eclipse Transit Timing Variation \\
    DE & Differential Evolution \\
    GA & Genetic Algorithm \\
    PSO & Particle Swarm Optimization \\
    ACF & Auto-Correlation Function \\
    AMD & Angular Momentum Deficit \\
    NAMD & Normalized AMD \\
    GP & Gaussian Process \\
    MEGNO & Mean Exponential Growth factor of Nearby Orbits \\

    \hline
\end{longtblr}

% --- Bibliography Section ---
{
    \footnotesize
    \bibliography{references}

@ARTICLE{Adams1847MmRAS..16..427A,
       author = {{Adams}, J.~C.},
        title = "{An Explanation of the Observed Irregularities in the Motion of Uranus, on the Hypothesis of Disturbances caused by a more Distant Planet; with a Determination of the Mass, Orbit, and Position of the Disturbing Body.}",
      journal = {\memras},
         year = 1847,
        month = jan,
       volume = {16},
        pages = {427},
       adsurl = {https://ui.adsabs.harvard.edu/abs/1847MmRAS..16..427A}
}

@ARTICLE{LeVerrier1877AnPar..14....1L,
       author = {{Le Verrier}, Urbain J.},
        title = "{Tables du mouvement de Neptune fondees sur la comparaison de la theorie avec les observations}",
      journal = {Annales de l'Observatoire de Paris},
         year = 1877,
        month = jan,
       volume = {14},
        pages = {1-97},
       adsurl = {https://ui.adsabs.harvard.edu/abs/1877AnPar..14....1L}
}

@BOOK{NR1992nrfa.book.....P,
       author = {{Press}, William H. and {Teukolsky}, Saul A. and {Vetterling}, William T. and {Flannery}, Brian P.},
        title = "{Numerical recipes in FORTRAN. The art of scientific computing}",
         year = 1992,
       adsurl = {https://ui.adsabs.harvard.edu/abs/1992nrfa.book.....P}
}

@ARTICLE{Agol2005MNRAS.359..567A,
       author = {{Agol}, Eric and {Steffen}, Jason and {Sari}, Re'em and {Clarkson}, Will},
        title = "{On detecting terrestrial planets with timing of giant planet transits}",
      journal = {\mnras},
         year = 2005,
        month = may,
       volume = {359},
       number = {2},
        pages = {567-579},
          doi = {10.1111/j.1365-2966.2005.08922.x},
       adsurl = {https://ui.adsabs.harvard.edu/abs/2005MNRAS.359..567A}
}

@ARTICLE{HolmanMurray2005Sci...307.1288H,
       author = {{Holman}, Matthew J. and {Murray}, Norman W.},
        title = "{The Use of Transit Timing to Detect Terrestrial-Mass Extrasolar Planets}",
      journal = {Science},
         year = 2005,
        month = feb,
       volume = {307},
       number = {5713},
        pages = {1288-1291},
          doi = {10.1126/science.1107822},
archivePrefix = {arXiv},
       eprint = {astro-ph/0412028},
 primaryClass = {astro-ph},
       adsurl = {https://ui.adsabs.harvard.edu/abs/2005Sci...307.1288H}
}

@ARTICLE{Holman2010Sci...330...51H,
       author = {{Holman}, Matthew J. and {Fabrycky}, Daniel C. and {Ragozzine}, Darin and {Ford}, Eric B. and {Steffen}, Jason H. and {Welsh}, William F. and {Lissauer}, Jack J. and {Latham}, David W. and {Marcy}, Geoffrey W. and {Walkowicz}, Lucianne M. and {Batalha}, Natalie M. and {Jenkins}, Jon M. and {Rowe}, Jason F. and {Cochran}, William D. and {Fressin}, Francois and {Torres}, Guillermo and {Buchhave}, Lars A. and {Sasselov}, Dimitar D. and {Borucki}, William J. and {Koch}, David G. and {Basri}, Gibor and {Brown}, Timothy M. and {Caldwell}, Douglas A. and {Charbonneau}, David and {Dunham}, Edward W. and {Gautier}, Thomas N. and {Geary}, John C. and {Gilliland}, Ronald L. and {Haas}, Michael R. and {Howell}, Steve B. and {Ciardi}, David R. and {Endl}, Michael and {Fischer}, Debra and {F{\"u}r{\'e}sz}, G{\'a}bor and {Hartman}, Joel D. and {Isaacson}, Howard and {Johnson}, John A. and {MacQueen}, Phillip J. and {Moorhead}, Althea V. and {Morehead}, Robert C. and {Orosz}, Jerome A.},
        title = "{Kepler-9: A System of Multiple Planets Transiting a Sun-Like Star, Confirmed by Timing Variations}",
      journal = {Science},
         year = 2010,
        month = oct,
       volume = {330},
       number = {6000},
        pages = {51},
          doi = {10.1126/science.1195778},
       adsurl = {https://ui.adsabs.harvard.edu/abs/2010Sci...330...51H}
}

@ARTICLE{Borsato2019MNRAS.484.3233B,
       author = {{Borsato}, L. and {Malavolta}, L. and {Piotto}, G. and {Buchhave}, L.~A. and {Mortier}, A. and {Rice}, K. and {Collier Cameron}, A. and {Coffinet}, A. and {Sozzetti}, A. and {Charbonneau}, D. and {Cosentino}, R. and {Dumusque}, X. and {Figueira}, P. and {Latham}, D.~W. and {Lopez-Morales}, M. and {Mayor}, M. and {Micela}, G. and {Molinari}, E. and {Pepe}, F. and {Phillips}, D. and {Poretti}, E. and {Udry}, S. and {Watson}, C.},
        title = "{HARPS-N radial velocities confirm the low densities of the Kepler-9 planets}",
      journal = {\mnras},
         year = 2019,
        month = apr,
       volume = {484},
       number = {3},
        pages = {3233-3243},
          doi = {10.1093/mnras/stz181},
archivePrefix = {arXiv},
       eprint = {1901.05471},
 primaryClass = {astro-ph.EP},
       adsurl = {https://ui.adsabs.harvard.edu/abs/2019MNRAS.484.3233B}
}

@INCOLLECTION{2025haex.book....2A,
       author = {{Agol}, Eric and {Fabrycky}, Daniel},
        title = "{Transit Timing and Duration Variations for the Discovery and Characterization of Exoplanets in the TESS Era}",
    booktitle = {Handbook of Exoplanets},
         year = 2025,
        pages = {2},
          doi = {10.1007/978-3-319-30648-3_7-2},
       adsurl = {https://ui.adsabs.harvard.edu/abs/2025haex.book....2A}
}

@INCOLLECTION{Winn2010exop.book...55W,
       author = {{Winn}, J.~N.},
        title = "{Exoplanet Transits and Occultations}",
    booktitle = {Exoplanets},
         year = 2010,
       editor = {{Seager}, S.},
        pages = {55-77},
          doi = {10.48550/arXiv.1001.2010},
       adsurl = {https://ui.adsabs.harvard.edu/abs/2010exop.book...55W}
}

@INCOLLECTION{Fabrycky2010exop.book..217F,
       author = {{Fabrycky}, D.~C.},
        title = "{Non-Keplerian Dynamics of Exoplanets}",
    booktitle = {Exoplanets},
         year = 2010,
       editor = {{Seager}, S.},
        pages = {217-238},
       adsurl = {https://ui.adsabs.harvard.edu/abs/2010exop.book..217F}
}

@ARTICLE{Ballard2011ApJ...743..200B,
       author = {{Ballard}, Sarah and {Fabrycky}, Daniel and {Fressin}, Francois and {Charbonneau}, David and {Desert}, Jean-Michel and {Torres}, Guillermo and {Marcy}, Geoffrey and {Burke}, Christopher J. and {Isaacson}, Howard and {Henze}, Christopher and {Steffen}, Jason H. and {Ciardi}, David R. and {Howell}, Steven B. and {Cochran}, William D. and {Endl}, Michael and {Bryson}, Stephen T. and {Rowe}, Jason F. and {Holman}, Matthew J. and {Lissauer}, Jack J. and {Jenkins}, Jon M. and {Still}, Martin and {Ford}, Eric B. and {Christiansen}, Jessie L. and {Middour}, Christopher K. and {Haas}, Michael R. and {Li}, Jie and {Hall}, Jennifer R. and {McCauliff}, Sean and {Batalha}, Natalie M. and {Koch}, David G. and {Borucki}, William J.},
        title = "{The Kepler-19 System: A Transiting 2.2 R $_{{\ensuremath{\oplus}}}$ Planet and a Second Planet Detected via Transit Timing Variations}",
      journal = {\apj},
         year = 2011,
        month = dec,
       volume = {743},
       number = {2},
          eid = {200},
        pages = {200},
          doi = {10.1088/0004-637X/743/2/200},
archivePrefix = {arXiv},
       eprint = {1109.1561},
 primaryClass = {astro-ph.EP},
       adsurl = {https://ui.adsabs.harvard.edu/abs/2011ApJ...743..200B}
}

@ARTICLE{Nesvorny2013ApJ...777....3N,
       author = {{Nesvorn{\'y}}, David and {Kipping}, David and {Terrell}, Dirk and {Hartman}, Joel and {Bakos}, G{\'a}sp{\'a}r {\'A}. and {Buchhave}, Lars A.},
        title = "{KOI-142, The King of Transit Variations, is a Pair of Planets near the 2:1 Resonance}",
      journal = {\apj},
         year = 2013,
        month = nov,
       volume = {777},
       number = {1},
          eid = {3},
        pages = {3},
          doi = {10.1088/0004-637X/777/1/3},
archivePrefix = {arXiv},
       eprint = {1304.4283},
 primaryClass = {astro-ph.EP},
       adsurl = {https://ui.adsabs.harvard.edu/abs/2013ApJ...777....3N}
}

@ARTICLE{Weiss2020AJ....159..242W,
       author = {{Weiss}, Lauren M. and {Fabrycky}, Daniel C. and {Agol}, Eric and {Mills}, Sean M. and {Howard}, Andrew W. and {Isaacson}, Howard and {Petigura}, Erik A. and {Fulton}, Benjamin and {Hirsch}, Lea and {Sinukoff}, Evan},
        title = "{The Discovery of the Long-Period, Eccentric Planet Kepler-88 d and System Characterization with Radial Velocities and Photodynamical Analysis}",
      journal = {\aj},
         year = 2020,
        month = may,
       volume = {159},
       number = {5},
          eid = {242},
        pages = {242},
          doi = {10.3847/1538-3881/ab88ca},
archivePrefix = {arXiv},
       eprint = {1909.02427},
 primaryClass = {astro-ph.EP},
       adsurl = {https://ui.adsabs.harvard.edu/abs/2020AJ....159..242W}
}

@ARTICLE{McKee2023AJ....165..236M,
       author = {{McKee}, Brendan J. and {Montet}, Benjamin T.},
        title = "{Transit Depth Variations Reveal TOI-216 b to be a Super-puff}",
      journal = {\aj},
         year = 2023,
        month = jun,
       volume = {165},
       number = {6},
          eid = {236},
        pages = {236},
          doi = {10.3847/1538-3881/accd66},
archivePrefix = {arXiv},
       eprint = {2212.07450},
 primaryClass = {astro-ph.EP},
       adsurl = {https://ui.adsabs.harvard.edu/abs/2023AJ....165..236M}
}

@ARTICLE{Agol2016ApJ...818..177A,
       author = {{Agol}, Eric and {Deck}, Katherine},
        title = "{Transit Timing to First Order in Eccentricity}",
      journal = {\apj},
         year = 2016,
        month = feb,
       volume = {818},
       number = {2},
          eid = {177},
        pages = {177},
          doi = {10.3847/0004-637X/818/2/177},
archivePrefix = {arXiv},
       eprint = {1509.01623},
 primaryClass = {astro-ph.EP},
       adsurl = {https://ui.adsabs.harvard.edu/abs/2016ApJ...818..177A}
}

@ARTICLE{CarterAgol2013ApJ...765..132C,
       author = {{Carter}, Joshua A. and {Agol}, Eric},
        title = "{The Quasiperiodic Automated Transit Search Algorithm}",
      journal = {\apj},
         year = 2013,
        month = mar,
       volume = {765},
       number = {2},
          eid = {132},
        pages = {132},
          doi = {10.1088/0004-637X/765/2/132},
archivePrefix = {arXiv},
       eprint = {1210.5136},
 primaryClass = {astro-ph.EP},
       adsurl = {https://ui.adsabs.harvard.edu/abs/2013ApJ...765..132C}
}

@BOOK{MuDe1999book,
   author = {{Murray}, C.~D. and {Dermott}, S.~F.},
    title = "{Solar system dynamics}",
booktitle = {Solar system dynamics by Murray, C.~D., 1999},
     year = 1999,
   adsurl = {http://adsabs.harvard.edu/abs/1999ssd..book.....M},
publisher = {Cambridge University Press}
}

@ARTICLE{Malavolta2017AJ....153..224M,
       author = {{Malavolta}, Luca and {Borsato}, Luca and {Granata}, Valentina and {Piotto}, Giampaolo and {Lopez}, Eric and {Vanderburg}, Andrew and {Figueira}, Pedro and {Mortier}, Annelies and {Nascimbeni}, Valerio and {Affer}, Laura and {Bonomo}, Aldo S. and {Bouchy}, Francois and {Buchhave}, Lars A. and {Charbonneau}, David and {Collier Cameron}, Andrew and {Cosentino}, Rosario and {Dressing}, Courtney D. and {Dumusque}, Xavier and {Fiorenzano}, Aldo F.~M. and {Harutyunyan}, Avet and {Haywood}, Rapha{\"e}lle D. and {Johnson}, John Asher and {Latham}, David W. and {Lopez-Morales}, Mercedes and {Lovis}, Christophe and {Mayor}, Michel and {Micela}, Giusi and {Molinari}, Emilio and {Motalebi}, Fatemeh and {Pepe}, Francesco and {Phillips}, David F. and {Pollacco}, Don and {Queloz}, Didier and {Rice}, Ken and {Sasselov}, Dimitar and {S{\'e}gransan}, Damien and {Sozzetti}, Alessandro and {Udry}, St{\'e}phane and {Watson}, Chris},
        title = "{The Kepler-19 System: A Thick-envelope Super-Earth with Two Neptune-mass Companions Characterized Using Radial Velocities and Transit Timing Variations}",
      journal = {\aj},
         year = 2017,
        month = may,
       volume = {153},
       number = {5},
          eid = {224},
        pages = {224},
          doi = {10.3847/1538-3881/aa6897},
archivePrefix = {arXiv},
       eprint = {1703.06885},
 primaryClass = {astro-ph.EP},
       adsurl = {https://ui.adsabs.harvard.edu/abs/2017AJ....153..224M}
}

@ARTICLE{Boue2012MNRAS.422L..57B,
       author = {{Bou{\'e}}, G. and {Oshagh}, M. and {Montalto}, M. and {Santos}, N.~C.},
        title = "{Degeneracy in the characterization of non-transiting planets from transit timing variations}",
      journal = {\mnras},
         year = 2012,
        month = may,
       volume = {422},
       number = {1},
        pages = {L57-L61},
          doi = {10.1111/j.1745-3933.2012.01236.x},
archivePrefix = {arXiv},
       eprint = {1201.2080},
 primaryClass = {astro-ph.EP},
       adsurl = {https://ui.adsabs.harvard.edu/abs/2012MNRAS.422L..57B}
}

@ARTICLE{Lithwick2012ApJ...761..122L,
       author = {{Lithwick}, Yoram and {Xie}, Jiwei and {Wu}, Yanqin},
        title = "{Extracting Planet Mass and Eccentricity from TTV Data}",
      journal = {\apj},
         year = 2012,
        month = dec,
       volume = {761},
       number = {2},
          eid = {122},
        pages = {122},
          doi = {10.1088/0004-637X/761/2/122},
archivePrefix = {arXiv},
       eprint = {1207.4192},
 primaryClass = {astro-ph.EP},
       adsurl = {https://ui.adsabs.harvard.edu/abs/2012ApJ...761..122L}
}

@ARTICLE{Borsato2014AA...571A..38B,
       author = {{Borsato}, L. and {Marzari}, F. and {Nascimbeni}, V. and {Piotto}, G. and {Granata}, V. and {Bedin}, L.~R. and {Malavolta}, L.},
        title = "{TRADES: A new software to derive orbital parameters from observed transit times and radial velocities. Revisiting Kepler-11 and Kepler-9}",
      journal = {\aap},
         year = 2014,
        month = nov,
       volume = {571},
          eid = {A38},
        pages = {A38},
          doi = {10.1051/0004-6361/201424080},
archivePrefix = {arXiv},
       eprint = {1408.2844},
 primaryClass = {astro-ph.EP},
       adsurl = {https://ui.adsabs.harvard.edu/abs/2014AA...571A..38B}
}

@ARTICLE{Barros2014AA...561L...1B,
       author = {{Barros}, S.~C.~C. and {D{\'\i}az}, R.~F. and {Santerne}, A. and {Bruno}, G. and {Deleuil}, M. and {Almenara}, J. -M. and {Bonomo}, A.~S. and {Bouchy}, F. and {Damiani}, C. and {H{\'e}brard}, G. and {Montagnier}, G. and {Moutou}, C.},
        title = "{SOPHIE velocimetry of Kepler transit candidates. X. KOI-142 c: first radial velocity confirmation of a non-transiting exoplanet discovered by transit timing}",
      journal = {\aap},
         year = 2014,
        month = jan,
       volume = {561},
          eid = {L1},
        pages = {L1},
          doi = {10.1051/0004-6361/201323067},
archivePrefix = {arXiv},
       eprint = {1311.4335},
 primaryClass = {astro-ph.EP},
       adsurl = {https://ui.adsabs.harvard.edu/abs/2014AA...561L...1B}
}

@ARTICLE{Kipping2019MNRAS.486.4980K,
       author = {{Kipping}, David and {Nesvorn{\'y}}, David and {Hartman}, Joel and {Torres}, Guillermo and {Bakos}, Gaspar and {Jansen}, Tiffany and {Teachey}, Alex},
        title = "{A resonant pair of warm giant planets revealed by TESS}",
      journal = {\mnras},
         year = 2019,
        month = jul,
       volume = {486},
       number = {4},
        pages = {4980-4986},
          doi = {10.1093/mnras/stz1141},
archivePrefix = {arXiv},
       eprint = {1902.03900},
 primaryClass = {astro-ph.EP},
       adsurl = {https://ui.adsabs.harvard.edu/abs/2019MNRAS.486.4980K}
}

@ARTICLE{Dawson2021AJ....161..161D,
       author = {{Dawson}, Rebekah I. and {Huang}, Chelsea X. and {Brahm}, Rafael and {Collins}, Karen A. and {Hobson}, Melissa J. and {Jord{\'a}n}, Andr{\'e}s and {Dong}, Jiayin and {Korth}, Judith and {Trifonov}, Trifon and {Abe}, Lyu and {Agabi}, Abdelkrim and {Bruni}, Ivan and {Butler}, R. Paul and {Barbieri}, Mauro and {Collins}, Kevin I. and {Conti}, Dennis M. and {Crane}, Jeffrey D. and {Crouzet}, Nicolas and {Dransfield}, Georgina and {Evans}, Phil and {Espinoza}, N{\'e}stor and {Gan}, Tianjun and {Guillot}, Tristan and {Henning}, Thomas and {Lissauer}, Jack J. and {Jensen}, Eric L.~N. and {Sainte}, Wenceslas Marie and {M{\'e}karnia}, Djamel and {Myers}, Gordon and {Nandakumar}, Sangeetha and {Relles}, Howard M. and {Sarkis}, Paula and {Torres}, Pascal and {Shectman}, Stephen and {Schmider}, Fran{\c{c}}ois-Xavier and {Shporer}, Avi and {Stockdale}, Chris and {Teske}, Johanna and {Triaud}, Amaury H.~M.~J. and {Wang}, Sharon Xuesong and {Ziegler}, Carl and {Ricker}, G. and {Vanderspek}, R. and {Latham}, David W. and {Seager}, S. and {Winn}, J. and {Jenkins}, Jon M. and {Bouma}, L.~G. and {Burt}, Jennifer A. and {Charbonneau}, David and {Levine}, Alan M. and {McDermott}, Scott and {McLean}, Brian and {Rose}, Mark E. and {Vanderburg}, Andrew and {Wohler}, Bill},
        title = "{Precise Transit and Radial-velocity Characterization of a Resonant Pair: The Warm Jupiter TOI-216c and Eccentric Warm Neptune TOI-216b}",
      journal = {\aj},
         year = 2021,
        month = apr,
       volume = {161},
       number = {4},
          eid = {161},
        pages = {161},
          doi = {10.3847/1538-3881/abd8d0},
archivePrefix = {arXiv},
       eprint = {2102.06754},
 primaryClass = {astro-ph.EP},
       adsurl = {https://ui.adsabs.harvard.edu/abs/2021AJ....161..161D}
}

@ARTICLE{Nesvorny2014ApJ...790...58N,
       author = {{Nesvorn{\'y}}, David and {Vokrouhlick{\'y}}, David},
        title = "{The Effect of Conjunctions on the Transit Timing Variations of Exoplanets}",
      journal = {\apj},
         year = 2014,
        month = jul,
       volume = {790},
       number = {1},
          eid = {58},
        pages = {58},
          doi = {10.1088/0004-637X/790/1/58},
archivePrefix = {arXiv},
       eprint = {1405.7433},
 primaryClass = {astro-ph.EP},
       adsurl = {https://ui.adsabs.harvard.edu/abs/2014ApJ...790...58N}
}

@ARTICLE{Deck2013ApJ...774..129D,
       author = {{Deck}, Katherine M. and {Payne}, Matthew and {Holman}, Matthew J.},
        title = "{First-order Resonance Overlap and the Stability of Close Two-planet Systems}",
      journal = {\apj},
         year = 2013,
        month = sep,
       volume = {774},
       number = {2},
          eid = {129},
        pages = {129},
          doi = {10.1088/0004-637X/774/2/129},
archivePrefix = {arXiv},
       eprint = {1307.8119},
 primaryClass = {astro-ph.EP},
       adsurl = {https://ui.adsabs.harvard.edu/abs/2013ApJ...774..129D}
}

@ARTICLE{Deck2014ApJ...787..132D,
       author = {{Deck}, Katherine M. and {Agol}, Eric and {Holman}, Matthew J. and {Nesvorn{\'y}}, David},
        title = "{TTVFast: An Efficient and Accurate Code for Transit Timing Inversion Problems}",
      journal = {\apj},
         year = 2014,
        month = jun,
       volume = {787},
       number = {2},
          eid = {132},
        pages = {132},
          doi = {10.1088/0004-637X/787/2/132},
archivePrefix = {arXiv},
       eprint = {1403.1895},
 primaryClass = {astro-ph.EP},
       adsurl = {https://ui.adsabs.harvard.edu/abs/2014ApJ...787..132D}
}

@ARTICLE{Deck2015ApJ...802..116D,
       author = {{Deck}, Katherine M. and {Agol}, Eric},
        title = "{Measurement of Planet Masses with Transit Timing Variations Due to Synodic {\textquotedblleft}Chopping{\textquotedblright} Effects}",
      journal = {\apj},
         year = 2015,
        month = apr,
       volume = {802},
       number = {2},
          eid = {116},
        pages = {116},
          doi = {10.1088/0004-637X/802/2/116},
archivePrefix = {arXiv},
       eprint = {1411.0004},
 primaryClass = {astro-ph.EP},
       adsurl = {https://ui.adsabs.harvard.edu/abs/2015ApJ...802..116D}
}

@ARTICLE{Deck2016ApJ...821...96D,
       author = {{Deck}, Katherine M. and {Agol}, Eric},
        title = "{Transit Timing Variations for Planets near Eccentricity-type Mean Motion Resonances}",
      journal = {\apj},
         year = 2016,
        month = apr,
       volume = {821},
       number = {2},
          eid = {96},
        pages = {96},
          doi = {10.3847/0004-637X/821/2/96},
archivePrefix = {arXiv},
       eprint = {1509.08460},
 primaryClass = {astro-ph.EP},
       adsurl = {https://ui.adsabs.harvard.edu/abs/2016ApJ...821...96D}
}

@ARTICLE{Nesvorny2016ApJ...823...72N,
       author = {{Nesvorn{\'y}}, David and {Vokrouhlick{\'y}}, David},
        title = "{Dynamics and Transit Variations of Resonant Exoplanets}",
      journal = {\apj},
         year = 2016,
        month = jun,
       volume = {823},
       number = {2},
          eid = {72},
        pages = {72},
          doi = {10.3847/0004-637X/823/2/72},
archivePrefix = {arXiv},
       eprint = {1603.07306},
 primaryClass = {astro-ph.EP},
       adsurl = {https://ui.adsabs.harvard.edu/abs/2016ApJ...823...72N}
}

@ARTICLE{Nesvorny2008ApJ...688..636N,
       author = {{Nesvorn{\'y}}, David and {Morbidelli}, Alessandro},
        title = "{Mass and Orbit Determination from Transit Timing Variations of Exoplanets}",
      journal = {\apj},
         year = 2008,
        month = nov,
       volume = {688},
       number = {1},
        pages = {636-646},
          doi = {10.1086/592230},
       adsurl = {https://ui.adsabs.harvard.edu/abs/2008ApJ...688..636N}
}

@ARTICLE{Hadden2016ApJ...828...44H,
       author = {{Hadden}, Sam and {Lithwick}, Yoram},
        title = "{Numerical and Analytical Modeling of Transit Timing Variations}",
      journal = {\apj},
         year = 2016,
        month = sep,
       volume = {828},
       number = {1},
          eid = {44},
        pages = {44},
          doi = {10.3847/0004-637X/828/1/44},
archivePrefix = {arXiv},
       eprint = {1510.02476},
 primaryClass = {astro-ph.EP},
       adsurl = {https://ui.adsabs.harvard.edu/abs/2016ApJ...828...44H}
}

@ARTICLE{Linial2018ApJ...860...16L,
       author = {{Linial}, Itai and {Gilbaum}, Shmuel and {Sari}, Re'em},
        title = "{Modal Decomposition of TTV: Inferring Planet Masses and Eccentricities}",
      journal = {\apj},
         year = 2018,
        month = jun,
       volume = {860},
       number = {1},
          eid = {16},
        pages = {16},
          doi = {10.3847/1538-4357/aac21b},
archivePrefix = {arXiv},
       eprint = {1802.07731},
 primaryClass = {astro-ph.EP},
       adsurl = {https://ui.adsabs.harvard.edu/abs/2018ApJ...860...16L}
}

@ARTICLE{Almenara2015MNRAS.453.2644A,
       author = {{Almenara}, J.~M. and {D{\'\i}az}, R.~F. and {Mardling}, R. and {Barros}, S.~C.~C. and {Damiani}, C. and {Bruno}, G. and {Bonfils}, X. and {Deleuil}, M.},
        title = "{Absolute masses and radii determination in multiplanetary systems without stellar models}",
      journal = {\mnras},
         year = 2015,
        month = nov,
       volume = {453},
       number = {3},
        pages = {2644-2652},
          doi = {10.1093/mnras/stv1735},
archivePrefix = {arXiv},
       eprint = {1508.06596},
 primaryClass = {astro-ph.EP},
       adsurl = {https://ui.adsabs.harvard.edu/abs/2015MNRAS.453.2644A}
}

@ARTICLE{Carter2011Sci...331..562C,
       author = {{Carter}, Joshua A. and {Fabrycky}, Daniel C. and {Ragozzine}, Darin and {Holman}, Matthew J. and {Quinn}, Samuel N. and {Latham}, David W. and {Buchhave}, Lars A. and {Van Cleve}, Jeffrey and {Cochran}, William D. and {Cote}, Miles T. and {Endl}, Michael and {Ford}, Eric B. and {Haas}, Michael R. and {Jenkins}, Jon M. and {Koch}, David G. and {Li}, Jie and {Lissauer}, Jack J. and {MacQueen}, Phillip J. and {Middour}, Christopher K. and {Orosz}, Jerome A. and {Rowe}, Jason F. and {Steffen}, Jason H. and {Welsh}, William F.},
        title = "{KOI-126: A Triply Eclipsing Hierarchical Triple with Two Low-Mass Stars}",
      journal = {Science},
         year = 2011,
        month = feb,
       volume = {331},
       number = {6017},
        pages = {562},
          doi = {10.1126/science.1201274},
archivePrefix = {arXiv},
       eprint = {1102.0562},
 primaryClass = {astro-ph.SR},
       adsurl = {https://ui.adsabs.harvard.edu/abs/2011Sci...331..562C}
}

@book{john1992adaptation,
  title={Adaptation in Natural and Artificial Systems: An Introductory Analysis with Applications to Biology, Control, and Artificial Intelligence},
  author={John H. Holland},
  isbn={9780585038445},
  series={Bradford book},
  url={https://books.google.it/books?id=cyV7nQEACAAJ},
  year={1992},
  publisher={MIT Press}
}

@ARTICLE{Charbonneau1995ApJS..101..309C,
       author = {{Charbonneau}, P.},
        title = "{Genetic Algorithms in Astronomy and Astrophysics}",
      journal = {\apjs},
         year = 1995,
        month = dec,
       volume = {101},
        pages = {309},
          doi = {10.1086/192242},
       adsurl = {https://ui.adsabs.harvard.edu/abs/1995ApJS..101..309C}
}

@article{StornP97,
  author = {Storn, Rainer and Price, Kenneth V.},
  ee = {https://www.wikidata.org/entity/Q60670715},
  journal = {J. Glob. Optim.},
  number = 4,
  pages = {341-359},
  title = {Differential Evolution - A Simple and Efficient Heuristic for global Optimization over Continuous Spaces.},
  url = {http://dblp.uni-trier.de/db/journals/jgo/jgo11.html\#StornP97},
  volume = 11,
  year = 1997
}

@INPROCEEDINGS{PSO_KennedyEberheart1995,
    author={Kennedy, J. and Eberhart, R.},
    booktitle={Neural Networks, 1995. Proceedings., IEEE International Conference on},
    title={Particle swarm optimization},
    year={1995},
    volume={4},
    pages={1942-1948 vol.4},
    doi={10.1109/ICNN.1995.488968},
}

@book{DorigoStutzle2004ANT,
    author = {Dorigo, Marco and Stützle, Thomas},
    title = {Ant Colony Optimization},
    publisher = {The MIT Press},
    year = {2004},
    month = {06},
    isbn = {9780262256032},
    doi = {10.7551/mitpress/1290.001.0001},
    url = {https://doi.org/10.7551/mitpress/1290.001.0001},
}

@book{yang2010nature,
  title={Nature-inspired Metaheuristic Algorithms},
  author={Yang, X.S.},
  isbn={9781905986286},
  year={2010},
  publisher={Luniver Press}
}

@article{Levenberg1944,
 ISSN = {0033569X, 15524485},
 URL = {http://www.jstor.org/stable/43633451},
 author = {Kenneth Levenberg},
 journal = {Quarterly of Applied Mathematics},
 number = {2},
 pages = {164--168},
 publisher = {Brown University},
 title = {A METHOD FOR THE SOLUTION OF CERTAIN NON-LINEAR PROBLEMS IN LEAST SQUARES},
 urldate = {2024-11-13},
 volume = {2},
 year = {1944}
}

@article{Marquardt1963,
  title={An algorithm for least-squares estimation of nonlinear parameters},
  author={Marquardt, Donald W.},
  journal={Journal of the Society for Industrial and Applied Mathematics},
  volume={11},
  number={2},
  pages={431--441},
  year={1963},
  publisher={SIAM}
}

@article{Nelder-Mead1965,
    author = {Nelder, J. A. and Mead, R.},
    title = {A Simplex Method for Function Minimization},
    journal = {The Computer Journal},
    volume = {7},
    number = {4},
    pages = {308-313},
    year = {1965},
    month = {01},
    issn = {0010-4620},
    doi = {10.1093/comjnl/7.4.308},
    url = {https://doi.org/10.1093/comjnl/7.4.308},
    eprint = {https://academic.oup.com/comjnl/article-pdf/7/4/308/1013182/7-4-308.pdf},
}

@article{broyden1970,
  title={The convergence of a class of double-rank minimization algorithms 1. General considerations},
  author={Broyden, Charles G.},
  journal={IMA Journal of Applied Mathematics},
  volume={6},
  number={1},
  pages={76--90},
  year={1970},
  publisher={Oxford University Press}
}

@article{fletcher1970,
  title={A new approach to variable metric algorithms},
  author={Fletcher, Roger},
  journal={The Computer Journal},
  volume={13},
  number={3},
  pages={317--322},
  year={1970},
  publisher={Oxford University Press}
}

@article{goldfarb1970,
  title={A family of variable-metric methods derived by variational means},
  author={Goldfarb, Donald},
  journal={Mathematics of Computation},
  volume={24},
  number={109},
  pages={23--26},
  year={1970},
  publisher={American Mathematical Society}
}

@article{shanno1970,
  title={Conditioning of quasi-Newton methods for function minimization},
  author={Shanno, David F.},
  journal={Mathematics of Computation},
  volume={24},
  number={111},
  pages={647--656},
  year={1970},
  publisher={American Mathematical Society}
}

@article{Hestenes1952MethodsOC,
  title={Methods of conjugate gradients for solving linear systems},
  author={Magnus R. Hestenes and Eduard Stiefel},
  journal={Journal of research of the National Bureau of Standards},
  year={1952},
  volume={49},
  pages={409-435},
  url={https://api.semanticscholar.org/CorpusID:2207234}
}

@article{Powell1964,
    author = {Powell, M. J. D.},
    title = {An efficient method for finding the minimum of a function of several variables without calculating derivatives},
    journal = {The Computer Journal},
    volume = {7},
    number = {2},
    pages = {155-162},
    year = {1964},
    month = {01},
    issn = {0010-4620},
    doi = {10.1093/comjnl/7.2.155},
    url = {https://doi.org/10.1093/comjnl/7.2.155},
    eprint = {https://academic.oup.com/comjnl/article-pdf/7/2/155/959784/070155.pdf},
}

@ARTICLE{Veras2011ApJ...727...74V,
       author = {{Veras}, Dimitri and {Ford}, Eric B. and {Payne}, Matthew J.},
        title = "{Quantifying the Challenges of Detecting Unseen Planetary Companions with Transit Timing Variations}",
      journal = {\apj},
         year = 2011,
        month = feb,
       volume = {727},
       number = {2},
          eid = {74},
        pages = {74},
          doi = {10.1088/0004-637X/727/2/74},
archivePrefix = {arXiv},
       eprint = {1011.1466},
 primaryClass = {astro-ph.EP},
       adsurl = {https://ui.adsabs.harvard.edu/abs/2011ApJ...727...74V}
}

@misc{Hadden2019TTV2Fast2Furious,
       author = {{Hadden}, Sam},
        title = "{shadden/TTV2Fast2Furious: First release of TTV2Fast2Furious}",
         year = 2019,
        month = jul,
          eid = {10.5281/zenodo.3356829},
          doi = {10.5281/zenodo.3356829},
      version = {v1.0.0},
    publisher = {Zenodo},
       adsurl = {https://ui.adsabs.harvard.edu/abs/2019zndo...3356829H}
}

@ARTICLE{Korth2023AA...675A.115K,
       author = {{Korth}, J. and {Gandolfi}, D. and {{\v{S}}ubjak}, J. and {Howard}, S. and {Ataiee}, S. and {Collins}, K.~A. and {Quinn}, S.~N. and {Mustill}, A.~J. and {Guillot}, T. and {Lodieu}, N. and {Smith}, A.~M.~S. and {Esposito}, M. and {Rodler}, F. and {Muresan}, A. and {Abe}, L. and {Albrecht}, S.~H. and {Alqasim}, A. and {Barkaoui}, K. and {Beck}, P.~G. and {Burke}, C.~J. and {Butler}, R.~P. and {Conti}, D.~M. and {Collins}, K.~I. and {Crane}, J.~D. and {Dai}, F. and {Deeg}, H.~J. and {Evans}, P. and {Grziwa}, S. and {Hatzes}, A.~P. and {Hirano}, T. and {Horne}, K. and {Huang}, C.~X. and {Jenkins}, J.~M. and {Kab{\'a}th}, P. and {Kielkopf}, J.~F. and {Knudstrup}, E. and {Latham}, D.~W. and {Livingston}, J. and {Luque}, R. and {Mathur}, S. and {Murgas}, F. and {Osborne}, H.~L.~M. and {Palle}, E. and {Persson}, C.~M. and {Rodriguez}, J.~E. and {Rose}, M. and {Rowden}, P. and {Schwarz}, R.~P. and {Seager}, S. and {Serrano}, L.~M. and {Sha}, L. and {Shectman}, S.~A. and {Shporer}, A. and {Srdoc}, G. and {Stockdale}, C. and {Tan}, T. -G. and {Teske}, J.~K. and {Van Eylen}, V. and {Vanderburg}, A. and {Vanderspek}, R. and {Wang}, S.~X. and {Winn}, J.~N.},
        title = "{TOI-1130: A photodynamical analysis of a hot Jupiter in resonance with an inner low-mass planet}",
      journal = {\aap},
         year = 2023,
        month = jul,
       volume = {675},
          eid = {A115},
        pages = {A115},
          doi = {10.1051/0004-6361/202244617},
archivePrefix = {arXiv},
       eprint = {2305.15565},
 primaryClass = {astro-ph.EP},
       adsurl = {https://ui.adsabs.harvard.edu/abs/2023AA...675A.115K}
}

@ARTICLE{Agol2021MNRAS.507.1582A,
       author = {{Agol}, Eric and {Hernandez}, David M. and {Langford}, Zachary},
        title = "{A differentiable N-body code for transit timing and dynamical modelling - I. Algorithm and derivatives}",
      journal = {\mnras},
         year = 2021,
        month = oct,
       volume = {507},
       number = {2},
        pages = {1582-1605},
          doi = {10.1093/mnras/stab2044},
archivePrefix = {arXiv},
       eprint = {2106.02188},
 primaryClass = {astro-ph.EP},
       adsurl = {https://ui.adsabs.harvard.edu/abs/2021MNRAS.507.1582A}
}

@ARTICLE{Luger2017ApJ85194L,
       author = {{Luger}, Rodrigo and {Lustig-Yaeger}, Jacob and {Agol}, Eric},
        title = "{Planet-Planet Occultations in TRAPPIST-1 and Other Exoplanet Systems}",
      journal = {\apj},
         year = 2017,
        month = dec,
       volume = {851},
       number = {2},
          eid = {94},
        pages = {94},
          doi = {10.3847/1538-4357/aa9c43},
archivePrefix = {arXiv},
       eprint = {1711.05739},
 primaryClass = {astro-ph.EP},
       adsurl = {https://ui.adsabs.harvard.edu/abs/2017ApJ...851...94L}
}

@misc{exostriker2019ascl.soft06004T,
       author = {{Trifonov}, Trifon},
        title = "{The Exo-Striker: Transit and radial velocity interactive fitting tool for orbital analysis and N-body simulations}",
 howpublished = {Astrophysics Source Code Library, record ascl:1906.004},
         year = 2019,
        month = jun,
          eid = {ascl:1906.004},
       adsurl = {https://ui.adsabs.harvard.edu/abs/2019ascl.soft06004T}
}

@ARTICLE{phodymm2016Natur.533..509M,
       author = {{Mills}, Sean M. and {Fabrycky}, Daniel C. and {Migaszewski}, Cezary and {Ford}, Eric B. and {Petigura}, Erik and {Isaacson}, Howard},
        title = "{A resonant chain of four transiting, sub-Neptune planets}",
      journal = {\nat},
         year = 2016,
        month = may,
       volume = {533},
       number = {7604},
        pages = {509-512},
          doi = {10.1038/nature17445},
archivePrefix = {arXiv},
       eprint = {1612.07376},
 primaryClass = {astro-ph.EP},
       adsurl = {https://ui.adsabs.harvard.edu/abs/2016Natur.533..509M}
}

@ARTICLE{Yoffe2021ApJ...908..114Y,
       author = {{Yoffe}, Gideon and {Ofir}, Aviv and {Aharonson}, Oded},
        title = "{A Simplified Photodynamical Model for Planetary Mass Determination in Low-eccentricity Multitransiting Systems}",
      journal = {\apj},
         year = 2021,
        month = feb,
       volume = {908},
       number = {1},
          eid = {114},
        pages = {114},
          doi = {10.3847/1538-4357/abc87a},
archivePrefix = {arXiv},
       eprint = {2011.04404},
 primaryClass = {astro-ph.EP},
       adsurl = {https://ui.adsabs.harvard.edu/abs/2021ApJ...908..114Y}
}

@ARTICLE{GoodmanWeare2010CAMCS...5...65G,
       author = {{Goodman}, Jonathan and {Weare}, Jonathan},
        title = "{Ensemble samplers with affine invariance}",
      journal = {Communications in Applied Mathematics and Computational Science},
         year = 2010,
        month = jan,
       volume = {5},
       number = {1},
        pages = {65-80},
          doi = {10.2140/camcos.2010.5.65},
       adsurl = {https://ui.adsabs.harvard.edu/abs/2010CAMCS...5...65G}
}

@ARTICLE{Gibbs_sampler,
  author={Geman, Stuart and Geman, Donald},
  journal={IEEE Transactions on Pattern Analysis and Machine Intelligence}, 
  title={Stochastic Relaxation, Gibbs Distributions, and the Bayesian Restoration of Images}, 
  year={1984},
  volume={PAMI-6},
  number={6},
  pages={721-741},
  doi={10.1109/TPAMI.1984.4767596}
}

@ARTICLE{Foreman2013,
       author = {{Foreman-Mackey}, Daniel and {Hogg}, David W. and {Lang}, Dustin and {Goodman}, Jonathan},
        title = "{emcee: The MCMC Hammer}",
      journal = {\pasp},
         year = 2013,
        month = mar,
       volume = {125},
       number = {925},
        pages = {306},
          doi = {10.1086/670067},
archivePrefix = {arXiv},
       eprint = {1202.3665},
 primaryClass = {astro-ph.IM},
       adsurl = {https://ui.adsabs.harvard.edu/abs/2013PASP..125..306F}
}

@ARTICLE{DFM2019JOSS....4.1864F,
       author = {{Foreman-Mackey}, Daniel and {Farr}, Will and {Sinha}, Manodeep and
         {Archibald}, Anne and {Hogg}, David and {Sanders}, Jeremy and
         {Zuntz}, Joe and {Williams}, Peter and {Nelson}, Andrew and
         {de Val-Borro}, Miguel and {Erhardt}, Tobias and {Pashchenko}, Ilya and
         {Pla}, Oriol},
        title = "{emcee v3: A Python ensemble sampling toolkit for affine-invariant MCMC}",
      journal = {The Journal of Open Source Software},
         year = 2019,
        month = nov,
       volume = {4},
       number = {43},
          eid = {1864},
        pages = {1864},
          doi = {10.21105/joss.01864},
archivePrefix = {arXiv},
       eprint = {1911.07688},
 primaryClass = {astro-ph.IM},
       adsurl = {https://ui.adsabs.harvard.edu/abs/2019JOSS....4.1864F}
}

@ARTICLE{Feroz2009MNRAS.398.1601F,
       author = {{Feroz}, F. and {Hobson}, M.~P. and {Bridges}, M.},
        title = "{MULTINEST: an efficient and robust Bayesian inference tool for cosmology and particle physics}",
      journal = {\mnras},
         year = 2009,
        month = oct,
       volume = {398},
       number = {4},
        pages = {1601-1614},
          doi = {10.1111/j.1365-2966.2009.14548.x},
archivePrefix = {arXiv},
       eprint = {0809.3437},
 primaryClass = {astro-ph},
       adsurl = {https://ui.adsabs.harvard.edu/abs/2009MNRAS.398.1601F}
}

@ARTICLE{Handley2015MNRAS.450L..61H,
       author = {{Handley}, W.~J. and {Hobson}, M.~P. and {Lasenby}, A.~N.},
        title = "{polychord: nested sampling for cosmology.}",
      journal = {\mnras},
         year = 2015,
        month = jun,
       volume = {450},
        pages = {L61-L65},
          doi = {10.1093/mnrasl/slv047},
archivePrefix = {arXiv},
       eprint = {1502.01856},
 primaryClass = {astro-ph.CO},
       adsurl = {https://ui.adsabs.harvard.edu/abs/2015MNRAS.450L..61H}
}

@Article{terBraak2008,
    author={ter Braak, Cajo J. F.
    and Vrugt, Jasper A.},
    title={Differential Evolution Markov Chain with snooker updater and fewer chains},
    journal={Statistics and Computing},
    year={2008},
    month={Dec},
    day={01},
    volume={18},
    number={4},
    pages={435-446},
    issn={1573-1375},
    doi={10.1007/s11222-008-9104-9},
    url={https://doi.org/10.1007/s11222-008-9104-9}
}

@ARTICLE{DENelson2014ApJS..210...11N,
       author = {{Nelson}, Benjamin and {Ford}, Eric B. and {Payne}, Matthew J.},
        title = "{RUN DMC: An Efficient, Parallel Code for Analyzing Radial Velocity Observations Using N-body Integrations and Differential Evolution Markov Chain Monte Carlo}",
      journal = {\apjs},
         year = 2014,
        month = jan,
       volume = {210},
       number = {1},
          eid = {11},
        pages = {11},
          doi = {10.1088/0067-0049/210/1/11},
archivePrefix = {arXiv},
       eprint = {1311.5229},
 primaryClass = {astro-ph.EP},
       adsurl = {https://ui.adsabs.harvard.edu/abs/2014ApJS..210...11N}
}

@ARTICLE{Metropolis1953JChPh..21.1087M,
       author = {{Metropolis}, Nicholas and {Rosenbluth}, Arianna W. and {Rosenbluth}, Marshall N. and {Teller}, Augusta H. and {Teller}, Edward},
        title = "{Equation of State Calculations by Fast Computing Machines}",
      journal = {\jcp},
         year = 1953,
        month = jun,
       volume = {21},
       number = {6},
        pages = {1087-1092},
          doi = {10.1063/1.1699114},
       adsurl = {https://ui.adsabs.harvard.edu/abs/1953JChPh..21.1087M}
}

@ARTICLE{Hastings1970Bimka..57...97H,
       author = {{Hastings}, W.~K.},
        title = "{Monte Carlo Sampling Methods using Markov Chains and their Applications}",
      journal = {Biometrika},
         year = 1970,
        month = apr,
       volume = {57},
       number = {1},
        pages = {97-109},
          doi = {10.1093/biomet/57.1.97},
       adsurl = {https://ui.adsabs.harvard.edu/abs/1970Bimka..57...97H}
}

@ARTICLE{gelman1992,
       author = {{Gelman}, Andrew and {Rubin}, Donald B.},
        title = "{Inference from Iterative Simulation Using Multiple Sequences}",
      journal = {Statistical Science},
         year = 1992,
        month = jan,
       volume = {7},
        pages = {457-472},
          doi = {10.1214/ss/1177011136},
       adsurl = {https://ui.adsabs.harvard.edu/abs/1992StaSc...7..457G}
}

@TechReport{geweke1991,
  author={John F. Geweke},
  title={{Evaluating the accuracy of sampling-based approaches to the calculation of posterior moments}},
  year=1991,
  institution={Federal Reserve Bank of Minneapolis},
  type={Staff Report},
  url={https://ideas.repec.org/p/fip/fedmsr/148.html},
  number={148},
  doi={},
}

@ARTICLE{Eastman2013PASP..125...83E,
       author = {{Eastman}, Jason and {Gaudi}, B. Scott and {Agol}, Eric},
        title = "{EXOFAST: A Fast Exoplanetary Fitting Suite in IDL}",
      journal = {\pasp},
         year = 2013,
        month = jan,
       volume = {125},
       number = {923},
        pages = {83},
          doi = {10.1086/669497},
archivePrefix = {arXiv},
       eprint = {1206.5798},
 primaryClass = {astro-ph.IM},
       adsurl = {https://ui.adsabs.harvard.edu/abs/2013PASP..125...83E}
}

@ARTICLE{Ford2006ApJ...642..505F,
       author = {{Ford}, Eric B.},
        title = "{Improving the Efficiency of Markov Chain Monte Carlo for Analyzing the Orbits of Extrasolar Planets}",
      journal = {\apj},
         year = 2006,
        month = may,
       volume = {642},
       number = {1},
        pages = {505-522},
          doi = {10.1086/500802},
archivePrefix = {arXiv},
       eprint = {astro-ph/0512634},
 primaryClass = {astro-ph},
       adsurl = {https://ui.adsabs.harvard.edu/abs/2006ApJ...642..505F}
}

@ARTICLE{Anderson2011ApJ...726L..19A,
       author = {{Anderson}, D.~R. and {Collier Cameron}, A. and {Hellier}, C. and {Lendl}, M. and {Maxted}, P.~F.~L. and {Pollacco}, D. and {Queloz}, D. and {Smalley}, B. and {Smith}, A.~M.~S. and {Todd}, I. and {Triaud}, A.~H.~M.~J. and {West}, R.~G. and {Barros}, S.~C.~C. and {Enoch}, B. and {Gillon}, M. and {Lister}, T.~A. and {Pepe}, F. and {S{\'e}gransan}, D. and {Street}, R.~A. and {Udry}, S.},
        title = "{WASP-30b: A 61 M $_{Jup}$ Brown Dwarf Transiting a V = 12, F8 Star}",
      journal = {\apjl},
         year = 2011,
        month = jan,
       volume = {726},
       number = {2},
          eid = {L19},
        pages = {L19},
          doi = {10.1088/2041-8205/726/2/L19},
archivePrefix = {arXiv},
       eprint = {1010.3006},
 primaryClass = {astro-ph.SR},
       adsurl = {https://ui.adsabs.harvard.edu/abs/2011ApJ...726L..19A}
}

@ARTICLE{Sozzetti2007ApJ...664.1190S,
       author = {{Sozzetti}, Alessandro and {Torres}, Guillermo and {Charbonneau}, David and {Latham}, David W. and {Holman}, Matthew J. and {Winn}, Joshua N. and {Laird}, John B. and {O'Donovan}, Francis T.},
        title = "{Improving Stellar and Planetary Parameters of Transiting Planet Systems: The Case of TrES-2}",
      journal = {\apj},
         year = 2007,
        month = aug,
       volume = {664},
       number = {2},
        pages = {1190-1198},
          doi = {10.1086/519214},
archivePrefix = {arXiv},
       eprint = {0704.2938},
 primaryClass = {astro-ph},
       adsurl = {https://ui.adsabs.harvard.edu/abs/2007ApJ...664.1190S}
}

@BOOK{Perryman2018exha.book.....P,
       author = {{Perryman}, Michael},
        title = "{The Exoplanet Handbook}",
         year = 2018,
       adsurl = {https://ui.adsabs.harvard.edu/abs/2018exha.book.....P}
}

@ARTICLE{Sharma2017ARAA..55..213S,
       author = {{Sharma}, Sanjib},
        title = "{Markov Chain Monte Carlo Methods for Bayesian Data Analysis in Astronomy}",
      journal = {\araa},
         year = 2017,
        month = aug,
       volume = {55},
       number = {1},
        pages = {213-259},
          doi = {10.1146/annurev-astro-082214-122339},
archivePrefix = {arXiv},
       eprint = {1706.01629},
 primaryClass = {astro-ph.IM},
       adsurl = {https://ui.adsabs.harvard.edu/abs/2017ARAA..55..213S}
}

@ARTICLE{Feroz2008MNRAS.384..449F,
       author = {{Feroz}, F. and {Hobson}, M.~P.},
        title = "{Multimodal nested sampling: an efficient and robust alternative to Markov Chain Monte Carlo methods for astronomical data analyses}",
      journal = {\mnras},
         year = 2008,
        month = feb,
       volume = {384},
       number = {2},
        pages = {449-463},
          doi = {10.1111/j.1365-2966.2007.12353.x},
archivePrefix = {arXiv},
       eprint = {0704.3704},
 primaryClass = {astro-ph},
       adsurl = {https://ui.adsabs.harvard.edu/abs/2008MNRAS.384..449F}
}

@ARTICLE{Hogg2018ApJS..236...11H,
       author = {{Hogg}, David W. and {Foreman-Mackey}, Daniel},
        title = "{Data Analysis Recipes: Using Markov Chain Monte Carlo}",
      journal = {\apjs},
         year = 2018,
        month = may,
       volume = {236},
       number = {1},
          eid = {11},
        pages = {11},
          doi = {10.3847/1538-4365/aab76e},
archivePrefix = {arXiv},
       eprint = {1710.06068},
 primaryClass = {astro-ph.IM},
       adsurl = {https://ui.adsabs.harvard.edu/abs/2018ApJS..236...11H}
}

@INCOLLECTION{Neal2011hmcm.book..113N,
       author = {{Neal}, Radford},
        title = "{MCMC Using Hamiltonian Dynamics}",
    booktitle = {Handbook of Markov Chain Monte Carlo},
         year = 2011,
        pages = {113-162},
          doi = {10.1201/b10905},
       adsurl = {https://ui.adsabs.harvard.edu/abs/2011hmcm.book..113N}
}

@ARTICLE{Betancourt2017arXiv170102434B,
       author = {{Betancourt}, Michael},
        title = "{A Conceptual Introduction to Hamiltonian Monte Carlo}",
      journal = {arXiv e-prints},
         year = 2017,
        month = jan,
          eid = {arXiv:1701.02434},
        pages = {arXiv:1701.02434},
          doi = {10.48550/arXiv.1701.02434},
archivePrefix = {arXiv},
       eprint = {1701.02434},
 primaryClass = {stat.ME},
       adsurl = {https://ui.adsabs.harvard.edu/abs/2017arXiv170102434B}
}

@ARTICLE{Higson2019DynamicNested,
  title    = "Dynamic nested sampling: an improved algorithm for parameter
              estimation and evidence calculation",
  author   = "Higson, Edward and Handley, Will and Hobson, Michael and Lasenby,
              Anthony",
  journal  = "Statistics and Computing",
  volume   =  29,
  number   =  5,
  pages    = "891--913",
  month    =  sep,
  year     =  2019
}

@ARTICLE{Handley2015MNRAS.453.4384H,
       author = {{Handley}, W.~J. and {Hobson}, M.~P. and {Lasenby}, A.~N.},
        title = "{POLYCHORD: next-generation nested sampling}",
      journal = {\mnras},
         year = 2015,
        month = nov,
       volume = {453},
       number = {4},
        pages = {4384-4398},
          doi = {10.1093/mnras/stv1911},
archivePrefix = {arXiv},
       eprint = {1506.00171},
 primaryClass = {astro-ph.IM},
       adsurl = {https://ui.adsabs.harvard.edu/abs/2015MNRAS.453.4384H}
}

@ARTICLE{Leleu2021AA...655A..66L_rivers,
       author = {{Leleu}, A. and {Chatel}, G. and {Udry}, S. and {Alibert}, Y. and {Delisle}, J. -B. and {Mardling}, R.},
        title = "{Alleviating the transit timing variation bias in transit surveys. I. RIVERS: Method and detection of a pair of resonant super-Earths around Kepler-1705}",
      journal = {\aap},
         year = 2021,
        month = nov,
       volume = {655},
          eid = {A66},
        pages = {A66},
          doi = {10.1051/0004-6361/202141471},
archivePrefix = {arXiv},
       eprint = {2111.06825},
 primaryClass = {astro-ph.EP},
       adsurl = {https://ui.adsabs.harvard.edu/abs/2021AA...655A..66L}
}

@ARTICLE{Beauge2006MNRAS.365.1160B,
       author = {{Beaug{\'e}}, C. and {Michtchenko}, T.~A. and {Ferraz-Mello}, S.},
        title = "{Planetary migration and extrasolar planets in the 2/1 mean-motion resonance}",
      journal = {\mnras},
         year = 2006,
        month = feb,
       volume = {365},
       number = {4},
        pages = {1160-1170},
          doi = {10.1111/j.1365-2966.2005.09779.x},
archivePrefix = {arXiv},
       eprint = {astro-ph/0404166},
 primaryClass = {astro-ph},
       adsurl = {https://ui.adsabs.harvard.edu/abs/2006MNRAS.365.1160B}
}

@ARTICLE{Chambers2009AREPS..37..321C,
       author = {{Chambers}, John E.},
        title = "{Planetary Migration: What Does It Mean for Planet Formation?}",
      journal = {Annual Review of Earth and Planetary Sciences},
         year = 2009,
        month = may,
       volume = {37},
       number = {1},
        pages = {321-344},
          doi = {10.1146/annurev.earth.031208.100122},
       adsurl = {https://ui.adsabs.harvard.edu/abs/2009AREPS..37..321C}
}

@ARTICLE{Lykawka2013ApJ...773...65L,
       author = {{Lykawka}, Patryk Sofia and {Ito}, Takashi},
        title = "{Terrestrial Planet Formation during the Migration and Resonance Crossings of the Giant Planets}",
      journal = {\apj},
         year = 2013,
        month = aug,
       volume = {773},
       number = {1},
          eid = {65},
        pages = {65},
          doi = {10.1088/0004-637X/773/1/65},
archivePrefix = {arXiv},
       eprint = {1306.3287},
 primaryClass = {astro-ph.EP},
       adsurl = {https://ui.adsabs.harvard.edu/abs/2013ApJ...773...65L}
}

@ARTICLE{Oshagh2013AA...556A..19O,
       author = {{Oshagh}, M. and {Santos}, N.~C. and {Boisse}, I. and {Bou{\'e}}, G. and {Montalto}, M. and {Dumusque}, X. and {Haghighipour}, N.},
        title = "{Effect of stellar spots on high-precision transit light-curve}",
      journal = {\aap},
         year = 2013,
        month = aug,
       volume = {556},
          eid = {A19},
        pages = {A19},
          doi = {10.1051/0004-6361/201321309},
archivePrefix = {arXiv},
       eprint = {1306.0739},
 primaryClass = {astro-ph.EP},
       adsurl = {https://ui.adsabs.harvard.edu/abs/2013AA...556A..19O}
}

@ARTICLE{Ioannidis2016AA...585A..72I,
       author = {{Ioannidis}, P. and {Huber}, K.~F. and {Schmitt}, J.~H.~M.~M.},
        title = "{How do starspots influence the transit timing variations of exoplanets? Simulations of individual and consecutive transits}",
      journal = {\aap},
         year = 2016,
        month = jan,
       volume = {585},
          eid = {A72},
        pages = {A72},
          doi = {10.1051/0004-6361/201527184},
archivePrefix = {arXiv},
       eprint = {1510.03276},
 primaryClass = {astro-ph.EP},
       adsurl = {https://ui.adsabs.harvard.edu/abs/2016AA...585A..72I}
}

@ARTICLE{Leonardi2024AA...686A..84L,
       author = {{Leonardi}, P. and {Nascimbeni}, V. and {Granata}, V. and {Malavolta}, L. and {Borsato}, L. and {Biazzo}, K. and {Lanza}, A.~F. and {Desidera}, S. and {Piotto}, G. and {Nardiello}, D. and {Damasso}, M. and {Cunial}, A. and {Bedin}, L.~R.},
        title = "{TASTE. V. A new ground-based investigation of orbital decay in the ultra-hot Jupiter WASP-12b}",
      journal = {\aap},
         year = 2024,
        month = jun,
       volume = {686},
          eid = {A84},
        pages = {A84},
          doi = {10.1051/0004-6361/202348363},
archivePrefix = {arXiv},
       eprint = {2402.12120},
 primaryClass = {astro-ph.EP},
       adsurl = {https://ui.adsabs.harvard.edu/abs/2024AA...686A..84L}
}

@ARTICLE{Rasio1996ApJ...470.1187R,
       author = {{Rasio}, F.~A. and {Tout}, C.~A. and {Lubow}, S.~H. and {Livio}, M.},
        title = "{Tidal Decay of Close Planetary Orbits}",
      journal = {\apj},
         year = 1996,
        month = oct,
       volume = {470},
        pages = {1187},
          doi = {10.1086/177941},
archivePrefix = {arXiv},
       eprint = {astro-ph/9605059},
 primaryClass = {astro-ph},
       adsurl = {https://ui.adsabs.harvard.edu/abs/1996ApJ...470.1187R}
}

@ARTICLE{Levrard2009ApJ...692L...9L,
       author = {{Levrard}, B. and {Winisdoerffer}, C. and {Chabrier}, G.},
        title = "{Falling Transiting Extrasolar Giant Planets}",
      journal = {\apjl},
         year = 2009,
        month = feb,
       volume = {692},
       number = {1},
        pages = {L9-L13},
          doi = {10.1088/0004-637X/692/1/L9},
archivePrefix = {arXiv},
       eprint = {0901.2048},
 primaryClass = {astro-ph.EP},
       adsurl = {https://ui.adsabs.harvard.edu/abs/2009ApJ...692L...9L}
}

@ARTICLE{Latham2011ApJ...732L..24L,
       author = {{Latham}, David W. and {Rowe}, Jason F. and {Quinn}, Samuel N. and {Batalha}, Natalie M. and {Borucki}, William J. and {Brown}, Timothy M. and {Bryson}, Stephen T. and {Buchhave}, Lars A. and {Caldwell}, Douglas A. and {Carter}, Joshua A. and {Christiansen}, Jessie L. and {Ciardi}, David R. and {Cochran}, William D. and {Dunham}, Edward W. and {Fabrycky}, Daniel C. and {Ford}, Eric B. and {Gautier}, III, Thomas N. and {Gilliland}, Ronald L. and {Holman}, Matthew J. and {Howell}, Steve B. and {Ibrahim}, Khadeejah A. and {Isaacson}, Howard and {Jenkins}, Jon M. and {Koch}, David G. and {Lissauer}, Jack J. and {Marcy}, Geoffrey W. and {Quintana}, Elisa V. and {Ragozzine}, Darin and {Sasselov}, Dimitar and {Shporer}, Avi and {Steffen}, Jason H. and {Welsh}, William F. and {Wohler}, Bill},
        title = "{A First Comparison of Kepler Planet Candidates in Single and Multiple Systems}",
      journal = {\apjl},
         year = 2011,
        month = may,
       volume = {732},
       number = {2},
          eid = {L24},
        pages = {L24},
          doi = {10.1088/2041-8205/732/2/L24},
archivePrefix = {arXiv},
       eprint = {1103.3896},
 primaryClass = {astro-ph.EP},
       adsurl = {https://ui.adsabs.harvard.edu/abs/2011ApJ...732L..24L}
}

@ARTICLE{Steffen2012PNAS..109.7982S,
       author = {{Steffen}, J.~H. and {Ragozzine}, D. and {Fabrycky}, D.~C. and {Carter}, J.~A. and {Ford}, E.~B. and {Holman}, M.~J. and {Rowe}, J.~F. and {Welsh}, W.~F. and {Borucki}, W.~J. and {Boss}, A.~P. and {Ciardi}, D.~R. and {Quinn}, S.~N.},
        title = "{Kepler constraints on planets near hot Jupiters}",
      journal = {Proceedings of the National Academy of Science},
         year = 2012,
        month = may,
       volume = {109},
       number = {21},
        pages = {7982-7987},
          doi = {10.1073/pnas.1120970109},
archivePrefix = {arXiv},
       eprint = {1205.2309},
 primaryClass = {astro-ph.EP},
       adsurl = {https://ui.adsabs.harvard.edu/abs/2012PNAS..109.7982S}
}

@ARTICLE{Huang2016ApJ...825...98H,
       author = {{Huang}, Chelsea and {Wu}, Yanqin and {Triaud}, Amaury H.~M.~J.},
        title = "{Warm Jupiters Are Less Lonely than Hot Jupiters: Close Neighbors}",
      journal = {\apj},
         year = 2016,
        month = jul,
       volume = {825},
       number = {2},
          eid = {98},
        pages = {98},
          doi = {10.3847/0004-637X/825/2/98},
archivePrefix = {arXiv},
       eprint = {1601.05095},
 primaryClass = {astro-ph.EP},
       adsurl = {https://ui.adsabs.harvard.edu/abs/2016ApJ...825...98H}
}

@ARTICLE{Hellier2012MNRAS.426..739H,
       author = {{Hellier}, Coel and {Anderson}, D.~R. and {Collier Cameron}, A. and {Doyle}, A.~P. and {Fumel}, A. and {Gillon}, M. and {Jehin}, E. and {Lendl}, M. and {Maxted}, P.~F.~L. and {Pepe}, F. and {Pollacco}, D. and {Queloz}, D. and {S{\'e}gransan}, D. and {Smalley}, B. and {Smith}, A.~M.~S. and {Southworth}, J. and {Triaud}, A.~H.~M.~J. and {Udry}, S. and {West}, R.~G.},
        title = "{Seven transiting hot Jupiters from WASP-South, Euler and TRAPPIST: WASP-47b, WASP-55b, WASP-61b, WASP-62b, WASP-63b, WASP-66b and WASP-67b}",
      journal = {\mnras},
         year = 2012,
        month = oct,
       volume = {426},
       number = {1},
        pages = {739-750},
          doi = {10.1111/j.1365-2966.2012.21780.x},
archivePrefix = {arXiv},
       eprint = {1204.5095},
 primaryClass = {astro-ph.EP},
       adsurl = {https://ui.adsabs.harvard.edu/abs/2012MNRAS.426..739H}
}

@ARTICLE{Nascimbeni2023AA...673A..42N,
       author = {{Nascimbeni}, V. and {Borsato}, L. and {Zingales}, T. and {Piotto}, G. and {Pagano}, I. and {Beck}, M. and {Broeg}, C. and {Ehrenreich}, D. and {Hoyer}, S. and {Majidi}, F.~Z. and {Granata}, V. and {Sousa}, S.~G. and {Wilson}, T.~G. and {Van Grootel}, V. and {Bonfanti}, A. and {Salmon}, S. and {Mustill}, A.~J. and {Delrez}, L. and {Alibert}, Y. and {Alonso}, R. and {Anglada}, G. and {B{\'a}rczy}, T. and {Barrado}, D. and {Barros}, S.~C.~C. and {Baumjohann}, W. and {Beck}, T. and {Benz}, W. and {Bergomi}, M. and {Billot}, N. and {Bonfils}, X. and {Brandeker}, A. and {Cabrera}, J. and {Charnoz}, S. and {Collier Cameron}, A. and {Csizmadia}, Sz. and {Cubillos}, P.~E. and {Davies}, M.~B. and {Deleuil}, M. and {Deline}, A. and {Demangeon}, O.~D.~S. and {Demory}, B. -O. and {Erikson}, A. and {Fortier}, A. and {Fossati}, L. and {Fridlund}, M. and {Gandolfi}, D. and {Gillon}, M. and {G{\"u}del}, M. and {Isaak}, K.~G. and {Kiss}, L.~L. and {Laskar}, J. and {Lecavelier des Etangs}, A. and {Lendl}, M. and {Lovis}, C. and {Luque}, R. and {Magrin}, D. and {Maxted}, P.~F.~L. and {Mordasini}, C. and {Olofsson}, G. and {Ottensamer}, R. and {Pall{\'e}}, E. and {Peter}, G. and {Piazza}, D. and {Pollacco}, D. and {Queloz}, D. and {Ragazzoni}, R. and {Rando}, N. and {Ratti}, F. and {Rauer}, H. and {Ribas}, I. and {Santos}, N.~C. and {Scandariato}, G. and {S{\'e}gransan}, D. and {Simon}, A.~E. and {Smith}, A.~M.~S. and {Steinberger}, M. and {Steller}, M. and {Szab{\'o}}, Gy. M. and {Thomas}, N. and {Udry}, S. and {Venturini}, J. and {Walton}, N.~A. and {Wolter}, D.},
        title = "{A new dynamical modeling of the WASP-47 system with CHEOPS observations}",
      journal = {\aap},
         year = 2023,
        month = may,
       volume = {673},
          eid = {A42},
        pages = {A42},
          doi = {10.1051/0004-6361/202245486},
archivePrefix = {arXiv},
       eprint = {2302.01352},
 primaryClass = {astro-ph.EP},
       adsurl = {https://ui.adsabs.harvard.edu/abs/2023AA...673A..42N}
}

@ARTICLE{Canas2019ApJ...870L..17C,
       author = {{Ca{\~n}as}, Caleb I. and {Wang}, Songhu and {Mahadevan}, Suvrath and {Bender}, Chad F. and {De Lee}, Nathan and {Fleming}, Scott W. and {Garc{\'\i}a-Hern{\'a}ndez}, D.~A. and {Hearty}, Fred R. and {Majewski}, Steven R. and {Roman-Lopes}, Alexandre and {Schneider}, Donald P. and {Stassun}, Keivan G.},
        title = "{Kepler-730: A Hot Jupiter System with a Close-in, Transiting, Earth-sized Planet}",
      journal = {\apjl},
         year = 2019,
        month = jan,
       volume = {870},
       number = {2},
          eid = {L17},
        pages = {L17},
          doi = {10.3847/2041-8213/aafa1e},
archivePrefix = {arXiv},
       eprint = {1812.08358},
 primaryClass = {astro-ph.EP},
       adsurl = {https://ui.adsabs.harvard.edu/abs/2019ApJ...870L..17C}
}

@ARTICLE{Huang2020ApJ...892L...7H,
       author = {{Huang}, Chelsea X. and {Quinn}, Samuel N. and {Vanderburg}, Andrew and {Becker}, Juliette and {Rodriguez}, Joseph E. and {Pozuelos}, Francisco J. and {Gandolfi}, Davide and {Zhou}, George and {Mann}, Andrew W. and {Collins}, Karen A. and {Crossfield}, Ian and {Barkaoui}, Khalid and {Collins}, Kevin I. and {Fridlund}, Malcolm and {Gillon}, Micha{\"e}l and {Gonzales}, Erica J. and {G{\"u}nther}, Maximilian N. and {Henry}, Todd J. and {Howell}, Steve B. and {James}, Hodari-Sadiki and {Jao}, Wei-Chun and {Jehin}, Emmanu{\"e}l and {Jensen}, Eric L.~N. and {Kane}, Stephen R. and {Lissauer}, Jack J. and {Matthews}, Elisabeth and {Matson}, Rachel A. and {Paredes}, Leonardo A. and {Schlieder}, Joshua E. and {Stassun}, Keivan G. and {Shporer}, Avi and {Sha}, Lizhou and {Tan}, Thiam-Guan and {Georgieva}, Iskra and {Mathur}, Savita and {Palle}, Enric and {Persson}, Carina M. and {Van Eylen}, Vincent and {Ricker}, George R. and {Vanderspek}, Roland K. and {Latham}, David W. and {Winn}, Joshua N. and {Seager}, S. and {Jenkins}, Jon M. and {Burke}, Christopher J. and {Goeke}, Robert F. and {Rinehart}, Stephen and {Rose}, Mark E. and {Ting}, Eric B. and {Torres}, Guillermo and {Wong}, Ian},
        title = "{TESS Spots a Hot Jupiter with an Inner Transiting Neptune}",
      journal = {\apjl},
         year = 2020,
        month = mar,
       volume = {892},
       number = {1},
          eid = {L7},
        pages = {L7},
          doi = {10.3847/2041-8213/ab7302},
archivePrefix = {arXiv},
       eprint = {2003.10852},
 primaryClass = {astro-ph.EP},
       adsurl = {https://ui.adsabs.harvard.edu/abs/2020ApJ...892L...7H}
}

@ARTICLE{Borsato2022ExA....53..635B,
       author = {{Borsato}, Luca and {Nascimbeni}, Valerio and {Piotto}, Giampaolo and {Szab{\'o}}, Gyula},
        title = "{Exploiting the transit timing capabilities of Ariel}",
      journal = {Experimental Astronomy},
         year = 2022,
        month = apr,
       volume = {53},
       number = {2},
        pages = {635-653},
          doi = {10.1007/s10686-021-09737-5},
archivePrefix = {arXiv},
       eprint = {2103.09239},
 primaryClass = {astro-ph.EP},
       adsurl = {https://ui.adsabs.harvard.edu/abs/2022ExA....53..635B}
}

@ARTICLE{Borsato2024AA...689A..52B,
       author = {{Borsato}, L. and {Degen}, D. and {Leleu}, A. and {Hooton}, M.~J. and {Egger}, J.~A. and {Bekkelien}, A. and {Brandeker}, A. and {Collier Cameron}, A. and {G{\"u}nther}, M.~N. and {Nascimbeni}, V. and {Persson}, C.~M. and {Bonfanti}, A. and {Wilson}, T.~G. and {Correia}, A.~C.~M. and {Zingales}, T. and {Guillot}, T. and {Triaud}, A.~H.~M.~J. and {Piotto}, G. and {Gandolfi}, D. and {Abe}, L. and {Alibert}, Y. and {Alonso}, R. and {B{\'a}rczy}, T. and {Navascues}, D. Barrado and {Barros}, S.~C.~C. and {Baumjohann}, W. and {Beck}, T. and {Bendjoya}, P. and {Benz}, W. and {Billot}, N. and {Broeg}, C. and {Busch}, M. -D. and {Csizmadia}, Sz. and {Cubillos}, P.~E. and {Davies}, M.~B. and {Deleuil}, M. and {Deline}, A. and {Delrez}, L. and {Demangeon}, O.~D.~S. and {Demory}, B. -O. and {Derekas}, A. and {Edwards}, B. and {Ehrenreich}, D. and {Erikson}, A. and {Fortier}, A. and {Fossati}, L. and {Fridlund}, M. and {Gazeas}, K. and {Gillon}, M. and {G{\"u}del}, M. and {Heitzmann}, A. and {Helling}, Ch. and {Hoyer}, S. and {Isaak}, K.~G. and {Kiss}, L.~L. and {Korth}, J. and {Lam}, K.~W.~F. and {Laskar}, J. and {Lecavelier des Etangs}, A. and {Lendl}, M. and {Magrin}, D. and {Marafatto}, L. and {Maxted}, P.~F.~L. and {Mecina}, M. and {M{\'e}karnia}, D. and {Mordasini}, C. and {Mura}, D. and {Olofsson}, G. and {Ottensamer}, R. and {Pagano}, I. and {Pall{\'e}}, E. and {Peter}, G. and {Pollacco}, D. and {Queloz}, D. and {Ragazzoni}, R. and {Rando}, N. and {Ratti}, F. and {Rauer}, H. and {Ribas}, I. and {Salmon}, S. and {Santos}, N.~C. and {Scandariato}, G. and {S{\'e}gransan}, D. and {Simon}, A.~E. and {Smith}, A.~M.~S. and {Sousa}, S.~G. and {Stalport}, M. and {Suarez}, O. and {Sulis}, S. and {Szab{\'o}}, Gy. M. and {Udry}, S. and {Van Grootel}, V. and {Venturini}, J. and {Villaver}, E. and {Walton}, N.~A. and {Wolter}, D.},
        title = "{Characterisation of the warm-Jupiter TOI-1130 system with CHEOPS and a photo-dynamical approach}",
      journal = {\aap},
         year = 2024,
        month = sep,
       volume = {689},
          eid = {A52},
        pages = {A52},
          doi = {10.1051/0004-6361/202450974},
archivePrefix = {arXiv},
       eprint = {2407.06097},
 primaryClass = {astro-ph.EP},
       adsurl = {https://ui.adsabs.harvard.edu/abs/2024AA...689A..52B}
}

@ARTICLE{Grieves2024arXiv240615986G,
       author = {{Grieves}, N. and {Bouchy}, F. and {Armstrong}, D.~J. and {Akinsanmi}, B. and {Psaridi}, A. and {Ulmer-Moll}, S. and {Frensch}, Y.~G.~C. and {Helled}, R. and {Muller}, S. and {Knierim}, H. and {Santos}, N.~C. and {Adibekyan}, V. and {Battley}, M.~P. and {Unger}, N. and {Chaverot}, G. and {Parc}, L. and {Bayliss}, D. and {Dumusque}, X. and {Hawthorn}, F. and {Figueira}, P. and {Keniger}, M.~A.~F. and {Lillo-Box}, J. and {Nielsen}, L.~D. and {Osborn}, A. and {Sousa}, S.~G. and {Strom}, P. and {Udry}, S.},
        title = "{Refining the WASP-132 multi-planetary system: discovery of a cold giant planet and mass measurement of a hot super-Earth}",
      journal = {arXiv e-prints},
         year = 2024,
        month = jun,
          eid = {arXiv:2406.15986},
        pages = {arXiv:2406.15986},
          doi = {10.48550/arXiv.2406.15986},
archivePrefix = {arXiv},
       eprint = {2406.15986},
 primaryClass = {astro-ph.EP},
       adsurl = {https://ui.adsabs.harvard.edu/abs/2024arXiv240615986G}
}

@ARTICLE{Sha2023MNRAS.524.1113S,
       author = {{Sha}, Lizhou and {Vanderburg}, Andrew M. and {Huang}, Chelsea X. and {Armstrong}, David J. and {Brahm}, Rafael and {Giacalone}, Steven and {Wood}, Mackenna L. and {Collins}, Karen A. and {Nielsen}, Louise D. and {Hobson}, Melissa J. and {Ziegler}, Carl and {Howell}, Steve B. and {Torres-Miranda}, Pascal and {Mann}, Andrew W. and {Zhou}, George and {Delgado-Mena}, Elisa and {Rojas}, Felipe I. and {Abe}, Lyu and {Trifonov}, Trifon and {Adibekyan}, Vardan and {Sousa}, S{\'e}rgio G. and {Fajardo-Acosta}, Sergio B. and {Guillot}, Tristan and {Howard}, Saburo and {Littlefield}, Colin and {Hawthorn}, Faith and {Schmider}, Fran{\c{c}}ois-Xavier and {Eberhardt}, Jan and {Tan}, Thiam-Guan and {Osborn}, Ares and {Schwarz}, Richard P. and {Str{\o}m}, Paul and {Jord{\'a}n}, Andr{\'e}s and {Wang}, Gavin and {Henning}, Thomas and {Massey}, Bob and {Law}, Nicholas and {Stockdale}, Chris and {Furlan}, Elise and {Srdoc}, Gregor and {Wheatley}, Peter J. and {Barrado Navascu{\'e}s}, David and {Lissauer}, Jack J. and {Stassun}, Keivan G. and {Ricker}, George R. and {Vanderspek}, Roland K. and {Latham}, David W. and {Winn}, Joshua N. and {Seager}, Sara and {Jenkins}, Jon M. and {Barclay}, Thomas and {Bouma}, Luke G. and {Christiansen}, Jessie L. and {Guerrero}, Natalia and {Rose}, Mark E.},
        title = "{TESS spots a mini-neptune interior to a hot saturn in the TOI-2000 system}",
      journal = {\mnras},
         year = 2023,
        month = sep,
       volume = {524},
       number = {1},
        pages = {1113-1138},
          doi = {10.1093/mnras/stad1666},
archivePrefix = {arXiv},
       eprint = {2209.14396},
 primaryClass = {astro-ph.EP},
       adsurl = {https://ui.adsabs.harvard.edu/abs/2023MNRAS.524.1113S}
}

@ARTICLE{Maciejewski2023MNRAS.525L..43M,
       author = {{Maciejewski}, G. and others},
        title = "{A hot super-Earth planet in the WASP-84 planetary system}",
      journal = {\mnras},
         year = 2023,
        month = oct,
       volume = {525},
       number = {1},
        pages = {L43-L49},
          doi = {10.1093/mnrasl/slad078},
archivePrefix = {arXiv},
       eprint = {2305.09177},
 primaryClass = {astro-ph.EP},
       adsurl = {https://ui.adsabs.harvard.edu/abs/2023MNRAS.525L..43M}
}

@ARTICLE{Mantovan2022MNRAS.516.4432M,
       author = {{Mantovan}, Giacomo and {Montalto}, Marco and {Piotto}, Giampaolo and {Wilson}, Thomas G. and {Collier Cameron}, Andrew and {Majidi}, Fatemeh Zahra and {Borsato}, Luca and {Granata}, Valentina and {Nascimbeni}, Valerio},
        title = "{Validation of TESS exoplanet candidates orbiting solar analogues in the all-sky PLATO input catalogue}",
      journal = {\mnras},
         year = 2022,
        month = nov,
       volume = {516},
       number = {3},
        pages = {4432-4447},
          doi = {10.1093/mnras/stac2451},
archivePrefix = {arXiv},
       eprint = {2208.12276},
 primaryClass = {astro-ph.EP},
       adsurl = {https://ui.adsabs.harvard.edu/abs/2022MNRAS.516.4432M}
}

@ARTICLE{Mantovan2024AA...682A.129M,
       author = {{Mantovan}, G. and {Malavolta}, L. and {Desidera}, S. and {Zingales}, T. and {Borsato}, L. and {Piotto}, G. and {Maggio}, A. and {Locci}, D. and {Polychroni}, D. and {Turrini}, D. and {Baratella}, M. and {Biazzo}, K. and {Nardiello}, D. and {Stassun}, K. and {Nascimbeni}, V. and {Benatti}, S. and {John}, A. Anna and {Watkins}, C. and {Bieryla}, A. and {Lissauer}, J.~J. and {Twicken}, J.~D. and {Lanza}, A.~F. and {Winn}, J.~N. and {Messina}, S. and {Montalto}, M. and {Sozzetti}, A. and {Boffin}, H. and {Cheryasov}, D. and {Strakhov}, I. and {Murgas}, F. and {D'Arpa}, M. and {Barkaoui}, K. and {Benni}, P. and {Bignamini}, A. and {Bonomo}, A.~S. and {Borsa}, F. and {Cabona}, L. and {Cameron}, A.~C. and {Claudi}, R. and {Cochran}, W. and {Collins}, K.~A. and {Damasso}, M. and {Dong}, J. and {Endl}, M. and {Fukui}, A. and {F{\H{u}}r{\'e}sz}, G. and {Gandolfi}, D. and {Ghedina}, A. and {Jenkins}, J. and {Kab{\'a}th}, P. and {Latham}, D.~W. and {Lorenzi}, V. and {Luque}, R. and {Maldonado}, J. and {McLeod}, K. and {Molinaro}, M. and {Narita}, N. and {Nowak}, G. and {Orell-Miquel}, J. and {Pall{\'e}}, E. and {Parviainen}, H. and {Pedani}, M. and {Quinn}, S.~N. and {Relles}, H. and {Rowden}, P. and {Scandariato}, G. and {Schwarz}, R. and {Seager}, S. and {Shporer}, A. and {Vanderburg}, A. and {Wilson}, T.~G.},
        title = "{The GAPS programme at TNG. XLIX. TOI-5398, the youngest compact multi-planet system composed of an inner sub-Neptune and an outer warm Saturn}",
      journal = {\aap},
         year = 2024,
        month = feb,
       volume = {682},
          eid = {A129},
        pages = {A129},
          doi = {10.1051/0004-6361/202347472},
archivePrefix = {arXiv},
       eprint = {2310.16888},
 primaryClass = {astro-ph.EP},
       adsurl = {https://ui.adsabs.harvard.edu/abs/2024AA...682A.129M}
}

@ARTICLE{Bonfanti2021AA...646A.157B,
       author = {{Bonfanti}, A. and {Delrez}, L. and {Hooton}, M.~J. and {Wilson}, T.~G. and {Fossati}, L. and {Alibert}, Y. and {Hoyer}, S. and {Mustill}, A.~J. and {Osborn}, H.~P. and {Adibekyan}, V. and {Gandolfi}, D. and {Salmon}, S. and {Sousa}, S.~G. and {Tuson}, A. and {Van Grootel}, V. and {Cabrera}, J. and {Nascimbeni}, V. and {Maxted}, P.~F.~L. and {Barros}, S.~C.~C. and {Billot}, N. and {Bonfils}, X. and {Borsato}, L. and {Broeg}, C. and {Davies}, M.~B. and {Deleuil}, M. and {Demangeon}, O.~D.~S. and {Fridlund}, M. and {Lacedelli}, G. and {Lendl}, M. and {Persson}, C. and {Santos}, N.~C. and {Scandariato}, G. and {Szab{\'o}}, Gy. M. and {Collier Cameron}, A. and {Udry}, S. and {Benz}, W. and {Beck}, M. and {Ehrenreich}, D. and {Fortier}, A. and {Isaak}, K.~G. and {Queloz}, D. and {Alonso}, R. and {Asquier}, J. and {Bandy}, T. and {B{\'a}rczy}, T. and {Barrado}, D. and {Barrag{\'a}n}, O. and {Baumjohann}, W. and {Beck}, T. and {Bekkelien}, A. and {Bergomi}, M. and {Brandeker}, A. and {Busch}, M. -D. and {Cessa}, V. and {Charnoz}, S. and {Chazelas}, B. and {Corral Van Damme}, C. and {Demory}, B. -O. and {Erikson}, A. and {Farinato}, J. and {Futyan}, D. and {Garcia Mu{\~n}oz}, A. and {Gillon}, M. and {Guedel}, M. and {Guterman}, P. and {Hasiba}, J. and {Heng}, K. and {Hernandez}, E. and {Kiss}, L. and {Kuntzer}, T. and {Laskar}, J. and {Lecavelier des Etangs}, A. and {Lovis}, C. and {Magrin}, D. and {Malvasio}, L. and {Marafatto}, L. and {Michaelis}, H. and {Munari}, M. and {Olofsson}, G. and {Ottacher}, H. and {Ottensamer}, R. and {Pagano}, I. and {Pall{\'e}}, E. and {Peter}, G. and {Piazza}, D. and {Piotto}, G. and {Pollacco}, D. and {Ragazzoni}, R. and {Rando}, N. and {Ratti}, F. and {Rauer}, H. and {Ribas}, I. and {Rieder}, M. and {Rohlfs}, R. and {Safa}, F. and {Salatti}, M. and {S{\'e}gransan}, D. and {Simon}, A.~E. and {Smith}, A.~M.~S. and {Sordet}, M. and {Steller}, M. and {Thomas}, N. and {Tschentscher}, M. and {Van Eylen}, V. and {Viotto}, V. and {Walter}, I. and {Walton}, N.~A. and {Wildi}, F. and {Wolter}, D.},
        title = "{CHEOPS observations of the HD 108236 planetary system: a fifth planet, improved ephemerides, and planetary radii}",
      journal = {\aap},
         year = 2021,
        month = feb,
       volume = {646},
          eid = {A157},
        pages = {A157},
          doi = {10.1051/0004-6361/202039608},
archivePrefix = {arXiv},
       eprint = {2101.00663},
 primaryClass = {astro-ph.EP},
       adsurl = {https://ui.adsabs.harvard.edu/abs/2021AA...646A.157B}
}

@ARTICLE{Berardo2019AJ....157..185B,
       author = {{Berardo}, David and {Crossfield}, Ian J.~M. and {Werner}, Michael and {Petigura}, Erik and {Christiansen}, Jessie and {Ciardi}, David R. and {Dressing}, Courtney and {Fulton}, Benjamin J. and {Gorjian}, Varoujan and {Greene}, Thomas P. and {Hardegree-Ullman}, Kevin and {Kane}, Stephen R. and {Livingston}, John and {Morales}, Farisa and {Schlieder}, Joshua E.},
        title = "{Revisiting the HIP 41378 System with K2 and Spitzer}",
      journal = {\aj},
         year = 2019,
        month = may,
       volume = {157},
       number = {5},
          eid = {185},
        pages = {185},
          doi = {10.3847/1538-3881/ab100c},
archivePrefix = {arXiv},
       eprint = {1809.11116},
 primaryClass = {astro-ph.EP},
       adsurl = {https://ui.adsabs.harvard.edu/abs/2019AJ....157..185B}
}

@ARTICLE{Sulis2024AA...686L..18S,
       author = {{Sulis}, S. and {Borsato}, L. and {Grouffal}, S. and {Osborn}, H.~P. and {Santerne}, A. and {Brandeker}, A. and {G{\"u}nther}, M.~N. and {Heitzmann}, A. and {Lendl}, M. and {Fridlund}, M. and {Gandolfi}, D. and {Alibert}, Y. and {Alonso}, R. and {B{\'a}rczy}, T. and {Barrado Navascues}, D. and {Barros}, S.~C.~C. and {Baumjohann}, W. and {Beck}, T. and {Benz}, W. and {Bergomi}, M. and {Billot}, N. and {Bonfanti}, A. and {Broeg}, C. and {Collier Cameron}, A. and {Corral van Damme}, C. and {Correia}, A.~C.~M. and {Csizmadia}, Sz. and {Cubillos}, P.~E. and {Davies}, M.~B. and {Deleuil}, M. and {Deline}, A. and {Delrez}, L. and {Demangeon}, O.~D.~S. and {Demory}, B. -O. and {Derekas}, A. and {Edwards}, B. and {Ehrenreich}, D. and {Erikson}, A. and {Fortier}, A. and {Fossati}, L. and {Gazeas}, K. and {Gillon}, M. and {G{\"u}del}, M. and {Helling}, Ch. and {Hoyer}, S. and {Isaak}, K.~G. and {Kiss}, L.~L. and {Korth}, J. and {Lam}, K.~W.~F. and {Laskar}, J. and {Lecavelier des Etangs}, A. and {Magrin}, D. and {Maxted}, P.~F.~L. and {Mordasini}, C. and {Nascimbeni}, V. and {Olofsson}, G. and {Ottensamer}, R. and {Pagano}, I. and {Pall{\'e}}, E. and {Peter}, G. and {Piazza}, D. and {Piotto}, G. and {Pollacco}, D. and {Queloz}, D. and {Ragazzoni}, R. and {Rando}, N. and {Rauer}, H. and {Ribas}, I. and {Santos}, N.~C. and {Scandariato}, G. and {S{\'e}gransan}, D. and {Simon}, A.~E. and {Smith}, A.~M.~S. and {Sousa}, S.~G. and {Stalport}, M. and {Steinberger}, M. and {Szab{\'o}}, Gy. M. and {Tuson}, A. and {Udry}, S. and {Ulmer-Moll}, S. and {Van Grootel}, V. and {Venturini}, J. and {Villaver}, E. and {Walton}, N.~A. and {Wilson}, T.~G. and {Wolter}, D. and {Zingales}, T.},
        title = "{HIP 41378 observed by CHEOPS: Where is planet d?}",
      journal = {\aap},
         year = 2024,
        month = jun,
       volume = {686},
          eid = {L18},
        pages = {L18},
          doi = {10.1051/0004-6361/202449689},
archivePrefix = {arXiv},
       eprint = {2405.20077},
 primaryClass = {astro-ph.EP},
       adsurl = {https://ui.adsabs.harvard.edu/abs/2024AA...686L..18S}
}

@ARTICLE{Rowe2014ApJ...784...45R,
       author = {{Rowe}, Jason F. and {Bryson}, Stephen T. and {Marcy}, Geoffrey W. and {Lissauer}, Jack J. and {Jontof-Hutter}, Daniel and {Mullally}, Fergal and {Gilliland}, Ronald L. and {Issacson}, Howard and {Ford}, Eric and {Howell}, Steve B. and {Borucki}, William J. and {Haas}, Michael and {Huber}, Daniel and {Steffen}, Jason H. and {Thompson}, Susan E. and {Quintana}, Elisa and {Barclay}, Thomas and {Still}, Martin and {Fortney}, Jonathan and {Gautier}, III, T.~N. and {Hunter}, Roger and {Caldwell}, Douglas A. and {Ciardi}, David R. and {Devore}, Edna and {Cochran}, William and {Jenkins}, Jon and {Agol}, Eric and {Carter}, Joshua A. and {Geary}, John},
        title = "{Validation of Kepler's Multiple Planet Candidates. III. Light Curve Analysis and Announcement of Hundreds of New Multi-planet Systems}",
      journal = {\apj},
         year = 2014,
        month = mar,
       volume = {784},
       number = {1},
          eid = {45},
        pages = {45},
          doi = {10.1088/0004-637X/784/1/45},
archivePrefix = {arXiv},
       eprint = {1402.6534},
 primaryClass = {astro-ph.EP},
       adsurl = {https://ui.adsabs.harvard.edu/abs/2014ApJ...784...45R}
}

@ARTICLE{Xie2014ApJS..210...25X,
       author = {{Xie}, Ji-Wei},
        title = "{Transit Timing Variation of Near-resonance Planetary Pairs. II. Confirmation of 30 Planets in 15 Multiple-planet Systems}",
      journal = {\apjs},
         year = 2014,
        month = feb,
       volume = {210},
       number = {2},
          eid = {25},
        pages = {25},
          doi = {10.1088/0067-0049/210/2/25},
archivePrefix = {arXiv},
       eprint = {1309.2329},
 primaryClass = {astro-ph.EP},
       adsurl = {https://ui.adsabs.harvard.edu/abs/2014ApJS..210...25X}
}

@ARTICLE{Fabrycky2012ApJ...750..114F,
       author = {{Fabrycky}, Daniel C. and {Ford}, Eric B. and {Steffen}, Jason H. and {Rowe}, Jason F. and {Carter}, Joshua A. and {Moorhead}, Althea V. and {Batalha}, Natalie M. and {Borucki}, William J. and {Bryson}, Steve and {Buchhave}, Lars A. and {Christiansen}, Jessie L. and {Ciardi}, David R. and {Cochran}, William D. and {Endl}, Michael and {Fanelli}, Michael N. and {Fischer}, Debra and {Fressin}, Francois and {Geary}, John and {Haas}, Michael R. and {Hall}, Jennifer R. and {Holman}, Matthew J. and {Jenkins}, Jon M. and {Koch}, David G. and {Latham}, David W. and {Li}, Jie and {Lissauer}, Jack J. and {Lucas}, Philip and {Marcy}, Geoffrey W. and {Mazeh}, Tsevi and {McCauliff}, Sean and {Quinn}, Samuel and {Ragozzine}, Darin and {Sasselov}, Dimitar and {Shporer}, Avi},
        title = "{Transit Timing Observations from Kepler. IV. Confirmation of Four Multiple-planet Systems by Simple Physical Models}",
      journal = {\apj},
         year = 2012,
        month = may,
       volume = {750},
       number = {2},
          eid = {114},
        pages = {114},
          doi = {10.1088/0004-637X/750/2/114},
archivePrefix = {arXiv},
       eprint = {1201.5415},
 primaryClass = {astro-ph.EP},
       adsurl = {https://ui.adsabs.harvard.edu/abs/2012ApJ...750..114F}
}

@ARTICLE{Lissauer2012ApJ...750..112L,
       author = {{Lissauer}, Jack J. and {Marcy}, Geoffrey W. and {Rowe}, Jason F. and {Bryson}, Stephen T. and {Adams}, Elisabeth and {Buchhave}, Lars A. and {Ciardi}, David R. and {Cochran}, William D. and {Fabrycky}, Daniel C. and {Ford}, Eric B. and {Fressin}, Francois and {Geary}, John and {Gilliland}, Ronald L. and {Holman}, Matthew J. and {Howell}, Steve B. and {Jenkins}, Jon M. and {Kinemuchi}, Karen and {Koch}, David G. and {Morehead}, Robert C. and {Ragozzine}, Darin and {Seader}, Shawn E. and {Tanenbaum}, Peter G. and {Torres}, Guillermo and {Twicken}, Joseph D.},
        title = "{Almost All of Kepler's Multiple-planet Candidates Are Planets}",
      journal = {\apj},
         year = 2012,
        month = may,
       volume = {750},
       number = {2},
          eid = {112},
        pages = {112},
          doi = {10.1088/0004-637X/750/2/112},
archivePrefix = {arXiv},
       eprint = {1201.5424},
 primaryClass = {astro-ph.EP},
       adsurl = {https://ui.adsabs.harvard.edu/abs/2012ApJ...750..112L}
}

@ARTICLE{Marcy2014ApJS..210...20M,
       author = {{Marcy}, Geoffrey W. and {Isaacson}, Howard and {Howard}, Andrew W. and {Rowe}, Jason F. and {Jenkins}, Jon M. and {Bryson}, Stephen T. and {Latham}, David W. and {Howell}, Steve B. and {Gautier}, III, Thomas N. and {Batalha}, Natalie M. and {Rogers}, Leslie and {Ciardi}, David and {Fischer}, Debra A. and {Gilliland}, Ronald L. and {Kjeldsen}, Hans and {Christensen-Dalsgaard}, J{\o}rgen and {Huber}, Daniel and {Chaplin}, William J. and {Basu}, Sarbani and {Buchhave}, Lars A. and {Quinn}, Samuel N. and {Borucki}, William J. and {Koch}, David G. and {Hunter}, Roger and {Caldwell}, Douglas A. and {Van Cleve}, Jeffrey and {Kolbl}, Rea and {Weiss}, Lauren M. and {Petigura}, Erik and {Seager}, Sara and {Morton}, Timothy and {Johnson}, John Asher and {Ballard}, Sarah and {Burke}, Chris and {Cochran}, William D. and {Endl}, Michael and {MacQueen}, Phillip and {Everett}, Mark E. and {Lissauer}, Jack J. and {Ford}, Eric B. and {Torres}, Guillermo and {Fressin}, Francois and {Brown}, Timothy M. and {Steffen}, Jason H. and {Charbonneau}, David and {Basri}, Gibor S. and {Sasselov}, Dimitar D. and {Winn}, Joshua and {Sanchis-Ojeda}, Roberto and {Christiansen}, Jessie and {Adams}, Elisabeth and {Henze}, Christopher and {Dupree}, Andrea and {Fabrycky}, Daniel C. and {Fortney}, Jonathan J. and {Tarter}, Jill and {Holman}, Matthew J. and {Tenenbaum}, Peter and {Shporer}, Avi and {Lucas}, Philip W. and {Welsh}, William F. and {Orosz}, Jerome A. and {Bedding}, T.~R. and {Campante}, T.~L. and {Davies}, G.~R. and {Elsworth}, Y. and {Handberg}, R. and {Hekker}, S. and {Karoff}, C. and {Kawaler}, S.~D. and {Lund}, M.~N. and {Lundkvist}, M. and {Metcalfe}, T.~S. and {Miglio}, A. and {Silva Aguirre}, V. and {Stello}, D. and {White}, T.~R. and {Boss}, Alan and {Devore}, Edna and {Gould}, Alan and {Prsa}, Andrej and {Agol}, Eric and {Barclay}, Thomas and {Coughlin}, Jeff and {Brugamyer}, Erik and {Mullally}, Fergal and {Quintana}, Elisa V. and {Still}, Martin and {Thompson}, Susan E. and {Morrison}, David and {Twicken}, Joseph D. and {D{\'e}sert}, Jean-Michel and {Carter}, Josh and {Crepp}, Justin R. and {H{\'e}brard}, Guillaume and {Santerne}, Alexandre and {Moutou}, Claire and {Sobeck}, Charlie and {Hudgins}, Douglas and {Haas}, Michael R. and {Robertson}, Paul and {Lillo-Box}, Jorge and {Barrado}, David},
        title = "{Masses, Radii, and Orbits of Small Kepler Planets: The Transition from Gaseous to Rocky Planets}",
      journal = {\apjs},
         year = 2014,
        month = feb,
       volume = {210},
       number = {2},
          eid = {20},
        pages = {20},
          doi = {10.1088/0067-0049/210/2/20},
archivePrefix = {arXiv},
       eprint = {1401.4195},
 primaryClass = {astro-ph.EP},
       adsurl = {https://ui.adsabs.harvard.edu/abs/2014ApJS..210...20M}
}

@ARTICLE{Steffen2013MNRAS.428.1077S,
       author = {{Steffen}, Jason H. and {Fabrycky}, Daniel C. and {Agol}, Eric and {Ford}, Eric B. and {Morehead}, Robert C. and {Cochran}, William D. and {Lissauer}, Jack J. and {Adams}, Elisabeth R. and {Borucki}, William J. and {Bryson}, Steve and {Caldwell}, Douglas A. and {Dupree}, Andrea and {Jenkins}, Jon M. and {Robertson}, Paul and {Rowe}, Jason F. and {Seader}, Shawn and {Thompson}, Susan and {Twicken}, Joseph D.},
        title = "{Transit timing observations from Kepler - VII. Confirmation of 27 planets in 13 multiplanet systems via transit timing variations and orbital stability}",
      journal = {\mnras},
         year = 2013,
        month = jan,
       volume = {428},
       number = {2},
        pages = {1077-1087},
          doi = {10.1093/mnras/sts090},
archivePrefix = {arXiv},
       eprint = {1208.3499},
 primaryClass = {astro-ph.EP},
       adsurl = {https://ui.adsabs.harvard.edu/abs/2013MNRAS.428.1077S}
}

@ARTICLE{Freudenthal2019AA...628A.108F,
       author = {{Freudenthal}, J. and {von Essen}, C. and {Ofir}, A. and {Dreizler}, S. and {Agol}, E. and {Wedemeyer}, S. and {Morris}, B.~M. and {Becker}, A.~C. and {Deeg}, H.~J. and {Hoyer}, S. and {Mallonn}, M. and {Poppenhaeger}, K. and {Herrero}, E. and {Ribas}, I. and {Boumis}, P. and {Liakos}, A.},
        title = "{Kepler Object of Interest Network. III. Kepler-82f: a new non-transiting 21 M$_{{\ensuremath{\oplus}}}$ planet from photodynamical modelling}",
      journal = {\aap},
         year = 2019,
        month = aug,
       volume = {628},
          eid = {A108},
        pages = {A108},
          doi = {10.1051/0004-6361/201935879},
archivePrefix = {arXiv},
       eprint = {1907.06534},
 primaryClass = {astro-ph.EP},
       adsurl = {https://ui.adsabs.harvard.edu/abs/2019AA...628A.108F}
}

@ARTICLE{Xie2013ApJS..208...22X,
       author = {{Xie}, Ji-Wei},
        title = "{Transit Timing Variation of Near-resonance Planetary Pairs: Confirmation of 12 Multiple-planet Systems}",
      journal = {\apjs},
         year = 2013,
        month = oct,
       volume = {208},
       number = {2},
          eid = {22},
        pages = {22},
          doi = {10.1088/0067-0049/208/2/22},
archivePrefix = {arXiv},
       eprint = {1208.3312},
 primaryClass = {astro-ph.EP},
       adsurl = {https://ui.adsabs.harvard.edu/abs/2013ApJS..208...22X}
}

@ARTICLE{Lissauer2013ApJ...770..131L,
       author = {{Lissauer}, Jack J. and {Jontof-Hutter}, Daniel and {Rowe}, Jason F. and {Fabrycky}, Daniel C. and {Lopez}, Eric D. and {Agol}, Eric and {Marcy}, Geoffrey W. and {Deck}, Katherine M. and {Fischer}, Debra A. and {Fortney}, Jonathan J. and {Howell}, Steve B. and {Isaacson}, Howard and {Jenkins}, Jon M. and {Kolbl}, Rea and {Sasselov}, Dimitar and {Short}, Donald R. and {Welsh}, William F.},
        title = "{All Six Planets Known to Orbit Kepler-11 Have Low Densities}",
      journal = {\apj},
         year = 2013,
        month = jun,
       volume = {770},
       number = {2},
          eid = {131},
        pages = {131},
          doi = {10.1088/0004-637X/770/2/131},
archivePrefix = {arXiv},
       eprint = {1303.0227},
 primaryClass = {astro-ph.EP},
       adsurl = {https://ui.adsabs.harvard.edu/abs/2013ApJ...770..131L}
}

@ARTICLE{MacDonald2016AJ....152..105M,
       author = {{MacDonald}, Mariah G. and {Ragozzine}, Darin and {Fabrycky}, Daniel C. and {Ford}, Eric B. and {Holman}, Matthew J. and {Isaacson}, Howard T. and {Lissauer}, Jack J. and {Lopez}, Eric D. and {Mazeh}, Tsevi and {Rogers}, Leslie and {Rowe}, Jason F. and {Steffen}, Jason H. and {Torres}, Guillermo},
        title = "{A Dynamical Analysis of the Kepler-80 System of Five Transiting Planets}",
      journal = {\aj},
         year = 2016,
        month = oct,
       volume = {152},
       number = {4},
          eid = {105},
        pages = {105},
          doi = {10.3847/0004-6256/152/4/105},
archivePrefix = {arXiv},
       eprint = {1607.07540},
 primaryClass = {astro-ph.EP},
       adsurl = {https://ui.adsabs.harvard.edu/abs/2016AJ....152..105M}
}

@ARTICLE{Beard2024AJ....167...70B,
       author = {{Beard}, Corey and {Robertson}, Paul and {Dai}, Fei and {Holcomb}, Rae and {Lubin}, Jack and {Akana Murphy}, Joseph M. and {Batalha}, Natalie M. and {Blunt}, Sarah and {Crossfield}, Ian and {Dressing}, Courtney and {Fulton}, Benjamin and {Howard}, Andrew W. and {Huber}, Dan and {Isaacson}, Howard and {Kane}, Stephen R. and {Nowak}, Grzegorz and {Petigura}, Erik A. and {Roy}, Arpita and {Rubenzahl}, Ryan A. and {Weiss}, Lauren M. and {Barrena}, Rafael and {Behmard}, Aida and {Brinkman}, Casey L. and {Carleo}, Ilaria and {Chontos}, Ashley and {Dalba}, Paul A. and {Fetherolf}, Tara and {Giacalone}, Steven and {Hill}, Michelle L. and {Kawauchi}, Kiyoe and {Korth}, Judith and {Luque}, Rafael and {MacDougall}, Mason G. and {Mayo}, Andrew W. and {Mo{\v{c}}nik}, Teo and {Morello}, Giuseppe and {Murgas}, Felipe and {Orell-Miquel}, Jaume and {Palle}, Enric and {Polanski}, Alex S. and {Rice}, Malena and {Scarsdale}, Nicholas and {Tyler}, Dakotah and {Van Zandt}, Judah},
        title = "{The TESS-Keck Survey. XVII. Precise Mass Measurements in a Young, High-multiplicity Transiting Planet System Using Radial Velocities and Transit Timing Variations}",
      journal = {\aj},
         year = 2024,
        month = feb,
       volume = {167},
       number = {2},
          eid = {70},
        pages = {70},
          doi = {10.3847/1538-3881/ad1330},
archivePrefix = {arXiv},
       eprint = {2312.04635},
 primaryClass = {astro-ph.EP},
       adsurl = {https://ui.adsabs.harvard.edu/abs/2024AJ....167...70B}
}

@ARTICLE{Gillon2017Natur.542..456G,
       author = {{Gillon}, Micha{\"e}l and {Triaud}, Amaury H.~M.~J. and {Demory}, Brice-Olivier and {Jehin}, Emmanu{\"e}l and {Agol}, Eric and {Deck}, Katherine M. and {Lederer}, Susan M. and {de Wit}, Julien and {Burdanov}, Artem and {Ingalls}, James G. and {Bolmont}, Emeline and {Leconte}, Jeremy and {Raymond}, Sean N. and {Selsis}, Franck and {Turbet}, Martin and {Barkaoui}, Khalid and {Burgasser}, Adam and {Burleigh}, Matthew R. and {Carey}, Sean J. and {Chaushev}, Aleksander and {Copperwheat}, Chris M. and {Delrez}, Laetitia and {Fernandes}, Catarina S. and {Holdsworth}, Daniel L. and {Kotze}, Enrico J. and {Van Grootel}, Val{\'e}rie and {Almleaky}, Yaseen and {Benkhaldoun}, Zouhair and {Magain}, Pierre and {Queloz}, Didier},
        title = "{Seven temperate terrestrial planets around the nearby ultracool dwarf star TRAPPIST-1}",
      journal = {\nat},
         year = 2017,
        month = feb,
       volume = {542},
       number = {7642},
        pages = {456-460},
          doi = {10.1038/nature21360},
archivePrefix = {arXiv},
       eprint = {1703.01424},
 primaryClass = {astro-ph.EP},
       adsurl = {https://ui.adsabs.harvard.edu/abs/2017Natur.542..456G}
}

@ARTICLE{Agol2021PSJ.....2....1A,
       author = {{Agol}, Eric and {Dorn}, Caroline and {Grimm}, Simon L. and {Turbet}, Martin and {Ducrot}, Elsa and {Delrez}, Laetitia and {Gillon}, Micha{\"e}l and {Demory}, Brice-Olivier and {Burdanov}, Artem and {Barkaoui}, Khalid and {Benkhaldoun}, Zouhair and {Bolmont}, Emeline and {Burgasser}, Adam and {Carey}, Sean and {de Wit}, Julien and {Fabrycky}, Daniel and {Foreman-Mackey}, Daniel and {Haldemann}, Jonas and {Hernandez}, David M. and {Ingalls}, James and {Jehin}, Emmanuel and {Langford}, Zachary and {Leconte}, J{\'e}r{\'e}my and {Lederer}, Susan M. and {Luger}, Rodrigo and {Malhotra}, Renu and {Meadows}, Victoria S. and {Morris}, Brett M. and {Pozuelos}, Francisco J. and {Queloz}, Didier and {Raymond}, Sean N. and {Selsis}, Franck and {Sestovic}, Marko and {Triaud}, Amaury H.~M.~J. and {Van Grootel}, Valerie},
        title = "{Refining the Transit-timing and Photometric Analysis of TRAPPIST-1: Masses, Radii, Densities, Dynamics, and Ephemerides}",
      journal = {\psj},
         year = 2021,
        month = feb,
       volume = {2},
       number = {1},
          eid = {1},
        pages = {1},
          doi = {10.3847/PSJ/abd022},
archivePrefix = {arXiv},
       eprint = {2010.01074},
 primaryClass = {astro-ph.EP},
       adsurl = {https://ui.adsabs.harvard.edu/abs/2021PSJ.....2....1A}
}

@ARTICLE{Steffen2016MNRAS.457.4384S,
       author = {{Steffen}, Jason H.},
        title = "{Sensitivity bias in the mass-radius distribution from transit timing variations and radial velocity measurements}",
      journal = {\mnras},
         year = 2016,
        month = apr,
       volume = {457},
       number = {4},
        pages = {4384-4392},
          doi = {10.1093/mnras/stw241},
archivePrefix = {arXiv},
       eprint = {1510.04750},
 primaryClass = {astro-ph.EP},
       adsurl = {https://ui.adsabs.harvard.edu/abs/2016MNRAS.457.4384S}
}

@ARTICLE{Liang2021AJ....161..202L,
       author = {{Liang}, Yan and {Robnik}, Jakob and {Seljak}, Uro{\v{s}}},
        title = "{Kepler-90: Giant Transit-timing Variations Reveal a Super-puff}",
      journal = {\aj},
         year = 2021,
        month = apr,
       volume = {161},
       number = {4},
          eid = {202},
        pages = {202},
          doi = {10.3847/1538-3881/abe6a7},
archivePrefix = {arXiv},
       eprint = {2011.08515},
 primaryClass = {astro-ph.EP},
       adsurl = {https://ui.adsabs.harvard.edu/abs/2021AJ....161..202L}
}

@ARTICLE{Cochran2011ApJS..197....7C,
       author = {{Cochran}, William D. and {Fabrycky}, Daniel C. and {Torres}, Guillermo and {Fressin}, Fran{\c{c}}ois and {D{\'e}sert}, Jean-Michel and {Ragozzine}, Darin and {Sasselov}, Dimitar and {Fortney}, Jonathan J. and {Rowe}, Jason F. and {Brugamyer}, Erik J. and {Bryson}, Stephen T. and {Carter}, Joshua A. and {Ciardi}, David R. and {Howell}, Steve B. and {Steffen}, Jason H. and {Borucki}, William. J. and {Koch}, David G. and {Winn}, Joshua N. and {Welsh}, William F. and {Uddin}, Kamal and {Tenenbaum}, Peter and {Still}, M. and {Seager}, Sara and {Quinn}, Samuel N. and {Mullally}, F. and {Miller}, Neil and {Marcy}, Geoffrey W. and {MacQueen}, Phillip J. and {Lucas}, Phillip and {Lissauer}, Jack J. and {Latham}, David W. and {Knutson}, Heather and {Kinemuchi}, K. and {Johnson}, John A. and {Jenkins}, Jon M. and {Isaacson}, Howard and {Howard}, Andrew and {Horch}, Elliott and {Holman}, Matthew J. and {Henze}, Christopher E. and {Haas}, Michael R. and {Gilliland}, Ronald L. and {Gautier}, III, Thomas N. and {Ford}, Eric B. and {Fischer}, Debra A. and {Everett}, Mark and {Endl}, Michael and {Demory}, Brice-Oliver and {Deming}, Drake and {Charbonneau}, David and {Caldwell}, Douglas and {Buchhave}, Lars and {Brown}, Timothy M. and {Batalha}, Natalie},
        title = "{Kepler-18b, c, and d: A System of Three Planets Confirmed by Transit Timing Variations, Light Curve Validation, Warm-Spitzer Photometry, and Radial Velocity Measurements}",
      journal = {\apjs},
         year = 2011,
        month = nov,
       volume = {197},
       number = {1},
          eid = {7},
        pages = {7},
          doi = {10.1088/0067-0049/197/1/7},
archivePrefix = {arXiv},
       eprint = {1110.0820},
 primaryClass = {astro-ph.EP},
       adsurl = {https://ui.adsabs.harvard.edu/abs/2011ApJS..197....7C}
}

@ARTICLE{Dai2015ApJ...813L...9D,
       author = {{Dai}, Fei and {Winn}, Joshua N. and {Arriagada}, Pamela and {Butler}, R. Paul and {Crane}, Jeffrey D. and {Johnson}, John Asher and {Shectman}, Stephen A. and {Teske}, Johanna K. and {Thompson}, Ian B. and {Vanderburg}, Andrew and {Wittenmyer}, Robert A.},
        title = "{Doppler Monitoring of the WASP-47 Multiplanet System}",
      journal = {\apjl},
         year = 2015,
        month = nov,
       volume = {813},
       number = {1},
          eid = {L9},
        pages = {L9},
          doi = {10.1088/2041-8205/813/1/L9},
archivePrefix = {arXiv},
       eprint = {1510.03811},
 primaryClass = {astro-ph.EP},
       adsurl = {https://ui.adsabs.harvard.edu/abs/2015ApJ...813L...9D}
}

@ARTICLE{Szabo2011ApJ...727L..44S,
       author = {{Szab{\'o}}, Gy. M. and {Kiss}, L.~L.},
        title = "{A Short-period Censor of Sub-Jupiter Mass Exoplanets with Low Density}",
      journal = {\apjl},
         year = 2011,
        month = feb,
       volume = {727},
       number = {2},
          eid = {L44},
        pages = {L44},
          doi = {10.1088/2041-8205/727/2/L44},
archivePrefix = {arXiv},
       eprint = {1012.4791},
 primaryClass = {astro-ph.EP},
       adsurl = {https://ui.adsabs.harvard.edu/abs/2011ApJ...727L..44S}
}

@ARTICLE{Owen2018MNRAS.479.5012O,
       author = {{Owen}, James E. and {Lai}, Dong},
        title = "{Photoevaporation and high-eccentricity migration created the sub-Jovian desert}",
      journal = {\mnras},
         year = 2018,
        month = oct,
       volume = {479},
       number = {4},
        pages = {5012-5021},
          doi = {10.1093/mnras/sty1760},
archivePrefix = {arXiv},
       eprint = {1807.00012},
 primaryClass = {astro-ph.EP},
       adsurl = {https://ui.adsabs.harvard.edu/abs/2018MNRAS.479.5012O}
}

@ARTICLE{Eggleton2001ApJ...562.1012E,
       author = {{Eggleton}, Peter P. and others},
        title = "{Orbital Evolution in Binary and Triple Stars, with an Application to SS Lacertae}",
      journal = {\apj},
         year = 2001,
        month = dec,
       volume = {562},
       number = {2},
        pages = {1012-1030},
          doi = {10.1086/323843},
archivePrefix = {arXiv},
       eprint = {astro-ph/0104126},
 primaryClass = {astro-ph},
       adsurl = {https://ui.adsabs.harvard.edu/abs/2001ApJ...562.1012E}
}

@ARTICLE{Weidenschilling1996Natur.384..619W,
       author = {{Weidenschilling}, Stuart and others},
        title = "{Gravitational scattering as a possible origin for giant planets at small stellar distances}",
      journal = {\nat},
         year = 1996,
        month = dec,
       volume = {384},
       number = {6610},
        pages = {619-621},
          doi = {10.1038/384619a0},
       adsurl = {https://ui.adsabs.harvard.edu/abs/1996Natur.384..619W}
}

@ARTICLE{Wu2003ApJ...589..605W,
       author = {{Wu}, Y. and {Murray}, N.},
        title = "{Planet Migration and Binary Companions: The Case of HD 80606b}",
      journal = {\apj},
         year = 2003,
        month = may,
       volume = {589},
       number = {1},
        pages = {605-614},
          doi = {10.1086/374598},
archivePrefix = {arXiv},
       eprint = {astro-ph/0303010},
 primaryClass = {astro-ph},
       adsurl = {https://ui.adsabs.harvard.edu/abs/2003ApJ...589..605W}
}

@ARTICLE{WisdomHolman1991AJ....102.1528W,
       author = {{Wisdom}, Jack and {Holman}, Matthew},
        title = "{Symplectic maps for the N-body problem.}",
      journal = {\aj},
         year = 1991,
        month = oct,
       volume = {102},
        pages = {1528-1538},
          doi = {10.1086/115978},
       adsurl = {https://ui.adsabs.harvard.edu/abs/1991AJ....102.1528W}
}

@ARTICLE{Laskar1993PhyD...67..257L,
       author = {{Laskar}, Jacques},
        title = "{Frequency analysis for multi-dimensional systems. Global dynamics and diffusion}",
      journal = {Physica D Nonlinear Phenomena},
         year = 1993,
        month = aug,
       volume = {67},
       number = {1-3},
        pages = {257-281},
          doi = {10.1016/0167-2789(93)90210-R},
       adsurl = {https://ui.adsabs.harvard.edu/abs/1993PhyD...67..257L}
}

@ARTICLE{Laskar1990Icar...88..266L,
       author = {{Laskar}, J.},
        title = "{The chaotic motion of the solar system: A numerical estimate of the size of the chaotic zones}",
      journal = {\icarus},
         year = 1990,
        month = dec,
       volume = {88},
       number = {2},
        pages = {266-291},
          doi = {10.1016/0019-1035(90)90084-M},
       adsurl = {https://ui.adsabs.harvard.edu/abs/1990Icar...88..266L}
}

@ARTICLE{Cincotta2000AAS..147..205C,
       author = {{Cincotta}, P.~M. and {Sim{\'o}}, C.},
        title = "{Simple tools to study global dynamics in non-axisymmetric galactic potentials - I}",
      journal = {\aaps},
         year = 2000,
        month = dec,
       volume = {147},
        pages = {205-228},
          doi = {10.1051/aas:2000108},
       adsurl = {https://ui.adsabs.harvard.edu/abs/2000AAS..147..205C}
}

@INCOLLECTION{Reichl1993sptn.book...47R,
       author = {{Reichl}, Linda E.},
        title = "{Methods of Conservative Chaos Theory in Stochastic Physics}",
    booktitle = {Statistical Physics and Thermodynamics of Nonlinear Nonequilibrium Systems. Edited by MUSCHIK W ET AL. Published by World Scientific Publishing Co. Pte. Ltd},
         year = 1993,
       editor = {{Muschik}, W. and {et al.}},
        pages = {47-53},
          doi = {10.1142/9789814503648_0004},
       adsurl = {https://ui.adsabs.harvard.edu/abs/1993sptn.book...47R}
}

@article{Ketchum_2013,
    doi = {10.1088/0004-637X/762/2/71},
    url = {https://dx.doi.org/10.1088/0004-637X/762/2/71},
    year = {2012},
    month = {dec},
    publisher = {The American Astronomical Society},
    volume = {762},
    number = {2},
    pages = {71},
    author = {Jacob A. Ketchum and Fred C. Adams and Anthony M. Bloch},
    title = {MEAN MOTION RESONANCES IN EXOPLANET SYSTEMS: AN INVESTIGATION INTO NODDING BEHAVIOR},
    journal = {The Astrophysical Journal}
}

@Inbook{Correia2018,
author="Correia, Alexandre C. M.
and Delisle, Jean-Baptiste
and Laskar, Jacques",
editor="Deeg, Hans J.
and Belmonte, Juan Antonio",
title="Planets in Mean-Motion Resonances and the System Around HD45364",
bookTitle="Handbook of Exoplanets ",
year="2018",
publisher="Springer International Publishing",
address="Cham",
pages="2693--2711",
isbn="978-3-319-55333-7",
doi="10.1007/978-3-319-55333-7_12",
url="https://doi.org/10.1007/978-3-319-55333-7_12"
}

@ARTICLE{Laune2022MNRAS.517.4472L,
       author = {{Laune}, J.~T. and {Rodet}, Laetitia and {Lai}, Dong},
        title = "{Apsidal alignment and anti-alignment of planets in mean-motion resonance: disc-driven migration and eccentricity driving}",
      journal = {\mnras},
         year = 2022,
        month = dec,
       volume = {517},
       number = {3},
        pages = {4472-4488},
          doi = {10.1093/mnras/stac2914},
archivePrefix = {arXiv},
       eprint = {2206.04810},
 primaryClass = {astro-ph.EP},
       adsurl = {https://ui.adsabs.harvard.edu/abs/2022MNRAS.517.4472L}
}

@ARTICLE{Beauge2003ApJ...593.1124B,
       author = {{Beaug{\'e}}, C. and {Ferraz-Mello}, S. and {Michtchenko}, T.~A.},
        title = "{Extrasolar Planets in Mean-Motion Resonance: Apses Alignment and Asymmetric Stationary Solutions}",
      journal = {\apj},
         year = 2003,
        month = aug,
       volume = {593},
       number = {2},
        pages = {1124-1133},
          doi = {10.1086/376568},
archivePrefix = {arXiv},
       eprint = {astro-ph/0210577},
 primaryClass = {astro-ph},
       adsurl = {https://ui.adsabs.harvard.edu/abs/2003ApJ...593.1124B}
}

@ARTICLE{Cabrera2014ApJ...781...18C,
       author = {{Cabrera}, J. and {Csizmadia}, Sz. and {Lehmann}, H. and {Dvorak}, R. and {Gandolfi}, D. and {Rauer}, H. and {Erikson}, A. and {Dreyer}, C. and {Eigm{\"u}ller}, Ph. and {Hatzes}, A.},
        title = "{The Planetary System to KIC 11442793: A Compact Analogue to the Solar System}",
      journal = {\apj},
         year = 2014,
        month = jan,
       volume = {781},
       number = {1},
          eid = {18},
        pages = {18},
          doi = {10.1088/0004-637X/781/1/18},
archivePrefix = {arXiv},
       eprint = {1310.6248},
 primaryClass = {astro-ph.EP},
       adsurl = {https://ui.adsabs.harvard.edu/abs/2014ApJ...781...18C}
}

@ARTICLE{Schwarz2011MNRAS.414.2763S,
       author = {{Schwarz}, R. and {Haghighipour}, N. and {Eggl}, S. and {Pilat-Lohinger}, E. and {Funk}, B.},
        title = "{Prospects of the detection of circumbinary planets with Kepler and CoRoT using the variations of eclipse timing}",
      journal = {\mnras},
         year = 2011,
        month = jul,
       volume = {414},
       number = {3},
        pages = {2763-2770},
          doi = {10.1111/j.1365-2966.2011.18594.x},
archivePrefix = {arXiv},
       eprint = {1101.1994},
 primaryClass = {astro-ph.EP},
       adsurl = {https://ui.adsabs.harvard.edu/abs/2011MNRAS.414.2763S}
}

@ARTICLE{MarzariThebault2019Galax...7...84M,
       author = {{Marzari}, Francesco and {Thebault}, Philippe},
        title = "{Planets in Binaries: Formation and Dynamical Evolution}",
      journal = {Galaxies},
         year = 2019,
        month = oct,
       volume = {7},
       number = {4},
          eid = {84},
        pages = {84},
          doi = {10.3390/galaxies7040084},
archivePrefix = {arXiv},
       eprint = {2002.12006},
 primaryClass = {astro-ph.EP},
       adsurl = {https://ui.adsabs.harvard.edu/abs/2019Galax...7...84M}
}

@ARTICLE{Columba2023AA...675A.156C,
       author = {{Columba}, G. and {Danielski}, C. and {Dorozsmai}, A. and {Toonen}, S. and {Lopez Puertas}, M.},
        title = "{Statistics of Magrathea exoplanets beyond the main sequence. Simulating the long-term evolution of circumbinary giant planets with TRES}",
      journal = {\aap},
         year = 2023,
        month = jul,
       volume = {675},
          eid = {A156},
        pages = {A156},
          doi = {10.1051/0004-6361/202345843},
archivePrefix = {arXiv},
       eprint = {2305.07057},
 primaryClass = {astro-ph.EP},
       adsurl = {https://ui.adsabs.harvard.edu/abs/2023AA...675A.156C}
}

@ARTICLE{Eastman2010PASP..122..935E,
       author = {{Eastman}, Jason and {Siverd}, Robert and {Gaudi}, B. Scott},
        title = "{Achieving Better Than 1 Minute Accuracy in the Heliocentric and Barycentric Julian Dates}",
      journal = {\pasp},
         year = 2010,
        month = aug,
       volume = {122},
       number = {894},
        pages = {935},
          doi = {10.1086/655938},
archivePrefix = {arXiv},
       eprint = {1005.4415},
 primaryClass = {astro-ph.IM},
       adsurl = {https://ui.adsabs.harvard.edu/abs/2010PASP..122..935E}
}

@ARTICLE{Damasso2015AA...575A.111D,
       author = {{Damasso}, M. and {Biazzo}, K. and {Bonomo}, A.~S. and {Desidera}, S. and {Lanza}, A.~F. and {Nascimbeni}, V. and {Esposito}, M. and {Scandariato}, G. and {Sozzetti}, A. and {Cosentino}, R. and {Gratton}, R. and {Malavolta}, L. and {Rainer}, M. and {Gandolfi}, D. and {Poretti}, E. and {Zanmar Sanchez}, R. and {Ribas}, I. and {Santos}, N. and {Affer}, L. and {Andreuzzi}, G. and {Barbieri}, M. and {Bedin}, L.~R. and {Benatti}, S. and {Bernagozzi}, A. and {Bertolini}, E. and {Bonavita}, M. and {Borsa}, F. and {Borsato}, L. and {Boschin}, W. and {Calcidese}, P. and {Carbognani}, A. and {Cenadelli}, D. and {Christille}, J.~M. and {Claudi}, R.~U. and {Covino}, E. and {Cunial}, A. and {Giacobbe}, P. and {Granata}, V. and {Harutyunyan}, A. and {Lattanzi}, M.~G. and {Leto}, G. and {Libralato}, M. and {Lodato}, G. and {Lorenzi}, V. and {Mancini}, L. and {Martinez Fiorenzano}, A.~F. and {Marzari}, F. and {Masiero}, S. and {Micela}, G. and {Molinari}, E. and {Molinaro}, M. and {Munari}, U. and {Murabito}, S. and {Pagano}, I. and {Pedani}, M. and {Piotto}, G. and {Rosenberg}, A. and {Silvotti}, R. and {Southworth}, J.},
        title = "{The GAPS programme with HARPS-N at TNG. V. A comprehensive analysis of the XO-2 stellar and planetary systems}",
      journal = {\aap},
         year = 2015,
        month = mar,
       volume = {575},
          eid = {A111},
        pages = {A111},
          doi = {10.1051/0004-6361/201425332},
archivePrefix = {arXiv},
       eprint = {1501.01424},
 primaryClass = {astro-ph.SR},
       adsurl = {https://ui.adsabs.harvard.edu/abs/2015AA...575A.111D}
}

@INPROCEEDINGS{Sterken2005ASPC..335....3S,
       author = {{Sterken}, C.},
        title = "{The O-C Diagram: Basic Procedures}",
    booktitle = {The Light-Time Effect in Astrophysics: Causes and cures of the O-C diagram},
         year = 2005,
       editor = {{Sterken}, C.},
       series = {Astronomical Society of the Pacific Conference Series},
       volume = {335},
        month = jul,
        pages = {3},
       adsurl = {https://ui.adsabs.harvard.edu/abs/2005ASPC..335....3S}
}

@ARTICLE{Applegate1992ApJ...385..621A,
       author = {{Applegate}, James H.},
        title = "{A Mechanism for Orbital Period Modulation in Close Binaries}",
      journal = {\apj},
         year = 1992,
        month = feb,
       volume = {385},
        pages = {621},
          doi = {10.1086/170967},
       adsurl = {https://ui.adsabs.harvard.edu/abs/1992ApJ...385..621A}
}

@article{Brown-Sevilla10.1093/mnras/stab1843,
    author = {Brown-Sevilla, S B and Nascimbeni, V and Borsato, L and Tartaglia, L and Nardiello, D and Granata, V and Libralato, M and Damasso, M and Piotto, G and Pollacco, D and West, R G and Colombo, L S and Cunial, A and Piazza, G and Scaggiante, F},
    title = {A new photometric and dynamical study of the eclipsing binary star HW Virginis},
    journal = {Monthly Notices of the Royal Astronomical Society},
    volume = {506},
    number = {2},
    pages = {2122-2135},
    year = {2021},
    month = {07},
    issn = {0035-8711},
    doi = {10.1093/mnras/stab1843},
    url = {https://doi.org/10.1093/mnras/stab1843},
    eprint = {https://academic.oup.com/mnras/article-pdf/506/2/2122/39117678/stab1843.pdf},
}

@ARTICLE{Irwin1952ApJ...116..211I,
       author = {{Irwin}, John B.},
        title = "{The Determination of a Light-Time Orbit.}",
      journal = {\apj},
         year = 1952,
        month = jul,
       volume = {116},
        pages = {211},
          doi = {10.1086/145604},
       adsurl = {https://ui.adsabs.harvard.edu/abs/1952ApJ...116..211I}
}

@ARTICLE{Irwin1959AJ.....64..149I,
       author = {{Irwin}, John B.},
        title = "{Standard light-time curves}",
      journal = {\aj},
         year = 1959,
        month = may,
       volume = {64},
        pages = {149},
          doi = {10.1086/107913},
       adsurl = {https://ui.adsabs.harvard.edu/abs/1959AJ.....64..149I}
}

@ARTICLE{Borkovits2016MNRAS.455.4136B,
       author = {{Borkovits}, T. and {Hajdu}, T. and {Sztakovics}, J. and {Rappaport}, S. and {Levine}, A. and {B{\'\i}r{\'o}}, I.~B. and {Klagyivik}, P.},
        title = "{A comprehensive study of the Kepler triples via eclipse timing}",
      journal = {\mnras},
         year = 2016,
        month = feb,
       volume = {455},
       number = {4},
        pages = {4136-4165},
          doi = {10.1093/mnras/stv2530},
archivePrefix = {arXiv},
       eprint = {1510.08272},
 primaryClass = {astro-ph.SR},
       adsurl = {https://ui.adsabs.harvard.edu/abs/2016MNRAS.455.4136B}
}

@ARTICLE{Borkovits2015MNRAS.448..946B,
       author = {{Borkovits}, T. and {Rappaport}, S. and {Hajdu}, T. and {Sztakovics}, J.},
        title = "{Eclipse timing variation analyses of eccentric binaries with close tertiaries in the Kepler field}",
      journal = {\mnras},
         year = 2015,
        month = mar,
       volume = {448},
       number = {1},
        pages = {946-993},
          doi = {10.1093/mnras/stv015},
archivePrefix = {arXiv},
       eprint = {1412.5759},
 primaryClass = {astro-ph.SR},
       adsurl = {https://ui.adsabs.harvard.edu/abs/2015MNRAS.448..946B}
}

@ARTICLE{Borkovits2025AA...695A.209B,
       author = {{Borkovits}, T. and {Rappaport}, S.~A. and {Mitnyan}, T. and {B{\'\i}r{\'o}}, I.~B. and {Cs{\'a}nyi}, I. and {Forg{\'a}cs-Dajka}, E. and {Forr{\'o}}, A. and {Hajdu}, T. and {Seli}, B. and {Sztakovics}, J. and {G{\"o}bly{\"o}s}, A. and {P{\'a}l}, A.},
        title = "{Then and now: A new look at the eclipse timing variations of hierarchical triple star candidates in the primordial Kepler field, revisited by TESS}",
      journal = {\aap},
         year = 2025,
        month = mar,
       volume = {695},
          eid = {A209},
        pages = {A209},
          doi = {10.1051/0004-6361/202453616},
archivePrefix = {arXiv},
       eprint = {2502.09480},
 primaryClass = {astro-ph.SR},
       adsurl = {https://ui.adsabs.harvard.edu/abs/2025AA...695A.209B}
}

@ARTICLE{Marcadon2024ApJ...976..242M,
       author = {{Marcadon}, Fr{\'e}d{\'e}ric and {Pr{\v{s}}a}, Andrej},
        title = "{Precision Timing of Eclipsing Binaries from TESS Full Frame Images: Method and Performance}",
      journal = {\apj},
         year = 2024,
        month = dec,
       volume = {976},
       number = {2},
          eid = {242},
        pages = {242},
          doi = {10.3847/1538-4357/ad8571},
archivePrefix = {arXiv},
       eprint = {2403.07694},
 primaryClass = {astro-ph.SR},
       adsurl = {https://ui.adsabs.harvard.edu/abs/2024ApJ...976..242M}
}

@ARTICLE{Bours2016MNRAS.460.3873B,
       author = {{Bours}, M.~C.~P. and {Marsh}, T.~R. and {Parsons}, S.~G. and {Dhillon}, V.~S. and {Ashley}, R.~P. and {Bento}, J.~P. and {Breedt}, E. and {Butterley}, T. and {Caceres}, C. and {Chote}, P. and {Copperwheat}, C.~M. and {Hardy}, L.~K. and {Hermes}, J.~J. and {Irawati}, P. and {Kerry}, P. and {Kilkenny}, D. and {Littlefair}, S.~P. and {McAllister}, M.~J. and {Rattanasoon}, S. and {Sahman}, D.~I. and {Vu{\v{c}}kovi{\'c}}, M. and {Wilson}, R.~W.},
        title = "{Long-term eclipse timing of white dwarf binaries: an observational hint of a magnetic mechanism at work}",
      journal = {\mnras},
         year = 2016,
        month = aug,
       volume = {460},
       number = {4},
        pages = {3873-3887},
          doi = {10.1093/mnras/stw1203},
archivePrefix = {arXiv},
       eprint = {1606.00780},
 primaryClass = {astro-ph.SR},
       adsurl = {https://ui.adsabs.harvard.edu/abs/2016MNRAS.460.3873B}
}

@ARTICLE{Siegel2021AJ....161..290S,
       author = {{Siegel}, Jared C. and {Fabrycky}, Daniel},
        title = "{Resonant Chains of Exoplanets: Libration Centers for Three-body Angles}",
      journal = {\aj},
         year = 2021,
        month = jun,
       volume = {161},
       number = {6},
          eid = {290},
        pages = {290},
          doi = {10.3847/1538-3881/abf8a6},
archivePrefix = {arXiv},
       eprint = {2104.14665},
 primaryClass = {astro-ph.EP},
       adsurl = {https://ui.adsabs.harvard.edu/abs/2021AJ....161..290S}
}

@ARTICLE{Borucki2010Sci...327..977B,
       author = {{Borucki}, William J. and {Koch}, David and {Basri}, Gibor and {Batalha}, Natalie and {Brown}, Timothy and {Caldwell}, Douglas and {Caldwell}, John and {Christensen-Dalsgaard}, J{\o}rgen and {Cochran}, William D. and {DeVore}, Edna and {Dunham}, Edward W. and {Dupree}, Andrea K. and {Gautier}, Thomas N. and {Geary}, John C. and {Gilliland}, Ronald and {Gould}, Alan and {Howell}, Steve B. and {Jenkins}, Jon M. and {Kondo}, Yoji and {Latham}, David W. and {Marcy}, Geoffrey W. and {Meibom}, S{\o}ren and {Kjeldsen}, Hans and {Lissauer}, Jack J. and {Monet}, David G. and {Morrison}, David and {Sasselov}, Dimitar and {Tarter}, Jill and {Boss}, Alan and {Brownlee}, Don and {Owen}, Toby and {Buzasi}, Derek and {Charbonneau}, David and {Doyle}, Laurance and {Fortney}, Jonathan and {Ford}, Eric B. and {Holman}, Matthew J. and {Seager}, Sara and {Steffen}, Jason H. and {Welsh}, William F. and {Rowe}, Jason and {Anderson}, Howard and {Buchhave}, Lars and {Ciardi}, David and {Walkowicz}, Lucianne and {Sherry}, William and {Horch}, Elliott and {Isaacson}, Howard and {Everett}, Mark E. and {Fischer}, Debra and {Torres}, Guillermo and {Johnson}, John Asher and {Endl}, Michael and {MacQueen}, Phillip and {Bryson}, Stephen T. and {Dotson}, Jessie and {Haas}, Michael and {Kolodziejczak}, Jeffrey and {Van Cleve}, Jeffrey and {Chandrasekaran}, Hema and {Twicken}, Joseph D. and {Quintana}, Elisa V. and {Clarke}, Bruce D. and {Allen}, Christopher and {Li}, Jie and {Wu}, Haley and {Tenenbaum}, Peter and {Verner}, Ekaterina and {Bruhweiler}, Frederick and {Barnes}, Jason and {Prsa}, Andrej},
        title = "{Kepler Planet-Detection Mission: Introduction and First Results}",
      journal = {Science},
         year = 2010,
        month = feb,
       volume = {327},
       number = {5968},
        pages = {977},
          doi = {10.1126/science.1185402},
       adsurl = {https://ui.adsabs.harvard.edu/abs/2010Sci...327..977B}
}

@ARTICLE{Ricker2015JATIS...1a4003R,
       author = {{Ricker}, George R. and {Winn}, Joshua N. and {Vanderspek}, Roland and {Latham}, David W. and {Bakos}, G{\'a}sp{\'a}r {\'A}. and {Bean}, Jacob L. and {Berta-Thompson}, Zachory K. and {Brown}, Timothy M. and {Buchhave}, Lars and {Butler}, Nathaniel R. and {Butler}, R. Paul and {Chaplin}, William J. and {Charbonneau}, David and {Christensen-Dalsgaard}, J{\o}rgen and {Clampin}, Mark and {Deming}, Drake and {Doty}, John and {De Lee}, Nathan and {Dressing}, Courtney and {Dunham}, Edward W. and {Endl}, Michael and {Fressin}, Francois and {Ge}, Jian and {Henning}, Thomas and {Holman}, Matthew J. and {Howard}, Andrew W. and {Ida}, Shigeru and {Jenkins}, Jon M. and {Jernigan}, Garrett and {Johnson}, John Asher and {Kaltenegger}, Lisa and {Kawai}, Nobuyuki and {Kjeldsen}, Hans and {Laughlin}, Gregory and {Levine}, Alan M. and {Lin}, Douglas and {Lissauer}, Jack J. and {MacQueen}, Phillip and {Marcy}, Geoffrey and {McCullough}, Peter R. and {Morton}, Timothy D. and {Narita}, Norio and {Paegert}, Martin and {Palle}, Enric and {Pepe}, Francesco and {Pepper}, Joshua and {Quirrenbach}, Andreas and {Rinehart}, Stephen A. and {Sasselov}, Dimitar and {Sato}, Bun'ei and {Seager}, Sara and {Sozzetti}, Alessandro and {Stassun}, Keivan G. and {Sullivan}, Peter and {Szentgyorgyi}, Andrew and {Torres}, Guillermo and {Udry}, Stephane and {Villasenor}, Joel},
        title = "{Transiting Exoplanet Survey Satellite (TESS)}",
      journal = {Journal of Astronomical Telescopes, Instruments, and Systems},
         year = 2015,
        month = jan,
       volume = {1},
          eid = {014003},
        pages = {014003},
          doi = {10.1117/1.JATIS.1.1.014003},
       adsurl = {https://ui.adsabs.harvard.edu/abs/2015JATIS...1a4003R}
}

@ARTICLE{Howell2014PASP..126..398H,
       author = {{Howell}, Steve B. and {Sobeck}, Charlie and {Haas}, Michael and {Still}, Martin and {Barclay}, Thomas and {Mullally}, Fergal and {Troeltzsch}, John and {Aigrain}, Suzanne and {Bryson}, Stephen T. and {Caldwell}, Doug and {Chaplin}, William J. and {Cochran}, William D. and {Huber}, Daniel and {Marcy}, Geoffrey W. and {Miglio}, Andrea and {Najita}, Joan R. and {Smith}, Marcie and {Twicken}, J.~D. and {Fortney}, Jonathan J.},
        title = "{The K2 Mission: Characterization and Early Results}",
      journal = {\pasp},
         year = 2014,
        month = apr,
       volume = {126},
       number = {938},
        pages = {398},
          doi = {10.1086/676406},
archivePrefix = {arXiv},
       eprint = {1402.5163},
 primaryClass = {astro-ph.IM},
       adsurl = {https://ui.adsabs.harvard.edu/abs/2014PASP..126..398H}
}

@ARTICLE{Benz2021ExA....51..109B,
       author = {{Benz}, W. and {Broeg}, C. and {Fortier}, A. and {Rando}, N. and {Beck}, T. and {Beck}, M. and {Queloz}, D. and {Ehrenreich}, D. and {Maxted}, P.~F.~L. and {Isaak}, K.~G. and {Billot}, N. and {Alibert}, Y. and {Alonso}, R. and {Ant{\'o}nio}, C. and {Asquier}, J. and {Bandy}, T. and {B{\'a}rczy}, T. and {Barrado}, D. and {Barros}, S.~C.~C. and {Baumjohann}, W. and {Bekkelien}, A. and {Bergomi}, M. and {Biondi}, F. and {Bonfils}, X. and {Borsato}, L. and {Brandeker}, A. and {Busch}, M. -D. and {Cabrera}, J. and {Cessa}, V. and {Charnoz}, S. and {Chazelas}, B. and {Collier Cameron}, A. and {Corral Van Damme}, C. and {Cortes}, D. and {Davies}, M.~B. and {Deleuil}, M. and {Deline}, A. and {Delrez}, L. and {Demangeon}, O. and {Demory}, B.~O. and {Erikson}, A. and {Farinato}, J. and {Fossati}, L. and {Fridlund}, M. and {Futyan}, D. and {Gandolfi}, D. and {Garcia Munoz}, A. and {Gillon}, M. and {Guterman}, P. and {Gutierrez}, A. and {Hasiba}, J. and {Heng}, K. and {Hernandez}, E. and {Hoyer}, S. and {Kiss}, L.~L. and {Kovacs}, Z. and {Kuntzer}, T. and {Laskar}, J. and {Lecavelier des Etangs}, A. and {Lendl}, M. and {L{\'o}pez}, A. and {Lora}, I. and {Lovis}, C. and {L{\"u}ftinger}, T. and {Magrin}, D. and {Malvasio}, L. and {Marafatto}, L. and {Michaelis}, H. and {de Miguel}, D. and {Modrego}, D. and {Munari}, M. and {Nascimbeni}, V. and {Olofsson}, G. and {Ottacher}, H. and {Ottensamer}, R. and {Pagano}, I. and {Palacios}, R. and {Pall{\'e}}, E. and {Peter}, G. and {Piazza}, D. and {Piotto}, G. and {Pizarro}, A. and {Pollaco}, D. and {Ragazzoni}, R. and {Ratti}, F. and {Rauer}, H. and {Ribas}, I. and {Rieder}, M. and {Rohlfs}, R. and {Safa}, F. and {Salatti}, M. and {Santos}, N.~C. and {Scandariato}, G. and {S{\'e}gransan}, D. and {Simon}, A.~E. and {Smith}, A.~M.~S. and {Sordet}, M. and {Sousa}, S.~G. and {Steller}, M. and {Szab{\'o}}, G.~M. and {Szoke}, J. and {Thomas}, N. and {Tschentscher}, M. and {Udry}, S. and {Van Grootel}, V. and {Viotto}, V. and {Walter}, I. and {Walton}, N.~A. and {Wildi}, F. and {Wolter}, D.},
        title = "{The CHEOPS mission}",
      journal = {Experimental Astronomy},
         year = 2021,
        month = feb,
       volume = {51},
       number = {1},
        pages = {109-151},
          doi = {10.1007/s10686-020-09679-4},
archivePrefix = {arXiv},
       eprint = {2009.11633},
 primaryClass = {astro-ph.IM},
       adsurl = {https://ui.adsabs.harvard.edu/abs/2021ExA....51..109B}
}

@ARTICLE{Rauer2025ExA....59...26R,
       author = {{Rauer}, Heike and {Aerts}, Conny and {Cabrera}, Juan and {Deleuil}, Magali and {Erikson}, Anders and {Gizon}, Laurent and {Goupil}, Mariejo and {Heras}, Ana and {Walloschek}, Thomas and {Lorenzo-Alvarez}, Jose and {Marliani}, Filippo and {Martin-Garcia}, C{\'e}sar and {Mas-Hesse}, J. Miguel and {O'Rourke}, Laurence and {Osborn}, Hugh and {Pagano}, Isabella and {Piotto}, Giampaolo and {Pollacco}, Don and {Ragazzoni}, Roberto and {Ramsay}, Gavin and {Udry}, St{\'e}phane and {Appourchaux}, Thierry and {Benz}, Willy and {Brandeker}, Alexis and {G{\"u}del}, Manuel and {Janot-Pacheco}, Eduardo and {Kabath}, Petr and {Kjeldsen}, Hans and {Min}, Michiel and {Santos}, Nuno and {Smith}, Alan and {Suarez}, Juan-Carlos and {Werner}, Stephanie C. and {Aboudan}, Alessio and {Abreu}, Manuel and {Acu{\~n}a}, Lorena and {Adams}, Moritz and {Adibekyan}, Vardan and {Affer}, Laura and {Agneray}, Fran{\c{c}}ois and {Agnor}, Craig and {Aguirre B{\o}rsen-Koch}, Victor and {Ahmed}, Saad and {Aigrain}, Suzanne and {Al-Bahlawan}, Ashraf and {Alcacera Gil}, Ma de los Angeles and {Alei}, Eleonora and {Alencar}, Silvia and {Alexander}, Richard and {Alfonso-Garz{\'o}n}, Julia and {Alibert}, Yann and {Allende Prieto}, Carlos and {Almeida}, Leonardo and {Alonso Sobrino}, Roi and {Altavilla}, Giuseppe and {Althaus}, Christian and {Alvarez Trujillo}, Luis Alonso and {Amarsi}, Anish and {Ammler-von Eiff}, Matthias and {Am{\^o}res}, Eduardo and {Andrade}, Laerte and {Antoniadis-Karnavas}, Alexandros and {Ant{\'o}nio}, Carlos and {Aparicio del Moral}, Beatriz and {Appolloni}, Matteo and {Arena}, Claudio and {Armstrong}, David and {Aroca Aliaga}, Jose and {Asplund}, Martin and {Audenaert}, Jeroen and {Auricchio}, Natalia and {Avelino}, Pedro and {Baeke}, Ann and {Bailli{\'e}}, Kevin and {Balado}, Ana and {Ballber Balaguer{\'o}}, Pau and {Balestra}, Andrea and {Ball}, Warrick and {Ballans}, Herve and {Ballot}, Jerome and {Barban}, Caroline and {Barbary}, Ga{\"e}le and {Barbieri}, Mauro and {Barcel{\'o} Forteza}, Sebasti{\`a} and {Barker}, Adrian and {Barklem}, Paul and {Barnes}, Sydney and {Barrado Navascues}, David and {Barragan}, Oscar and {Baruteau}, Cl{\'e}ment and {Basu}, Sarbani and {Baudin}, Frederic and {Baumeister}, Philipp and {Bayliss}, Daniel and {Bazot}, Michael and {Beck}, Paul G. and {Belkacem}, Kevin and {Bellinger}, Earl and {Benatti}, Serena and {Benomar}, Othman and {B{\'e}rard}, Diane and {Bergemann}, Maria and {Bergomi}, Maria and {Bernardo}, Pierre and {Biazzo}, Katia and {Bignamini}, Andrea and {Bigot}, Lionel and {Billot}, Nicolas and {Binet}, Martin and {Biondi}, David and {Biondi}, Federico and {Birch}, Aaron C. and {Bitsch}, Bertram and {Bluhm Ceballos}, Paz Victoria and {B{\'o}di}, Attila and {Bogn{\'a}r}, Zs{\'o}fia and {Boisse}, Isabelle and {Bolmont}, Emeline and {Bonanno}, Alfio and {Bonavita}, Mariangela and {Bonfanti}, Andrea and {Bonfils}, Xavier and {Bonito}, Rosaria and {Bonomo}, Aldo Stefano and {B{\"o}rner}, Anko and {Boro Saikia}, Sudeshna and {Borreguero Mart{\'\i}n}, Elisa and {Borsa}, Francesco and {Borsato}, Luca and {Bossini}, Diego and {Bouchy}, Francois and {Bou{\'e}}, Gwena{\"e}l and {Boufleur}, Rodrigo and {Boumier}, Patrick and {Bourrier}, Vincent and {Bowman}, Dominic M. and {Bozzo}, Enrico and {Bradley}, Louisa and {Bray}, John and {Bressan}, Alessandro and {Breton}, Sylvain and {Brienza}, Daniele and {Brito}, Ana and {Brogi}, Matteo and {Brown}, Beverly and {Brown}, David J.~A. and {Brun}, Allan Sacha and {Bruno}, Giovanni and {Bruns}, Michael and {Buchhave}, Lars A. and {Bugnet}, Lisa and {Buldgen}, Ga{\"e}l and {Burgess}, Patrick and {Busatta}, Andrea and {Busso}, Giorgia and {Buzasi}, Derek and {Caballero}, Jos{\'e} A. and {Cabral}, Alexandre and {Cabrero Gomez}, Juan-Francisco and {Calderone}, Flavia and {Cameron}, Robert and {Cameron}, Andrew and {Campante}, Tiago and {Campos Gestal}, N{\'e}stor and {Canto Martins}, Bruno Leonardo and {Cara}, Christophe and {Carone}, Ludmila and {Carrasco}, Josep Manel and {Casagrande}, Luca and {Casewell}, Sarah L. and {Cassisi}, Santi and {Castellani}, Marco and {Castro}, Matthieu and {Catala}, Claude and {Catal{\'a}n Fern{\'a}ndez}, Irene and {Catelan}, M{\'a}rcio and {Cegla}, Heather and {Cerruti}, Chiara and {Cessa}, Virginie and {Chadid}, Merieme and {Chaplin}, William and {Charpinet}, Stephane and {Chiappini}, Cristina and {Chiarucci}, Simone and {Chiavassa}, Andrea and {Chinellato}, Simonetta and {Chirulli}, Giovanni and {Christensen-Dalsgaard}, J{\o}rgen and {Church}, Ross and {Claret}, Antonio and {Clarke}, Cathie and {Claudi}, Riccardo and {Clermont}, Lionel and {Coelho}, Hugo and {Coelho}, Joao and {Cogato}, Fabrizio and {Colom{\'e}}, Josep and {Condamin}, Mathieu and {Conde Garc{\'\i}a}, Fernando and {Conseil}, Simon},
        title = "{The PLATO mission}",
      journal = {Experimental Astronomy},
         year = 2025,
        month = jun,
       volume = {59},
       number = {3},
          eid = {26},
        pages = {26},
          doi = {10.1007/s10686-025-09985-9},
archivePrefix = {arXiv},
       eprint = {2406.05447},
 primaryClass = {astro-ph.IM},
       adsurl = {https://ui.adsabs.harvard.edu/abs/2025ExA....59...26R}
}

@ARTICLE{Tinetti2018ExA....46..135T,
       author = {{Tinetti}, Giovanna and {Drossart}, Pierre and {Eccleston}, Paul and {Hartogh}, Paul and {Heske}, Astrid and {Leconte}, J{\'e}r{\'e}my and {Micela}, Giusi and {Ollivier}, Marc and {Pilbratt}, G{\"o}ran and {Puig}, Ludovic and {Turrini}, Diego and {Vandenbussche}, Bart and {Wolkenberg}, Paulina and {Beaulieu}, Jean-Philippe and {Buchave}, Lars A. and {Ferus}, Martin and {Griffin}, Matt and {Guedel}, Manuel and {Justtanont}, Kay and {Lagage}, Pierre-Olivier and {Machado}, Pedro and {Malaguti}, Giuseppe and {Min}, Michiel and {N{\o}rgaard-Nielsen}, Hans Ulrik and {Rataj}, Mirek and {Ray}, Tom and {Ribas}, Ignasi and {Swain}, Mark and {Szabo}, Robert and {Werner}, Stephanie and {Barstow}, Joanna and {Burleigh}, Matt and {Cho}, James and {Coud{\'e} du Foresto}, Vincent and {Coustenis}, Athena and {Decin}, Leen and {Encrenaz}, Therese and {Galand}, Marina and {Gillon}, Michael and {Helled}, Ravit and {Morales}, Juan Carlos and {Garc{\'\i}a Mu{\~n}oz}, Antonio and {Moneti}, Andrea and {Pagano}, Isabella and {Pascale}, Enzo and {Piccioni}, Giuseppe and {Pinfield}, David and {Sarkar}, Subhajit and {Selsis}, Franck and {Tennyson}, Jonathan and {Triaud}, Amaury and {Venot}, Olivia and {Waldmann}, Ingo and {Waltham}, David and {Wright}, Gillian and {Amiaux}, Jerome and {Augu{\`e}res}, Jean-Louis and {Berth{\'e}}, Michel and {Bezawada}, Naidu and {Bishop}, Georgia and {Bowles}, Neil and {Coffey}, Deirdre and {Colom{\'e}}, Josep and {Crook}, Martin and {Crouzet}, Pierre-Elie and {Da Peppo}, Vania and {Sanz}, Isabel Escudero and {Focardi}, Mauro and {Frericks}, Martin and {Hunt}, Tom and {Kohley}, Ralf and {Middleton}, Kevin and {Morgante}, Gianluca and {Ottensamer}, Roland and {Pace}, Emanuele and {Pearson}, Chris and {Stamper}, Richard and {Symonds}, Kate and {Rengel}, Miriam and {Renotte}, Etienne and {Ade}, Peter and {Affer}, Laura and {Alard}, Christophe and {Allard}, Nicole and {Altieri}, Francesca and {Andr{\'e}}, Yves and {Arena}, Claudio and {Argyriou}, Ioannis and {Aylward}, Alan and {Baccani}, Cristian and {Bakos}, Gaspar and {Banaszkiewicz}, Marek and {Barlow}, Mike and {Batista}, Virginie and {Bellucci}, Giancarlo and {Benatti}, Serena and {Bernardi}, Pernelle and {B{\'e}zard}, Bruno and {Blecka}, Maria and {Bolmont}, Emeline and {Bonfond}, Bertrand and {Bonito}, Rosaria and {Bonomo}, Aldo S. and {Brucato}, John Robert and {Brun}, Allan Sacha and {Bryson}, Ian and {Bujwan}, Waldemar and {Casewell}, Sarah and {Charnay}, Bejamin and {Pestellini}, Cesare Cecchi and {Chen}, Guo and {Ciaravella}, Angela and {Claudi}, Riccardo and {Cl{\'e}dassou}, Rodolphe and {Damasso}, Mario and {Damiano}, Mario and {Danielski}, Camilla and {Deroo}, Pieter and {Di Giorgio}, Anna Maria and {Dominik}, Carsten and {Doublier}, Vanessa and {Doyle}, Simon and {Doyon}, Ren{\'e} and {Drummond}, Benjamin and {Duong}, Bastien and {Eales}, Stephen and {Edwards}, Billy and {Farina}, Maria and {Flaccomio}, Ettore and {Fletcher}, Leigh and {Forget}, Fran{\c{c}}ois and {Fossey}, Steve and {Fr{\"a}nz}, Markus and {Fujii}, Yuka and {Garc{\'\i}a-Piquer}, {\'A}lvaro and {Gear}, Walter and {Geoffray}, Herv{\'e} and {G{\'e}rard}, Jean Claude and {Gesa}, Lluis and {Gomez}, H. and {Graczyk}, Rafa{\l} and {Griffith}, Caitlin and {Grodent}, Denis and {Guarcello}, Mario Giuseppe and {Gustin}, Jacques and {Hamano}, Keiko and {Hargrave}, Peter and {Hello}, Yann and {Heng}, Kevin and {Herrero}, Enrique and {Hornstrup}, Allan and {Hubert}, Benoit and {Ida}, Shigeru and {Ikoma}, Masahiro and {Iro}, Nicolas and {Irwin}, Patrick and {Jarchow}, Christopher and {Jaubert}, Jean and {Jones}, Hugh and {Julien}, Queyrel and {Kameda}, Shingo and {Kerschbaum}, Franz and {Kervella}, Pierre and {Koskinen}, Tommi and {Krijger}, Matthijs and {Krupp}, Norbert and {Lafarga}, Marina and {Landini}, Federico and {Lellouch}, Emanuel and {Leto}, Giuseppe and {Luntzer}, A. and {Rank-L{\"u}ftinger}, Theresa and {Maggio}, Antonio and {Maldonado}, Jesus and {Maillard}, Jean-Pierre and {Mall}, Urs and {Marquette}, Jean-Baptiste and {Mathis}, Stephane and {Maxted}, Pierre and {Matsuo}, Taro and {Medvedev}, Alexander and {Miguel}, Yamila and {Minier}, Vincent and {Morello}, Giuseppe and {Mura}, Alessandro and {Narita}, Norio and {Nascimbeni}, Valerio and {Nguyen Tong}, N. and {Noce}, Vladimiro and {Oliva}, Fabrizio and {Palle}, Enric and {Palmer}, Paul and {Pancrazzi}, Maurizio and {Papageorgiou}, Andreas and {Parmentier}, Vivien and {Perger}, Manuel and {Petralia}, Antonino and {Pezzuto}, Stefano and {Pierrehumbert}, Ray and {Pillitteri}, Ignazio},
        title = "{A chemical survey of exoplanets with ARIEL}",
      journal = {Experimental Astronomy},
         year = 2018,
        month = nov,
       volume = {46},
       number = {1},
        pages = {135-209},
          doi = {10.1007/s10686-018-9598-x},
       adsurl = {https://ui.adsabs.harvard.edu/abs/2018ExA....46..135T}
}

@ARTICLE{Nascimbeni2025AA...694A.313N,
       author = {{Nascimbeni}, V. and {Piotto}, G. and {Cabrera}, J. and {Montalto}, M. and {Marinoni}, S. and {Marrese}, P.~M. and {Aerts}, C. and {Altavilla}, G. and {Benatti}, S. and {B{\"o}rner}, A. and {Deleuil}, M. and {Desidera}, S. and {Gizon}, L. and {Goupil}, M.~J. and {Granata}, V. and {Heras}, A.~M. and {Magrin}, D. and {Malavolta}, L. and {Mas-Hesse}, J.~M. and {Osborn}, H.~P. and {Pagano}, I. and {Paproth}, C. and {Pollacco}, D. and {Prisinzano}, L. and {Ragazzoni}, R. and {Ramsay}, G. and {Rauer}, H. and {Tkachenko}, A. and {Udry}, S.},
        title = "{The PLATO field selection process: II. Characterization of LOPS2, the first long-pointing field}",
      journal = {\aap},
         year = 2025,
        month = feb,
       volume = {694},
          eid = {A313},
        pages = {A313},
          doi = {10.1051/0004-6361/202452325},
archivePrefix = {arXiv},
       eprint = {2501.07687},
 primaryClass = {astro-ph.EP},
       adsurl = {https://ui.adsabs.harvard.edu/abs/2025AA...694A.313N}
}

@ARTICLE{Gardner2023PASP..135f8001G,
       author = {{Gardner}, Jonathan P. and {Mather}, John C. and {Abbott}, Randy and {Abell}, James S. and {Abernathy}, Mark and {Abney}, Faith E. and {Abraham}, John G. and {Abraham}, Roberto and {Abul-Huda}, Yasin M. and {Acton}, Scott and {Adams}, Cynthia K. and {Adams}, Evan and {Adler}, David S. and {Adriaensen}, Maarten and {Aguilar}, Jonathan Albert and {Ahmed}, Mansoor and {Ahmed}, Nasif S. and {Ahmed}, Tanjira and {Albat}, R{\"u}deger and {Albert}, Lo{\"\i}c and {Alberts}, Stacey and {Aldridge}, David and {Allen}, Mary Marsha and {Allen}, Shaune S. and {Altenburg}, Martin and {Altunc}, Serhat and {Alvarez}, Jose Lorenzo and {{\'A}lvarez-M{\'a}rquez}, Javier and {Alves de Oliveira}, Catarina and {Ambrose}, Leslie L. and {Anandakrishnan}, Satya M. and {Andersen}, Gregory C. and {Anderson}, Harry James and {Anderson}, Jay and {Anderson}, Kristen and {Anderson}, Sara M. and {Aprea}, Julio and {Archer}, Benita J. and {Arenberg}, Jonathan W. and {Argyriou}, Ioannis and {Arribas}, Santiago and {Artigau}, {\'E}tienne and {Arvai}, Amanda Rose and {Atcheson}, Paul and {Atkinson}, Charles B. and {Averbukh}, Jesse and {Aymergen}, Cagatay and {Bacinski}, John J. and {Baggett}, Wayne E. and {Bagnasco}, Giorgio and {Baker}, Lynn L. and {Balzano}, Vicki Ann and {Banks}, Kimberly A. and {Baran}, David A. and {Barker}, Elizabeth A. and {Barrett}, Larry K. and {Barringer}, Bruce O. and {Barto}, Allison and {Bast}, William and {Baudoz}, Pierre and {Baum}, Stefi and {Beatty}, Thomas G. and {Beaulieu}, Mathilde and {Bechtold}, Kathryn and {Beck}, Tracy and {Beddard}, Megan M. and {Beichman}, Charles and {Bellagama}, Larry and {Bely}, Pierre and {Berger}, Timothy W. and {Bergeron}, Louis E. and {Bernier}, Antoine-Darveau and {Bertch}, Maria D. and {Beskow}, Charlotte and {Betz}, Laura E. and {Biagetti}, Carl P. and {Birkmann}, Stephan and {Bjorklund}, Kurt F. and {Blackwood}, James D. and {Blazek}, Ronald Paul and {Blossfeld}, Stephen and {Bluth}, Marcel and {Boccaletti}, Anthony and {Boegner}, Jr., Martin E. and {Bohlin}, Ralph C. and {Boia}, John Joseph and {B{\"o}ker}, Torsten and {Bonaventura}, N. and {Bond}, Nicholas A. and {Bosley}, Kari Ann and {Boucarut}, Rene A. and {Bouchet}, Patrice and {Bouwman}, Jeroen and {Bower}, Gary and {Bowers}, Ariel S. and {Bowers}, Charles W. and {Boyce}, Leslye A. and {Boyer}, Christine T. and {Boyer}, Martha L. and {Boyer}, Michael and {Boyer}, Robert and {Bradley}, Larry D. and {Brady}, Gregory R. and {Brandl}, Bernhard R. and {Brannen}, Judith L. and {Breda}, David and {Bremmer}, Harold G. and {Brennan}, David and {Bresnahan}, Pamela A. and {Bright}, Stacey N. and {Broiles}, Brian J. and {Bromenschenkel}, Asa and {Brooks}, Brian H. and {Brooks}, Keira J. and {Brown}, Bob and {Brown}, Bruce and {Brown}, Thomas M. and {Bruce}, Barry W. and {Bryson}, Jonathan G. and {Bujanda}, Edwin D. and {Bullock}, Blake M. and {Bunker}, A.~J. and {Bureo}, Rafael and {Burt}, Irving J. and {Bush}, James Aaron and {Bushouse}, Howard A. and {Bussman}, Marie C. and {Cabaud}, Olivier and {Cale}, Steven and {Calhoon}, Charles D. and {Calvani}, Humberto and {Canipe}, Alicia M. and {Caputo}, Francis M. and {Cara}, Mihai and {Carey}, Larkin and {Case}, Michael Eli and {Cesari}, Thaddeus and {Cetorelli}, Lee D. and {Chance}, Don R. and {Chandler}, Lynn and {Chaney}, Dave and {Chapman}, George N. and {Charlot}, S. and {Chayer}, Pierre and {Cheezum}, Jeffrey I. and {Chen}, Bin and {Chen}, Christine H. and {Cherinka}, Brian and {Chichester}, Sarah C. and {Chilton}, Zachary S. and {Chittiraibalan}, Dharini and {Clampin}, Mark and {Clark}, Charles R. and {Clark}, Kerry W. and {Clark}, Stephanie M. and {Claybrooks}, Edward E. and {Cleveland}, Keith A. and {Cohen}, Andrew L. and {Cohen}, Lester M. and {Col{\'o}n}, Knicole D. and {Coleman}, Benee L. and {Colina}, Luis and {Comber}, Brian J. and {Comeau}, Thomas M. and {Comer}, Thomas and {Conde Reis}, Alain and {Connolly}, Dennis C. and {Conroy}, Kyle E. and {Contos}, Adam R. and {Contreras}, James and {Cook}, Neil J. and {Cooper}, James L. and {Cooper}, Rachel Aviva and {Correia}, Michael F. and {Correnti}, Matteo and {Cossou}, Christophe and {Costanza}, Brian F. and {Coulais}, Alain and {Cox}, Colin R. and {Coyle}, Ray T. and {Cracraft}, Misty M. and {Crew}, Keith A. and {Curtis}, Gary J. and {Cusveller}, Bianca and {Da Costa Maciel}, Cleyciane and {Dailey}, Christopher T. and {Daugeron}, Fr{\'e}d{\'e}ric and {Davidson}, Greg S. and {Davies}, James E. and {Davis}, Katherine Anne and {Davis}, Michael S. and {Day}, Ratna and {de Chambure}, Daniel and {de Jong}, Pauline and {De Marchi}, Guido and {Dean}, Bruce H. and {Decker}, John E. and {Delisa}, Amy S. and {Dell}, Lawrence C. and {Dellagatta}, Gail},
        title = "{The James Webb Space Telescope Mission}",
      journal = {\pasp},
         year = 2023,
        month = jun,
       volume = {135},
       number = {1048},
          eid = {068001},
        pages = {068001},
          doi = {10.1088/1538-3873/acd1b5},
archivePrefix = {arXiv},
       eprint = {2304.04869},
 primaryClass = {astro-ph.IM},
       adsurl = {https://ui.adsabs.harvard.edu/abs/2023PASP..135f8001G}
}

@ARTICLE{Wheatley2018MNRAS.475.4476W,
       author = {{Wheatley}, Peter J. and {West}, Richard G. and {Goad}, Michael R. and {Jenkins}, James S. and {Pollacco}, Don L. and {Queloz}, Didier and {Rauer}, Heike and {Udry}, St{\'e}phane and {Watson}, Christopher A. and {Chazelas}, Bruno and {Eigm{\"u}ller}, Philipp and {Lambert}, Gregory and {Genolet}, Ludovic and {McCormac}, James and {Walker}, Simon and {Armstrong}, David J. and {Bayliss}, Daniel and {Bento}, Joao and {Bouchy}, Fran{\c{c}}ois and {Burleigh}, Matthew R. and {Cabrera}, Juan and {Casewell}, Sarah L. and {Chaushev}, Alexander and {Chote}, Paul and {Csizmadia}, Szil{\'a}rd and {Erikson}, Anders and {Faedi}, Francesca and {Foxell}, Emma and {G{\"a}nsicke}, Boris T. and {Gillen}, Edward and {Grange}, Andrew and {G{\"u}nther}, Maximilian N. and {Hodgkin}, Simon T. and {Jackman}, James and {Jord{\'a}n}, Andr{\'e}s and {Louden}, Tom and {Metrailler}, Lionel and {Moyano}, Maximiliano and {Nielsen}, Louise D. and {Osborn}, Hugh P. and {Poppenhaeger}, Katja and {Raddi}, Roberto and {Raynard}, Liam and {Smith}, Alexis M.~S. and {Soto}, Maritza and {Titz-Weider}, Ruth},
        title = "{The Next Generation Transit Survey (NGTS)}",
      journal = {\mnras},
         year = 2018,
        month = apr,
       volume = {475},
       number = {4},
        pages = {4476-4493},
          doi = {10.1093/mnras/stx2836},
archivePrefix = {arXiv},
       eprint = {1710.11100},
 primaryClass = {astro-ph.EP},
       adsurl = {https://ui.adsabs.harvard.edu/abs/2018MNRAS.475.4476W}
}

@article{narita2019muscat2,
  title={MuSCAT2: four-color simultaneous camera for the 1.52-m Telescopio Carlos S{\'a}nchez},
  author={Narita, Norio and Fukui, Akihiko and Kusakabe, Nobuhiko and Watanabe, Noriharu and Palle, Enric and Parviainen, Hannu and Monta{\~n}{\'e}s-Rodr{\'\i}guez, Pilar and Murgas, Felipe and Monelli, Matteo and Aguiar, Marta and others},
  journal={Journal of Astronomical Telescopes, Instruments, and Systems},
  volume={5},
  number={1},
  pages={015001--015001},
  year={2019},
  publisher={Society of Photo-Optical Instrumentation Engineers}
}

@INPROCEEDINGS{Narita2020SPIE11447E..5KN,
       author = {{Narita}, Norio and {Fukui}, Akihiko and {Yamamuro}, Tomoyasu and {Harbeck}, Daniel and {Bowman}, Mark and {Elphick}, Mark and {Nation}, Jon and {Armstrong}, J.~D. and {Han}, Jacqueline and {Abe}, Shunichi and {Ikoma}, Masahiro and {Isogai}, Keisuke and {Kawauchi}, Kiyoe and {Kurita}, Seiya and {Kusakabe}, Nobuhiko and {de Leon}, Jerome and {Livingston}, John and {Mori}, Mayuko and {Nishiumi}, Taku and {Tamura}, Motohide and {Watanabe}, Noriharu and {Volgenau}, Nikolaus and {Heinrich-Josties}, Elisabeth and {Foale}, Steve and {Daily}, Matt and {McCully}, Curtis and {Kirby}, Annie and {Smith}, Cary and {Haworth}, Brian and {Conway}, Patrick and {Storrie-Lombardi}, Lisa and {Rosing}, Wayne and {Chatelain}, Joey and {Bachelet}, Etienne and {Johnson}, Marshall and {Rabus}, Markus},
        title = "{MuSCAT3: a 4-color simultaneous camera for the 2m Faulkes Telescope North}",
    booktitle = {Ground-based and Airborne Instrumentation for Astronomy VIII},
         year = 2020,
       editor = {{Evans}, Christopher J. and {Bryant}, Julia J. and {Motohara}, Kentaro},
       series = {Society of Photo-Optical Instrumentation Engineers (SPIE) Conference Series},
       volume = {11447},
        month = dec,
          eid = {114475K},
        pages = {114475K},
          doi = {10.1117/12.2559947},
       adsurl = {https://ui.adsabs.harvard.edu/abs/2020SPIE11447E..5KN}
}

@ARTICLE{Guillot2015AN....336..638G,
       author = {{Guillot}, T. and {Abe}, L. and {Agabi}, A. and {Rivet}, J. -P. and {Daban}, J. -B. and {M{\'e}karnia}, D. and {Aristidi}, E. and {Schmider}, F. -X. and {Crouzet}, N. and {Gon{\c{c}}alves}, I. and {Gouvret}, C. and {Ottogalli}, S. and {Faradji}, H. and {Blanc}, P. -E. and {Bondoux}, E. and {Valbousquet}, F.},
        title = "{Thermalizing a telescope in Antarctica - analysis of ASTEP observations}",
      journal = {Astronomische Nachrichten},
         year = 2015,
        month = sep,
       volume = {336},
       number = {7},
        pages = {638},
          doi = {10.1002/asna.201512174},
archivePrefix = {arXiv},
       eprint = {1506.06009},
 primaryClass = {astro-ph.IM},
       adsurl = {https://ui.adsabs.harvard.edu/abs/2015AN....336..638G}
}

@ARTICLE{Mekarnia2016MNRAS.463...45M,
       author = {{M{\'e}karnia}, D. and {Guillot}, T. and {Rivet}, J. -P. and {Schmider}, F. -X. and {Abe}, L. and {Gon{\c{c}}alves}, I. and {Agabi}, A. and {Crouzet}, N. and {Fruth}, T. and {Barbieri}, M. and {Bayliss}, D.~D.~R. and {Zhou}, G. and {Aristidi}, E. and {Szulagyi}, J. and {Daban}, J. -B. and {Fante{\"\i}-Caujolle}, Y. and {Gouvret}, C. and {Erikson}, A. and {Rauer}, H. and {Bouchy}, F. and {Gerakis}, J. and {Bouchez}, G.},
        title = "{Transiting planet candidates with ASTEP 400 at Dome C, Antarctica}",
      journal = {\mnras},
         year = 2016,
        month = nov,
       volume = {463},
       number = {1},
        pages = {45-62},
          doi = {10.1093/mnras/stw1934},
       adsurl = {https://ui.adsabs.harvard.edu/abs/2016MNRAS.463...45M}
}

@ARTICLE{Nascimbeni2011AA...527A..85N,
       author = {{Nascimbeni}, V. and {Piotto}, G. and {Bedin}, L.~R. and {Damasso}, M.},
        title = "{TASTE: The Asiago Search for Transit timing variations of Exoplanets. I.  Overview and improved parameters for HAT-P-3b and HAT-P-14b}",
      journal = {\aap},
         year = 2011,
        month = mar,
       volume = {527},
          eid = {A85},
        pages = {A85},
          doi = {10.1051/0004-6361/201015199},
archivePrefix = {arXiv},
       eprint = {1011.6395},
 primaryClass = {astro-ph.EP},
       adsurl = {https://ui.adsabs.harvard.edu/abs/2011AA...527A..85N}
}

@ARTICLE{Speagle2020MNRAS.493.3132S,
       author = {{Speagle}, Joshua S.},
        title = "{DYNESTY: a dynamic nested sampling package for estimating Bayesian posteriors and evidences}",
      journal = {\mnras},
         year = 2020,
        month = apr,
       volume = {493},
       number = {3},
        pages = {3132-3158},
          doi = {10.1093/mnras/staa278},
archivePrefix = {arXiv},
       eprint = {1904.02180},
 primaryClass = {astro-ph.IM},
       adsurl = {https://ui.adsabs.harvard.edu/abs/2020MNRAS.493.3132S}
}

@article{Skilling10.1214/06-BA127,
    author = {John Skilling},
    title = {{Nested sampling for general Bayesian computation}},
    volume = {1},
    journal = {Bayesian Analysis},
    number = {4},
    publisher = {International Society for Bayesian Analysis},
    pages = {833 -- 859},
    year = {2006},
    doi = {10.1214/06-BA127},
    URL = {https://doi.org/10.1214/06-BA127}
}

@INPROCEEDINGS{Skilling2004AIPC..735..395S,
       author = {{Skilling}, John},
        title = "{Nested Sampling}",
    booktitle = {Bayesian Inference and Maximum Entropy Methods in Science and Engineering: 24th International Workshop on Bayesian Inference and Maximum Entropy Methods in Science and Engineering},
         year = 2004,
       editor = {{Fischer}, Rainer and {Preuss}, Roland and {Toussaint}, Udo Von},
       series = {American Institute of Physics Conference Series},
       volume = {735},
        month = nov,
    publisher = {AIP},
        pages = {395-405},
          doi = {10.1063/1.1835238},
       adsurl = {https://ui.adsabs.harvard.edu/abs/2004AIPC..735..395S}
}

@ARTICLE{Buchner2023StSur..17..169B,
       author = {{Buchner}, Johannes},
        title = "{Nested Sampling Methods}",
      journal = {Statistics Surveys},
         year = 2023,
        month = jan,
       volume = {17},
        pages = {169-215},
          doi = {10.1214/23-SS144},
archivePrefix = {arXiv},
       eprint = {2101.09675},
 primaryClass = {stat.CO},
       adsurl = {https://ui.adsabs.harvard.edu/abs/2023StSur..17..169B}
}

@article{nautilus,
    author = {Lange, Johannes U},
    title = "{nautilus: boosting Bayesian importance nested sampling with deep learning}",
    journal = {Monthly Notices of the Royal Astronomical Society},
    volume = {525},
    number = {2},
    pages = {3181-3194},
    year = {2023},
    month = {08},
    doi = {10.1093/mnras/stad2441},
    url = {https://doi.org/10.1093/mnras/stad2441},
    eprint = {https://academic.oup.com/mnras/article-pdf/525/2/3181/51331635/stad2441.pdf},
}

@ARTICLE{Gladman1993Icar..106..247G,
       author = {{Gladman}, Brett},
        title = "{Dynamics of Systems of Two Close Planets}",
      journal = {\icarus},
         year = 1993,
        month = nov,
       volume = {106},
       number = {1},
        pages = {247-263},
          doi = {10.1006/icar.1993.1169},
       adsurl = {https://ui.adsabs.harvard.edu/abs/1993Icar..106..247G}
}

@ARTICLE{Laskar1997AA...317L..75L,
       author = {{Laskar}, J.},
        title = "{Large scale chaos and the spacing of the inner planets.}",
      journal = {\aap},
         year = 1997,
        month = jan,
       volume = {317},
        pages = {L75-L78},
       adsurl = {https://ui.adsabs.harvard.edu/abs/1997AA...317L..75L}
}

@ARTICLE{Laskar2000PhRvL..84.3240L,
       author = {{Laskar}, J.},
        title = "{On the Spacing of Planetary Systems}",
      journal = {\prl},
         year = 2000,
        month = apr,
       volume = {84},
       number = {15},
        pages = {3240-3243},
          doi = {10.1103/PhysRevLett.84.3240},
       adsurl = {https://ui.adsabs.harvard.edu/abs/2000PhRvL..84.3240L}
}

@ARTICLE{LaskarPetit2017AA...605A..72L,
       author = {{Laskar}, J. and {Petit}, A.~C.},
        title = "{AMD-stability and the classification of planetary systems}",
      journal = {\aap},
         year = 2017,
        month = sep,
       volume = {605},
          eid = {A72},
        pages = {A72},
          doi = {10.1051/0004-6361/201630022},
archivePrefix = {arXiv},
       eprint = {1703.07125},
 primaryClass = {astro-ph.EP},
       adsurl = {https://ui.adsabs.harvard.edu/abs/2017AA...605A..72L}
}

@ARTICLE{Petit2018AA...617A..93P,
       author = {{Petit}, Antoine C. and {Laskar}, Jacques and {Bou{\'e}}, Gwena{\"e}l},
        title = "{Hill stability in the AMD framework}",
      journal = {\aap},
         year = 2018,
        month = sep,
       volume = {617},
          eid = {A93},
        pages = {A93},
          doi = {10.1051/0004-6361/201833088},
archivePrefix = {arXiv},
       eprint = {1806.08869},
 primaryClass = {astro-ph.EP},
       adsurl = {https://ui.adsabs.harvard.edu/abs/2018AA...617A..93P}
}

@ARTICLE{Chambers1999MNRAS.304..793C,
       author = {{Chambers}, J.~E.},
        title = "{A hybrid symplectic integrator that permits close encounters between massive bodies}",
      journal = {\mnras},
         year = 1999,
        month = apr,
       volume = {304},
       number = {4},
        pages = {793-799},
          doi = {10.1046/j.1365-8711.1999.02379.x},
       adsurl = {https://ui.adsabs.harvard.edu/abs/1999MNRAS.304..793C}
}

@book{NR-Press1996,
 author = {Press, William H. and Teukolsky, Saul A. and Vetterling, William T. and Flannery, Brian P.},
 title = {Numerical recipes in Fortran 90 (2nd ed.): the art of parallel scientific computing},
 year = {1996},
 isbn = {0-521-57439-0},
 source = {diskette for IBM PC 3.5, v 2.06, \$39.95, ISBN 0-521-57440-4; Unix CD-ROM, v 2.06, \$149.95, ISBN 0-521-57607-5; Windows, DOS, Macintosh CD-ROM, v 2.06, \$89.95, ISBN 0-521-57608-3},
 publisher = {Cambridge University Press},
 address = {New York, NY, USA},
}

@ARTICLE{rebound,
       author = {{Rein}, H. and {Liu}, S. -F.},
        title = "{REBOUND: an open-source multi-purpose N-body code for collisional dynamics}",
      journal = {\aap},
         year = 2012,
        month = jan,
       volume = {537},
          eid = {A128},
        pages = {A128},
          doi = {10.1051/0004-6361/201118085},
archivePrefix = {arXiv},
       eprint = {1110.4876},
 primaryClass = {astro-ph.EP},
       adsurl = {https://ui.adsabs.harvard.edu/abs/2012AA...537A.128R}
}

@ARTICLE{reboundwhfast,
       author = {{Rein}, Hanno and {Tamayo}, Daniel},
        title = "{WHFAST: a fast and unbiased implementation of a symplectic Wisdom-Holman integrator for long-term gravitational simulations}",
      journal = {\mnras},
         year = 2015,
        month = sep,
       volume = {452},
       number = {1},
        pages = {376-388},
          doi = {10.1093/mnras/stv1257},
archivePrefix = {arXiv},
       eprint = {1506.01084},
 primaryClass = {astro-ph.EP},
       adsurl = {https://ui.adsabs.harvard.edu/abs/2015MNRAS.452..376R}
}

@ARTICLE{Baran2015AA...577A.146B,
       author = {{Baran}, A.~S. and {Zola}, S. and {Blokesz}, A. and {{\O}stensen}, R.~H. and {Silvotti}, R.},
        title = "{Detection of a planet in the sdB + M dwarf binary system 2M 1938+4603}",
      journal = {\aap},
         year = 2015,
        month = may,
       volume = {577},
          eid = {A146},
        pages = {A146},
          doi = {10.1051/0004-6361/201425392},
       adsurl = {https://ui.adsabs.harvard.edu/abs/2015A&A...577A.146B}
}

@ARTICLE{Potter2011MNRAS.416.2202P,
       author = {{Potter}, Stephen B. and {Romero-Colmenero}, Encarni and {Ramsay}, Gavin and {Crawford}, Steven and {Gulbis}, Amanda and {Barway}, Sudhanshu and {Zietsman}, Ewald and {Kotze}, Marissa and {Buckley}, David A.~H. and {O'Donoghue}, Darragh and {Siegmund}, O.~H.~W. and {McPhate}, J. and {Welsh}, B.~Y. and {Vallerga}, John},
        title = "{Possible detection of two giant extrasolar planets orbiting the eclipsing polar UZ Fornacis}",
      journal = {\mnras},
         year = 2011,
        month = sep,
       volume = {416},
       number = {3},
        pages = {2202-2211},
          doi = {10.1111/j.1365-2966.2011.19198.x},
archivePrefix = {arXiv},
       eprint = {1106.1404},
 primaryClass = {astro-ph.SR},
       adsurl = {https://ui.adsabs.harvard.edu/abs/2011MNRAS.416.2202P}
}

@ARTICLE{Pulley2022MNRAS.514.5725P,
       author = {{Pulley}, D. and {Sharp}, I.~D. and {Mallett}, J. and {von Harrach}, S.},
        title = "{Eclipse timing variations in post-common envelope binaries: Are they a reliable indicator of circumbinary companions?}",
      journal = {\mnras},
         year = 2022,
        month = aug,
       volume = {514},
       number = {4},
        pages = {5725-5738},
          doi = {10.1093/mnras/stac1676},
archivePrefix = {arXiv},
       eprint = {2206.06919},
 primaryClass = {astro-ph.SR},
       adsurl = {https://ui.adsabs.harvard.edu/abs/2022MNRAS.514.5725P}
}

@ARTICLE{Chambers2001Icar..152..205C,
       author = {{Chambers}, J.~E.},
        title = "{Making More Terrestrial Planets}",
      journal = {\icarus},
         year = 2001,
        month = aug,
       volume = {152},
       number = {2},
        pages = {205-224},
          doi = {10.1006/icar.2001.6639},
       adsurl = {https://ui.adsabs.harvard.edu/abs/2001Icar..152..205C}
}

@ARTICLE{Turrini2020AA...636A..53T,
       author = {{Turrini}, D. and {Zinzi}, A. and {Belinchon}, J.~A.},
        title = "{Normalized angular momentum deficit: a tool for comparing the violence of the dynamical histories of planetary systems}",
      journal = {\aap},
         year = 2020,
        month = apr,
       volume = {636},
          eid = {A53},
        pages = {A53},
          doi = {10.1051/0004-6361/201936301},
archivePrefix = {arXiv},
       eprint = {2003.05366},
 primaryClass = {astro-ph.EP},
       adsurl = {https://ui.adsabs.harvard.edu/abs/2020A&A...636A..53T}
}

@ARTICLE{2023AA...677A..33B,
       author = {{Bonomo}, A.~S. and {Dumusque}, X. and {Massa}, A. and {Mortier}, A. and {Bongiolatti}, R. and {Malavolta}, L. and {Sozzetti}, A. and {Buchhave}, L.~A. and {Damasso}, M. and {Haywood}, R.~D. and {Morbidelli}, A. and {Latham}, D.~W. and {Molinari}, E. and {Pepe}, F. and {Poretti}, E. and {Udry}, S. and {Affer}, L. and {Boschin}, W. and {Charbonneau}, D. and {Cosentino}, R. and {Cretignier}, M. and {Ghedina}, A. and {Lega}, E. and {L{\'o}pez-Morales}, M. and {Margini}, M. and {Mart{\'\i}nez Fiorenzano}, A.~F. and {Mayor}, M. and {Micela}, G. and {Pedani}, M. and {Pinamonti}, M. and {Rice}, K. and {Sasselov}, D. and {Tronsgaard}, R. and {Vanderburg}, A.},
        title = "{Cold Jupiters and improved masses in 38 Kepler and K2 small planet systems from 3661 HARPS-N radial velocities. No excess of cold Jupiters in small planet systems}",
      journal = {\aap},
         year = 2023,
        month = sep,
       volume = {677},
          eid = {A33},
        pages = {A33},
          doi = {10.1051/0004-6361/202346211},
archivePrefix = {arXiv},
       eprint = {2304.05773},
 primaryClass = {astro-ph.EP},
       adsurl = {https://ui.adsabs.harvard.edu/abs/2023AA...677A..33B}
}

@ARTICLE{2025AJ....170..146S,
       author = {{Shaw}, David E. and {Weiss}, Lauren M. and {Agol}, Eric and {Collins}, Karen A. and {Barkaoui}, Khalid and {Watkins}, Cristilyn N. and {Schwarz}, Richard P. and {Relles}, Howard M. and {Stockdale}, Chris and {Kielkopf}, John F. and {Rodriguez Frustaglia}, Fabian and {Bieryla}, Allyson and {Gregorio}, Joao and {Mitchem}, Owen and {Linnenkohl}, Katherine and {Popowicz}, Adam and {Narita}, Norio and {Fukui}, Akihiko and {Gillon}, Micha{\"e}l and {Sefako}, Ramotholo and {Shporer}, Avi and {Lark}, Adam and {Heying}, Amelie and {Khan}, Isa and {Chen}, Beibei and {Carden}, Kylee and {Terndrup}, Donald M. and {Taylor}, Robert and {Crocker}, Dasha and {Ballard}, Sarah and {Fabrycky}, Daniel C.},
        title = "{Updated Masses for the Gas Giants in the Eight-planet Kepler-90 System Via Transit-timing Variation and Radial Velocity Observations}",
      journal = {\aj},
         year = 2025,
        month = sep,
       volume = {170},
       number = {3},
          eid = {146},
        pages = {146},
          doi = {10.3847/1538-3881/ade67b},
archivePrefix = {arXiv},
       eprint = {2507.13588},
 primaryClass = {astro-ph.EP},
       adsurl = {https://ui.adsabs.harvard.edu/abs/2025AJ....170..146S}
}

@ARTICLE{2014ApJ...784...44L,
       author = {{Lissauer}, Jack J. and {Marcy}, Geoffrey W. and {Bryson}, Stephen T. and {Rowe}, Jason F. and {Jontof-Hutter}, Daniel and {Agol}, Eric and {Borucki}, William J. and {Carter}, Joshua A. and {Ford}, Eric B. and {Gilliland}, Ronald L. and et al.},
        title = "{Validation of Kepler's Multiple Planet Candidates. II. Refined Statistical Framework and Descriptions of Systems of Special Interest}",
      journal = {\apj},
         year = 2014,
        month = mar,
       volume = {784},
       number = {1},
          eid = {44},
        pages = {44},
          doi = {10.1088/0004-637X/784/1/44},
archivePrefix = {arXiv},
       eprint = {1402.6352},
 primaryClass = {astro-ph.EP},
       adsurl = {https://ui.adsabs.harvard.edu/abs/2014ApJ...784...44L}
}

@ARTICLE{2019AJ....158...32K,
       author = {{Kostov}, Veselin B. and {Schlieder}, Joshua E. and {Barclay}, Thomas and {Quintana}, Elisa V. and {Col{\'o}n}, Knicole D. and {Brande}, Jonathan and {Collins}, Karen A. and {Feinstein}, Adina D. and {Hadden}, Samuel and {Kane}, Stephen R. and {Kreidberg}, Laura and {Kruse}, Ethan and {Lam}, Christopher and {Matthews}, Elisabeth and {Montet}, Benjamin T. and {Pozuelos}, Francisco J. and {Stassun}, Keivan G. and {Winters}, Jennifer G. and {Ricker}, George and {Vanderspek}, Roland and {Latham}, David and {Seager}, Sara and {Winn}, Joshua and {Jenkins}, Jon M. and {Afanasev}, Dennis and {Armstrong}, James J.~D. and {Arney}, Giada and {Boyd}, Patricia and {Barentsen}, Geert and {Barkaoui}, Khalid and {Batalha}, Natalie E. and {Beichman}, Charles and {Bayliss}, Daniel and {Burke}, Christopher and {Burdanov}, Artem and {Cacciapuoti}, Luca and {Carson}, Andrew and {Charbonneau}, David and {Christiansen}, Jessie and {Ciardi}, David and {Clampin}, Mark and {Collins}, Kevin I. and {Conti}, Dennis M. and {Coughlin}, Jeffrey and {Covone}, Giovanni and {Crossfield}, Ian and {Delrez}, Laetitia and {Domagal-Goldman}, Shawn and {Dressing}, Courtney and {Ducrot}, Elsa and {Essack}, Zahra and {Everett}, Mark E. and {Fauchez}, Thomas and {Foreman-Mackey}, Daniel and {Gan}, Tianjun and {Gilbert}, Emily and {Gillon}, Micha{\"e}l and {Gonzales}, Erica and {Hamann}, Aaron and {Hedges}, Christina and {Hocutt}, Hannah and {Hoffman}, Kelsey and {Horch}, Elliott P. and {Horne}, Keith and {Howell}, Steve and {Hynes}, Shane and {Ireland}, Michael and {Irwin}, Jonathan M. and {Isopi}, Giovanni and {Jensen}, Eric L.~N. and {Jehin}, Emmanu{\"e}l and {Kaltenegger}, Lisa and {Kielkopf}, John F. and {Kopparapu}, Ravi and {Lewis}, Nikole and {Lopez}, Eric and {Lissauer}, Jack J. and {Mann}, Andrew W. and {Mallia}, Franco and {Mandell}, Avi and {Matson}, Rachel A. and {Mazeh}, Tsevi and {Monsue}, Teresa and {Moran}, Sarah E. and {Moran}, Vickie and {Morley}, Caroline V. and {Morris}, Brett and {Muirhead}, Philip and {Mukai}, Koji and {Mullally}, Susan and {Mullally}, Fergal and {Murray}, Catriona and {Narita}, Norio and {Palle}, Enric and {Pidhorodetska}, Daria and {Quinn}, David and {Relles}, Howard and {Rinehart}, Stephen and {Ritsko}, Matthew and {Rodriguez}, Joseph E. and {Rowden}, Pamela and {Rowe}, Jason F. and {Sebastian}, Daniel and {Sefako}, Ramotholo and {Shahaf}, Sahar and {Shporer}, Avi and {Ta{\~n}{\'o}n Reyes}, Naylynn and {Tenenbaum}, Peter and {Ting}, Eric B. and {Twicken}, Joseph D. and {van Belle}, Gerard T. and {Vega}, Laura and {Volosin}, Jeffrey and {Walkowicz}, Lucianne M. and {Youngblood}, Allison},
        title = "{The L 98-59 System: Three Transiting, Terrestrial-size Planets Orbiting a Nearby M Dwarf}",
      journal = {\aj},
         year = 2019,
        month = jul,
       volume = {158},
       number = {1},
          eid = {32},
        pages = {32},
          doi = {10.3847/1538-3881/ab2459},
archivePrefix = {arXiv},
       eprint = {1903.08017},
 primaryClass = {astro-ph.EP},
       adsurl = {https://ui.adsabs.harvard.edu/abs/2019AJ....158...32K}
}

@ARTICLE{2025AJ....170..154C,
       author = {{Cadieux}, Charles and {L'Heureux}, Alexandrine and {Piaulet-Ghorayeb}, Caroline and {Doyon}, Ren{\'e} and {Artigau}, {\'E}tienne and {Cook}, Neil J. and {Coulombe}, Louis-Philippe and {Roy}, Pierre-Alexis and {Lafreni{\`e}re}, David and {Lamontagne}, Pierrot and {Radica}, Michael and {Benneke}, Bj{\"o}rn and {Ahrer}, Eva-Maria and {Weisserman}, Drew and {Cloutier}, Ryan},
        title = "{Detailed Architecture of the L 98-59 System and Confirmation of a Fifth Planet in the Habitable Zone}",
      journal = {\aj},
         year = 2025,
        month = sep,
       volume = {170},
       number = {3},
          eid = {154},
        pages = {154},
          doi = {10.3847/1538-3881/adef59},
archivePrefix = {arXiv},
       eprint = {2507.09343},
 primaryClass = {astro-ph.EP},
       adsurl = {https://ui.adsabs.harvard.edu/abs/2025AJ....170..154C}
}

@ARTICLE{2012MNRAS.420.2580S,
       author = {{Southworth}, John and {Bruni}, I. and {Mancini}, L. and {Gregorio}, J.},
        title = "{Refined physical properties of the HAT-P-13 planetary system}",
      journal = {\mnras},
         year = 2012,
        month = mar,
       volume = {420},
       number = {3},
        pages = {2580-2587},
          doi = {10.1111/j.1365-2966.2011.20230.x},
archivePrefix = {arXiv},
       eprint = {1111.5432},
 primaryClass = {astro-ph.EP},
       adsurl = {https://ui.adsabs.harvard.edu/abs/2012MNRAS.420.2580S}
}

@ARTICLE{2008ApJ...680.1450P,
       author = {{P{\'a}l}, A. and {Bakos}, G. {\'A}. and {Torres}, G. and {Noyes}, R.~W. and {Latham}, D.~W. and {Kov{\'a}cs}, G{\'e}za and {Marcy}, G.~W. and {Fischer}, D.~A. and {Butler}, R.~P. and {Sasselov}, D.~D. and et al.},
        title = "{HAT-P-7b: An Extremely Hot Massive Planet Transiting a Bright Star in the Kepler Field}",
      journal = {\apj},
         year = 2008,
        month = jun,
       volume = {680},
       number = {2},
        pages = {1450-1456},
          doi = {10.1086/588010},
archivePrefix = {arXiv},
       eprint = {0803.0746},
 primaryClass = {astro-ph},
       adsurl = {https://ui.adsabs.harvard.edu/abs/2008ApJ...680.1450P}
}

@ARTICLE{2025ApJ...986..117Y,
       author = {{Yang}, Eritas and {Su}, Yubo and {Winn}, Joshua N.},
        title = "{A Third Star in the HAT-P-7 System and a New Dynamical Pathway to Misaligned Hot Jupiters}",
      journal = {\apj},
         year = 2025,
        month = jun,
       volume = {986},
       number = {2},
          eid = {117},
        pages = {117},
          doi = {10.3847/1538-4357/add5f7},
archivePrefix = {arXiv},
       eprint = {2505.07927},
 primaryClass = {astro-ph.EP},
       adsurl = {https://ui.adsabs.harvard.edu/abs/2025ApJ...986..117Y}
}

@ARTICLE{2014AA...572A...2B,
       author = {{Bonomo}, A.~S. and {Sozzetti}, A. and {Lovis}, C. and {Malavolta}, L. and {Rice}, K. and {Buchhave}, L.~A. and {Sasselov}, D. and {Cameron}, A.~C. and {Latham}, D.~W. and {Molinari}, E. and {Pepe}, F. and {Udry}, S. and {Affer}, L. and {Charbonneau}, D. and {Cosentino}, R. and {Dressing}, C.~D. and {Dumusque}, X. and {Figueira}, P. and {Fiorenzano}, A.~F.~M. and {Gettel}, S. and {Harutyunyan}, A. and {Haywood}, R.~D. and {Horne}, K. and {Lopez-Morales}, M. and {Mayor}, M. and {Micela}, G. and {Motalebi}, F. and {Nascimbeni}, V. and {Phillips}, D.~F. and {Piotto}, G. and {Pollacco}, D. and {Queloz}, D. and {S{\'e}gransan}, D. and {Szentgyorgyi}, A. and {Watson}, C.},
        title = "{Characterization of the planetary system Kepler-101 with HARPS-N. A hot super-Neptune with an Earth-sized low-mass companion}",
      journal = {\aap},
         year = 2014,
        month = dec,
       volume = {572},
          eid = {A2},
        pages = {A2},
          doi = {10.1051/0004-6361/201424617},
archivePrefix = {arXiv},
       eprint = {1409.4592},
 primaryClass = {astro-ph.EP},
       adsurl = {https://ui.adsabs.harvard.edu/abs/2014AA...572A...2B}
}

@ARTICLE{2025AJ....169...90O,
       author = {{Ofir}, Aviv and {Yoffe}, Gideon and {Aharonson}, Oded},
        title = "{Planetary Mass Determinations from a Simplified Photodynamical Model{\textemdash}Application to the Complete Kepler Dataset}",
      journal = {\aj},
         year = 2025,
        month = feb,
       volume = {169},
       number = {2},
          eid = {90},
        pages = {90},
          doi = {10.3847/1538-3881/ad91a7},
archivePrefix = {arXiv},
       eprint = {2410.11401},
 primaryClass = {astro-ph.EP},
       adsurl = {https://ui.adsabs.harvard.edu/abs/2025AJ....169...90O}
}

@ARTICLE{2023AA...669A.117L,
       author = {{Leleu}, A. and {Delisle}, J.-B. and {Udry}, S. and {Mardling}, R. and {Turbet}, M. and {Egger}, J.~A. and {Alibert}, Y. and {Chatel}, G. and {Eggenberger}, P. and {Stalport}, M.},
        title = "{Removing biases on the density of sub-Neptunes characterised via transit timing variations. Update on the mass-radius relationship of 34 Kepler planets}",
      journal = {\aap},
         year = 2023,
        month = jan,
       volume = {669},
          eid = {A117},
        pages = {A117},
          doi = {10.1051/0004-6361/202244132},
archivePrefix = {arXiv},
       eprint = {2207.07456},
 primaryClass = {astro-ph.EP},
       adsurl = {https://ui.adsabs.harvard.edu/abs/2023A&A...669A.117L}
}

@ARTICLE{2020AA...634A..29J,
       author = {{Jofr{\'e}}, E. and {Almenara}, J.~M. and {Petrucci}, R. and {D{\'\i}az}, R.~F. and {G{\'o}mez Maqueo Chew}, Y. and {Martioli}, E. and {Ram{\'\i}rez}, I. and {Garc{\'\i}a}, L. and {Saffe}, C. and {Canul}, E.~F. and {Buccino}, A. and {G{\'o}mez}, M. and {Moreno Hilario}, E.},
        title = "{Gemini-GRACES high-quality spectra of Kepler evolved stars with transiting planets. I. Detailed characterization of multi-planet systems Kepler-278 and Kepler-391}",
      journal = {\aap},
         year = 2020,
        month = feb,
       volume = {634},
          eid = {A29},
        pages = {A29},
          doi = {10.1051/0004-6361/201936446},
archivePrefix = {arXiv},
       eprint = {1912.10278},
 primaryClass = {astro-ph.EP},
       adsurl = {https://ui.adsabs.harvard.edu/abs/2020A&A...634A..29J}
}

@ARTICLE{2024ApJS..270....8W,
       author = {{Weiss}, Lauren M. and {Isaacson}, Howard and {Howard}, Andrew W. and {Fulton}, Benjamin J. and {Petigura}, Erik A. and {Fabrycky}, Daniel and {Jontof-Hutter}, Daniel and {Steffen}, Jason H. and {Schlichting}, Hilke E. and {Wright}, Jason T. and {Beard}, Corey and {Brinkman}, Casey L. and {Chontos}, Ashley and {Giacalone}, Steven and {Hill}, Michelle L. and {Kosiarek}, Molly R. and {MacDougall}, Mason G. and {Mo{\v{c}}nik}, Teo and {Polanski}, Alex S. and {Turtelboom}, Emma V. and {Tyler}, Dakotah and {Van Zandt}, Judah},
        title = "{The Kepler Giant Planet Search. I. A Decade of Kepler Planet-host Radial Velocities from W. M. Keck Observatory}",
      journal = {\apjs},
         year = 2024,
        month = jan,
       volume = {270},
       number = {1},
          eid = {8},
        pages = {8},
          doi = {10.3847/1538-4365/ad0cab},
archivePrefix = {arXiv},
       eprint = {2304.00071},
 primaryClass = {astro-ph.EP},
       adsurl = {https://ui.adsabs.harvard.edu/abs/2024ApJS..270....8W}
}

@ARTICLE{2013Sci...342..331H,
       author = {{Huber}, Daniel and {Carter}, Joshua A. and {Barbieri}, Mauro and {Miglio}, Andrea and {Deck}, Katherine M. and {Fabrycky}, Daniel C. and {Montet}, Benjamin T. and {Buchhave}, Lars A. and {Chaplin}, William J. and {Hekker}, Saskia and {Montalb{\'a}n}, Josefina and {Sanchis-Ojeda}, Roberto and {Basu}, Sarbani and {Bedding}, Timothy R. and {Campante}, Tiago L. and {Christensen-Dalsgaard}, J{\o}rgen and {Elsworth}, Yvonne P. and {Stello}, Dennis and {Arentoft}, Torben and {Ford}, Eric B. and {Gilliland}, Ronald L. and {Handberg}, Rasmus and {Howard}, Andrew W. and {Isaacson}, Howard and {Johnson}, John Asher and {Karoff}, Christoffer and {Kawaler}, Steven D. and {Kjeldsen}, Hans and {Latham}, David W. and {Lund}, Mikkel N. and {Lundkvist}, Mia and {Marcy}, Geoffrey W. and {Metcalfe}, Travis S. and {Silva Aguirre}, Victor and {Winn}, Joshua N.},
        title = "{Stellar Spin-Orbit Misalignment in a Multiplanet System}",
      journal = {Science},
         year = 2013,
        month = oct,
       volume = {342},
       number = {6156},
        pages = {331-334},
          doi = {10.1126/science.1242066},
archivePrefix = {arXiv},
       eprint = {1310.4503},
 primaryClass = {astro-ph.EP},
       adsurl = {https://ui.adsabs.harvard.edu/abs/2013Sci...342..331H}
}

@ARTICLE{2014AA...561A.103O,
       author = {{Ofir}, Aviv and {Dreizler}, Stefan and {Zechmeister}, Mathias and {Husser}, Tim-Oliver},
        title = "{An independent planet search in the Kepler dataset. II. An extremely low-density super-Earth mass planet around Kepler-87}",
      journal = {\aap},
         year = 2014,
        month = jan,
       volume = {561},
          eid = {A103},
        pages = {A103},
          doi = {10.1051/0004-6361/201220935},
archivePrefix = {arXiv},
       eprint = {1310.2064},
 primaryClass = {astro-ph.EP},
       adsurl = {https://ui.adsabs.harvard.edu/abs/2014A&A...561A.103O}
}

@ARTICLE{2014AA...562A.109L,
       author = {{Lillo-Box}, J. and {Barrado}, D. and {Moya}, A. and {Montesinos}, B. and {Montalb{\'a}n}, J. and {Bayo}, A. and {Barbieri}, M. and {R{\'e}gulo}, C. and {Mancini}, L. and {Bouy}, H. and {Henning}, T.},
        title = "{Kepler-91b: a planet at the end of its life. Planet and giant host star properties via light-curve variations}",
      journal = {\aap},
         year = 2014,
        month = feb,
       volume = {562},
          eid = {A109},
        pages = {A109},
          doi = {10.1051/0004-6361/201322001},
archivePrefix = {arXiv},
       eprint = {1312.3943},
 primaryClass = {astro-ph.EP},
       adsurl = {https://ui.adsabs.harvard.edu/abs/2014A&A...562A.109L}
}

@ARTICLE{2016ApSS.361...17B,
       author = {{Budding}, E. and {P{\"u}sk{\"u}ll{\"u}}, {\c{C}}. and {Rhodes}, M.~D. and {Demircan}, O. and {Erdem}, A.},
        title = "{Analysis of the exoplanet containing system Kepler-91}",
      journal = {\apss},
         year = 2016,
        month = jan,
       volume = {361},
          eid = {17},
        pages = {17},
          doi = {10.1007/s10509-015-2564-4},
archivePrefix = {arXiv},
       eprint = {1507.02060},
 primaryClass = {astro-ph.EP},
       adsurl = {https://ui.adsabs.harvard.edu/abs/2016Ap&SS.361...17B}
}

@ARTICLE{2022AJ....163...61D,
       author = {{Dalba}, Paul A. and {Kane}, Stephen R. and {Dragomir}, Diana and {Villanueva}, Steven and {Collins}, Karen A. and {Jacobs}, Thomas Lee and {LaCourse}, Daryll M. and {Gagliano}, Robert and {Kristiansen}, Martti H. and {Omohundro}, Mark and et al.},
        title = "{The TESS-Keck Survey. VIII. Confirmation of a Transiting Giant Planet on an Eccentric 261 Day Orbit with the Automated Planet Finder Telescope}",
      journal = {\aj},
         year = 2022,
        month = feb,
       volume = {163},
       number = {2},
          eid = {61},
        pages = {61},
          doi = {10.3847/1538-3881/ac415b},
archivePrefix = {arXiv},
       eprint = {2201.04146},
 primaryClass = {astro-ph.EP},
       adsurl = {https://ui.adsabs.harvard.edu/abs/2022AJ....163...61D}
}

@ARTICLE{2022RNAAS...6...76D,
       author = {{Dalba}, Paul A. and {Jacobs}, Thomas Lee and {Omohundro}, Mark and {Gagliano}, Robert and {Jursich}, Jay and {Kristiansen}, Martti H. and {LaCourse}, Daryll M. and {Schwengeler}, Hans M. and {Terentev}, Ivan A.},
        title = "{The Refined Transit Ephemeris of TOI-2180 b}",
      journal = {Research Notes of the American Astronomical Society},
         year = 2022,
        month = apr,
       volume = {6},
       number = {4},
          eid = {76},
        pages = {76},
          doi = {10.3847/2515-5172/ac64fd},
       adsurl = {https://ui.adsabs.harvard.edu/abs/2022RNAAS...6...76D}
}

@ARTICLE{2024ApJS..271...16D,
       author = {{Dalba}, Paul A. and {Kane}, Stephen R. and {Isaacson}, Howard and {Fulton}, Benjamin and {Howard}, Andrew W. and {Schwieterman}, Edward W. and {Thorngren}, Daniel P. and {Fortney}, Jonathan and {Vowell}, Noah and {Beard}, Corey and et al.},
        title = "{Giant Outer Transiting Exoplanet Mass (GOT 'EM) Survey. IV. Long-term Doppler Spectroscopy for 11 Stars Thought to Host Cool Giant Exoplanets}",
      journal = {\apjs},
         year = 2024,
        month = mar,
       volume = {271},
       number = {1},
          eid = {16},
        pages = {16},
          doi = {10.3847/1538-4365/ad18c3},
archivePrefix = {arXiv},
       eprint = {2401.03021},
 primaryClass = {astro-ph.EP},
       adsurl = {https://ui.adsabs.harvard.edu/abs/2024ApJS..271...16D}
}

@ARTICLE{Dai2024AJ....168..239D,
       author = {{Dai}, Fei and {Goldberg}, Max and {Batygin}, Konstantin and {van Saders}, Jennifer and {Chiang}, Eugene and {Choksi}, Nick and {Li}, Rixin and {Petigura}, Erik A. and {Gilbert}, Gregory J. and {Millholland}, Sarah C. and {Dai}, Yuan-Zhe and {Bouma}, Luke and {Weiss}, Lauren M. and {Winn}, Joshua N.},
        title = "{The Prevalence of Resonance Among Young, Close-in Planets}",
      journal = {\aj},
         year = 2024,
        month = dec,
       volume = {168},
       number = {6},
          eid = {239},
        pages = {239},
          doi = {10.3847/1538-3881/ad83a6},
archivePrefix = {arXiv},
       eprint = {2406.06885},
 primaryClass = {astro-ph.EP},
       adsurl = {https://ui.adsabs.harvard.edu/abs/2024AJ....168..239D}
}

@ARTICLE{Murillo2026AJ....171...63L,
       author = {{Lopez Murillo}, Ana Isabel and {Mann}, Andrew W. and {Barber}, Madyson G. and {Vanderburg}, Andrew and {Thao}, Pa Chia and {Boyle}, Andrew W.},
        title = "{Searching for Transit Timing Variations in Young Transiting Systems}",
      journal = {\aj},
         year = 2026,
        month = feb,
       volume = {171},
       number = {2},
          eid = {63},
        pages = {63},
          doi = {10.3847/1538-3881/ae231a},
archivePrefix = {arXiv},
       eprint = {2512.06035},
 primaryClass = {astro-ph.EP},
       adsurl = {https://ui.adsabs.harvard.edu/abs/2026AJ....171...63L}
}

@INPROCEEDINGS{2013prpl.conf2K066G,
       author = {{Gillon}, Micha{\"e}l and {Jehin}, Emmanu{\"e}l and {Delrez}, Laetitia and {Magain}, Pierre and {Opitom}, Cyrielle and {Sohy}, Sandrine},
        title = "{SPECULOOS: Search for habitable Planets EClipsing ULtra-cOOl Stars}",
    booktitle = {Protostars and Planets VI Posters},
         year = 2013,
        month = jul,
       adsurl = {https://ui.adsabs.harvard.edu/abs/2013prpl.conf2K066G}
}

@ARTICLE{2024AJ....168..294M,
       author = {{Masuda}, Kento and {Libby-Roberts}, Jessica E. and {Livingston}, John H. and {Stevenson}, Kevin B. and {Gao}, Peter and {Vissapragada}, Shreyas and {Fu}, Guangwei and {Han}, Te and {Greklek-McKeon}, Michael and {Mahadevan}, Suvrath and et al.},
        title = "{A Fourth Planet in the Kepler-51 System Revealed by Transit Timing Variations}",
      journal = {\aj},
         year = 2024,
        month = dec,
       volume = {168},
       number = {6},
          eid = {294},
        pages = {294},
          doi = {10.3847/1538-3881/ad83d3},
archivePrefix = {arXiv},
       eprint = {2410.01625},
 primaryClass = {astro-ph.EP},
       adsurl = {https://ui.adsabs.harvard.edu/abs/2024AJ....168..294M}
}

@ARTICLE{2025MNRAS.540..106L,
       author = {{Langford}, Zachary and {Agol}, Eric},
        title = "{A differentiable N-body code for transit timing and dynamical modelling - II. Photodynamics}",
      journal = {\mnras},
         year = 2025,
        month = jun,
       volume = {540},
       number = {1},
        pages = {106-127},
          doi = {10.1093/mnras/staf687},
archivePrefix = {arXiv},
       eprint = {2410.03874},
 primaryClass = {astro-ph.EP},
       adsurl = {https://ui.adsabs.harvard.edu/abs/2025MNRAS.540..106L}
}

@ARTICLE{2026AA...706A.222B,
       author = {{Bocchieri}, A. and {Zak}, J. and {Turrini}, D.},
        title = "{ExoNAMD: Leveraging the spin-orbit angle to constrain the dynamics of multi-planet systems}",
      journal = {\aap},
         year = 2026,
        month = feb,
       volume = {706},
          eid = {A222},
        pages = {A222},
          doi = {10.1051/0004-6361/202557279},
archivePrefix = {arXiv},
       eprint = {2512.06126},
 primaryClass = {astro-ph.EP},
       adsurl = {https://ui.adsabs.harvard.edu/abs/2026A&A...706A.222B}
}
}
\end{document}